Дубовиченко С.Б.

# Избранные методы ядерной астрофизики

**с е р и я с ы**

# ҚАЗАҚСТАНДАҒЫ ҒАРЫШТЫҚ ЗЕРТТЕУЛЕР

**с е р и я**

# КАЗАХСТАНСКИЕ КОСМИЧЕСКИЕ ИССЛЕДОВАНИЯ

**s e r i e s**

# KAZAKHSTAN SPACE RESEARCH

***Бұл кітап Қазақстан Республикасында алғашқы астрономиялық бақылаулардың басталуына 70 жыл толуына арналады***

***Книга посвящается 70-ти летию первых астрономических наблюдений в Республике Казахстан***

***The book is dedicated to the seventieth anniversary of the first astronomical observations in the Republic of Kazakhstan***

*Қазақстандағы* Том 9

*Ғарыштық*

*Зерттеулер*

***ҚР Ұлттық ғарыш агенттігі***
***Ұлттық ғарыштық зерттеулер мен технологиялар орталығы***
***В.Г. Фесенков атындағы астрофизика институты***
***Халықаралық ақпараттандыру академиясы ҚР***

# Дубовиченко С.Б.

# *ЯДРОЛЫҚ АСТРФИЗИКАНЫҢ ІРІКТЕЛГЕН ӘДІСТЕРІ*

**Ядролық астрофизиканың іріктелген есептерінде ядролық физиканың әдістері**

Басылымға ұсынған "ҰҒЗТО" ҚР ҰҒА ғылыми техникалық кеңесі,
"В.Г. Фесенков атындағы астрофизика институты"
"ҰҒЗТО" ҚР ҰҒА, және АХА (ҚР) Президиумы

***Пікір жазғандар***

Ф.-м.ғ.д., профессор **Мухамеджанов А.М.** (ТУ, Техас, АҚШ),
ҚР мемлекеттік сыйлығының лауреаты, академик ҚР ҰИА, ф.-м.ғ.д.,
профессор **Данаев Н.Т.** (ҚазҰУ, Алматы, ҚР),
ҚР мемлекеттік сыйлығының лауреаты, ф.-м.ғ.д.,
профессор **Дуйсебаев А.Д.** (ЯФИ, Алматы, ҚР),
ҚР ХААН академигі, ҚР және РФ ф.-м.ғ.д.,
профессор **Чечин Л.М.** (ФИФА, Алматы, ҚР),
ЖРА корреспондент-мүшесі (РФ), ф.-м.ғ.д.,
профессор **Буртебаев Н.Т.** (ЯФИ, Алматы, ҚР)

***Ғылыми редакторы***

ҚР және РФ ф.-м.ғ. докторы, профессор **Буркова Н.А.**
(ҚазҰУ, Алматы, ҚР)

***Ғылыми кеңесшісі***

ҚР ҰҒА академигі, ф.-м.ғ.д., профессор **Боос Э.Г.** (ИФТ, Алматы, ҚР)

Бұл кітап өте төмен және төмен энергиядағы процестердің және жеңіл ядролы атомдардағы ядролық астрофизикадағы және ядролық физиканың жеке сұрақтарына арналады. Ядролық астрофизикада қарастырылатын термоядролық процестердің ядролық мінездемелерін есептейтін кейбір әдістер келтірілген. Күнде, жұлдыздарда және Әлемдегі термоядролық процестердің сипаттамаларының аймағындағы ядролық астрофизиканың кейбір есептерінің тікелей қолданылатын шешімдерінен алынған нәтижелер.

Кітап үш бөлімнен тұрады, соңғы бес-жеті жылдың ішіндегі шамамен үш ондық жарияланған ғылыми мақалалардың нәтижелерінің негізінде жазылды. Біріншісі - квант бөлшектерінің континуумы немесе байланыс күйі үшін ядролық мінездемелерін есептейтін кейбір жалпы әдістерін сипаттайды. Екіншісі - өте төмен және төмен энергиялы $^4$He$^{12}$C және p$^3$He, p$^6$Li, p$^{12}$C, n$^{12}$C, p$^{13}$C, $^4$He$^4$He жүйелерінің ядролық серпімді шашырауының фазалық талдауының нәтижелері және компьютерлік бағдарламалардың әдістеріне арналады. Үшінші бөлімде үш денелі моделдің негізінде алынған кейбір жеңіл атомдардың, атап айтқанда $^7$Li, $^9$Be және $^{11}$B тексеруге қолданылатын, серпімді шашырауының фазаларын анықтау негізінде, Әлемдегі термоядролық процестерді сипаттауға байланысты ядролық астрофизиканың есептерінде қолданылатын жұпты кластерaралық потенциалдар нәтижелері келтірілген.

Кітап ядролы-физика саласындағы ҒЗИ және ЖОО PhD докторанттарына және аспиранттарға, жоғарғы курс студенттеріне пайдалы болуы мүмкін.

***К**азахстанские* Том 9

***К**осмические*

***И**сследования*

***Национальное космическое агентство РК***
***Национальный центр космических исследований и технологий***
***Астрофизический институт им. В.Г. Фесенкова***
***Международная Академия Информатизации РК***

**Дубовиченко С.Б.**

# *ИЗБРАННЫЕ МЕТОДЫ ЯДЕРНОЙ АСТРОФИЗИКИ*

**Методы ядерной физики в избранных задачах ядерной астрофизики**

Рекомендовано к изданию научно-техническими советами
АО "НЦКИТ" НКА РК, департамента "Астрофизический институт
им. В.Г. Фесенкова" (АФИФ) "НЦКИТ" НКА РК и Президиумом МАИН РК

***Рецензенты***
Доктор ф.-м.н., профессор **Мухамеджанов А.М.** (Техасский университет, Техас, США),
Лауреат государственной премии РК, академик НИА РК, д.ф.-м.н., профессор **Данаев Н.Т.** (КазНУ, Алматы, РК),
Лауреат государственной премии РК, д.ф.-м.н., профессор **Дуйсебаев А.Д.** (ИЯФ, Алматы, РК),
академик МАИН РК, д.ф.-м.н. РК и РФ, профессор **Чечин Л.М.** (АФИФ, Алматы, РК),
член-корр. РАЕ (РФ), д.ф.-м.н., профессор **Буртебаев Н.Т.** (ИЯФ, Алматы, РК)

***Научный редактор***
Д.ф.-м.н. РК и РФ, профессор **Буркова Н.А.** (КазНУ, Алматы, РК)

***Научный консультант***
Академик НАН РК, д.ф.-м.н., профессор **Боос Э.Г.** (ФТИ, Алматы, РК)

Книга посвящена отдельным вопросам ядерной физики и ядерной астрофизики легких атомных ядер и процессов с ними при низких и сверхнизких энергиях. Приводятся некоторые методы расчета ядерных характеристик термоядерных процессов, рассматриваемых в ядерной астрофизике. Полученные результаты непосредственно применимы к решению некоторых задач ядерной астрофизики в области описания термоядерных процессов на Солнце, звездах и Вселенной.

Книга основана на результатах, примерно, трех-четырех десятках научных статей, опубликованных, в основном, за последние пять-семь лет и состоит из трех разделов. Первый из них посвящен описанию общих методов расчета некоторых ядерных характеристик для связанных состояний или континуума квантовых частиц. Второй – методам, компьютерным программам и результатам фазового анализа упругого рассеяния ядерных $p^3He$, $p^6Li$, $p^{12}C$, $n^{12}C$, $p^{13}C$, $^4He^4He$ и $^4He^{12}C$ систем при низких энергиях. В третьем разделе приводятся результаты, полученные на основе трехтельных моделей некоторых легких атомных ядер, а именно, $^7Li$, $^9Be$ и $^{11}B$, которые используются для проверки, определяемых на основе фаз упругого рассеяния, парных межкластерных потенциалов, используемых затем в задачах ядерной астрофизики, связанных с описанием термоядерных процессов Вселенной.

Книга может быть полезна студентам старших курсов, аспирантам и PhD докторантам ВУЗов и НИИ астрофизического и ядерно-физического профиля.

***National Space Agency of the Republic of Kazakhstan***
***National Center of the Space Research and Technology***
***V.G. Fessenkov Astrophysical Institute***
***International Informatization Academy of the Republic of Kazakhstan***

**Dubovichenko S.B.**

# *SELECTED METHODS of NUCLEAR ASTROPHYSICS*

**Nuclear Physics Methods in Selected Nuclear Astrophysics Problems**

The book covers the certain questions of nuclear physics and nuclear astrophysics of light atomic nuclei and their processes at low and ultralow energies. Some methods of calculation of nuclear characteristics of the thermonuclear processes considered in nuclear astrophysics are given here. The obtained results are directly applicable to the solution of certain nuclear astrophysics problems in the field of description of the thermonuclear processes in the Sun, the stars and the Universe.

The book is based on the results of approximately three-four tens of scientific papers generally published in recent five-seven years and consists of three sections. The first one covers the description of the general methods of calculation of certain nuclear characteristics for the bound states or the continuum of quantum particles. The second section deals with the methods, the computer programs and the results of the phase shift analysis of elastic scattering in $p^3He$, $p^6Li$, $p^{12}C$, $n^{12}C$, $p^{13}C$, $^4He^4He$ and $^4He^{12}C$ nuclear systems at low and ultralow energies. The results obtained on the basis of three-body models of certain light atomic nuclei are given in the third section, notably the $^7Li$, $^9Be$ and $^{11}B$ nuclei which are used for examination of the conjugated intercluster potentials determined on the basis of the phase shifts of elastic scattering and using then in the nuclear astrophysics problems connected with the description of the thermonuclear processes in the Universe.

The book will be useful for advanced students, postgraduate students and PhD doctoral candidates in the universities and research institutes in the field of astrophysics and nuclear physics.

# ОГЛАВЛЕНИЕ
### Content

# АЛҒЫ СӨЗ

***Автор***

*Күн және жұлдыздардағы термоядролық реакцияларды және ғарыштық құбылыстарды зерттеуге қазіргі кездегі ядролық физиканың табысты нәтижелерін қолдану бақылаулар мен теорияны сапалы келісетін білімдерін құрастыруға мүмкіндік берді, жұлдыздардың эволюциясы мен құрылымын, Әлемдегі химиялық элементтердің таралуын түсіндіру [1]*

Бұл кітапта төмен және өте төмен термоядролық процесстерді зерттеулердің аймағында кейбір жаңа табыстарды, яғни астрофизикалық энергияларды және олардың ортақ заңдарын талдау әдістері және қазіргі ядролық физикасы қағидаларына сүйене отырып бейнелеуге талпындық.

Ядролық астрофизиканың іс жүзінде кез келген есептері әдеттегідей, өте төмен энергиялы және термоядролық процестерде қатысатын жеңіл атом ядролары ядролық астрофизиканың нақтылы мәселелермен байланысты. Мысалы, Күн және жұлдыздардың термоядролық реакцияларының астрофизикалық сипаттамаларына [1,2] өте төмен энергиялы ядролық физика деген түсінікті қатыстырмау мүмкін емес. Мысалы үшін, өз негізінде, ядролық реакциялы жоғары және өте жоғары энергиялы болатын, Үлкен жарылыс болғандағы процесстерді талдау кезінде, қазіргі кездегі элементар бөлшектер физикасының әдістерінің бірі "Стандартты Моделді" қолданбау мүмкін емес [3,4,5]. Басқаша айтқанда, астрономиялық құбылыстарды және астрономиялық объектілердің физикалық қасиеттерін физика заңдарын және жекеленген ядролық физиканы қолданбау мүмкін емес!

Біз бұл кітапта, Күн және жұлдыздардағы термоядролық реакцияларды сипаттамасы беру үшін, яғни кейбір ядролық реакцияларды төмен және өте төмен энергияларда қолданылатын астрофизикалық энергиялардың ядролық физикасының нәтижелері қараймыз. Бұл кітап негізінен соңғы бес-жеті жыл ішінде Ресейде, Европада, АҚШ, ТМД және Қазақстанда жарияланған шамамен отыз ғылыми мақалалардан және үш бөлімнен тұрады.

Біріншісі жалпы есептеу әдістерін шолуға арналған, кейбір ядролық сипаттамалар үшін, берілген потенциалдық өзара байланыстағы бөлшектер жүйесінің толқындық функцияларын, байланған – күйлер немесе континуумдерде табу үшін.

Екінші әдісте, астрофизикалық энергиядағы $^{4}He^{12}C$ және ядролық бөлшектердің серпімді шашырауының фазалық талдауының нәтижелері $p^{3}He$, $p^{6}Li$, $p^{12}C$, $n^{12}C$, $p^{13}C$, $^{4}He^{4}He$ және компьютерлік бағдарламалар қамтылған. Бұл нәтижелер үздіксіз және дискретті спектрлерде жұп кластерaралық өзара байланысу потенциалдарын құрастыру үшін қолданылады. Бұл алынған потенциалдар сайып келгенде Әлемнің термоядролық процесстерінің кейбір негізгі сипаттамаларын есептеулері үшін қолданылады [3].

Үшінші бөлімде кейбір жеңіл атом ядроларының үш – денелік модельдерінің нәтижелері беріледі, атап айтқанда, $^{7}Li$, $^{9}Be$ и $^{11}B$, жұп кластерaралық серпімді шашырауының фазаларының негізінде алынатын потенциалдарын тексеруге мүмкіндік береді.

Қазіргі кезде ғарышқа ұшудың мүмкіндіктерін барлық жағынан бақылаумен анықтайды, ол бүкіл ғарыш ғылымының дамуына үлкен нәтижелер бергенін атып өтуіміз қажет. Сонымен қатар, бүгінгі уақытта қазақсандық астрофизиктердің әлемдік астрономия ғылымына үлесін көруге және атап айтуымыз қажет.

Автор, осы монографияны әзірлеуде, атап айтқанда Қазақстан Республикасының халықаралық деңгейін күшейтуге қатысты екі ең маңызды тарихи оқиғаларды айғақ ретінде айта алады:

– алғашқы ғарышкер Ю.А.Гагариннің Жерден ”Восток” ғарыш кемесімен ”Байқоңыр” ғарыш айлағынан ұшқанына 50 жыл толу тойымен байланысты.

және

– отандық ғылымның дамыуына жаңа қарқын немесе белсенділік берген, біздің мемлекетіміздің тәуелсіздігіне 20-жыл толуына байланысты.

# ПРЕДИСЛОВИЕ

*Применение достижений современной ядерной физики к изучению космических явлений и термоядерных реакций на Солнце и звездах позволило построить качественно согласующиеся с наблюдениями теорию образования, строения и эволюции звезд, объяснить распространенность химических элементов во Вселенной [1].*

В этой книге представлены новые достижения в области исследований термоядерных процессов при низких и сверхнизких, т.е. астрофизических энергиях и методов их анализа с точки зрения общих законов и принципов современной ядерной физики.

Практически любые задачи ядерной астрофизики связаны с определенными проблемами ядерной физики, обычно, сверхнизких энергий и, как правило, легких атомных ядер, участвующих в термоядерных процессах [1,2]. Например, невозможно рассматривать астрофизические характеристики термоядерных реакций на Солнце и звездах, не привлекая для этого понятий и представлений ядерной физики сверхнизких энергий [3]. Невозможно анализировать процессы, протекавшие при Большом Взрыве [4], которые, в своей основе, являются ядерными реакциями при высоких и сверхвысоких энергиях, не привлекая для этого модели и методы современной физики элементарных частиц, например, "Стандартной Модели" [5]. Иначе говоря, невозможно рассматривать астрономические явления и физические свойства астрономических объектов, не привлекая для этого законы физики, в целом, и ядерной физики, в частности!

В книге будут рассмотрены некоторые методы и ре-зультаты ядерной физики астрофизических энергий, которые используются для описания термоядерных реакций на Солнце и звездах, т.е. некоторых ядерных реакций при низких и сверхнизких энергиях. Книга основана на результатах, примерно, трех-четырех десятков научных статей, опубликованных, в основном, за последние пять-семь лет в России, Европе, США, СНГ и Казахстане и состоит из трех разделов.

Первый из них посвящен описанию общих методов расчета некоторых ядерных характеристик для связанных состояний и континуума, которые используются для нахождения волновой функции системы частиц при заданных потенциалах взаимодействия.

Второй содержит методы, компьютерные программы и результаты фазового анализа упругого рассеяния ядерных частиц $p^3He$, $p^6Li$, $p^{12}C$, $n^{12}C$, $p^{13}C$, $^4He^4He$ и $^4He^{12}C$ при астрофизических энергиях. Эти результаты используются для построения парных межкластерных потенциалов взаимодействия в непрерывном и дискретном спектре. Полученные таким образом потенциалы используются далее для расчетов некоторых основных характеристик термоядерных процессов Вселенной [3].

В третьем разделе приводятся результаты трехтельных моделей некоторых легких атомных ядер, а именно, $^7Li$, $^9Be$ и $^{11}B$, которые позволяют проверить, получаемые на основе фаз упругого рассеяния, парные межкластерные потенциалы.

Вместе со сказанным выше, следует подчеркнуть, что эти результаты стали возможными на фоне общих успехов в развитии всей космической науки, которая в наше время во многом определяется наблюдательными возможностями пилотируемой космонавтики. Кроме того, сегодня все более заметным становится и вклад казахстанских астрофизиков в мировую астрономическую науку.

Поэтому автор специально констатирует тот факт,

что подготовка данной монографии была стимулирована двумя важнейшими историческими событиями, связанными с укреплением международного имиджа Республики Казахстан, а именно:

– 50-летием запуска с космодрома ”Байконур” космического корабля ”Восток”, пилотируемого первым космонавтом Земли Ю.А.Гагариным

и

– 20-летием независимости нашего государства, давшего новый импульс развитию отечественной науки.

# PREFACE

*The application of the achievement in the modern nuclear physics to study of cosmic phenomena and thermonuclear reactions in the Sun and the stars allows one to create the theory of formation, construction and evolution of stars that is qualitatively conform with measurements and explain the abundance of chemical elements in the Universe [1].*

The new achievements in sphere of thermonuclear processes at low and ultralow energies, i.e. astrophysical energies and methods of their analysis in terms of modern nuclear physics general laws and concepts was done in this book.

Practically, any nuclear astrophysics problems are connected with the certain nuclear physics problems, usually, as a rule, with light atomic nuclei at ultralow energy region in thermonuclear processes [1,2]. It is impossible, for example, to consider astrophysical characteristics of thermonuclear reactions in the Sun and the stars, without using the ideas and conceptions of ultralow energy nuclear physics for this consideration [3]. It is impossible to analyze the astrophysical processes during the Big Bang [4], which, basically, are the nuclear reactions at high and ultrahigh energies, without using the models and methods of modern elementary-particle physics, for example, “Standard Model” [5]. In other words, it is impossible to consider astronomical phenomena and physical properties of astronomical objects without using laws of physics, in whole, and nuclear physics, particularly.

In this book, we will consider some methods and results of nuclear physics of astrophysical energies using for the description of thermonuclear reactions in the Sun and the stars, i.e. certain nuclear reactions at low and ultralow energies. The book is based on the results of about three-four tens of scientific papers that were published in Europe, USA, Russia, CIS and Kazakhstan and consists of three sections.

The first of them is devoted to the description of the general calculation methods of nuclear characteristics for bound states or continuum, which are used for finding of wave function of system of particles at the specified potentials.

The second contains methods, computer programs and results of the phase shift analysis of the elastic scattering of nuclear particles in $p^3He$, $p^6Li$, $p^{12}C$, $n^{12}C$, $p^{13}C$, $^4He^4He$ and $^4He^{12}C$ systems at astrophysical energies. These results are used for the construction of pair intercluster interaction potentials in continuous and discrete spectrum. The potentials, obtained in such a way, are used further for the calculations of some fundamental characteristics of thermonuclear processes in the Universe [3].

The results of three-body models of some light atomic nuclei, specifically $^7Li$, $^9Be$ and $^{11}B$, are given in the third section. They allow us to check the pair intercluster potentials, obtained on the basis of the phase shifts of elastic scattering.

It should be note, together with the said above, that these results become possible on the background of general successes in the development of the space science, which, in the present time, is determined by the observation facilities of the manned space exploration. In addition, the contribution of the astrophysicists of Kazakhstan to the world astronomical science becomes more appreciable for today.

Therefore, the author specially states the fact that the preparation of this monograph was stimulated by the two significant historical events, connected with the strengthening of the international image of the Republic of Kazakhstan, namely:

– The fiftieth anniversary of the launching of the spaceship “Vostok“ manned by the first spaceman of the Earth Yuri Gagarin from the spaceport “Baikonur”

and

– The twentieth anniversary of the independence of our state, which gives a new impetus to the development of the science of our country.

# КІРІСПЕ

*Ядролық реакциялардың қималары көптеген тәжирибелік мәліметерт бойынша төмен энергиядағы ядролық шашырау ядроның құрылымының негізгі ақппарат көзі және ядролар арасындағы және олардың фрагменттерінің ықтималдылығына кластерлеу қасиетінде және өзара байланыстыратын механизімдерінде [9].*

Табиғи термоядролық синтез ағатын термоядролық процесстердің басты сипаттамаларының бірі, астрофизикалық *S*-фактор болып табылады – бұл реакцияның ағу жылдамдығын анықтайтын фактор. Сірә, ядролық астрофизиканың негізгі есебі S-фактордың өлшемін анықтау болып табылады деп санауға болады, яғни бұл нөлге ұмтылатын энергиялар. Бұл есептің шешуін тәжірибелік және теориялық ядролық физиканың мүмкіндіктерінен шығатын бірнеше әдістермен анықтау мүмкін. Солардың бірі тек қана ядролық реакциялардың толық қималарының өлшемдерінің өте төмен энергияларының тәжірибелік әдістеріне негізделген, екінші, нақтылы атап айтқанда, анықталған теория негізіндегі астрофизика аймағындағы ядролық процесстері туралы.

Негізінен, астрофизикалық S-факторды тәжірибе түрінде анықтауға болады, бірақ қазіргі кезде өзара байланысатын жеңіл ядролардың көпшілігі үшін әртүрлі термоядролық процестерге қатысатын энергиялары шамамен, 0.1÷1.0 МэВ аймағында болу керек. Тәжірибелік өлшемдердің қателері жүзге дейін сирек жетеді және одан көп пайызға жетеді [6]. Алайда, астрофизикалық есептеулері үшін,мысалға, жұлдыздар эволюциясының моделін нақытылауда және дамытуда, мүмкін, және барлық Біздің Әлемнің [4,8] қазіргі кезеңдегі дамуының мәні мен және де ең кішкентай мүмкін болатын қателермен, 0.1 ден 100 кэВ, әдеттегідей энергиялары жұлдыздың ортасындағы температуралары $10^6 \div 10^9$ К болатын аймақтағы талаптарды қажет етеді [7].

Астрофизикалық *S*-фактордың шамаларын анықтау әдістерінің бірі, нөлдік энергияның, яғни 1 кэВ шамасында және одан да төмен, бұл тәжірибеде анықталатын, төмен энергиялар аймақтарындағы оның мәндерінің экстраполяциясы болады. Бұл қолданылған әдіс ең алдымен кейбір термоядролық реакциялардың қимасының өлшемінің орындауынан кейін орындалады. Бірақ, бұл астрофизикалық S-факторды анықтауының үлкен тәжірибелік қателері мұндай нәтижелердің құндылығы айтарлықтай төмендетеді, жүргізілген экстраполяциялардың үлкен мәнсіздікке алып келеді [6].

Екінші, өзекті және артықшылығы бар жол, астрофизикалық *S*-фактордың теориялық есептеулерінде болады – нақтылы ядролық модельдеудің негізінде радиациялық қармап алуды кейбір термоядролық реакцияның факторы [3]. Бұл әдіс ядролық процесстердің кейбір үлгісі егер оларда болатын энергиялардың аймағындағы тәжірибелердің мәліметтерін сипаттауға дұрыс мүмкіндік берген, анық жорамалдау негізделген, ол астрофизикалық S-фактордың болжамды үлгісін алып береді, шамамен 1 кэВ, бірақ қазіргі кезде тікелей тәжірибелік өлшеу мүмкіндігі жоқ.

Кез келген ядролық модельді ортақ қағида бойынша қазіргі ядролық физика және кванттық механиканың нақтылы микроскоптық дәлелдеуінен жасалынады, бұл нөлдік энергияға тәжірибелік мәліметтерінің кәдімгі экстраполяциясы екінші жолдың артықшылығын көрсетеді [9]. Сондықтан, мұндай модльдің кейбір болжаушы мүмкіндіктерінің болуы нақты түрде үміттендіреді, әсіресе, тәжірибелік мәліметтері бар, энергия аймағында болжаушы нәтижелер алуға болатын және ол жаңа нәтижелерден еш айырмашылығы жоқ.

Ядролық модельді таңдауда, онда есептеулер үшін біз Юнга схемалар бойынша орбиталық күйлердің классификациясы бар жеңіл атом ядроларының потенциалдық кластер модельдері қолданылады [10-13]. Бұл модель қарапайым астрофизикалық есептеулерді, сипаттамаларды анықтауға арналған, мысалы, радиациялық қармау астрофизикалық *S*-фактор, электромагниттік өту кластерінің шашырау күйінен осы кластерлік каналдардан жеңіл ядролардың атамдарының байланыс күйіне өтуі [14].

Негізінен, осы кітап потенциалдық үш кластерлі модельдердің кейбір мүмкіндіктерінің демонстрациясы және серпімді шашыратудың фазаларын талдауды, нәтижелері кластерaралық кластер потенциалдарының құрастыруы үшін керек болатын серпімді шашыратудың фазалық талдауына бірнеше жеңіл ядролардың мысалында және үш бөлімдерден тұрады.

Біріншісінде олардың ішінен есептеулердің жалпы әдістері қаралады [9,15]. Екіншісінде – компьютерлік бағдарламалары және $p^3He$, $p^6Li$, $p^{12}C$, $n^{12}C$, $p^{13}C$, $^4He^4He$ серпімді шашыратуының фазалық талдауының нәтижелері және төмен және өте төмен энергиядағы $^4He^{12}C$ кластерлеріолар ядролық астрофизика есептеріндегі кластерaралық потенциалдарын құрастыруы үшін қолданылады[14,15]. Үшінші бөлімде үш денелі кейбір жеңіл атом ядроларының потенциалдық модельдері, атап айтқанда, $^7Li$, $^9Be$ және $^{11}B$ шашырау фазаларының суырылған тәжірибелерінің негізінде аралық кластер потенциалдарын құрастыруға нақтылы тесттер қызметін атқарады [16].

1-ші тарауда есептеудің айтылған әдістері ядролық қасиеттер және бұдан әрі қаралатын процесстердің әр түрлі мінездемелерінің анықтауындағы бір мәнсіздіктен құтылуға мүмкіндік береді. Мысалы, алгоритмдарды келтірсек, әртүрлі спинді кванттық бөлшектердің жүйелері үшін серпімді шашыратудың фазаларының пішінді мінездемелерін кітапта қаралатын есептеулері үшін компьютерлік бағдарламалар берілген [17].

2-ші тарауда фазалық талдаудың нәтижелері, яғни серпімді шашыратудың фазасын, алуға мүмкіндік береді, алдыңғы кітапта жартылай қарастырылған, болашақта әр түрлі астрофизикалық қосымшалар үшін қолданылалатын өзара байланысатын кластерлік потенциалдар [14].

3-ші тарауда кейбір үш денелі модельдерді қолдану кластераралық потенциалдарының фазалық талдауының негізінде қосымша ресми түрде мақұлдауы, және ядролық жүйелер және ядролық реакциялардың астрофизикалық мінездемелерімен Күн және жұлдыздарда ағатын төмен және өте төмен энергияларда есептеулерде олардың ары қарай қолданылғыштығын анықтауға мүмкіндік береді [1,3].

Кітапта төмен энергиялы ядролық физиканың негізгі алынатын жалғыз емес әдістері және термоядролық реакцияларды есептеу үшін содан соң астрофизикалық энергияларда қолданылатын нәтижелер көрсетілген. Сайып келгенде, төмен энергиялы ядролық физиканың әдістерінің негізінде ядролық астрофизиканың Әлемдегі ағатын нақтылы процестер қаралады.

Ядролық сипаттамалардың қалай алынатыны көрсетілген, астрофизикалық есептеулерді орындауы үшін қолданылатын кластераралық потенциалдарын құру үшін қолданылатын серпімді шашыратудың фазалары [14]. Олардың құрастыруының дұрыстығын көрсететін үш денелі есептеулердегі потенциалдарды тексерудің мысалдары келтірілген.

Бұл кітап, қазіргі ядролық астрофизикада қолданылатын барлық әдістерді және де термоядролық процесстерді жеткілікті түрде түсіндіруге талаптанбайды. Кітап, төмен энергиялы ядролық физиканың тек қана, кейбір әдістеріне және нәтижелеріне арналған, ол әдістер ядролық астрофизиканың нақтылы кейбір есептерінде, мысалға, Әлемдегі және жұлдыздардағы, Күндегі термоядролық реакцияларды түсіндіруде қолданылады.

# ВВЕДЕНИЕ

*Многочисленные экспериментальные данные по сечениям ядерных реакций и ядерного рассеяния при низких энергиях являются основным источником информации о структуре ядер, свойствах и механизмах взаимодействия между ядрами и их фрагментами, вероятности кластеризации таких ядер [9].*

Одной из главных характеристик термоядерных процессов, протекающих в природных термоядерных котлах – звездах или реакциях управляемого термоядерного синтеза, является астрофизический $S$-фактор, определяющий скорость протекания такой реакции. Поэтому можно, по-видимому, считать, что основной задачей ядерной астрофизики является определение формы $S$-фактора и его зависимости от энергии в области нулевой энергии, т.е. при энергиях, стремящихся к нулю. Решить эту задачу можно несколькими способами, вытекающими из возможностей экспериментальной и теоретической ядерной физики. Один из них основывается исключительно на экспериментальных методиках измерений полных сечений ядерных реакций при сверхнизких энергиях, а второй, на определенных теоретических, а именно, модельных представлениях о ядерных процессах в астрофизической области энергий.

В принципе, астрофизический $S$-фактор всегда можно определить экспериментально, но в настоящее время для большинства взаимодействующих легких ядер, которые участвуют в различных термоядерных процессах, это оказывается возможным только при энергиях, примерно, в области $0.1 \div 1.0$ МэВ. Причем ошибки экспериментальных измерений не редко доходят до ста и более процентов [6]. Однако, для реальных астрофизических расчетов, например, развития и уточнения модели эволюции звезд [7], а, возможно, и всей нашей Вселенной на современном этапе ее развития [4,8], его значения, причем, с минимально возможными ошибками, требуются при энергиях, как правило, в области от 0.1 до 100 кэВ, что соответствует температурам в центре звезды порядка $10^6$ К÷$10^9$ К.

Один из методов определения величины астрофизического *S*-фактора при нулевой энергии, т.е. энергии порядка 1 кэВ и меньше, это экстраполяция его значений из области, где он экспериментально определим, в область более низких энергий. Это обычный путь, который используется, в первую очередь, после выполнения измерения сечения некоторой термоядерной реакции. Однако большие экспериментальные ошибки определения *S*-фактора [6] приводят к большим неоднозначностям проводимых экстраполяций, что существенно снижает ценность таких результатов.

Второй, и, по-видимому, наиболее перспективный и предпочтительный путь, заключается в теоретических расчетах *S*-фактора некоторой термоядерной реакции, например, радиационного захвата на основе определенной ядерной модели [3]. Такой метод основан на вполне очевидном предположении, что если некоторая модель ядерных процессов позволяет правильно описать экспериментальные данные в той области энергий, где они имеются, то допустимо предположить, что она будет правильно передавать форму *S*-фактора и при более низких энергиях, порядка 1 кэВ, где на сегодняшний день прямые экспериментальные измерения его значений пока еще не возможны.

В этом и заключается определенное преимущество второго подхода [9] над обычной экстраполяцией экспериментальных данных к нулевой энергии, поскольку любая ядерная модель строится так, что имеет вполне определенное микроскопическое обоснование с точки зрения общих принципов современной ядерной физики и квантовой механики. Поэтому имеется вполне определенная надежда в существовании некоторых предсказательных возможностей такой модели, особенно, если области энергий, где имеются экспериментальные данные и в которой требуется получить новые, т.е., по сути, предсказательные результаты, существенно не отличаются.

Что касается выбора ядерной модели, то нами для подобных расчетов обычно используется потенциальная кластерная модель (ПКМ) легких атомных ядер с классификацией орбитальных состояний по схемам Юнга [10,11,12,13]. Такая модель предоставляет сравнительно много простых возможностей для выполнения различных расчетов астрофизических характеристик, например, астрофизического $S$-фактора или полных сечений радиационного захвата для электромагнитных переходов из состояний рассеяния кластеров на связанные состояния легких атомных ядер в этих кластерных каналах [14].

В целом, данная книга посвящена демонстрации некоторых возможностей потенциальных трехкластерных моделей и фазовому анализу упругого рассеяния, результаты которого требуются для построения межкластерных потенциалов по фазам рассеяния, на примере нескольких легких ядер и состоит из трех разделов.

В первом из них рассматриваются общие методы расчетов [9,15]. Во втором – методы, компьютерные программы и результаты фазового анализа упругого рассеяния $p^3He$, $p^6Li$, $p^{12}C$, $n^{12}C$, $p^{13}C$, $^4He^4He$ и $^4He^{12}C$ кластеров при низких и сверхнизких энергиях, которые используются для построения межкластерных потенциалов в задачах ядерной астрофизики [14,15]. И, наконец, в третьем разделе приведены результаты трехтельной, трехкластерной потенциальной модели некоторых легких атомных ядер, а именно, $^7Li$, $^9Be$ и $^{11}B$, которые служат определенным тестом качества построения межкластерных потенциалов на основе извлеченных из эксперимента фаз рассеяния [16].

Изложенные методы расчета (Раздел 1) позволяют избежать неоднозначности в определении различных характеристик ядерных свойств и процессов, которые рассматриваются далее. Привести алгоритмы, а в дальнейшем и компьютерные программы для расчетов, рассматриваемых в книге, модельных характеристик, например, фаз упругого рассеяния для систем квантовых частиц с разным спином [17].

Результаты фазового анализа (Раздел 2), т.е. фазы упругого рассеяния, позволяют получать, как уже говорилось, межкластерные потенциалы взаимодействия, которые могут использоваться в дальнейшем для различных астрофизических приложений, частично рассмотренных, например, в нашей предыдущей книге [14].

Использование некоторых трехтельных моделей (Раздел 3) дает возможность дополнительного апробирования полученных на основе фазового анализа межкластерных потенциалов, и выяснения их дальнейшей применимости в расчетах, связанных с астрофизическими характеристиками ядерных систем и ядерных реакций при низких и сверхнизких энергиях, которые протекают на Солнце и звездах [1,3].

Книга демонстрирует определенные, причем, не единственные методы ядерной физики низких энергий и, получаемые на их основе, результаты, которые используются затем для расчета характеристик термоядерных реакций при астрофизических энергиях [14]. Показано, как получать некоторые ядерные характеристики, в частности, фазы упругого рассеяния, использующиеся для построения межкластерных потенциалов, применяемых для выполнения астрофизических расчетов [14]. Приведены примеры проверки таких потенциалов в трехтельных расчетах, которые демонстрируют корректность их построения.

Конечно, данная книга не претендует на исчерпывающее изложение всех методов, используемых в современной ядерной астрофизике даже при объяснении термоядерных процессов. Она посвящена только некоторым методам и результатам ядерной физики низких энергий, которые могут быть непосредственно использованы для решения определенного круга задач ядерной астрофизики при описании термоядерных реакций на Солнце, звездах и Вселенной.

# INTRODUCTION

*The numerous experimental data on cross-sections of nuclear reactions and nuclear scattering at low energies are the main source of information about nuclear structure, properties and mechanisms of interaction between nuclei and their fragments, about probability of clusterization of such nuclei [9].*

One of the main characteristics of thermonuclear processes, flowing in the natural thermonuclear reactors – stars or in controlled thermonuclear fusion, is the astrophysical *S*-factor, determining the reaction rate. Therefore, it is possible, evidently, to consider that the main problem of nuclear astrophysics is the obtaining of the form of the *S*-factor and its dependence from energy in the zero energy range, i.e. at vanishing energies. This problem can be solved by few methods; they follow from the resources of experimental and theoretical nuclear physics. One of them is exclusively based on experimental procedures of measurement of total cross-sections of nuclear reactions at ultralow energies, and the second is based on certain theoretical, specifically model-based, representations of nuclear processes in the astrophysical energy range.

In principle, the astrophysical *S*-factor can be obtained experimentally, but, at present, for the majority of interacting light nuclei, taking place in different thermonuclear processes, it is possible only at the energy range about 100 keV÷1 MeV. At that, the errors of experimental measurements reach up to one hundred and more percents, quite often [6]. However, for real astrophysical calculations, for example for the developing and adjusting of the star evolution model [7] and, probably, for the whole Universe at the modern stage of its evolution [4,8], the values of astrophysical *S*-factor, with the minimally possible errors, are required at the energy range about 0.1-100 keV, which corresponds to the temperatures in the star core about $10^6$ K÷$10^9$ K.

One of the methods for obtaining the astrophysical *S*-factor at zero energy, i.e. the energy on the order of 1 keV and less, is the extrapolation of its values to lower energy range where it can be determined experimentally. It is the general way that is used, first of all, after carrying out the experimental measurements of cross-sections of certain thermonuclear reaction. But, the large experimental errors in obtaining of the *S*-factor [6] lead to the big ambiguities of the calculated extrapolations and this fact essentially reduces the importance of such results.

The second and, evidently, most preferable approach consists in theoretical calculations of the *S*-factor of some thermonuclear reaction, for example the radiative capture reaction, on the basis of certain nuclear model [3]. This method is based on the obvious assumption: if there is the certain nuclear model that correctly describes the experimental data of the astrophysical *S*-factor in that energy range where these data exist, then it is reasonably assume that this model will describe the form of the *S*-factor correctly at lower energies (about 1 keV); because of the fact that its direct experimental measurements can not be feasible so far, at present.

This is the certain advantage of the second approach [9] stated above over the simple data extrapolation to zero energy, because the using model has, as a rule, the certain microscopic justification with a view to the general principles of nuclear physics and quantum mechanics. Therefore, there is well defined hope in the existence of certain predictive resources of such model, especially, if there is no essential difference between the results obtaining in the energy ranges where experimental data exist and predictive results.

Concerning the nuclear model choice: we usually use the potential cluster model for light atomic nuclei with the classification of the orbital states according to Young's schemes [10-13]. This model gives comparatively a lot of simple possibilities for carrying out of different calculations of astrophysical characteristics, for example, the astrophysical *S*- factor of radiative capture for electromagnetic transitions from the states of cluster scattering to the bound states of light atomic nuclei in these cluster channels [14].

In the large, this book is devoted to the demonstration of some possibilities of potential three-cluster models and to the phase shift analysis of elastic scattering. The results of this analysis are needed for the construction of the intercluster potentials according to phase shifts of scattering by example few light nuclei. In addition, this book consists in three sections.

The general calculation methods [9,15] are considered in the first of them. In the second – methods, computer programs and results of the phase shift analysis of the elastic scattering in the $p^3He$, $p^6Li$, $p^{12}C$, $n^{12}C$, $p^{13}C$, $^4He^4He$ and $^4He^{12}C$ cluster systems at low and ultralow energies, which are used for the construction of the intercluster potentials for nuclear astrophysics problems [14,15]. Finally, the results of three-body (three-cluster) potential model of few light atomic nuclei, notably $^7Li$, $^9Be$ and $^{11}B$, which are in use as a certain quality test for the construction of the intercluster potentials on the basis of scattering phase shifts extracted from the experiment [16], are given in the third sections.

The given calculation methods (Section 1) allow to avoid ambiguities in the obtaining of different characteristics of nuclear properties and processes, which are considered further. The algorithms and, in what follows, computer programs for the calculation of the considered model characteristics, for example, phase shifts of elastic scattering for quantum particle systems with different spin [17], are given here.

The results of the phase shift analysis (Section 2), i.e. the phase shifts of elastic scattering allow to obtain, as it was said, the intercluster interaction potentials, which can be used in future for different astrophysical applications, partially considered, for example, in our previous book [14].

The usage of certain three-body models (Section 3) gives the possibility of the additional approbation of the obtained on the basis of phase shift analysis intercluster potentials, and clearing of their future applicability in calculations connected with astrophysical characteristics of nuclear systems and nuclear reactions at low and ultralow energies flowing in the Sun and the stars [1,3].

The book demonstrates the certain and not the unique methods of low energy nuclear physics, and the results, obtained on their basis, which are used further for the calculation of the thermonuclear reactions at astrophysical energies. Thus, the certain questions of nuclear astrophysics flowing in the Universe are considered on the basis of the methods of low energy nuclear physics.

It was shown how to obtain certain nuclear characteristics, particularly, phase shifts of elastic scattering, which are used for the construction of intercluster potentials applying for carrying out astrophysical calculations [14]. The examples, which demonstrate correctness of their construction, are given here for checking such potentials in three-body calculations.

Certainly, this book does not claim to exhausting description of the whole methods, using in modern nuclear astrophysics, even in the case of explanation of thermonuclear processes. It is devoted only to the certain methods and results of low energy nuclear physics, which can be directly used for solving of the certain scope of nuclear astrophysics problems of the thermonuclear reactions in the Sun, the stars and the Universe.

# I. МЕТОДЫ РАСЧЕТА
**Calculation methods**

*Большинство задач ядерной физики требуют знания волновой функции относительного движения частиц, которые участвуют в ядерных столкновениях или определяют связанное состояние ядра, т.е. являются внутренними фрагментами полной системы. Такую функцию можно найти из решений уравнения Шредингера для каждой конкретной физической задачи в дискретном или непрерывном спектре, если известен потенциал взаимодействия этих частиц [17].*

## *Введение*
## *Introduction*

В этом разделе кратко изложены некоторые математические методы и численные алгоритмы, используемые в фазовом анализе, при решении радиального уравнения Шредингера и в трехтельной вариационной задаче. В частности, описанные здесь методы, позволяют решать обобщенную матричную задачу на собственные значения и собственные функции с использованием альтернативного способа, который, особенно в трехтельной модели, приводит к устойчивому алгоритму решения.

Изложению этих методов предшествует краткий обзор некоторых основных результатов, полученных в потенциальной кластерной модели с запрещенными состояниями (ЗС) и классификацией орбитальных состояний по схемам Юнга. Использование подобных моделей обусловлено тем, что до настоящего времени не существует общей и достаточно законченной теории легких атомных ядер. Поэтому для анализа отдельных ядерных характеристик используются различные физические модели и методы [12,18,19,20,21,22,23]. В этой связи, большой интерес представляет изучение возможностей потенциальной кластерной модели, использующей, в том числе, межкластерные взаимодействия с запрещенными состояниями.

В ПКМ считается, что ядро состоит из двух бесструктурных, но не точечных фрагментов – кластеров, которым можно сопоставить свойства соответствующих ядерных частиц в свободном состоянии. Используемые потенциалы с ЗС позволяют эффективно учитывать принцип Паули в межкластерных взаимодействиях и не требуют явной антисимметризации полной волновой функции системы, что заметно упрощает все компьютерные вычисления [13,14,17,24].

Параметры этих потенциалов обычно согласованы с фазами упругого рассеяния соответствующих свободных частиц. В случае большой вероятности кластеризации рассматриваемых ядер ПКМ позволяет правильно воспроизвести основные характеристики их связанных состояний в этих кластерных каналах [12,13].

Именно эти общие представления используются далее при рассмотрении некоторых характеристик легких атомных ядер в трехкластерных конфигурациях. Они практически не отличаются от принципов, заложенных в двухкластерном варианте ПКМ, который использовался нами ранее при рассмотрении некоторых астрофизических аспектов термоядерных процессов на Солнце, звездах и Вселенной [14].

Что касается фазового анализа упругого рассеяния ядерных частиц при низких и астрофизических энергиях, то ситуация здесь более проста, чем в предыдущем случае, поскольку не требует знания ядерных межкластерных потенциалов взаимодействия. Поэтому далее, после описания основных математических методов расчета и некоторых численных алгоритмов, мы сразу перейдем к рассмотрению отдельных задач фазового анализа.

## *1.1 Обзор возможностей кластерной модели The review of the cluster model possibilities*

Многие свойства легких атомных ядер, которые участвуют в термоядерных процессах, хорошо описываются различными вариантами кластерных моделей и, по-видимому, наиболее распространенной на сегодняшний день является потенциальная кластерная модель. ПКМ основана на единых парных гамильтонианах взаимодействий в непрерывном и дискретном спектрах и строится на предположении, что рассматриваемые ядра с высокой степенью вероятности имеют двух- или трехкластерную структуру [12,13].

### 1.1.1 Основные принципы модели Basic model principles

Работоспособность ПКМ обусловлена тем, что во многих легких атомных ядрах вероятность образования нуклонных ассоциаций-кластеров и степень их обособления друг от друга действительно высоки. Это подтверждается многочисленными экспериментальными данными и результатами различных теоретических расчетов, полученными за последние 50-60 лет [11].

Для нахождения феноменологических потенциалов взаимодействия между каждой парой кластеров используются результаты фазового анализа экспериментальных данных по дифференциальным сечениям упругого рассеяния соответствующих свободных ядер [13]. Потенциалы процессов рассеяния строятся из условия наилучшего описания, полученных на основе этих данных, фаз упругого рассеяния ядерных частиц. Потенциалы связанного состояния (СС) кластеров строятся, как правило, на основе описания некоторых характеристик основного состояния (ОС) ядра, которое рассматривается в данном кластерном канале.

Рассматриваемая здесь потенциальная кластерная модель сравнительно проста, поскольку сводится к решению задачи двух тел, или, что эквивалентно – одного тела в поле силово-го центра. Конечно, могут возникнуть возражения, что такая модель совершенно не соответствует проблеме многих тел, задача которой является описание свойств системы, состоящей из $A$ нуклонов. В этой связи следует заметить, что одной из наиболее успешных моделей в теории атомного ядра за всю историю развития ядерной физики является модель ядерных оболочек (МО), которая математически представляет собой именно проблему движения одного тела в поле силового центра. Физические основания рассматриваемой здесь потенциальной кластерной модели восходят именно к оболочечной модели, или точнее, к удивительной связи между моделью оболочек и кластерной моделью, которая в литературе часто встречается под названием модели нуклонных ассоциаций (МНА) [9,11].

В МНА и ПКМ волновая функция (ВФ) ядра, состоящего из двух кластеров с числом нуклонов $A_1$ и $A_2$ ($A = A_1 + A_2$), имеет вид антисимметризованного произведения полностью антисимметричных внутренних волновых функций (ВФ) кластеров $\Psi(R_1)$ и $\Psi(R_2)$, умноженных на волновую функцию их относительного движения $\Phi(R = R_1 - R_2)$,

$$\Psi = \hat{A}\ \{\Psi(R_1)\Psi(R_2)\Phi(R)\}\ , \tag{B.1}$$

где $\hat{A}$ – оператор антисимметризации, который действует по отношению к перестановкам нуклонов из разных кластеров ядра, $R$ – межкластерное расстояние, $R_1$ и $R_2$ – радиус-векторы положения центра масс кластеров.

Обычно, кластерные волновые функции выбирают так, чтобы соответствовать основным состояниям свободных ядер, состоящих из $A_1$ и $A_2$ нуклонов. Эти волновые функции характеризуются специфическими квантовыми числами, включая схемы Юнга $\{f\}$, которые определяют перестановочную симметрию орбитальной части ВФ относительного движения кластеров. Кроме того, определенные выводы кластерной модели [9,11] приводят к понятию запрещенных принципом Паули состояний. Поэтому, некоторые полные волновые функции ядра $\Psi$ с определенным типом функций относительного движения $\Phi(R)$ обращаются в ноль при анти-симметризации по всем $A$ нуклонам.

Основное, т.е. реально существующее связанное в данном потенциале состояние кластерной системы, описывается волновой функцией с ненулевым, в общем случае, числом узлов. Таким образом, представление о запрещенных принципом Паули состояниях позволяет учесть многочастичный характер задачи в терминах двухчастичного потенциала взаимодействия между кластерами [9,11]. При этом на практике потенциал межкластерного взаимодействия выбирается так, чтобы правильно описать извлеченные из экспериментальных данных фазы упругого рассеяния кластеров в соответствующей парциальной волне и, предпочтительно, в состоянии с одной определенной схемой Юнга $\{f\}$ для пространственной части волновой функции $A$ нуклонов ядра [22].

Однако результаты фазового анализа в ограниченной области энергий, как правило, не позволяют однозначно восстановить потенциал взаимодействия. Поэтому дополнительным условием на потенциал СС является требование правильного воспроизведения энергии связи ядра в соответствующем кластерном канале и описание некоторых других статических ядерных характеристик.

Например, это может быть зарядовый радиус и асимптотическая константа (АК), причем характеристики связанных в ядре кластеров отождествляются с характеристиками соответствующих свободных легчайших ядер [9]. Это дополнительное требование, очевидно, является идеализацией, т.к. предполагает, что в основном состоянии ядро имеет 100%-ую кластеризацию. Поэтому успех данной потенциальной кластерной модели при описании системы из $A$ нуклонов в связанном состоянии определяется тем, насколько велика реальная кластеризация основного состояния такого ядра в двух- или трехчастичных каналах [9,11,22].

Однако некоторые ядерные характеристики отдельных, даже не кластерных, ядер могут быть обусловлены преимущественно одним определенным кластерным каналом, при малом вкладе других возможных кластерных конфигураций. В этом случае используемая одноканальная двух- или трех-кластерная модель позволяет идентифицировать доминирующий кластерный канал и выделить некоторые основные свойства такой ядерной системы, которые им обусловлены [9].

### 1.1.2 Развитие модели и основные результаты Model development and general results

Остановимся теперь вкратце на некоторых основных этапах развития ПКМ и ЗС. В начале 70-х годов прошлого века в работах [25,26,27] было показано, что фазы упругого рассеяния легких кластерных систем могут быть описаны на основе глубоких чисто притягивающих потенциалов Вудс-Саксоновского типа, которые содержат связанные запрещенные принципом Паули состояния. Поведение фаз рассеяния при нулевой энергии для таких взаимодействий подчиняется обобщенной теореме Левинсона [25-27,28,29], а фазы рассеяния при больших энергиях стремятся к нулю, оставаясь положительными. Радиальная ВФ разрешенных состояний (РС) потенциалов с ЗС осциллирует на малых расстояниях, а не вымирает, как это было для взаимодействий с кором. Такой подход можно рассматривать, как альтернативу часто используемой концепции отталкивающего кора. Кор вводится в потенциал взаимодействия кластеров для качественного учета принципа Паули без выполнения полной и явной антисимметризации ВФ.

Далее, например, в работах [28,30,31,32,33,34,35,36,37] были параметризованы межкластерные центральные гауссовы потенциалы взаимодействия, правильно воспроизводящие фазы упругого $^4$He$^2$H рассеяния при низких энергиях и содержащие запрещенные состояния. Показано, что на основе этих потенциалов в кластерной модели можно воспроизвести основные характеристики связанных состояний ядра $^6$Li, вероятность кластеризации которого в рассматриваемом двухкластерном канале сравнительно высока [22]. Все кластерные состояния в такой системе оказываются чистыми по орбитальным схемам Юнга [25-37] и потенциалы, полученные из фаз рассеяния, можно непосредственно применять для опи-сания характеристик основного состояния этого ядра.

Определенный успех одноканальной модели, основанной на таких потенциалах, обусловлен не только большой степенью кластеризации обсуждаемых ядер, но и тем, что в каждом состоянии кластеров существует только одна разрешенная орбитальная схема Юнга [12,13], определяющая симметрию этого состояния. Тем самым, достигается некое "единое" описание непрерывного и дискретного спектра, и потенциалы, полученные на основе экспериментальных фаз рассеяния, вполне успешно используются для описания различных характеристик ОС ядер лития.

Для более легких кластерных систем вида $N^2H$, $^2H^2H$, $N^3H$, $N^3He$, $^2H^3He$ и т.д. в состояниях рассеяния с минимальным спином уже возможно смешивание по орбитальным симметриям и ситуация оказывается более сложной. В состояниях с минимальным спином, в непрерывном спектре разрешены две орбитальные симметрии с различными схемами Юнга, в то время как связанным основным состояниям, по-видимому, соответствует только одна из этих схем [10-13,38,39,40,41,42,43,44,45,46,47,48,49].

Поэтому потенциалы, непосредственно полученные на основе экспериментальных фаз рассеяния, эффективно зависят от различных орбитальных схем и не могут в таком виде использоваться для описания характеристик основного состояния ядер. Из таких взаимодействий, необходимо выделять чистую компоненту, применимую уже при анализе характеристик СС.

В более тяжелых ядерных системах $N^6Li$, $N^7Li$ и $^2H^6Li$ также реализуется подобная ситуация [50,51,52,53,54], когда в некоторых случаях различные состояния оказываются смешанными по схемам Юнга. В указанных выше работах были получены чистые по схемам Юнга потенциалы взаимодействия для перечисленных выше тяжелых ядерных систем. Они, в основном, оказались способны правильно описывать как характеристики рассеяния, так и свойства связанных состояний соответствующих ядер.

Несмотря на определенные успехи такого подхода, обычно рассматривались только чисто центральные межкла-стерные взаимодействия. При рассмотрении $^4$He$^2$H системы в рамках потенциальной кластерной модели не учитывалась тензорная компонента, которая приводит к появлению $D$ волны в ВФ СС и рассеяния и позволяет рассматривать квадрупольный момент ядра $^6$Li. Под тензорным потенциалом здесь следует понимать взаимодействие, оператор которого зависит от взаимной ориентации полного спина системы и межкластерного расстояния. Математическая форма записи такого оператора полностью совпадает с оператором двухнуклонной задачи, поэтому потенциал, по аналогии, называется тензорным [55,56,57].

По-видимому, впервые тензорные потенциалы были использованы для описания $^2$H$^4$He взаимодействия в начале 80-х годов XX века в работе [55], где предпринята попытка ввести тензорную компоненту в оптический потенциал. Это позволило заметно улучшить качество описания дифференциальных сечений упругого рассеяния и поляризации. В работе [56] на основе "фолдинг" модели выполнены расчеты сечений и поляризаций, и учет тензорной компоненты потенциала позволил улучшить их описание. В дальнейшем такой подход был использован в работе [57], где “сверткой” нуклон-нуклонных потенциалов получены $^2$H$^4$He взаимодействия с тензорной компонентой. Показано, что в принципе удается правильно описать основные характеристики связанного состояния $^6$Li, включая правильный знак и порядок величины квадрупольного момента.

Однако, в работах [55,56] рассматривались только процессы рассеяния кластеров, а в [57] только характеристики СС ядра $^6$Li без анализа фаз упругого рассеяния. Тем не менее, гамильтониан взаимодействия должен быть единым и для процессов рассеяния и СС кластеров, как это было сделано в [25-37] в случае чисто центральных потенциалов. Большая вероятность кластеризации ядра $^6$Li в $^2$H$^4$He канал позволяет надеяться на корректность постановки такой задачи в потенциальной кластерной модели.

Поскольку ОС $^6$Li [10,25-37,58,59,60,61,62,63] сопоставляется орбитальная схема {42}, то в $S$ состоянии должно быть ЗС со схемой {6}. В тоже время в $D$ волне ЗС отсутст-вует, так как схема {42} совместима с орбитальным моментом $L$ = 2. Это значит, что ВФ $S$ состояния будет иметь узел, а ВФ для $D$ волны – безузловая. Такая классификация запрещенных и разрешенных состояний по схемам Юнга в целом позволяет определить общий вид ВФ СС кластерной $^2H^4He$ системы [10,28].

В работе [64], в рамках потенциальной кластерной модели, был получен именно такой, единый гамильтониан $^2H^4He$ взаимодействия, т.е. единый потенциал с тензорной компонентой и запрещенным в $S$ волне состоянием. Он удовлетворял всем перечисленным выше условиям и позволял описывать как характеристики рассеяния, т.е. ядерные фазы, так и свойства связанных состояний ядра $^6Li$, включая его квадрупольный момент [65].

Далее, в работах [66,67] понятие ЗС было распространено на синглетный нуклон-нуклонный (NN) потенциал, а затем и на его триплетную часть [68]. Впоследствии появился NN потенциал гауссова типа с тензорной компонентой и ОПЕП (потенциал однопионного обмена) [69], а несколько позже [70,71] удалось получить параметры NN потенциала с тензорной компонентой и запрещенными состояниями, волновая функция которого, как и предсказывалось ранее теоретически, содержала в дискретном спектре узел только в $S$ волне, а $D$ волна была безузловая [72].

Такой потенциальный подход в NN системе [71] оказался способен описывать практически все характеристики дейтрона и NN рассеяния при низких и средних энергиях. Кроме того, заметно улучшилось описание высокоэнергетических векторных и тензорных поляризаций в $e^2H$ рассеянии, по сравнению с известными ранее взаимодействиями подобного типа [69].

### 1.1.3 Представления и методы модели Model representations and methods

Прежде чем переходить к описанию конкретных методов расчета ядерных характеристик, остановимся на некоторых общих соображениях, используемых обычно при решении определенного круга задач ядерной физики и ядерной астрофизики. Известно, что ядерный или NN потенциал взаимодействия частиц в задачах рассеяния или связанных состояниях заведомо не известен, и определить его форму напрямую какими-либо способами не представляется возможным. Поэтому выбирается определенная форма его зависимости от расстояния (например, гауссова или экспоненциальная) и по некоторым ядерным характеристикам (обычно, это фазы ядерного рассеяния) на основе известных методов расчета (см., например, [73,74]) фиксируются его параметры, так чтобы он описывал эти характеристики. В дальнейшем такой потенциал можно применять для расчетов любых других ядерных характеристик, например, энергий связи рассматриваемых ядер и свойств их связанных состояний или сечений различных реакций, включая, термоядерные процессы при сверхнизких энергиях [14,17].

Когда, так или иначе, определен ядерный потенциал взаимодействия двух частиц, практически весь круг, рассмотренных выше физических задач, сводится к решению уравнения Шредингера или связанной системы этих уравнений в случае тензорных ядерных сил с определенными начальными и асимптотическими условиями. В принципе, это чисто математическая задача из области математического моделирования физических процессов и систем. Однако многие существующие методы ее решения [75,76,77,78,79,80, 81,82] не всегда приводят к устойчивой численной схеме, а обычно используемые алгоритмы часто дают недостаточную точность результатов, либо вообще приводят к переполнению в процессе работы компьютерных программ.

Поэтому далее мы определим общее направление и обозначим основные численные методы, которые приводят к наиболее устойчивым схемам решения рассматриваемых физических задач. Для квантовых задач такие решения основываются на уравнении Шредингера, которое позволяет получить ВФ системы ядерных частиц при известных потенциалах их взаимодействия. Решать уравнения Шредингера для связанных состояний и процессов рассеяния можно, например, методом Рунге-Кутта (РК) [83,84] или конечно-разностным методом (КРМ) [85].

Такие методы позволяют легко найти собственные волновые функции и собственные энергии квантовой системы частиц. Причем, если использовать, предложенную в [17], комбинацию численных и вариационных методов и контролировать, в некоторых случаях, точность решения уравнения или системы связанных уравнений Шредингера с помощью невязок [86], то процедура получения конечных результатов заметно упрощается. Затем, на основе полученных решений, т.е. волновых функций ядра, которые являются решениями исходных уравнений, вычисляются многочисленные ядерные характеристики, в том числе, фазы рассеяния и энергия связи атомных ядер в различных каналах.

И, наконец, переходя к непосредственному изложению избранных нами математических методов решения определенного круга задач ядерной физики низких и сверхнизких, т.е. астрофизических энергий, заметим, что в конечном итоге практически все рассматриваемые здесь физические проблемы в математическом плане сводятся к задачам вариационного характера.

В частности, при рассмотрении трехтельной модели, например, ядра $^7$Li [13,17,87] очень эффективно использовать альтернативный математический метод нахождения собственных значений обобщенной вариационной матричной задачи, рассматриваемой на основе уравнения Шредингера с использованием неортогонального вариационного базиса. Этот метод позволяет избавиться от неустойчивостей, иногда возникающих при численной реализации обычных вычислительных схем решения обобщенной вариационной задачи [17].

Кроме того, для фазового анализа рассеяния ядерных частиц применялись новые, предложенные нами еще в работе [17], алгоритмы реализации численных методов, которые используются для нахождения частных решений общей многопараметрической вариационной задачи для функционала $\chi^2$. Эта величина определяет точность описания экспериментальных данных на основе выбранного теоретического представления, т.е. некоторого функционала нескольких пере-менных.

## *1.2 Потенциалы и волновые функции Potentials and wave functions*

Межкластерные потенциалы взаимодействия для каждой парциальной волны, т.е. для заданного орбитального момента *L,* и точечным кулоновским членом могут быть выбраны в виде

$$V(r) = V_0\exp(-\gamma r^2) + V_1\exp(-\delta r^2) \qquad (1.2.1)$$

или

$$V(r) = V_0\exp(-\gamma r^2)\ . \qquad (1.2.2)$$

Здесь параметры $V_0$ и $V_1$ выражены в МэВ, $\gamma$ и $\delta$ имеют размерность Фм$^{-2}$, и являются параметрами потенциала, которые обычно находятся из условия наилучшего описания фаз упругого рассеяния, извлекаемых в процессе фазового анализа из экспериментальных данных по дифференциальным сечениям, т.е. угловым распределениям или функциям возбуждения.

Кулоновский потенциал при нулевом кулоновском радиусе $R_{\text{coul}} = 0$ записывался в форме

$$V_{\text{coul}}(\text{МэВ}) = 1.439975\cdot\frac{Z_1Z_2}{R}\ ,$$

где $R$ – относительное расстояние между частицами входного канала в Фм и $Z$ – заряды частиц в единицах элементарного заряда "$e$".

В некоторых случаях в кулоновский потенциал вводят кулоновский радиус $R_{\text{coul}}$, и тогда кулоновская часть размерности Фм$^{-2}$ принимает вид [88]

$$V_{\text{coul}}(r) = 2\mu\frac{m_0}{\hbar^2}\begin{cases} \dfrac{Z_1Z_2}{r} & r > R_\text{c} \\ Z_1Z_2\left(3-\dfrac{r^2}{R_\text{c}^2}\right)\Big/2R_\text{c} & r < R_\text{c} \end{cases}.$$

Поведение волновой функции связанных состояний, в том числе, основных состояний ядер в кластерных каналах на больших расстояниях характеризуется асимптотической константой $C_\text{w}$, которая определяется через функцию Уиттекера [89]

$$\chi_\text{L}(r) = \sqrt{2k_0}\, C_\text{w}W_{-\eta\text{L}+1/2}(2k_0r) \ , \qquad (1.2.3)$$

где $\chi_\text{L}(R)$ – численная волновая функция связанного состояния, получаемая из решения радиального уравнения Шредингера и нормированная на единицу, $W_{-\eta\text{L}+1/2}$ – функция Уиттекера связанного состояния, определяющая асимптотическое поведение ВФ и являющаяся решением того же уравнения без ядерного потенциала, т.е. на больших расстояниях $R$, $k_0$ – волновое число, обусловленное канальной энергией связи, $\eta$ – кулоновский параметр, определенный далее и $L$ – орбитальный момент связанного состояния.

Асимптотическая константа (АК или, как ее часто называют, асимптотический нормировочный коэффициент – АНК) является важной ядерной характеристикой, определяющей поведение "хвоста", т.е. асимптотики волновой функции на больших расстояниях. Во многих случаях ее значение для ядра $A$ в кластерном канале $b$+$c$ определяет значение астрофизического $S$-фактора для процесса радиационного захвата $b(c,\gamma)A$ [90]. Асимптотическая константа пропорциональна ядерной вершинной константе для виртуального процесса $A \to b+c$, которая является матричным элементом этого процесса на массовой поверхности [91].

Численная волновая функция $\chi_\text{L}(r)$ относительного движения кластеров является решением радиального уравнения

Шредингера вида

$$\chi''_L(r) + [\,k^2 - V(r) - V_{coul}(r) - L(L{+}1)/r^2]\chi_L(r) = 0 \;, \qquad (1.2.4)$$

где $V(r)$ – межкластерный ядерный потенциал (1.2.1) или (1.2.2), приведенный к размерности Фм$^{-2}$, т.е. умноженный на $2\mu\frac{m_0}{\hbar^2}$, $V_{coul}(r)$ – кулоновский потенциал размерности Фм$^{-2}$, $k$ – волновое число, определяемое энергией $E$ взаимодействия частиц и равное $k^2 = 2\mu\frac{m_0}{\hbar^2}E$, а константа $\frac{\hbar^2}{m_0}$ обычно принимается равной 41.4686 МэВ·Фм$^2$.

В данных расчетах обычно задавались целые значения масс частиц, а кулоновский параметр $\eta = \frac{\mu Z_1 Z_2 e^2}{\hbar^2 k}$ представлялся в виде

$$\eta = 3.44476\cdot 10^{-2}\,\frac{\mu Z_1 Z_2}{k}\,,$$

где $k$ – волновое число, выраженное в Фм$^{-1}$, $\mu$ – приведенная масса в а.е.м., $Z$ – заряды частиц в единицах элементарного заряда "е".

## 1.3 Методы фазового анализа
## Phase shifts methods

Зная экспериментальные дифференциальные сечения упругого рассеяния всегда можно найти некоторый набор параметров, называемых фазами рассеяния $\delta^{J}_{S,L}$, позволяющий с определенной точностью описать поведение этих сечений. Качество описания экспериментальных данных на основе некоторой теоретической функции (функционала нескольких переменных) можно оценить по методу $\chi^2$, который представляется в виде [88]

$$\chi^2 = \frac{1}{N}\sum_{i=1}^{N}\left[\frac{\sigma_i^t(\theta) - \sigma_i^e(\theta)}{\Delta\sigma_i^e(\theta)}\right]^2 = \frac{1}{N}\sum_{i=1}^{N}\chi_i^2 \quad , \qquad (1.3.1)$$

где $\sigma^e$ и $\sigma^t$ – экспериментальное и теоретическое, т.е. рассчитанное при некоторых заданных значениях фаз $\delta^J_{S,L}$ сечение упругого рассеяния ядерных частиц для $i$ – го угла рассеяния, $\Delta\sigma^e$ – ошибка экспериментальных сечений для этого угла и $N$ – число измерений.

Выражения, описывающие дифференциальные сечения, являются разложением некоторого функционала $d\sigma(\theta)/d\Omega$ в числовой ряд [88], и нужно найти такие вариационные параметры разложения $\delta^J_{S,L}$ , которые наилучшим образом описывают его поведение. Поскольку выражения для дифференциальных сечений обычно являются точными [88], то при увеличении членов разложения $L$ до бесконечности величина $\chi^2$ должна стремиться к нулю. Именно этот критерий использовался для выбора определенного набора фаз, приводящего к минимуму $\chi^2$, который мог бы претендовать на роль глобального минимума данной многопараметрической вариационной задачи [92]. Более подробно методы и критерии фазового анализа, использованные в данных расчетах, приведены в работах [17,92].

Таким образом, для поиска фаз рассеяния по экспериментальным сечениям выполнялась процедура минимизации функционала $\chi^2$, как функции определенного числа переменных, каждая из которых является фазой $\delta^J_{S,L}$ определенной парциальной волны. Для решения этой задачи ищется минимум $\chi^2$ в некоторой ограниченной области значений этих переменных. Но и в этой области можно найти множество локальных минимумов $\chi^2$ с величиной порядка единицы. Выбор наименьшего из них позволяет надеяться, что он будет соответствовать глобальному минимуму, который является решением такой вариационной задачи.

Изложенные критерии и методы использовались нами для выполнения фазового анализа в $^4$He$^4$He, p$^3$He, p$^6$Li, p$^{12}$C,

$n^{12}C$, $p^{13}C$ и $^4He^{12}C$ системах при низких энергиях, которые важны для астрофизических задач. Выражения для вычисления дифференциальных сечений упругого рассеяния, требуемые для выполнения фазового анализа в указанных выше системах, приведены далее в соответствующих разделах книги.

## *1.4 Некоторые численные методы*
## *Some numerical methods*

Конечно-разностные методы, которые являются модификацией методов [85], и содержат учет кулоновских взаимодействий, вариационные методы (ВМ) решения уравнения Шредингера и другие вычислительные методы, используемые в данных расчетах ядерных характеристик, подробно описаны в [17]. Поэтому здесь только вкратце перечислим основные моменты, связанные с численными методами компьютерных вычислений.

Во всех расчетах, полученных конечно-разностным и вариационным методом [17], в конце области стабилизации асимптотической константы, т.е. примерно на 10÷20 Фм, численная или вариационная волновая функция заменялась функцией Уиттекера (1.2.3) с учетом найденной ранее асимптотической константы. Численное интегрирование в любых матричных элементах проводилось на интервале от 0 до 25÷30 Фм. При этом был использован метод Симпсона [76], который дает хорошие результаты для плавных и слабо осциллирующих функций при задании нескольких сотен шагов на период [17]. ВФ при рассматриваемых здесь низких и сверхнизких энергиях вполне удовлетворяют указанным требованиям.

Для выполнения настоящих расчетов были переписаны и модифицированы наши компьютерные программы, основанные на конечно-разностном методе [17], для расчета полных сечений радиационного захвата и характеристик связанных состояний ядер с языка Turbo Basic на современную версию языка Fortran-90, которая имеет заметно больше возможно-стей. Это позволило существенно поднять скорость и точность всех вычислений, в том числе, энергии связи ядра в двухчастичном канале. Теперь, например, точность вычисления кулоновских волновых функций для процессов рассеяния, контролируемая по величине Вронскиана, и точность поиска корня детерминанта в КРМ [17], определяющая точность поиска энергии связи, находятся на уровне $10^{-15}\div10^{-20}$. Реальная абсолютная точность определения энергии связи в конечно-разностном методе для разных двухчастичных систем составила $10^{-6}\div10^{-8}$ МэВ.

Для проведения вариационных расчетов была переписана на Fortran и несколько модифицирована вариационная программа нахождения вариационных ВФ и энергии связи ядер в кластерных каналах, что позволило существенно поднять скорость поиска минимума многопараметрического функционала, который определяет энергию связи двух- и трехчастичных систем во всех рассматриваемых ядрах [17].

Данная программа по прежнему использует многопараметрический вариационный метод с разложением ВФ по неортогональному вариационному гауссову базису с независимым варьированием параметров. Модифицированы и переведены на Fortran-90 аналогичные вариационные программы, основанные на многопараметрическом вариационном методе и предназначенные для выполнения фазового анализа по дифференциальным сечениям упругого рассеяния ядерных частиц.

Для вычисления кулоновских функций рассеяния использовалось быстро сходящееся представление в виде цепных дробей [93], позволяющее получить их значения с высокой степенью точности в широком диапазоне переменных и с малыми затратами компьютерного времени [94]. Волновые кулоновские функции рассеяния имеют две составляющие – регулярную $F_L(\eta,\rho)$ и нерегулярную $G_L(\eta,\rho)$, которые являются линейно независимыми решениями радиального уравнения Шредингера с кулоновским потенциалом для процессов рассеяния [117]

$$\chi_L''(\rho)+\left(1-\frac{2\eta}{\rho}-\frac{L(L+1)}{\rho^2}\right)\chi_L(\rho)=0 \quad ,$$

где $\chi_L = F_L(\eta,\rho)$ или $G_L(\eta,\rho)$, $\rho = kr$, а $\eta$ – кулоновский параметр, определенный выше и $k$ – волновое число, определяемое энергией $E$ частиц.

Вронскианы этих кулоновских функций записываются [95]

$$W_1 = F_L' G_L - F_L G_L' = 1 ,$$

$$W_2 = F_{L-1} G_L - F_L G_{L-1} = \frac{L}{\sqrt{\eta^2+L^2}} .$$

Рекуррентные соотношения между ними представляются в форме

$$L[(L+1)^2+\eta^2]^{1/2}u_{L+1} = (2L+1)\left[\eta+\frac{L(L+1)}{\rho}\right]u_L-(L+1)[L^2+\eta^2]^{1/2}u_{L-1} \quad ,$$

$$(L+1)u_L' = \left[\frac{(L+1)^2}{\rho}+\eta\right]u_L-[(L+1)^2+\eta^2]^{1/2}u_{L+1} \quad ,$$

$$Lu_L' = [L^2+\eta^2]^{1/2}u_{L-1}-\left[\frac{L^2}{\rho}+\eta\right]u_L \quad ,$$

где $u_L = F_L(\eta,\rho)$ или $G_L(\eta,\rho)$.

Асимптотика таких функций при $\rho \to \infty$ может быть представлена в виде [96]

$$F_L = \sin(\rho-\eta\ln 2\rho-\pi L/2+\sigma_L) ,$$

$$G_L = \cos(\rho-\eta\ln 2\rho-\pi L/2+\sigma_L) .$$

Имеется достаточно много методов и приближений для вычисления кулоновских волновых функций рассеяния [97, 98,99,100,101,102,103], которые использовались, начиная с 30-х годов 20-го века. Однако, только в 70-х годах прошлого века появилось быстро сходящееся представление, позволяющее получать их значения с высокой степенью точности в широком диапазоне переменных и с малыми затратами компьютерного времени [93,94].

Кулоновские функции в таком методе представляются в виде бесконечных цепных дробей [104]

$$f_{\rm L} = F'_{\rm L} / F_{\rm L} = b_0 + \cfrac{a_1}{b_1 + \cfrac{a_2}{b_2 + \cfrac{a_3}{b_3 + \ldots}}} \quad ,$$

где

$$b_0 = (L+1)/\rho + \eta/(L+1) \quad ,$$

$$b_{\rm n} = \left[2(L+n)+1\right]\left[(L+n)(L+n+1)+\eta\rho\right] \quad ,$$

$$a_1 = -\rho[(L+1)^2+\eta^2](L+2)/(L+1) \quad ,$$

$$a_{\rm n} = -\rho^2[(L+n)^2+\eta^2]\left[(L+n)^2-1\right]$$

и

$$P_{\rm L} + iQ_{\rm L} = \frac{G'_{\rm L} + iF'_{\rm L}}{G_{\rm L} + iF_{\rm L}} = \frac{i}{\rho}\left( b_0 + \cfrac{a_1}{b_1 + \cfrac{a_2}{b_2 + \cfrac{a_3}{b_3 + \ldots}}} \right) \quad ,$$

где

$$b_0 = \rho - \eta \quad ,$$

$$b_n = 2(b_0 + in) \quad ,$$

$$a_n = -\eta^2 + n(n-1) - L(L+1) + i\eta(2n-1) \quad .$$

Используя приведенные выражения, можно получить связь между кулоновскими функциями и их производными [105]

$$F_L' = f_L F_L \ ,$$

$$G_L = (F_L' - P_L F_L)/Q_L = (f_L - P_L)F_L/Q_L \ ,$$

$$G_L' = P_L G_L - Q_L F_L = [P_L(f_L - P_L)/Q_L - Q_L]F_L \ .$$

Такой метод расчета оказывается применим в области $\rho \ge \eta + \sqrt{\eta^2 + L(L+1)}$, т.е. для $L = 0$ имеем $\rho > 2\eta$, и легко позволяет получить высокую точность благодаря быстрой сходимости цепных дробей. Поскольку кулоновский параметр $\eta$ обычно имеет величину порядка единицы, а орбитальный момент $L$ всегда можно положить равным нулю, то метод дает хорошие результаты уже при $\rho > 2$. Значения кулоновских функций для любых $L > 0$ всегда можно получить из рекуррентных соотношений.

Таким образом, задавая некоторое значение $F_L$ в точке $\rho$, находим все остальные функции и их производные с точностью до постоянного множителя, который определяется из вронскианов. Вычисления кулоновских функций по приведенным формулам и сравнение их с табличным материалом [95] показывает, что можно легко получить восемь-девять правильных знаков при расчетах с двойной точностью, если $\rho$ удовлетворяет приведенному выше условию.

Другой вид функций, который использовался нами, это функция Уиттекера, являющаяся решением уравнения Шре-дингера с кулоновским взаимодействием для связанных состояний [95]

$$\frac{d^2W(\mu,\nu,z)}{dz^2}-\left(\frac{1}{4}-\frac{\nu}{z}-\frac{1/4-\mu^2}{z^2}\right)W(\mu,\nu,z)=0 \quad .$$

Это уравнение можно привести к стандартному виду уравнения Шредингера

$$\frac{d^2\chi(k,L,r)}{dr^2}-\left(k^2+\frac{g}{r}+\frac{L(L+1)}{r^2}\right)\chi(k,L,r)=0 \quad ,$$

где $g=\dfrac{2\mu Z_1Z_2}{\eta^2}=2k\eta$, $\eta=\dfrac{\mu Z_1Z_2e^2}{\eta^2 k}$ – кулоновский параметр, приближенное выражение для которого приведено в п.п. 1.2, $z = 2kr$, $\nu=-\dfrac{g}{2k}=-\eta$ и $\mu = L+1/2$ , $k$ – волновое число взаимодействующих частиц.

Для нахождения численных значений функции Уиттекера обычно используют ее интегральное представление, которое имеет вид

$$W(\mu,\nu,z)=\frac{z^{\nu}e^{-z/2}}{\Gamma(1/2-\nu+\mu)}\int t^{\mu-\nu-1/2}(1+t/z)^{\mu+\nu-1/2}e^{-t}dt \quad .$$

Его можно привести к виду

$$W_{-\eta L+1/2}(z)=W(L+1/2,-\eta,z)=\frac{z^{-\eta}e^{-z/2}}{\Gamma(L+\eta+1)}\int t^{L+\eta}(1+t/z)^{L-\eta}e^{-t}dt$$

Легко видеть, что при $L = 1$ и $\eta = 1$ приведенный интеграл превращается в $\Gamma(3)$, который сокращается со знаменателем и остается простое выражение

$$W_{-1,1+1/2}(z) = W(1+1/2,-1,z) = \frac{e^{-z/2}}{z} \quad .$$

Такую запись можно использовать для контроля правильности вычислений функции Уиттекера при любых значениях $z$ для $L = 1$, $\eta = 1$ и $z = 2kr$. Более подробно вопросы, связанные с методами вычисления этих двух функций приведены в [17] и Приложении работы [14].

## 1.5 Обобщенная матричная задача на собственные значения The generalized matrix eigenvalue problem

При рассмотрении обобщенной матричной задачи на собственные значения и функции, которая получается после разложения ВФ по неортогональному гауссову базису, исходим из стандартного уравнения Шредингера в общем виде [106]

$$H\chi = E\chi \quad ,$$

где $H$ – гамильтониан и $E$ – энергия системы, $\chi$ – волновые функции.

Разлагая ВФ по некоторому, неортогональному в общем случае, вариационному базису [58]

$$\chi = \sum_{\mathrm{i}} C_{\mathrm{i}}\varphi_{\mathrm{i}} \quad ,$$

подставляя их в исходную систему, умножая ее слева на комплексно сопряженную $\varphi_{\mathrm{i}}^{*}$ базисную функцию и интегрируя по всем переменным, получим известную матричную систему вида [86,107]

$$(H - EL)C = 0 \quad , \qquad (1.5.1)$$

которая в общем случае является обобщенной матричной задачей для нахождения собственных значений и собственных функций [108,109]. Если разложение ВФ выполняется по ортогональному базису, матрица интегралов перекрывания $L$ превращается в единичную матрицу $I$, и мы имеем стандартную задачу на собственные значения, для решения которой существует множество методов [110].

Для решения обобщенной матричной задачи также имеются известные методы, приведенные, например, в книге [109]. Остановимся вначале на стандартном методе решения обобщенной матричной задачи для уравнения Шредингера, которая возникает при использовании неортогонального вариационного базиса в ядерной физике или ядерной астрофизике. Затем рассмотрим ее модификацию или альтернативный метод, который удобно применять для решения этой задачи при численных расчетах на современных компьютерах [17,108].

Итак, для определения спектра собственных значений энергии и собственных волновых функций в вариационном методе, при разложении ВФ по неортогональному гауссову базису [58-60], решается обобщенная матричная задача на собственные значения [109]

$$\sum_{i} (H_{ij} - E L_{ij}) C_i = 0 \quad , \qquad (1.5.2)$$

где $H$ – симметричная матрица гамильтониана, $L$ – матрица интегралов перекрывания, $E$ – собственные значения энергии и $C$ – собственные векторы задачи.

Представляя матрицу $L$ в виде произведения нижней $N$ и верхней $V$ треугольных матриц [109], после несложных преобразований переходим к обычной задаче на собственные значения

$$H'C' = EIC' \quad , \qquad (1.5.3)$$

или

$$(H' - EI)C' = 0 \quad ,$$

где

$$H' = N^{-1}HV^{-1} \quad , \qquad C' = VC \quad ,$$

и $V^{-1}$ и $N^{-1}$ обратные по отношению к $V$ и $N$ матрицы.

Далее находим матрицы $N$ и $V$, выполняя триангуляризацию симметричной матрицы $L$ [110], например, методом Халецкого [109]. Затем определяем обратные матрицы $N^{-1}$ и $V^{-1}$, например, методом Гаусса и вычисляем элементы матрицы $H' = N^{-1}HV^{-1}$. Далее находим полную диагональную по $E$ матрицу $(H' - EI)$ и вычисляем ее детерминант $\det(H' - EI)$ при некоторой энергии $E$.

Энергия, которая приводит к нулю детерминанта, является собственной энергией задачи, а соответствующие ей вектора $C'$ – это собственные вектора уравнения (1.5.3). Зная $C'$, не трудно найти и собственные вектора исходной задачи $C$ (1.5.1), поскольку матрица $V^{-1}$ уже известна. Описанный метод сведения обобщенной матричной задачи к обычной матричной задаче на собственные значения и функции называется методом ортогонализации по Шмидту [111].

Однако в некоторых задачах при некоторых значениях вариационных параметров процедура нахождения обратных матриц оказывается неустойчивой и при работе компьютерной программы выдается переполнение. Например, в двухтельных задачах для легких атомных ядер с одним вариационным параметром $\alpha_i$ в вариационной ВФ такой метод, как правило, достаточно устойчив и позволяет получать надежные результаты. Однако в трехтельной ядерной системе, когда вариационная ВФ представляется в виде [58-60]

$$\Phi_{l,\lambda}(r,R) = Nr^{\lambda}R^{l}\sum_{i} C_i \exp(-\alpha_i r^2 - \beta_i R^2) = N\sum_{i} C_i\Phi_i \quad , \qquad (1.5.4)$$

при некоторых значениях двух вариационных параметров $\alpha_i$ и $\beta_i$, метод нахождения обратных матриц иногда приводит к неустойчивости и переполнению при работе компьютерной программы [112], что представляет определенную проблему для решения задач такого типа.

Здесь выражение

$$\Phi_i = r^{\lambda} R^{1} \exp(-\alpha_i r^2 - \beta_i R^2)$$

называется базисной функцией.

Теперь рассмотрим альтернативный метод численного решения обобщенной матричной задачи на собственные значения, свободный от указанных трудностей и имеющий бóльшую скорость счета на компьютере. А именно, исходное матричное уравнение (1.5.1) или (1.5.2) есть однородная система линейных уравнений – она имеет нетривиальные решения, только если ее детерминант det($H$ - $EL$) равен нулю. Для численных методов, реализуемых на компьютере, не обязательно разлагать матрицу $L$ на треугольные матрицы и находить новую матрицу $H'$ и новые вектора $C'$, определяя обратные матрицы, как это было описано выше при использовании стандартного метода.

Можно сразу разлагать недиагональную, симметричную матрицу ($H$ - $EL$) на треугольные и численными методами в заданной области значений искать энергии, которые приводят к нулю ее детерминанта, т.е. являются собственными энергиями. В реальной физической задаче обычно не требуется искать все собственные значения и собственные функции. Нужно найти только 1÷2 собственных значения для определенной энергии системы и, как правило, это низшие их значения и соответствующие им собственные волновые функции.

Поэтому, например, методом Халецкого исходная матрица ($H$ - $EL$) разлагается на две треугольные, причем в главной диагонали верхней треугольной матрицы $V$ стоят единицы

$$A = H - EL = NV$$

и далее вычисляется ее детерминант при условии $\det(V) = 1$ [109]

$$D(E) = \det(A) = \det(N)\cdot\det(V) = \det(N) = \prod_{i=1}^{m} n_{ii}$$

по нулю, которого ищется нужное собственное значение *E,* т.е. значение энергии. Здесь *m* – размерность матриц, а детерминант треугольной матрицы *N* равен произведению ее диагональных элементов [109].

Таким образом, имеем довольно простую задачу поиска нуля функционала одной переменной

$$D(E) = 0 \ ,$$

численное решение которой не представляет большой сложности и может быть выполнено с любой точностью, например, методом половинного деления [86].

В результате, мы избавляемся от необходимости искать две обратные к *V* и *N* матрицы и выполнять несколько матричных умножений, чтобы вначале получить новую матрицу *H'*, а затем, конечную матрицу собственных векторов *C*. Отсутствие таких операций, особенно поиска обратных матриц, заметно увеличивает скорость счета на компьютере независимо от языка программирования, на котором реализовано решение этой задачи [73].

Для оценки точности решения, т.е. точности разложения исходной матрицы на две треугольные, использовано понятие невязок [109]. После разложения матрицы *A* на треугольные, вычисляется матрица невязок [109], как разность исходной матрицы *A* и матрицы

$$S = NV \ ,$$

где *N* и *V* найденные численные треугольные матрицы. Теперь берется разность по всем элементам с исходной матрицей *A*

$$A_N = S - A \quad .$$

Матрица $A_N$ невязок дает отклонение приближенной величины $NV$, найденной численными методами, от истинного значения каждого элемента исходной матрицы $A$. Можно выполнить суммирование всех элементов матрицы $A_N$ и получить численное значение невязки. Во всех приведенных в данной книге вариационных расчетах использовался описанный здесь метод, и максимальное значение любого элемента матрицы $A_N$ обычно не превышало величину $10^{-10}$.

Изложенный метод, который представляется вполне очевидным в численном исполнении, позволяет получить хорошую устойчивость алгоритма решения любых рассматриваемых задач и не приводит к переполнению при работе компьютерных программ [113]. Таким образом, описанный здесь альтернативный метод нахождения собственных значений обобщенной матричной задачи [17], рассматриваемой на основе вариационных методов решения уравнения Шредингера с использованием неортогонального вариационного базиса, избавляет нас от неустойчивостей, возникающих вместе с применением обычных методов решения такой математической задачи, т.е. обычного метода ортогонализации по Шмидту.

После нахождения собственного значения (обычно это первое или второе собственное значение с минимальной величиной) решаем известную систему уравнений для поиска собственных векторов $X$, которая имеет вид

$$AX = NVX = (H - EL)X = 0 \quad .$$

Такая система линейных уравнений относительно $N$ неизвестных $X$ может быть решена при $E$, которое равно собственному значению. Равенство нулю ее детерминанта означает линейную зависимость одного из уравнений системы, т.е. ее ранг $R$ меньше порядка системы $N$. Полагаем, что линейно зависимым является последнее $N$-е уравнение и, отбрасывая его, получаем систему ($N$ - 1) уравнений с $N$ неизвестными [114]

$a_{11}x_1 + a_{12}x_2 + a_{13}x_3 + \ldots\ldots + a_{1N}x_N = 0$
$a_{21}x_1 + a_{22}x_2 + a_{23}x_3 + \ldots\ldots + a_{2N}x_N = 0$
..............................................................
$a_{N-11}x_1 + a_{N-12}x_2 + a_{N-13}x_3 + \ldots\ldots + a_{N-1N}x_N = 0$

Принимаем $X_N = 1$, получаем систему ($N$ - 1) уравнений с ($N$ - 1) неизвестными и столбцом свободных членов, который получается из коэффициентов при $N$-ой неизвестной $a_{iN}$, где $i$ меняется от 1 до ($N$ - 1).

В матричном виде это можно записать

$$A'X' = D \quad , \tag{1.5.5}$$

где $A'$ – матрица размерности $N$ - 1, $X'$ – решение системы, $D$ – матрица свободных членов – $a_{iN}$. Решаем ее описанным далее методом, разлагая на две треугольные, и находим все $X'$ при $i = 1 \div (N - 1)$.

Теперь нам известны все решения исходной системы

$$(H - EL)X = 0$$

при $i = 1 \div N$.

Поскольку собственные вектора должны удовлетворять условию нормировки

$$N\sum_{\mathrm{i}} X_{\mathrm{i}}^2 = 1 \quad ,$$

можно найти эту нормировку и окончательно определить собственные вектора.

Для оценки точности решения системы можно использовать невязки, т.е. вычислять матрицу

$$B_N = (H - EL)X \quad ,$$

элементы которой должны быть близки к нулю при правильном определении всех $X$.

Пример программы реализующей этот метод решения матричного уравнения $A'X' = D$ после разложения матрицы на две треугольные $A = NV$ и нахождения всех решений уравнения $(H - EL)X = 0$ приведен в распечатке в третьей разделе.

В качестве примера рассмотрим теперь общий случай решения матричного уравнения стандартного вида и продемонстрируем применение метода Халецкого для решения подобных задач

$$Ax = b \quad ,$$

где $b$ и $x$ – матрицы столбцы размерности $N$, а $A$ – квадратная матрица размерности $N{\times}N$. Матрицу $A$ можно разложить на треугольные матрицы

$$A = BC \quad ,$$

где $B$ – нижняя треугольная матрица и $C$ – верхняя треугольная матрица, в главной диагонали которой стоят единицы. Нахождение нижней и верхней треугольных матриц выполняется по следующей схеме, называемой методом Халецкого [109]

$$b_{i1} = a_{i1} \quad , \qquad (1.5.6)$$

$$b_{ij} = a_{ij} - \sum_{k=1}^{j-1} b_{ik} c_{kj} \quad ,$$

где $i \geq j > 1$ и

$$c_{1j} = a_{1j} / b_{11} \quad ,$$

$$c_{ij} = \frac{1}{b_{ii}} \left( a_{ij} - \sum_{k=1}^{i-1} b_{ik} c_{kj} \right) \quad ,$$

при $1 < i < j$.

Такой метод позволяет определить и детерминант исходной матрицы $A$ [109]

$$\det(A) = \det(B)\det(C) \ .$$

Известно, что детерминант треугольной матрицы равен произведению ее диагональных элементов, а поскольку

$$\det(C) = 1 \ ,$$

то

$$\det(A) = \det(B) = (b_{11}b_{22}....b_{nn}) \ .$$

После разложения матрицы $A$ на треугольные, решение матричной системы можно записать в виде

$$By = b \ ,$$

$$Cx = y \ ,$$

где сами решения находятся из следующих простых выражений [109]

$$y_1 = a_{1,n+1}/b_{11} \ , \qquad (1.5.7)$$

$$y_i = \frac{a_{i,n+1} - \sum_{k=1}^{i-1} b_{ik} y_k}{b_{ii}}$$

при $i > 1$ и

$$x_n = y_n \ ,$$

$$x_i = y_i - \sum_{k=i+1}^{n} c_{ik} x_k$$

при $i < n$, где $a_{i,n+1}$ – элементы матрицы – столбца $b$ (здесь $i$ меняется от 1 до $N$ – размерности матрицы $A$). В такой задаче все треугольные матрицы и решения $X$ определяются вполне однозначно.

## 1.6 Общие принципы трехтельной модели General principles of the three-body model

Рассмотрим радиальное уравнение Шредингера с центральными ядерными силами для волновой функции системы трех частиц [115,116]

$$(H - E)\Phi_{l,\lambda}(r,R) = 0 \ . \qquad (1.6.1)$$

где

$$H = T + V \ ,$$

$$T = T_1 + T_2 = -\frac{\hbar^2}{2\mu}\Delta_r - \frac{\hbar^2}{2\mu_0}\Delta_R \ ,$$

$$V = V_{12} + V_{23} + V_{13} \ , \qquad (1.6.2)$$

$$\mu = \frac{m_2 m_3}{m_{23}} \ , \quad \mu_0 = \frac{m_1 m_{23}}{m} \ ,$$

$$m_{23} = m_2 + m_3 \ , \quad m = m_1 + m_2 + m_3 \ .$$

Здесь $m$ и $\mu$ – масса и приведенная масса частиц, $\Delta$ – оператор Лапласа, $\hbar$ – постоянная Планка, $T$ и $V$ – операторы кинетической и потенциальной энергии, $H$ – гамильтониан и $E$ – энергия системы.

Величина $r$ в такой записи определяет расстояние между частицами 2 и 3, которые находятся в основании треугольника из трех тел с орбитальным моментом $\lambda$, а $R$ – это расстояние между первой частицей, которая расположена в вершине треугольника и центром масс первых двух частиц с орбитальным моментом *l*.

Полная трехтельная волновая функция такой системы имеет вид [115]

$$\Psi(r,R) = \sum_{l,\lambda} \Phi_{l,\lambda}(r,R) Y_{LS}^{JM}(\hat{r},\hat{R}) \quad ,$$

а ее угловая часть $Y_{LS}^{JM}(\hat{r},\hat{R})$ записывается в обычной форме [58-60]

$$Y_{LS}^{JM}(\hat{r},\hat{R}) = \sum_{M_S M_L} \langle LM_L SM_S | JM \rangle Y_{LM_L}(\hat{r},\hat{R}) \chi_{SM_S}(\sigma) \; .$$

Здесь $\boldsymbol{L} = \boldsymbol{l} + \boldsymbol{\lambda}$ – полный орбитальный момент, $S$ – полный спин, $J$ – полный момент системы частиц, $M$ – их проекции, $Y_{LS}^{JM}$ – спин-угловая функция, $\Phi_{l,\lambda}$ – радиальная волновая функция (1.5.4), $r$ и $R$ – скалярные расстояния между частицами, $\hat{r}$ и $\hat{R}$ – углы между направлениями векторов **r** и **R** и осью $z$, $Y_{LM}$ – сферическая функция, $\chi_{SM}$ – спиновая функция системы, зависящая от спина $\sigma$ частиц, угловые скобки обозначают коэффициенты Клебша-Гордона [58-60].

В реальных расчетах при некоторых значениях вариационных параметров $\alpha_i$ и $\beta_i$ находим энергию системы, которая дает ноль детерминанта системы (1.6.1), а затем, варьируя эти параметры, проводим поиск минимума этой энергии. Затем увеличиваем размерность базиса $N$ и повторяем все вычисления, до тех пор, пока величина собственного значения, т.е. энергия связи $E_N$, на очередном шаге $N$ не станет отличаться от предыдущего значения $E_{N-1}$ на величину $\varepsilon$, которая обычно задается на уровне 0.1÷1.0% или 1÷2 кэВ. В соответствии с общими вариационными принципами [117] эта минимальная энергия и будет реальной энергией связи трехчастичной системы, т.е. энергией связи атомного ядра в рассматриваемой модели.

## 1.7 Вариационные методы трехтельной модели
## Variational methods for three-body model

Матричные элементы гамильтониана системы (1.6.2) и интегралов перекрывания, вычисленные по базисным функциям $\Phi_i$, (1.5.4) имеют вид [87,118]

$$T_{ij} = \frac{\pi}{16} N^2 \frac{h^2}{m_N} \frac{(2l+1)!!(2\lambda+1)!!}{2^{1+\lambda}} \alpha_{ij}^{-\lambda-1/2} \beta_{ij}^{-1-1/2} G_{ij} \quad ,$$

$$G_{ij} = \frac{B_{ij}(\alpha,\lambda)}{\mu\beta_{ij}} + \frac{B_{ij}(\beta,l)}{\mu_0\alpha_{ij}} \quad ,$$

$$B_{ij}(\delta,\nu) = \frac{\nu^2}{2\nu+1} + \frac{\delta_i\delta_j}{\delta_{ij}^2}(2\nu+3) - \nu \quad ,$$

$$L_{ij} = \frac{\pi}{16} N^2 \frac{(2l+1)!!(2\lambda+1)!!}{2^{1+\lambda}} \alpha_{ij}^{-\lambda-3/2} \beta_{ij}^{-1-3/2} \quad ,$$

$$N = \left( \sum_{ij} C_i C_j L_{ij} \right)^{-1/2} \quad ,$$

с потенциалами (cb – центробежный, coul – кулоновский)

$$[(V_{cb})_R]_{ij} = \frac{\pi}{16} N^2 \frac{h^2}{\mu_0} l(l+1) \frac{(2l-1)!!(2\lambda+1)!!}{2^{1+\lambda}} \alpha_{ij}^{-\lambda-3/2} \beta_{ij}^{-1-1/2} \quad ,$$

$$[(V_{cb})_r]_{ij} = \frac{\pi}{16} N^2 \frac{h^2}{\mu} \lambda(\lambda+1) \frac{(2\lambda-1)!!(2l+1)!!}{2^{1+\lambda}} \alpha_{ij}^{-\lambda-1/2} \beta_{ij}^{-1-3/2} \quad ,$$

$$[\{V_{coul}(23)\}_r]_{ij} = Z_2 Z_3 \frac{\pi}{16} N^2 \frac{2}{\sqrt{\pi}} \frac{(2l+1)!!}{2^l} \frac{\lambda!}{\alpha_{ij}^{\lambda+1} \beta_{ij}^{l+3/2}} \quad ,$$

$$[\{V_{\rm coul}(12)\}_{\rm R}]_{\rm ij} = Z_1 Z_2 \frac{\pi}{16} N^2 \frac{2}{\sqrt{\pi}} \frac{(2\lambda+1)!!}{2^\lambda} \frac{l!}{\beta_{\rm ij}^{l+1} \alpha_{\rm ij}^{\lambda+3/2}}$$

$$[\{V_{\rm coul}(13)\}_{\rm R}]_{\rm ij} = Z_1 Z_3 \frac{\pi}{16} N^2 \frac{2}{\sqrt{\pi}} \frac{(2\lambda+1)!!}{2^\lambda} \frac{l!}{\beta_{\rm ij}^{l+1} \alpha_{\rm ij}^{\lambda+3/2}}$$

$$(V_{23})_{\rm ij} = \frac{\pi}{16} N^2 V_{23} \frac{(2l+1)!!(2\lambda+1)!!}{2^{l+\lambda}} (\alpha_{\rm ij} + \gamma_{23})^{-\lambda-3/2} \beta_{\rm ij}^{-l-3/2} \quad ,$$

$$\alpha_{\rm ij} = \alpha_{\rm i} + \alpha_{\rm j} \ , \qquad \beta_{\rm ij} = \beta_{\rm i} + \beta_{\rm j} \ .$$

Далее, например, при значениях $l = 1$ и $\lambda = 0$ имеем следующие выражения для матричных элементов от ядерных потенциалов вида (1.2.1) или (1.2.2)

$$(V_{12})_{\rm ij} = \frac{\pi}{16} N^2 \frac{3V_{12}}{2A_{\rm ij}^{3/2}(\beta_{\rm ij} + \gamma_{12})} \left[ \frac{a^2 \gamma_{12}^2}{A_{\rm ij}} + 1 \right] \quad ,$$

где

$$A_{\rm ij} = \alpha_{\rm ij}\beta_{\rm ij} + \gamma_{12}(\alpha_{\rm ij} + a^2 \beta_{\rm ij}) \ , \quad a = m_3 / m_{23} \ .$$

В случае $l = 0$ и $\lambda = 0$ для этой части потенциала находим выражение

$$(V_{12})_{\rm ij} = \frac{\pi}{16} N^2 \frac{V_{12}}{A_{\rm ij}^{3/2}} \quad .$$

Здесь величина $\gamma_{12}$ является параметром ширины гауссова потенциала, а $V_{12}$ – его глубина между соответствующей парой частиц, в данном случае, 12 или 13.

Для случая произвольных $l$, когда $\lambda = 0$ можно получить выражение

$$(V_{12})_{\rm ij} = \frac{\pi}{16} N^2 V_{12} \frac{(2l+1)!!}{2^{\rm l}} \frac{d_{\rm ij}^{\rm l}}{A_{\rm ij}^{\rm l+3/2}} \ ,$$

где

$$d = \alpha_{\rm ij} + \gamma_{12} a^2 \ .$$

Среднеквадратичный массовый радиус ядра в такой модели представляется в виде [87,119]

$$< r^2 >_{\rm m} = m_1 / m < r^2 >_{\rm m1} + m_2 / m < r^2 >_{\rm m2} + m_3 / m < r^2 >_{\rm m3} + A / m$$

$$A = \frac{\pi}{16} N^2 \frac{(2l+1)!!(2\lambda+1)!!}{2^{\rm l+\lambda+1}} \sum_{\rm ij} C_{\rm i} C_{\rm j} \alpha_{\rm ij}^{-\lambda-3/2} \beta_{\rm ij}^{-\rm l-3/2} \begin{pmatrix} \dfrac{2\lambda+3}{\alpha_{\rm ij}} \mu + \\ + \dfrac{2l+3}{\beta_{\rm ij}} \mu_0 \end{pmatrix}$$

$$\mu = \frac{m_2 m_3}{m_{23}}, \quad \mu_0 = \frac{m_1 m_{23}}{m},$$

$$m_{23} = m_2 + m_3, \quad m = m_1 + m_2 + m_3 \quad .$$

Среднеквадратичный зарядовый радиус ядра в трехтельной модели имеет вид

$$< r^2 >_{\rm z} = Z_1 / Z < r^2 >_{\rm z1} + Z_2 / Z < r^2 >_{\rm z2} + Z_3 / Z < r^2 >_{\rm z3} + B / Z$$

Здесь величина $B$ выражена через моменты частиц, вариационные параметры и коэффициенты разложения ВФ следующим образом:

$$C = \frac{Z_1 m_{23}^2 + Z_{23} m_1^2}{m^2} \ , \quad Z_{23} = Z_2 + Z_3 \ ,$$

$$D = \frac{Z_2 m_3^2 + Z_3 m_2^2}{m_{23}^2} \ ,$$

$$E = \frac{m_1}{m m_{23}} (Z_3 m_2 - Z_2 m_3) \ .$$

$$B = \frac{\pi}{16} N^2 \frac{(2l+1)!!(2\lambda+1)!!}{2^{l+\lambda+1}} \sum_{ij} C_i C_j \alpha_{ij}^{-\lambda-3/2} \beta_{ij}^{-l-3/2} \begin{pmatrix} \frac{2l+3}{\beta_{ij}} C + \\ + \frac{2\lambda+3}{\alpha_{ij}} D \end{pmatrix} +$$

$$+ N^2 E (\lambda+1)!(l+1)! \sum_{ij} \frac{C_i C_j}{2\alpha_{ij}^{\lambda+2} \beta_{ij}^{l+2}} \ ,$$

В качестве зарядовых и массовых радиусов кластеров, а именно, протона, дейтрона, тритона и $\alpha$ частицы обычно принимались величины $\langle r \rangle_{mp} = \langle r \rangle_{zp} = 0.877$ Фм, $\langle r \rangle_{md} = \langle r \rangle_{zd} = 1.96$ Фм, $\langle r \rangle_{mt} = \langle r \rangle_{zt} = 1.70$ Фм, $\langle r \rangle_{m\alpha} = \langle r \rangle_{z\alpha} = 1.67$ Фм [120,121,122,123].

При поиске энергии связи ядра в трехтельной модели начальные значения вариационных параметров $\alpha_i$ и $\beta_i$ находились из линейной сетки вида

$$\alpha_i = i/30 \ , \qquad \beta_i = 2\alpha_i \ .$$

Затем проводилось независимое варьирование каждого из них так, чтобы минимизировать энергию системы с заданной точностью.

Для проверки предложенного метода расчета и компьютерной программы рассматривалась модельная задача для трех частиц, взаимодействующих в потенциале Афнана-Танга [124] с усреднением триплетных и синглетных состояний. Для энергии связи такой системы в работе [124] получено значение -7.74 МэВ, а в статьях [125], где использовался неортогональный вариационный метод с изменением параметров $\alpha$ и $\beta$ волновой функции на основе тангенциальной сетки, найдено -7.76 МэВ.

Нами, на основе изложенных выше методов, при независимом варьировании всех параметров и размерности базиса $N = 5$, получено -7.83 МэВ. Тем самым, энергия связи рассматриваемой системы изменилась примерно на 1% относительно результатов работ [124,125].

## *Заключение*
## *Conclusion*

В заключение этого раздела заметим, что существует множество различных математических методов решения дифференциальных уравнений второго порядка, которым является уравнение Шредингера. Однако в известной нам математической литературе обычно приводятся довольно абстрактные методы решений таких уравнений, которые бывает достаточно сложно применить для решения конкретного уравнения, типа уравнения Шредингера, в реальных физических задачах квантовой физики.

Поэтому данный раздел описывает некоторые основные математические методы, непосредственно применимые для нахождения волновых функций из решений уравнения Шредингера в задачах ядерной физики [17]. Рассматриваются вариационные методы решений, а также компьютерные программы, применимые в задачах непрерывного и дискретного спектров состояний двух или трех ядерных частиц [73,74].

# II. ФАЗОВЫЙ АНАЛИЗ
## Phase shifts analysis

*Процедура фазового анализа состоит в разложении полной амплитуды рассеяния в ряд по парциальным волнам или амплитудам и анализе, появляющихся при этом параметров, называемых фазами рассеяния. Такие фазы позволяют получать сведения о природе сильных взаимодействий, структуре резонансных состояний и общем строении атомного ядра [126].*

### *Введение*
### *Introduction*

Во многих задачах ядерной физики низких энергий и ядерной астрофизики требуется знание фаз упругого рассеяния, которые могут быть определены из дифференциальных сечений упругого рассеяния различных ядерных частиц [126]. Как мы уже говорили, фазы рассеяния используются, в частности, для построения межкластерных потенциалов взаимодействия в рассматриваемой здесь потенциальной кластерной модели.

Задача определения или извлечения ядерных фаз из сечений упругого рассеяния в математическом плане сводится к многопараметрической вариационной задаче. Когда известны экспериментальные сечения рассеяния ядерных частиц и математические выражения, полученные в квантовой механике, которые описывают эти сечения в зависимости от некоторых параметров $\delta_L$ – ядерных фаз рассеяния, возникает многопараметрическая вариационная задача нахождения этих параметров на заданном интервале значений.

Поскольку не существует общих методов решения многопараметрической вариационной задачи для поиска глобального минимума, мы можем надеяться найти только неко-торые локальные минимумы при каждой энергии и, исходя из физических соображений, отобрать те из них, которые могут являться решениями исходной задачи. Одним из критериев такого отбора является требование плавного поведения каждой парциальной ядерной фазы, как функции энергии в нерезонансной области и переход ее значения через 90° при резонансной энергии [126].

В разных ядерных системах в зависимости от энергии сталкивающихся частиц, число фаз упругого рассеяния может меняться от 1÷3 до 10÷20 [92]. Например, в $^4$He$^4$He рассеянии мы использовали до 20 парциальных волн, а в p$^{12}$C системе при низких энергиях, как это будет показано далее, всего одну парциальную $^2S$ фазу.

В этом разделе будут рассмотрены различные экспериментальные данные, методы расчетов дифференциальных сечений, включая компьютерные программы, и результаты для фазового анализа упругого $^4$He$^4$He, p$^3$He, p$^6$Li, p$^{12}$C, n$^{12}$C, p$^{13}$C и $^4$He$^{12}$C рассеяния. Для первой приведенной выше системы рассматриваются энергии до 50 МэВ, 11.5 МэВ во втором случае и примерно до 1÷1.5 МэВ для остальных процессов рассеяния, кроме последнего – в $^4$He$^{12}$C системе фазовый анализ выполнен в области 1.5÷6.5 МэВ.

## *2.1 Фазовый анализ упругого $^4$He$^4$He рассеяния* *Phase shifts analysis of the elastic $^4$He$^4$He scattering*

Рассмотрим вначале результаты измерений дифференциальных сечений упругого рассеяния и результаты фазового анализа, полученного из этих сечений, для $^4$He$^4$He системы при разных энергиях. Основные данные таких исследований относятся к области энергий до 120 МэВ, но не во всей этой области был выполнен последовательный фазовый анализ экспериментально измеренных дифференциальных сечений упругого рассеяния [127].

### 2.1.1 Обзор экспериментальных данных по упругому $^4$He$^4$He рассеянию Review of the experimental data on elastic $^4$He$^4$He scattering

Приведем краткий обзор экспериментальных данных и результатов фазового анализа упругого $^4$He$^4$He рассеяния, выполненных в разных работах:

1. Измерение дифференциальных сечений упругого рассеяния и фазовый анализ в области энергий 0.6÷3.0 МэВ (л.с.) выполнены в работе [128], где сечения и фазы приведены в таблицах, что очень удобно для их использования и в любой момент позволяет повторить все результаты по нахождению фаз рассеяния.

2. Область энергий 3.0÷5.0 МэВ была рассмотрена в работе [129], но фазы и сечения рассеяния приведены только на рисунках.

3. Энергии 3.8÷11.9 МэВ анализировались в работе [130], где результаты фазового анализа приведены в таблице, а дифференциальные сечения рассеяния показаны только на рисунках.

4. Очень аккуратные измерения дифференциальных се-

чений упругого рассеяния и фазовый анализ выполнены в работах [131,132], где для области энергий 12.3÷22.9 МэВ экспериментальные сечения и фазы рассеяния приведены в таблицах.

5. Область 12.9÷21.6 МэВ рассматривалась и в работе [133], но сечения и фазы рассеяния приведены в ней только на рисунках.

6. Очень хорошие данные для энергий 18.0÷29.5 МэВ приведены в работе [134], где сечения упругого рассеяния и результаты фазового анализа приведены в подробных таблицах.

7. При энергиях 23.1÷38.4 МэВ выполнены экспериментальные измерения сечений и проведен фазовый анализ в работе [135], однако в таблицах приведены только сечения рассеяния.

8. Измерение сечений и фазовый анализ в области 53÷120 МэВ были выполнены в работе [136], но там приведены только табличные фазы, а сечения рассеяния даны на рисунках.

9. Экспериментальные исследования сечений упругого рассеяния для энергий 36.8÷47.3 МэВ были проведены в работе [137], где в подробных таблицах даны результаты измерений дифференциальных сечений, но фазовый анализ этих данных вообще не проводился. Теоретические исследования некоторых энергий из этой области имеются только в работе [138], где выполнен поиск параметров оптического потенциала, а затем, расчетным путем, были получены фазы упругого $^4He^4He$ рассеяния. Качество оптической подгонки, сделанной в этой области энергий, оставляет желать много лучшего, что непосредственно видно из рисунков работы [138]. Кроме того, при извлечении фаз рассеяния из оптических потенциалов вычислялась только их действительная часть, а мнимая часть фаз считалась малой, находящейся на уровне $1^\circ \div 2^\circ$, что представляется не вполне оправданным для энергий в области 40 МэВ и выше.

10. Энергии 38.5, 49.9 и 51.1 МэВ были рассмотрены в работах [139,140,141] соответственно, где измерены диффе-

ренциальные сечения, но фазовый анализ этих данных не проводился. Вместо этого в работах [139,141] получены параметры оптических потенциалов, и в [141], на их основе, вычислены фазы упругого рассеяния.

Из приведенного обзора видно, что стандартный фазовый анализ экспериментальных данных, т.е. дифференциальных сечений упругого рассеяния в $^4$He$^4$He системе при энергиях 36.85÷51.1 МэВ, до сих пор не выполнен. Качество подгонки параметров оптических потенциалов для области энергий 23÷47 МэВ [138] вряд ли можно считать удовлетворительным, что неудивительно, поскольку все эти результаты были получены в 60-е годы, когда вычислительная техника и методы расчетов только начинали развиваться.

Поэтому представляется интересным выполнить точный и максимально полный фазовый анализ экспериментальных дифференциальных сечений упругого рассеяния в области энергий 37÷51 МэВ. Провести такой анализ позволяет табличное представление экспериментальных сечений упругого $^4$He$^4$He рассеяния в этой области энергий, приведенное в работах [137,139,140].

### 2.1.2 Методы фазового анализа упругого $^4$He$^4$He рассеяния<br>Methods for phase shift analysis of elastic $^4$He$^4$He scattering

Рассмотрим задачу фазового анализа и методы определения фаз из экспериментальных данных, т.е. определим способы и подходы, которые использовались в нашем фазовом анализе. Дифференциальное сечение упругого рассеяния определяется через фазы рассеяния тождественных частиц следующим образом [88]:

$$\frac{d\sigma(\theta)}{d\Omega} = \left| f(\theta) + f(\pi-\theta) \right|^2 \quad , \qquad (2.1.1)$$

где амплитуда рассеяния представляется в виде суммы кулоновской и ядерной амплитуд

$$f(\theta) = f_c(\theta) + f_N(\theta) \qquad (2.1.2)$$

и выражается через ядерные $\delta_l$ и кулоновские $\sigma_l$ фазы рассеяния [88]:

$$f_c(\theta) = -\left(\frac{\eta}{2k\sin^2(\theta/2)}\right)\exp\{i\eta\ln[\sin^{-2}(\theta/2)] + 2i\sigma_0\}, \qquad (2.1.3)$$

$$f_N(\theta) = \frac{1}{2ik}\sum_l (2l+1)\exp(2i\sigma_l)[S_l - 1]P_l(\cos\theta) \quad ,$$

где $k$ – волновое число относительного движения частиц $k^2 = 2\mu E/\hbar^2$, $E$ – энергия сталкивающихся частиц в центре масс, $\mu$ – приведенная масса, $\eta$ – кулоновский параметр, $\theta$ – угол рассеяния, $P_l(\cos\theta)$ – полиномы Лежандра.

Обычно ядерные фазы упругого рассеяния представляются в виде

$$\delta_l = \mathrm{Re}\delta_l + i\mathrm{Im}\delta_l \quad ,$$

тогда для матрицы рассеяния и параметра неупругости получим

$$S_l(k) = \eta_l(k)\exp[2i\mathrm{Re}\delta_l(k)] \quad ,$$

$$\eta_l(k) = \exp[-2\mathrm{Im}\delta_l(k)] \quad .$$

Суммирование в выражении (2.1.3) выполняется только по четным $l$, поскольку нечетные парциальные волны не дают вклада в полное сечение, и проводится до некоторого $L$, величина которого зависит от энергии.

Кулоновскую амплитуду рассеяния (2.1.3), используя выражение

$D = \sin^{-2}(\theta/2) = 2/[1 - \cos(\theta)]$ ,

можно записать в виде

$f_c = -\eta D/2k\,[\cos(C) + i\sin(C)]$ ,

где

$C = 2\sigma_0 + \eta\ln A$ .

Ядерная амплитуда может быть представлена в следующей форме

$$f_N = \frac{1}{2k}\sum_L \hat{L}\left\{\begin{matrix} [B\cos(2\sigma_L) + A\sin(2\sigma_L)] + \\ +i[B\sin(2\sigma_L) - A\cos(2\sigma_L)] \end{matrix}\right\} P_L(x), \qquad (2.1.4)$$

где

$x = \cos(\theta)$ , $\hat{L} = 2l + 1$ , $A = \eta_l\cos(2\delta_l) - 1$ , $B = \eta_l\sin(2\delta_l)$

зависят только от ядерных фаз, параметра неупругости и орбитального момента.

Кулоновские фазы рассеяния выражаются через Гамма функцию [88]

$$\sigma_l = \arg\{\Gamma(l + 1 + i\eta)\}$$

и удовлетворяют рекуррентному процессу

$$\sigma_l = \sigma_{l+1} - \mathrm{Arctg}\left(\frac{\eta}{l+1}\right) .$$

Откуда сразу можно получить следующее выражение для кулоновских фаз

$$\alpha_l = \sigma_l - \sigma_0 = \sum_{n=1}^{l} \mathrm{Arctg}\left(\frac{\eta}{n}\right) , \qquad \alpha_0 = 0 .$$

Величина $\alpha_l$ используется в преобразованных выражениях (2.1.3), если вынести общий множитель exp($2i\sigma_0$). Тогда $\sigma_l \to \alpha_l$ с $\alpha_0 = 0$, что избавляет нас от необходимости вычислять кулоновские фазы в явном виде, и кулоновская амплитуда принимает следующую форму

$$f_c(\theta) = -\left(\frac{\eta}{2k\sin^2(\theta/2)}\right)\exp\{i\eta\ln[\sin^{-2}(\theta/2)]\} .$$

В $^4He^4He$ задаче с нулевым спином набор фаз $\delta^J_{S,L}$, зависящий от полного момента $J$ и спина $S$, переходит в $\delta_l$. Поскольку $S = 0$, то полный момент равен орбитальному моменту $J = L$.

### 2.1.3 Проверка компьютерной программы Computer program check

Для выполнения данного фазового анализа была написана компьютерная программа на языке "Basic" для компилятора "Turbo Basic" фирмы “Borland International Inc.” [142], использующая режим двойной точности, которая затем переведена на Fortran-90. Программа тестировалась по выполненному ранее фазовому анализу из различных работ при разных энергиях [17]. Здесь мы приведем только некоторые из этих тестов.

Например, при энергии 6.47 МэВ (л.с.) в работе [130] приводятся следующие фазы: $\delta_0 = 79.5° \pm 2°$, $\delta_2 = 80.8° \pm 2°$. В нашем фазовом анализе для них получаются значения $\delta_0 = 80.43°$, $\delta_2 = 80.73°$ при $\chi^2 = 0.18$. Ошибка определения сечений из рисунков работы [130] принималась равной 10%, что и объясняет столь малую величину $\chi^2$.

При энергии 12.3 МэВ в работе [132] получено $\delta_0 = 29° \pm$

4°, $\delta_2$ = 103° ± 8°, $\delta_4$ = 3° ± 1.5°, а наш анализ дает $\delta_0$ = 28.37°, $\delta_2$ = 105.03°, $\delta_4$ = 2.62° при $\chi^2$ = 3.43.

В этой же работе для 17.8 МэВ найдено $\delta_0$ = 7° ± 2°, $\delta_2$ = 104° ± 4°, $\delta_4$ = 16.2° ± 2°, а в наших вычислениях мы приходим к значениям $\delta_0$ = 7.25°, $\delta_2$ = 103.93°, $\delta_4$ = 17.0° при $\chi^2$ = 0.46.

В работе [132] рассмотрена и энергия 22.9 МэВ, для которой получены следующие фазы: $\delta_0$ = 169.7° ± 2°, $\delta_2$ = 94.0° ± 2°, $\delta_4$ = 59.2° ± 2°, $\delta_6$ = 1.09°. Наш фазовый анализ этих данных дает величины $\delta_0$ = 169.30°, $\delta_2$ = 94.49°, $\delta_4$ = 59.55°, $\delta_6$ = 1.0° при $\chi^2$ = 1.46.

При энергиях 12.3, 17.8 и 22.9 МэВ дифференциальные сечения, их ошибки, фазы рассеяния и величина $\chi^2$ приведены в таблицах [132], но сравнивать можно только сами фазы рассеяния. Сравнивать $\chi^2$ достаточно сложно, потому что в работах [130,132] рассматривается не величина $\chi^2$, определенная выше (1.3.1), а ее производная, зависящая от некоторых констант, связанных с экспериментальной методикой.

Приведем более подробно вариант контрольного счета, который выполнен для упругого $^4He^4He$ рассеяния при энергии 29.5 МэВ. В работе [134], где даны экспериментальные сечения и результаты фазового анализа (см. табл.2.1.1), для среднего $\chi^2$ была получена величина 0.68, но методы ее расчета несколько отличаются от изложенных выше, поэтому значение 0.68 также нельзя напрямую сравнивать с нашими результатами.

Табл.2.1.1. Сравнение результатов фазового анализа из работы [134] и наших результатов при 29.5 МэВ.

| $\delta_L$, град. | Результаты [134] | Наши результаты |
|---|---|---|
| $\delta_0$ | 150,88 ±0.17 | 150.76 |
| $\delta_2$ | 86.90±0.13 | 86.61 |
| $\delta_4$ | 121.19±0.17 | 121.00 |
| $\delta_6$ | 2.20±0.11 | 2.16 |
| $\delta_8$ | 0.11±0.08 | 0.09 |

В результате наших расчетов с фазами из работы [134] для среднего $\chi^2$ по всем точкам получено 1.086. Если учесть весовые множители из [134] можно получить величину 0.6, вполне согласующуюся с результатами этой работы.

Далее нами выполнены подробные расчеты с минимизацией $\chi^2$ по нашей программе, приведенной далее, и сравнение результатов с экспериментальными данными [134]. Для среднего $\chi^2$ было получено 0.600 (вместо 1.086), т.е. наблюдается улучшение качества описания эксперимента почти в 2 раза при очень небольшом изменении значений самих фаз, которые показаны в табл.2.1.1.

В работе [134] получены данные и при энергии 25.5 МэВ, для которой найдены фазы $\delta_0 = 160,36° \pm 1.01°$, $\delta_2 = 89.37° \pm 1.54°$, $\delta_4 = 88.64° \pm 1.77°$, $\delta_6 = 1.61° \pm 0.39°$, $\delta_8 = 0.36° \pm 0.19°$. Наш расчет с этими фазами приводит к $\chi^2 = 2.127$. Выполняя дополнительное варьирование фаз рассеяния, получим заметное улучшение описания имеющихся данных с $\chi^2 = 0.886$. Для фаз получены следующие значения: $\delta_0 = 160.49°$, $\delta_2 = 89.00°$, $\delta_4 = 88.60°$, $\delta_6 = 1.41°$, $\delta_8 = 0.18°$, которые совпадают с результатами [134] в пределах, приведенных в этой работе, ошибок фаз рассеяния.

Небольшие отличия в фазах рассеяния могут быть обусловлены различными значениями констант или масс частиц, которые используются в таких расчетах. Например, можно использовать точные значения масс частиц или же их целые величины, а константа $\hbar^2 / m_0$ может быть равна 41.47 или, например, 41.4686 МэВ·Фм$^2$. Поэтому, в целом можно считать, что во всех рассмотренных выше случаях, которые можно считать контрольными, наши результаты, в пределах, приведенных в различных работах ошибок фаз, совпадают с данными, полученными ранее, разными авторами.

### 2.1.4 Результаты фазового анализа упругого $^4$He$^4$He рассеяния<br>Results of the $^4$He$^4$He scattering phase shifts analysis

Приведем теперь результаты нашего фазового анализа в области энергий 30÷40 МэВ, которая исследовалась в работе [135], где в таблицах приведены экспериментальные сечения упругого рассеяния, а результаты фазового анализа даны на рисунках.

В работе [135] рассматривалась энергия 30.3 МэВ, для которой получены фазы рассеяния 135 ± 5, 75 ± 5, 110 ± 5, -2 ± 1 (здесь и далее фазы приведены в градусах), в целом описывающие экспериментальные данные. Приведенные здесь ошибки определяют точность извлечения этих фаз из рисунков работы [135]. Используя их в качестве начальных и выполняя варьирование фаз с шестью парциальными волнами, получим $\chi^2 = 2.11$ с фазами 135.89, 76.51, 116.29, 0.0, которые вполне согласуются с результатами [135]. Видно, что достаточно четырех парциальных волн ($l$ = 0,2,4), чтобы сравнительно хорошо описать такие сечения. Для получения $\chi^2$ = 0.177 требуется уже $L$ = 16 и учет мнимой части фаз, но парциальный $\chi^2_i$ при 85 градусах остается порядка 1.6. И только увеличение $L$ до 20 позволяет получить $\chi^2$ = 0.108 со всеми парциальными $\chi^2_i$ меньше единицы и фазами

$$\delta_{Re} = 134.4998;\ 69.7085;\ 116.0319;\ 0.0;\ 0.0476;\ 2.5423;\ 0.8548;\ 4.4579;\ 0.1409;\ 0.0;\ 1.0187;$$

$$\delta_{Im} = 1.8530;\ 3.1955;\ 0.4002;\ 3.6621;\ 0.0;\ 4.0655;\ 1.5142;\ 3.7219;\ 2.1339;\ 0.0;\ 0.1611,$$

действительная часть которых в пределах ошибок вполне согласуется с результатами работы [135].

Приведенная точность фаз обусловлена тем, что их округление даже до сотых долей при $L$ порядка 20 может изменить $\chi^2$ в 2÷3 раза. На рис.2.1.1 показаны результаты расчетов сечений с этими фазами.

В работе [135] была рассмотрена также энергия 31.8 МэВ и на рисунках приведены фазы рассеяния 144 ± 5, 79 ± 5, 126 ± 5 при $l$ = 0, 2.4. Используя их в качестве начальных фаз, т.е. входных параметров для нашей программы, и выполняя варьирование, получим $\chi^2$ = 1.99. Для фаз рассеяния, приводящих к такому минимуму $\chi^2$, найдены величины $\delta_{Re}$ =

143.78, 78.39, 125.71, которые полностью совпадают с результатами работы [135].

Дальнейшее увеличение числа парциальных волн без учета мнимой части не приводит к заметному улучшению описания экспериментальных данных. И только если учесть мнимую часть фаз и выполнить варьирование с десятью парциальными волнами, можно получить улучшение величины $\chi^2$ до 1.10 со следующими фазами:

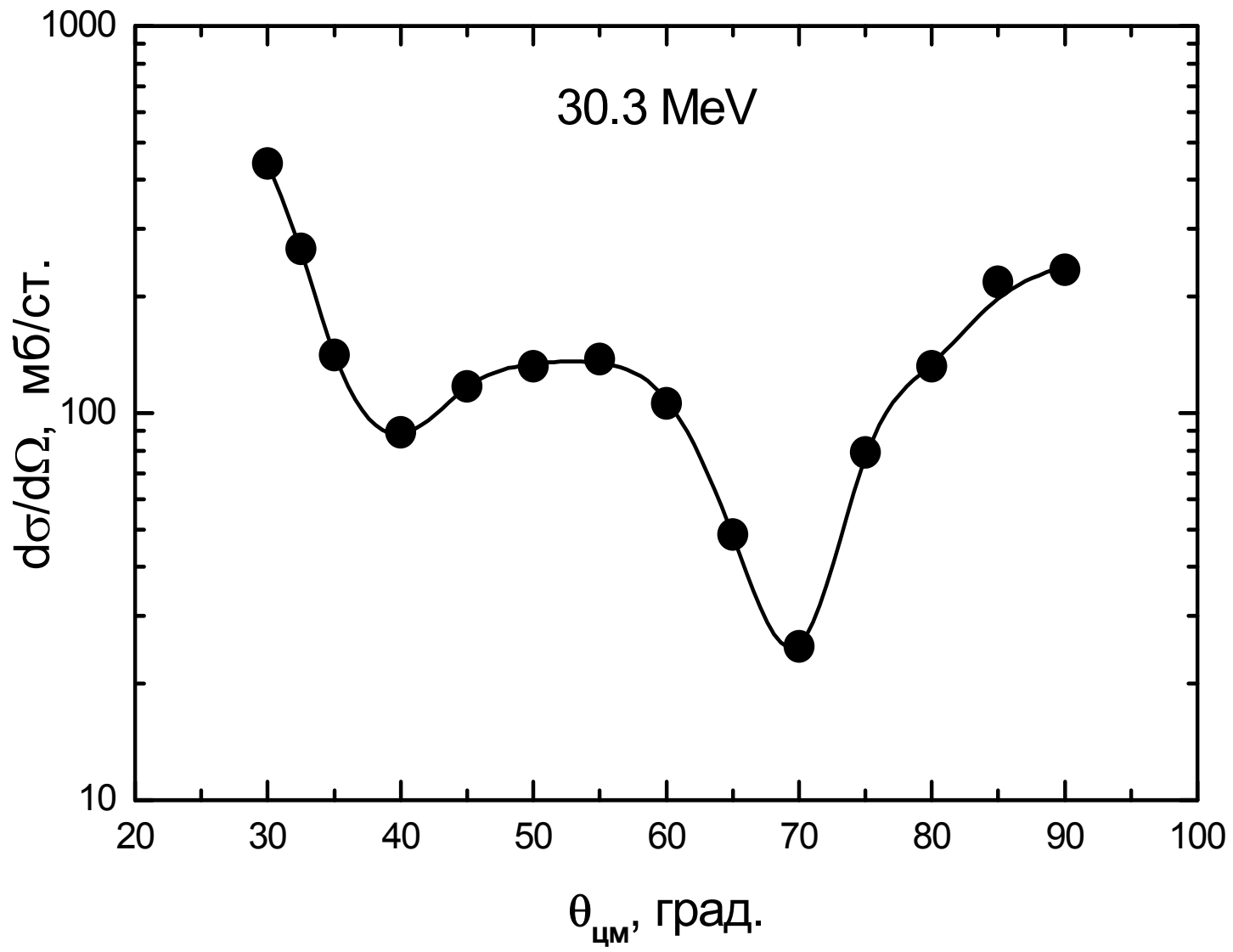

Рис.2.1.1. Дифференциальные сечения упругого рассеяния альфа-частиц на ядрах гелия при энергии 30.3 МэВ [135]. Точки – экспериментальные данные [135], сплошная кривая – расчет сечений с найденными фазами.

$\delta_{Re}$ = 142.89; 78.09; 125.80; 0.0; 0.14; 0.41;
$\delta_{Im}$ = 0.0; 0.0; 0.0; 0.0; 1.26; 0.59.

Однако и при таком значении $L$ некоторые парциальные $\chi^2_i$ заметно превышают единицу. И только при $L$ = 16 можно получить хорошее описание экспериментальных данных со средним $\chi^2$ = 0.088, всеми парциальными $\chi^2_i$ лежащими в об-ласти 0.003÷0.5, и фазами

$\delta_{Re}$ = 152.5263; 69.4016; 126.8901; 0.0; 1.3442; 4.5831; 0.0918; 2.1935; 0.1769;
$\delta_{Im}$ = 0.0; 2.0620; 5.3600; 1.4498; 0.2164; 5.6597; 2.2074; 2.8611; 0.4500.

Из этих результатов видно, что увеличение числа парциальных волн заметно сказывается на величине $S$ и $D$ фаз рассеяния, практически не меняя $G$ волну, а мнимая часть фаз при 30.3 и 31.8 МэВ уже оказывает заметное влияние на качество описания дифференциальных сечений, хотя мнимые части фаз сравнительно малы.

Энергия 34.2 МэВ также была рассмотрена в работе [135], где на рисунках приведены фазы рассеяния 145 ± 5, 65 ± 5, 145 ± 5, 5 ± 2 при $l$ = 0,2,4,6, с которыми нами получено $\chi^2$ = 6.4. Выполняя далее варьирование фаз, и включая восьмую парциальную волну, находим заметное улучшение описания экспериментальных данных с $\chi^2$ = 0.976. Для фаз рассеяния были получены следующие величины: $\delta_{Re}$ = 145.18, 69.10, 143.52, 4.20, 0.65, которые практически совпадают с данными [135] в первых шести парциальных волнах.

Дальнейшее увеличение числа парциальных волн до 20 и учет мнимой части фаз приводит к уменьшению $\chi^2$ только до 0.56 и уже не оказывает существенного влияния на качество описания экспериментальных данных, поскольку некоторые парциальные $\chi^2_i$ имеют величину порядка трех. И только увеличение числа парциальных волн до 26 приводит к достаточно гибкому вариационному базису, который дает среднее $\chi^2$ = 0.065 со всеми парциальными $\chi^2_i$ меньше 0.4 и фазами

$\delta_{Re}$ = 140.8989; 54.4378; 133.2759; 0.9489; 0.3164; 0.7580; 5.3713; 0.1052; 0.0; 0.3722; 2.6026; 0.1765; 0.9466; 3.0101;
$\delta_{Im}$ = 1.2698; 3.0833; 3.3667; 2.9079; 1.4433; 0.0; 3.2185; 0.0400; 1.8923; 0.3312; 1.6748; 0.6881; 0.0; 0.8276.

Здесь мы наблюдаем заметное изменение значений $D$ и

$G$ фаз рассеяния, по сравнению с результатами при $L = 8$.

Следующая энергия, которая была рассмотрена в работе [135] – это 35.1 МэВ и для нее были получены фазы 147 ± 5, 80 ± 5, 150 ± 5, 7 ± 2 при $l = 0,2,4,6$. Используя их в качестве начальных, нами выполнено варьирование при $L = 8$ без мнимой части. В результате получено $\chi^2 = 2.2$ с фазами 142.29, 76.95, 147.46, 5.42, 1.67, которые вполне согласуются с результатами [135] для первых шести значений $l$.

При увеличении $L$ до 16 и учете мнимой части фаз получается очень хорошее описание экспериментальных данных при $\chi^2 = 0.015$ с парциальными $\chi^2_i$ меньше 0.06 и фазами

$\delta_{Re}$ = 155.1029; 70.2311; 137.7721; 0.0; 1.0646; 3.1268; 1.9988; 0.4711; 1.7829;

$\delta_{Im}$ = 3.3855; 0.4609; 4.6500; 0.8327; 1.1338; 1.5133; 5.1306; 0.0; 3.3445.

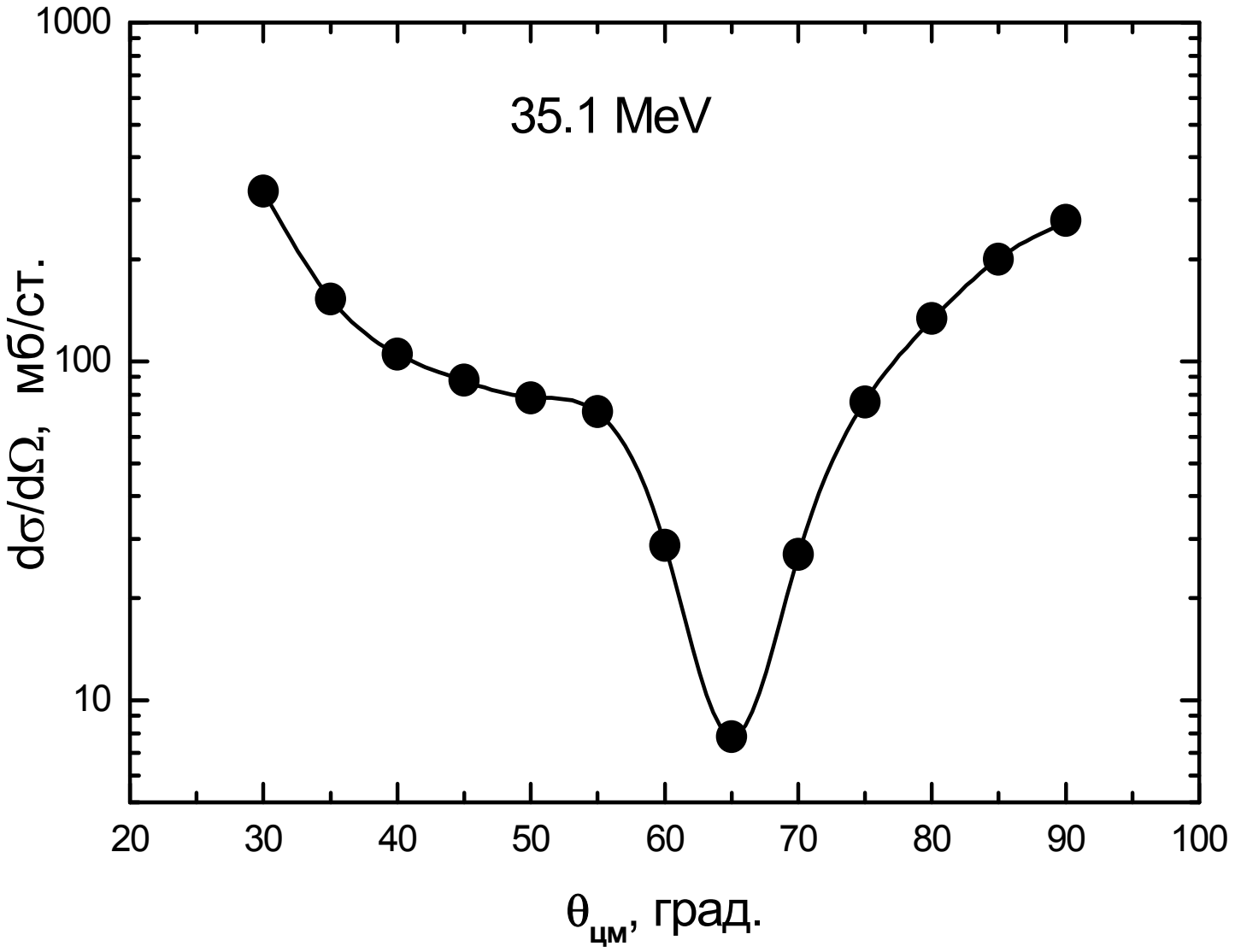

Рис.2.1.2. Дифференциальные сечения упругого рассеяния альфа-частиц на ядрах гелия при энергии 35.1 МэВ [135].
Точки – экспериментальные данные [135], сплошная кривая – расчет сечений с найденными фазами.

При увеличении числа парциальных волн наблюдается заметное изменение фаз рассеяния уже для трех первых *l*. Результаты расчетов дифференциальных сечений для этой энергии представлены на рис.2.1.2.

Далее в [135] была рассмотрена энергия 37.0 МэВ, для которой на рисунках приведены следующие фазы: 137 ± 5, 72 ± 5, 145 ± 5, -2 ± 2. Выполняя варьирование фаз с этими начальными условиями при $L$ = 6, имеем $\chi^2$ = 5.95 и фазы 135.07, 73.23, 147.97, 0.95, которые хорошо согласуются с данными работы [135]. При $L$ = 20 удается получить $\chi^2$ = 0.20, но некоторые парциальные $\chi^2_i$ остаются больше единицы, а именно, равными 1.5 при 85 градусах. И только увеличение $L$ до 24 позволяет получить практически нулевое $\chi^2$ = 0.003 при всех парциальных $\chi^2_i$ меньше 0.01 и фазы

$$\delta_{Re} = 111.5375; 55.9449; 138.4665; 8.1425; 0.0; 10.6324; 3.1546; 7.1735; 2.6095; 1.8621; 3.6626; 0.0; 0.7932;$$

$$\delta_{Im} = 3.5728; 1.5788; 10.0754; 6.7187; 0.0515; 7.7003; 7.8696; 9.2205; 1.9690; 0.0; 1.2802; 0.0562; 0.2180,$$

которые заметно отличаются от результатов работы [135] и наших вычислений при $L$ = 6 без учета мнимой части.

В работе [135] была рассмотрена и энергия 38.4 МэВ, для которой на рисунках приведены следующие фазы рассеяния: 135 ± 5, 75 ± 5, 170 ± 5, 5 ± 2. С такими фазами без варьирования их значений по нашей компьютерной программе получается $\chi^2$ = 19.5. Варьируя значения фаз, находим существенное улучшение описания эксперимента при шести парциальных волнах с $\chi^2$ = 1.39. Для фаз рассеяния были получены следующие значения: 137.65, 89.62, 175.80, 6.35. Здесь только *D* волна заметно отличается от результатов работы [135]. Увеличим теперь $L$ до 10 и учтем мнимую часть фаз. Величина $\chi^2$ уменьшается до 0.91 с фазами $\delta_{Re}$ = 134.09, 86.30, 170.78, 5.39, 1.25, 0.78 и $\delta_{Im}$ = 0.0, 0.0, 1.00, 0.49, 0.39, 0.62, но некоторые парциальные $\chi^2_i$ остаются намного больше единицы.

В спектрах ядра $^8$Be в области 19 МэВ имеется несколько узких уровней, которые должны оказывать определенное влияние на экспериментальные сечения при энергии 38.4 МэВ. Возможно, поэтому увеличение $L$ даже до 28 приводит к $\chi^2 = 0.207$ с почти всеми парциальными $\chi^2_i$ меньше единицы, за исключением одного угла при 66 градусах, при котором экспериментальное сечение равно 2.7 ± 0.5 мб, а расчетное не поднимается выше 2.0 мб, что приводит к величине $\chi^2_i$ около 2 и фазам

$\delta_{Re}$ = 129.8968; 76.4875; 167.6839; 1.7278; 0.0; 2.0189; 0.1684; 0.0051; 0.6075; 0.5257; 0.0056; 0.0211; 0.3073; 0.3494; 0.0575;
$\delta_{Im}$ = 4.2386; 2.3498; 0.5185; 0.3036; 1.6971; 0.0; 1.4852; 1.7773; 0.0747; 0.1783; 0.6247; 2.1599; 0.0; 1.0262; 0.5343.

Дальнейшее увеличение числа парциальных волн не приводит к улучшению описания экспериментальной точки при 66 градусах и практически не меняет величину $\chi^2$. Полученные таким образом фазы несколько отличаются от наших результатов при $L = 6 \div 10$, но вполне согласуются с результатами работы [135] для первых $l$.

В работе [139] в таблице приведены дифференциальные сечения при энергии 38.5 МэВ, но фазовый анализ этих данных не выполнялся. Используем в качестве начальных фазы, полученные в предыдущем случае при $L = 16$, находим $\chi^2 = 0.55$. Для $L = 24$ получаем $\chi^2 = 0.50$, и только при 26 парциальных волнах $\chi^2$ начинает уменьшаться, и оказывается равен 0.265, а для $L = 30$ $\chi^2$ достигает своего предела, равного 0.207 с фазами

$\delta_{Re}$ = 129.9036; 77.1998; 165.6615; 1.5187; 0.5671; 2.1129; 0.0601; 0.0; 0.6059; 0.4907; 0.0005; 0.0896; 0.3055; 0.4272; 0.1328; 0.0;
$\delta_{Im}$ = 3.8528; 2.4699; 1.1513; 0.9460; 1.4118; 0.0; 1.7083; 2.0442; 0.0; 0.2196; 0.8160; 2.4573; 0.0; 0.9449; 0.4563; 0.0.

И здесь, как и в предыдущем случае, все парциальные $\chi^2_i$ меньше единицы, за исключением того же угла при 66 граду-сах, при котором разница экспериментального и расчетного сечений приводит к величине $\chi^2_i$ около 2.

Далее, в первой из работ [143] на рисунках приведены данные для энергий 39, 40 и 41 МэВ – используем одну их них, а именно 40 МэВ, для фазового анализа. Принимая в качестве начальных фаз результаты предыдущего анализа, для $L = 12$ получим $\chi^2 = 0.22$ со следующими фазами:

$$\delta_{\mathrm{Re}} = 69.4967;\ 49.5392;\ 81.4271;\ 1.3593;\ 0.0;\ 0.9287;\ 0.0255;$$
$$\delta_{\mathrm{Im}} = 0.8975;\ 0.0;\ 4.5934;\ 7.1150;\ 1.2930;\ 0.0;\ 0.1762.$$

Дальнейшее увеличение числа парциальных волн не приводит к заметному уменьшению $\chi^2$. Полученные фазы рассеяния заметно отличаются от найденных ранее для энергии 38.4 МэВ и рассмотренной далее энергии 40.77 МэВ. Отметим, что в работе [138] для энергии 40 МэВ из оптических потенциалов были получены следующие фазы: 75.4, 22.3, 84.6, 5.1, 0.35, которые существенно отличаются от наших результатов только в *D* волне.

Из приведенного анализа видно, что в рассмотренной области энергий практически во всех случаях наши результаты при малом числе парциальных волн, в пределах приведенных ошибок, совпадают с данными, полученными ранее. Поскольку мы учитывали большее число парциальных волн, можно считать такие результаты уточнением известных данных по фазам упругого $^4$He$^4$He рассеяния в области 30÷40 МэВ.

Перейдем теперь к рассмотрению данных по дифференциальным сечениям в области энергий 40÷50 МэВ, для которых проводилась только подгонка оптических потенциалов, а фазовый анализ не выполнялся.

В работе [137] была рассмотрена энергия 40.77 МэВ и в таблицах приведены дифференциальные сечения упругого рассеяния. На основе этих данных в [138] была проведена подгонка параметров оптических потенциалов, а фазы рассеяния рассчитывались на их основе, причем в [138] приведена только их действительная часть. В результате были по-

лучены фазы 94.4, 21.8, 86.0, 5.4, 0.38 для $l$ = 0,2,4,6,8.

Используя их в качестве начальных и выполняя варьирование при $L$ = 8 и учете мнимой части фаз, получим $\delta_{Re}$ = 146.46, 44.09, 92.40, 9.48, 0.0 и $\delta_{Im}$ = 28.32, 0.02, 0.0, 1.78, 0.0 при $\chi^2$ = 5.0. Увеличение $L$ до 34 приводит нас к $\chi^2$ = 1.34, а при $L$ = 40 к $\chi^2$ = 1.21 с фазами рассеяния (некоторые парциальные $\chi^2_i$ заметно больше единицы)

$\delta_{Re}$ = 124.9455; 37.3915; 86.8874; 1.8977; 0.0; 1.1045; 1.9122; 0.0897; 4.1335; 2.7145; 4.0382; 0.3077; 0.0; 0.1285; 0.7936; 1.1646; 0.4322; 0.0; 0.1584; 0.0; 0.1347;
$\delta_{Im}$ = 18.7685; 4.3167; 0.0011; 1.5199; 0.0001; 1.4317; 1.9380; 0.9304; 0.0; 0.3225; 0.4839; 2.2038; 1.0627; 0.6151; 0.0822; 0.0; 0.5439; 0.4801; 0.1615; 0.1930; 0.0.

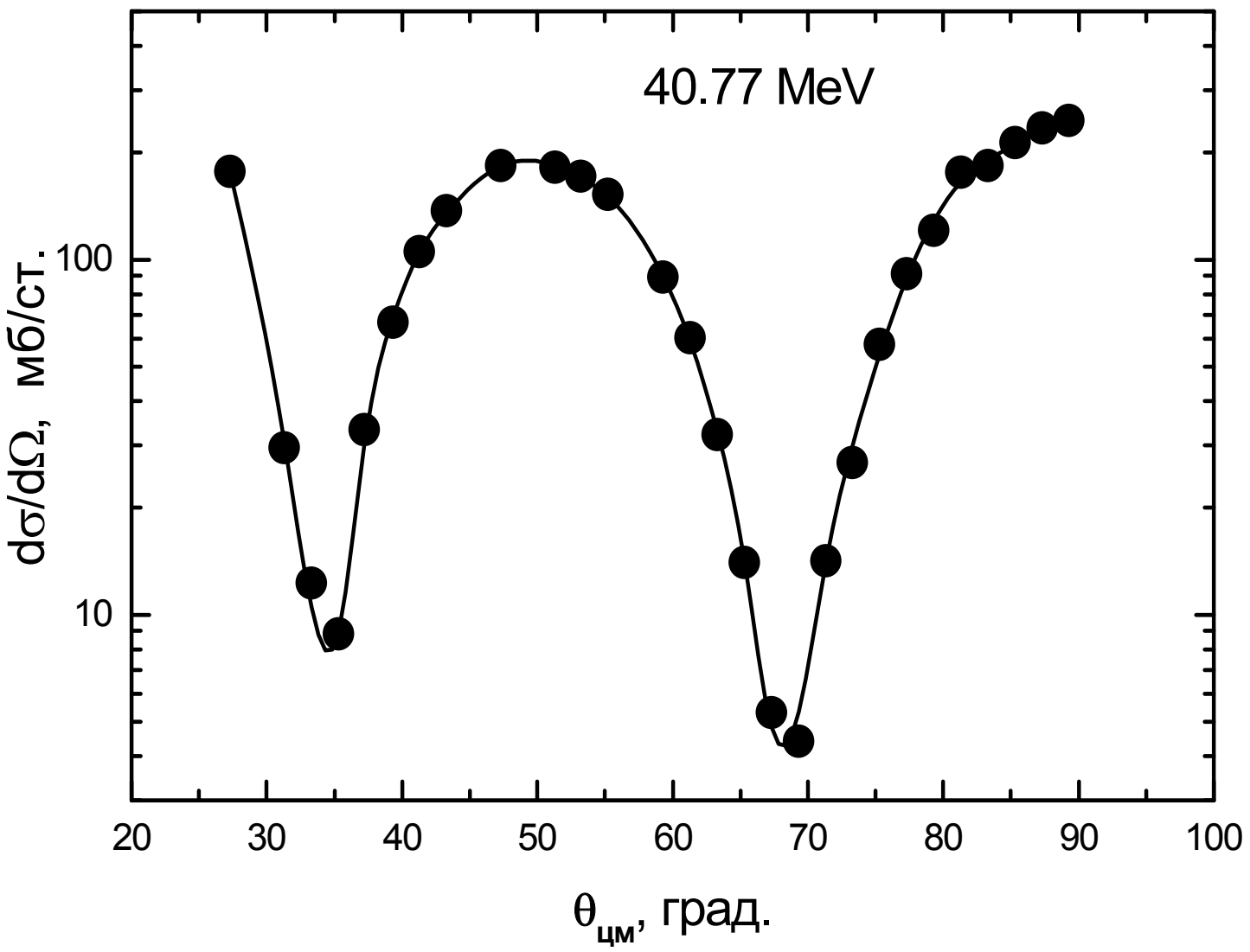

Рис.2.1.3. Дифференциальные сечения упругого рассеяния альфа-частиц на ядрах гелия при энергии 40.77 МэВ [137].
Точки – экспериментальные данные [137], сплошная кривая – расчет сечений с найденными фазами.

Результаты расчетов дифференциальных сечений при этой энергии показаны на рис.2.1.3.

Далее в работе [137] была рассмотрена энергия 41.9 МэВ, измеренные сечения приведены в таблице, а теоретический анализ вообще не проводился. С фазами рассеяния для предыдущей энергии и $L$ = 8, но при 41.9 МэВ, получаем очень большое значение $\chi^2$, равное 1300. Варьирование фаз с восемью парциальными волнами и учетом мнимой части приводит к следующему результату: $\delta_{Re}$ = 108.91, 56.36, 110.12, 12.53, 2.44 и $\delta_{Im}$ = 22.74 при $L$ = 0, а остальные мнимые фазы равны нулю, и величине $\chi^2$ = 10.3. Увеличение $L$ до 14 приводит к несколько лучшему результату $\chi^2$ = 5.66, при $L$ = 30 находим $\chi^2$ = 4.10, и только увеличение $L$ до 40 позволяет получить меньшее $\chi^2$ = 0.857 со следующими фазами рассеяния:

$\delta_{Re}$ = 105.8962; 53.0717; 103.3716; 16.0950; 0.0; 0.0710; 3.9094; 1.5391; 0.0; 0.5518; 1.4615; 0.0091; 0.4501; 0.0; 1.1544; 0.0631; 0.1885; 0.8750; 0.0000; 0.4869; 0.0;
$\delta_{Im}$ = 19.0395; 0.0005; 0.0; 0.2844; 0.0; 0.1207; 0.4153; 0.0; 0.2214; 0.1117; 0.0968; 0.0069; 0.5820; 0.2484; 0.2132; 0.3826; 0.3416; 0.3118; 0.0642; 0.1986; 0.1950.

И при этой энергии 41.9 МэВ, несмотря на сравнительно малое среднее $\chi^2$, некоторые парциальные $\chi^2_i$ оказываются больше единицы.

Следующая энергия, рассмотренная в работе [137] – это 44.41 МэВ. Используем в качестве начальных фаз результаты предыдущего анализа, тогда при $L$ = 8 получим величину $\chi^2$ = 10.85 с фазами $\delta_{Re}$ = 98.46, 73.53, 128.64, 17.86, 3.93 и $\delta_{Im}$ = 12.31, 3.97, 4.85, 0.0, 0.88. При $L$ = 20 получаем некоторое улучшение описания дифференциальных сечений с $\chi^2$ = 4.97, а, увеличивая $L$ до 30, находим $\chi^2$ = 0.97. Для $L$ = 34 можно получить еще некоторое улучшение $\chi^2$ = 0.68, и только увеличение $L$ до 40 приводит к заметному улучшению описания экспериментальных данных при $\chi^2$ = 0.48 и фазам

$\delta_{Re}$ = 117.3916; 72.1036; 115.9565; 16.7765; 3.3375; 0.3926;

0.0756; 0.0270; 0.0; 2.6021; 2.2519; 0.2051; 0.1094; 0.0091; 0.0; 0.2321; 0.3424; 0.0896; 0.0557; 0.0609; 0.0279;

$\delta_{\mathrm{Im}}$ = 30.1833; 9.4258; 3.4545; 0.1919; 1.4489; 2.9905; 1.8006; 1.2168; 0.0054; 0.1728; 0.3725; 0.0587; 0.3169; 0.1322; 0.0572; 0.0; 0.0583; 0.0817; 0.0583; 0.0316; 0.0383.

Но и здесь некоторые парциальные $\chi^2{}_i$ имеют величину несколько больше единицы.

В работе [137] приведены и данные по дифференциальным сечениям для энергии 47.1 МэВ. Измеренные сечения даны в таблицах, а извлечение фаз [138] из оптических потенциалов приводит к следующим действительным фазам рассеяния: 99, 51.8, 145.5, 18.7, 2.8. Наши вычисления с такими фазами дают $\chi^2$ = 156, а учет мнимой части фаз и их варьирование позволяет улучшить согласие с экспериментом почти на 1.5 порядка и получить $\chi^2$ = 2.63. Фазы рассеяния для этой энергии и $L$ = 8 оказались следующими: $\delta_{\mathrm{Re}}$ = 105.34, 55.05, 140.60, 18.85, 2.84 и $\delta_{\mathrm{Im}}$ = 5.48, 0.52, 0.78, 0.85, 0.03, действительная часть которых практически совпадает с результатами [138].

Увеличим теперь $L$ и посмотрим, сколько парциальных волн нужно учитывать для получения $\chi^2$ меньше единицы. При 14 парциальных волнах получаем $\chi^2$ = 1.33, для $L$ = 20 находим $\chi^2$ = 1.02, и только при 30 парциальных волнах имеем $\chi^2$ = 0.70 с фазами

$\delta_{\mathrm{Re}}$ = 106.1704; 51.6346; 134.1084; 17.2203; 2.3269; 0.3964; 0.6347; 0.0; 0.4220; 0.2338; 0.4530; 0.0; 0.0020; 0.0; 0.1307; 0.3588;

$\delta_{\mathrm{Im}}$ = 11.9789; 3.4758; 1.2031; 0.0; 0.0544; 1.0305; 1.1103; 0.5890; 0.2763; 0.1363; 0.4951; 0.2665; 0.2795; 0.0; 0.0; 0.1021.

И при этой энергии некоторые парциальные $\chi^2{}_i$ имеют величину несколько больше единицы.

Рассмотрим теперь энергию 51.1 МэВ. Сечения были измерены в работе [141] и приведены на рисунках, а фазовый анализ вообще не проводился. Поэтому используем в качест-ве начальных фаз результаты работы [136] при 53.4 МэВ, где для реальной части фаз получено $\delta_0$ = 104,8 ± 2.4, $\delta_2$ = 47.9 ± 1.7, $\delta_4$ = 137.9 ± 1.3, $\delta_6$ = 27.5 ± 0.6, $\delta_8$ = 2.0 ± 0.5. Для мнимой части соответственно найдено 12.1±3.1, 22.1±1.7, 16.3±1.1, 3.2±0.5, 0±0.4.

Выполняя варьирование этих фаз при $L$ = 10, получим $\chi^2$ = 1.12, а для самих фаз рассеяния находим $\delta_{Re}$ = 111.30, 54.76, 152.85, 24.97, 3.34, 0.03 и $\delta_{Im}$ = 14.03, 19.98, 23.36, 1.91, 0.27, 0.12. В качестве экспериментальных ошибок использовались ошибки определения сечений из рисунка работы [141], которые принимались равными 10%.

Увеличение $L$ до 20 позволяет получить несколько лучшее $\chi^2$ = 0.97 и фазы рассеяния

$\delta_{Re}$ = 110.8739; 55.0885; 151.8536; 24.9089; 3.2213; 0.0379; 0.0; 0.2313; 0.1331; 0.2534; 0.2644;
$\delta_{Im}$ = 14.9625; 20.3880; 23.6627; 1.8434; 0.3412; 0.1910; 0.0009; 0.0; 0.0; 0.0; 0.1214.

Эти фазы мало отличаются от наших результатов при 10 парциальных волнах, а увеличение числа парциальных волн до 30 приводит к $\chi^2$ = 0.56, что принципиально не меняет качество описания экспериментальных данных.

Отметим, что в работе [141] была выполнена подгонка оптического потенциала, а затем вычислены следующие фазы рассеяния $\delta_l$ и неупругости $\eta_l$

$\delta_{Re}$ = 111±4; 65±4; 163±4; 28±3; 4.2±0.6;
$\eta$ = 0.51±0.07; 0.51±0.07; 0.53±0.07; 0.855±0.03; 0.985±0.004; 0.998±0.001.

Как видно, их действительная часть довольно близка к результатам нашего фазового анализа.

И в заключение этого параграфа проведем фазовый анализ экспериментальных данных при энергии 49.9 МэВ [140,144], принимая в качестве начальных фаз результаты предыдущего анализа. При $L$ = 20 получим довольно боль-шую величину $\chi^2$ = 20.2, и только увеличивая $L$ до 30, находим заметное улучшение согласия расчета с экспериментальными данными с $\chi^2$ = 0.019, всеми парциальными $\chi^2_i$ меньше 0.2 и фазами

$\delta_{Re}$ = 127.5003; 34.3877; 141.3168; 8.0940; 0.0092; 0.3824; 1.8194; 0.7321; 4.4653; 2.4917; 0.0005; 0.9836; 0.1437; 0.0018; 0.5503; 0.3922;
$\delta_{Im}$ = 31.2338; 17.9109; 6.1097; 0.2166; 3.4990; 5.1733; 4.1288; 0.0; 1.8539; 1.1013; 0.0491; 0.9386; 0.3148; 0.0746; 0.2505; 0.1446.

Результаты расчета дифференциальных сечений с этими фазами показаны на рис.2.1.4.

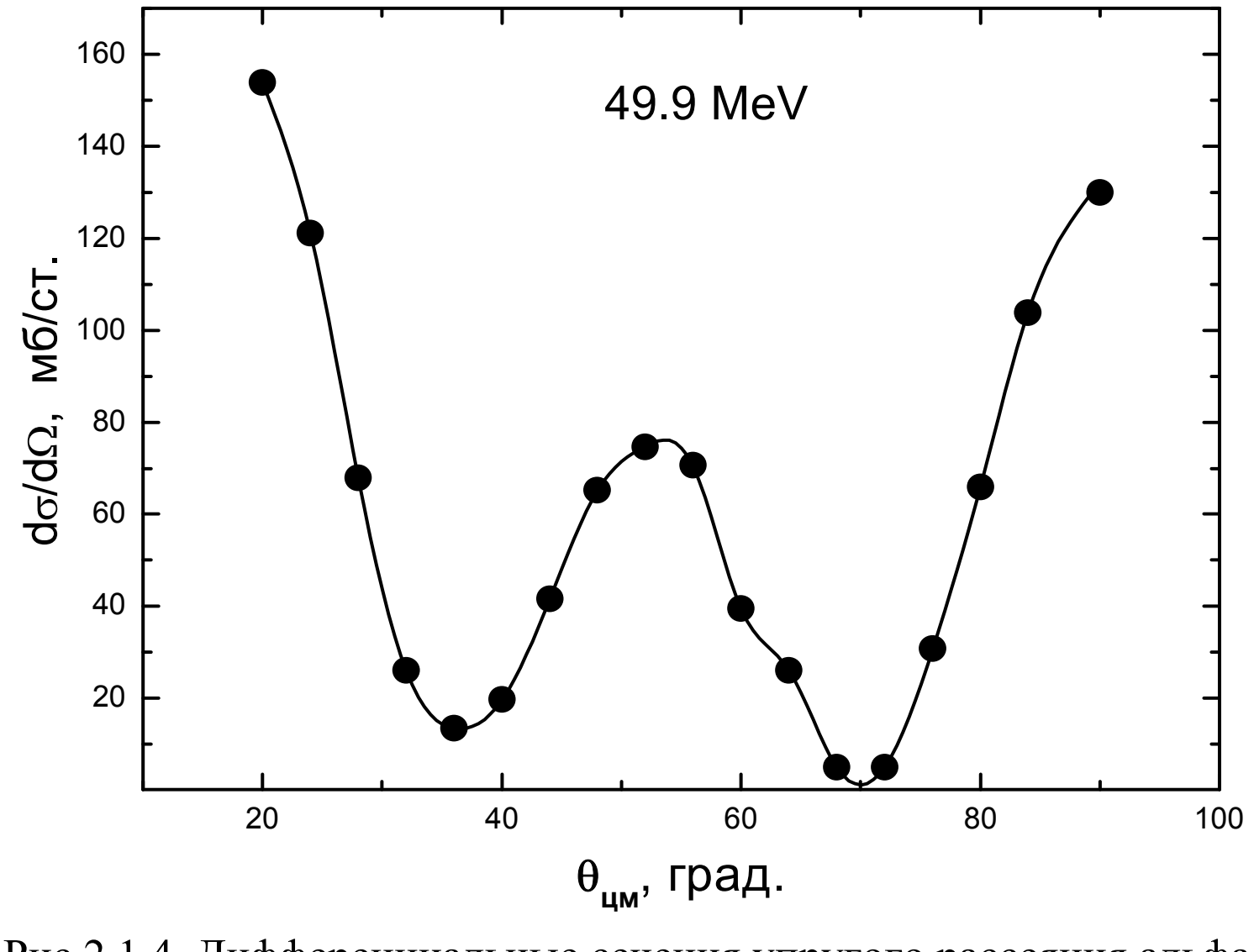

Рис.2.1.4. Дифференциальные сечения упругого рассеяния альфа-частиц на ядрах гелия при энергии 49.9 МэВ.
Точки – экспериментальные данные [140,144], сплошная кривая – расчет сечений с найденными фазами.

В сводных табл.2.1.2 и 2.1.3 приведены фазы рассеяния

(от 360 градусов при нулевой энергии, поскольку в $S$ волне имеется два запрещенных состояния [13], которые на рис.2.1.5÷2.1.8 приведены в более привычной форме – от 180 градусов), полученные в наших расчетах при рассмотренных здесь энергиях. В табл.2.1.2 даны результаты для $L$< 12 без учета мнимой части, а в табл.2.1.3 при $L$ > 12 с учетом мнимой компоненты фаз.

Почти для всех энергий имеется заметное различие этих фазовых анализов хотя бы в одной парциальной фазе – при анализе с небольшим числом парциальных волн, т.е. когда $L$ = 4÷10 и при максимальном учете высших парциальных волн, когда $L$ достигает 20÷40.

На рис.2.1.5÷2.1.8 крестиками приведены фазы из табл.2.1.2, а треугольниками из табл.2.1.3. Приведены также известные на сегодняшний день результаты других фазовых анализов при энергиях в области 20÷60 МэВ.

Табл.2.1.2. Фазы рассеяния (в град.) и величина $\chi^2$ для $L < 12$.
(При $E$ = 49.9 МэВ приведены результаты для $L$ = 20).

| $E_{лаб}$, МэВ | $\delta_0$ | $\delta_2$ | $\delta_4$ | $\delta_6$ | $\delta_8$ | $\chi^2$ |
|---|---|---|---|---|---|---|
| 29.5 | 150.7 | 86.4 | 120.7 | 2.0 | 0.01 | 0.6 |
| 30.3 | 135.9 | 76.5 | 116.3 | 0.0 | 0.0 | 2.1 |
| 31.8 | 142.9 | 78.1 | 125.8 | 0.0 | 0.1 | 1.1 |
| 34.2 | 145.2 | 69.1 | 143.5 | 4.2 | 0.7 | 1.0 |
| 35.1 | 142.3 | 77.0 | 147.5 | 5.4 | 1.7 | 2.2 |
| 37.0 | 135.1 | 73.2 | 148.0 | 1.0 | 0.0 | 6.0 |
| 38,4 | 137.7 | 89.6 | 175.8 | 6.4 | 0.0 | 1.4 |
| 40.77 | 146.5 | 44.1 | 92.4 | 9.5 | 0.0 | 5.0 |
| 41.9 | 108.9 | 56.4 | 110.1 | 12.5 | 2.4 | 10.3 |
| 44.41 | 98.5 | 73.5 | 128.6 | 17.9 | 3.9 | 10.9 |
| 47.1 | 105.3 | 55.1 | 140.6 | 18.9 | 2.8 | 2.6 |
| 49.9 | 128.5 | 26.1 | 138.5 | 6.9 | 0.0 | 20.2 |
| 51.1 | 111.3 | 54.8 | 152.8 | 25.0 | 3.3 | 1.1 |

Табл.2.1.3. Фазы рассеяния (в град.) и величина $\chi^2$ для $L > 12$. (При $E = 49.9$ МэВ приведены результаты для $L = 30$).

| $E_{лаб}$, МэВ | | $\delta_0$ | $\delta_2$ | $\delta_4$ | $\delta_6$ | $\delta_8$ | $\chi^2$ |
|---|---|---|---|---|---|---|---|
| 30.3 | Re $\delta_L$ | 134.5 | 69.7 | 116.0 | 0.0 | 0.04 | 0.11 |
| | Im $\delta_L$ | 1.8 | 3.2 | 0.4 | 3.7 | 0.0 | |
| 31.8 | Re $\delta_L$ | 152.5 | 69.4 | 126.9 | 0.0 | 1.3 | 0.09 |
| | Im $\delta_L$ | 0.0 | 2.1 | 5.4 | 1.5 | 0.2 | |
| 34.2 | Re $\delta_L$ | 140.9 | 54.4 | 133.3 | 0.9 | 0.3 | 0.06 |
| | Im $\delta_L$ | 1.3 | 3.1 | 3.4 | 2.9 | 1.4 | |
| 35.1 | Re $\delta_L$ | 155.1 | 70.2 | 137.8 | 0.0 | 1.1 | 0.015 |
| | Im $\delta_L$ | 3.4 | 0.5 | 4.6 | 0.8 | 1.1 | |
| 37.0 | Re $\delta_L$ | 111.5 | 55.9 | 138.5 | 8.1 | 0.0 | 0.003 |
| | Im $\delta_L$ | 3.6 | 1.6 | 10.1 | 6.7 | 0.1 | |
| 38.4 | Re $\delta_L$ | 129.9 | 76.5 | 167.7 | 1.7 | 0.0 | 0.21 |
| | Im $\delta_L$ | 4.2 | 2.3 | 0.5 | 0.3 | 1.7 | |
| 40.0 | Re $\delta_L$ | 69.5 | 49.5 | 81.4 | 1.4 | 0.0 | 0.22 |
| | Im $\delta_L$ | 0.9 | 0.0 | 4.6 | 7.1 | 1.3 | |
| 40.77 | Re $\delta_L$ | 124.9 | 37.4 | 86.9 | 1.9 | 0.0 | 1.21 |
| | Im $\delta_L$ | 18.8 | 4.3 | 0.0 | 1.5 | 0.0 | |
| 41.9 | Re $\delta_L$ | 105.9 | 53.1 | 103.4 | 16.1 | 0.0 | 0.86 |
| | Im $\delta_L$ | 19.0 | 0.0 | 0.0 | 0.3 | 0.0 | |
| 44.41 | Re $\delta_L$ | 117.4 | 72.1 | 116.0 | 16.8 | 3.3 | 0.48 |
| | Im $\delta_L$ | 30.2 | 9.4 | 3.4 | 0.2 | 1.4 | |
| 47.1 | Re $\delta_L$ | 106.2 | 51.6 | 134.1 | 17.2 | 2.3 | 0.70 |
| | Im $\delta_L$ | 12.0 | 3.5 | 1.2 | 0.0 | 0.0 | |
| 49.9 | Re $\delta_L$ | 127.5 | 34.4 | 141.3 | 8.1 | 0.0 | 0.02 |
| | Im $\delta_L$ | 31.2 | 17.9 | 6.1 | 0.2 | 3.5 | |
| 51.1 | Re $\delta_L$ | 110.9 | 55.1 | 151.8 | 24.9 | 3.2 | 0.97 |
| | Im $\delta_L$ | 15.0 | 20.4 | 23.7 | 1.8 | 0.3 | |

Рассмотрим теперь, как полученные фазы рассеяния согласуются с энергетическими уровнями, присутствующими в ядре $^8$Be [145]. В спектрах этого ядра при энергиях 19.86, 20.1 и 20.2 МэВ находятся уровни с ширинами 0.7, примерно 1.1 и около 1.0 МэВ и моментами $4^+$, $2^+$ и $0^+$ при нулевом изоспине, которые могут присутствовать, как резонансы в $^4$He$^4$He системе.

В лабораторной системе они попадают в область энергий 39.7÷40.4 МэВ и влияют на поведение фаз рассеяния при 40.0 и 40.77 МэВ, приводя к их скачку в парциальных волнах с такими же орбитальными моментами 0, 2 и 4, как это показано на рис.2.1.5÷2.1.7.

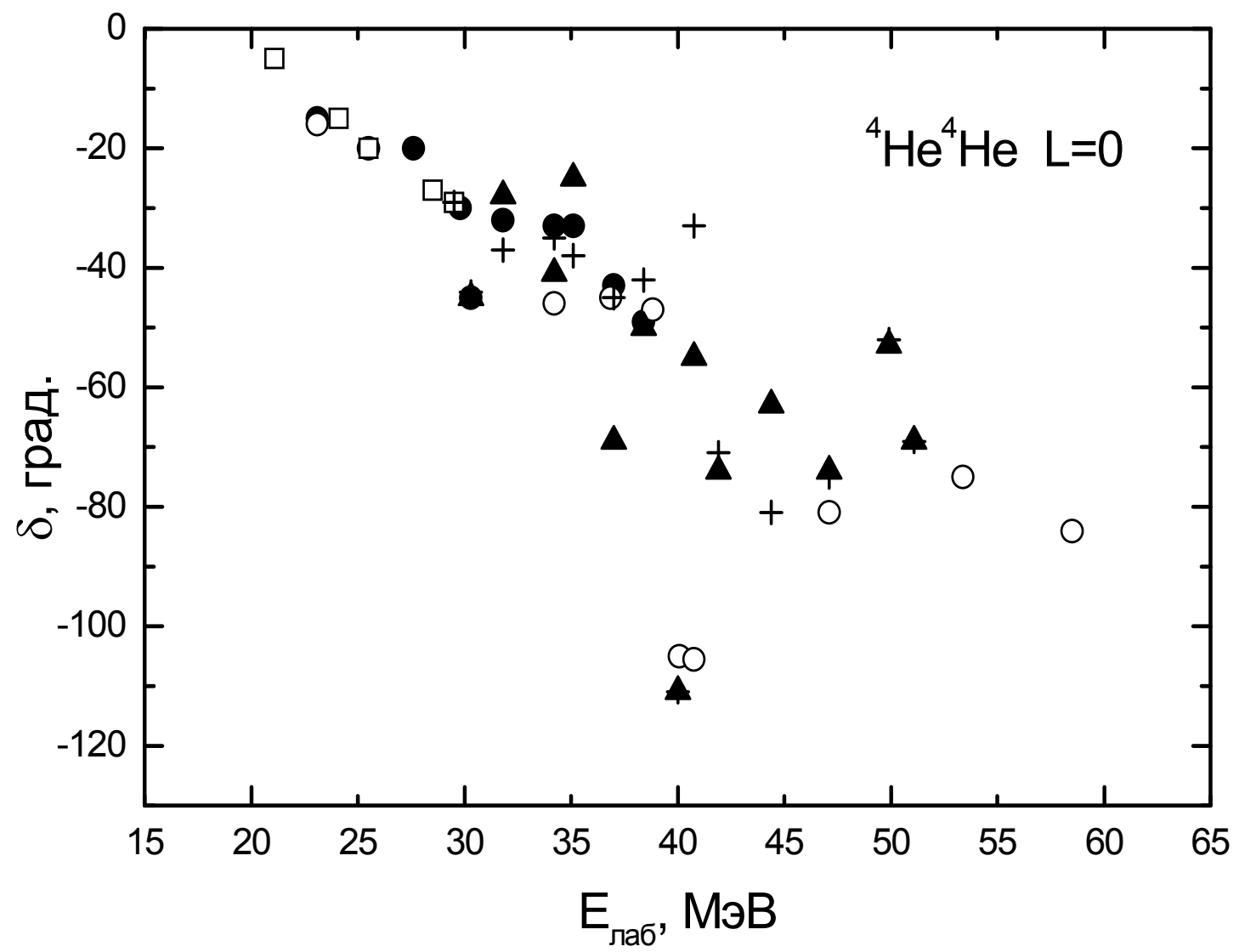

Рис.2.1.5. Фазы упругого $^4$He$^4$He рассеяния при $L$ = 0.
Точки: ○ – данные работ [138,141] при 51.1 МэВ и [136] при энергии больше 53 МэВ; • – результаты работы [135] при 23÷38 МэВ, которые определялись из рисунков; □ – данные [134] в области энергий 20÷30 МэВ; + и ▲– наши результаты.

Этот скачок фаз очень хорошо виден в $S$ и $G$ волнах, а в $G$ волне можно видеть даже подъем фазы, который наблюдается до 47 МэВ. В наших расчетах нет столь явного провала фазы в $D$ волне рассеяния при 40 МэВ, какой имеется в результатах работы [138], но, тем не менее, он также присутствует и виден на рис.2.1.6.

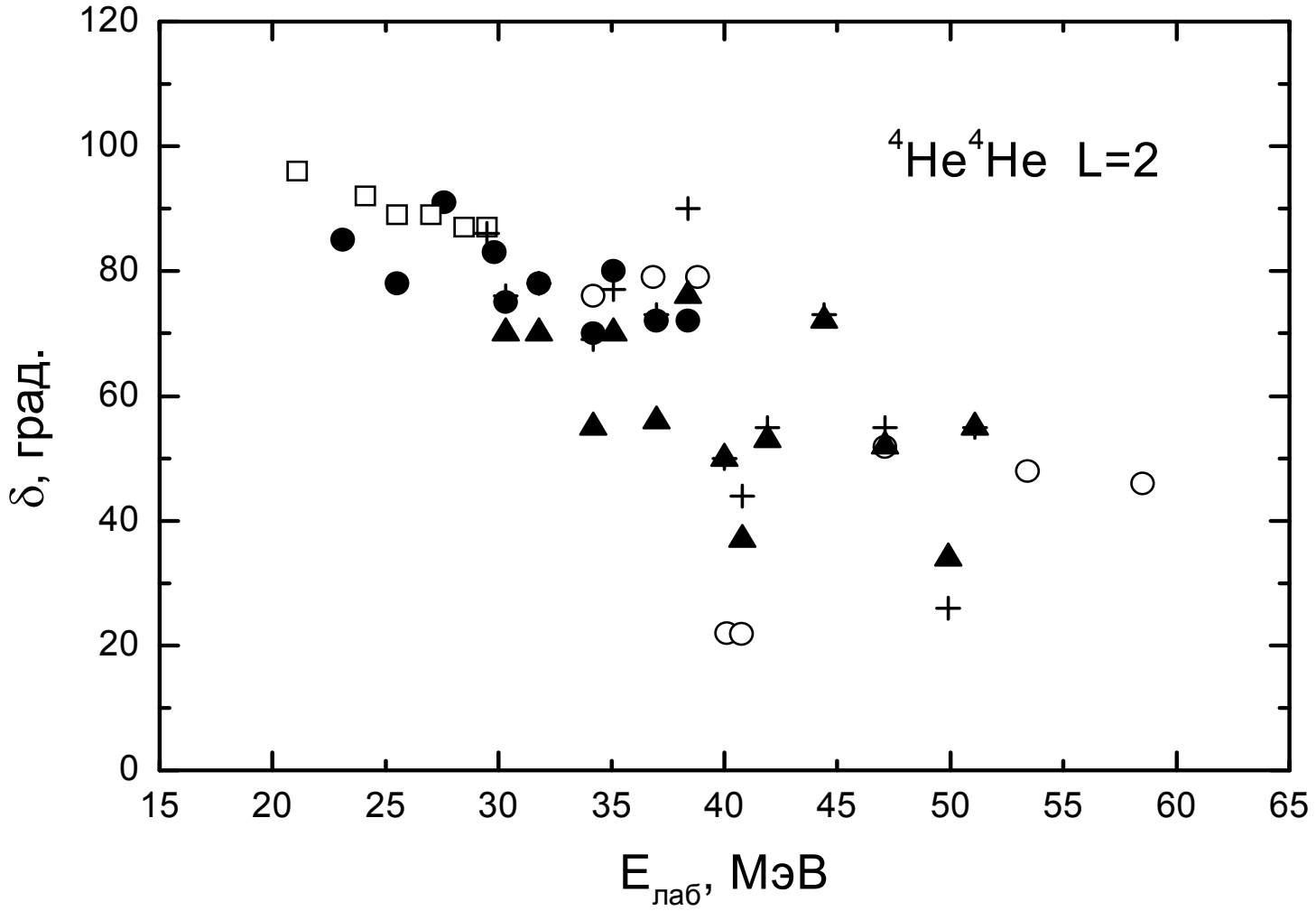

Рис.2.1.6. Фазы упругого $^4He^4He$ рассеяния при $L = 2$.
Точки: ○ – данные работ [138,141] при 51.1 МэВ и [136] при энергии больше 53 МэВ; • – результаты работы [135] при 23÷38 МэВ, которые определялись из рисунков; □ – данные [134] в области энергий 20÷30 МэВ; + и ▲– наши результаты.

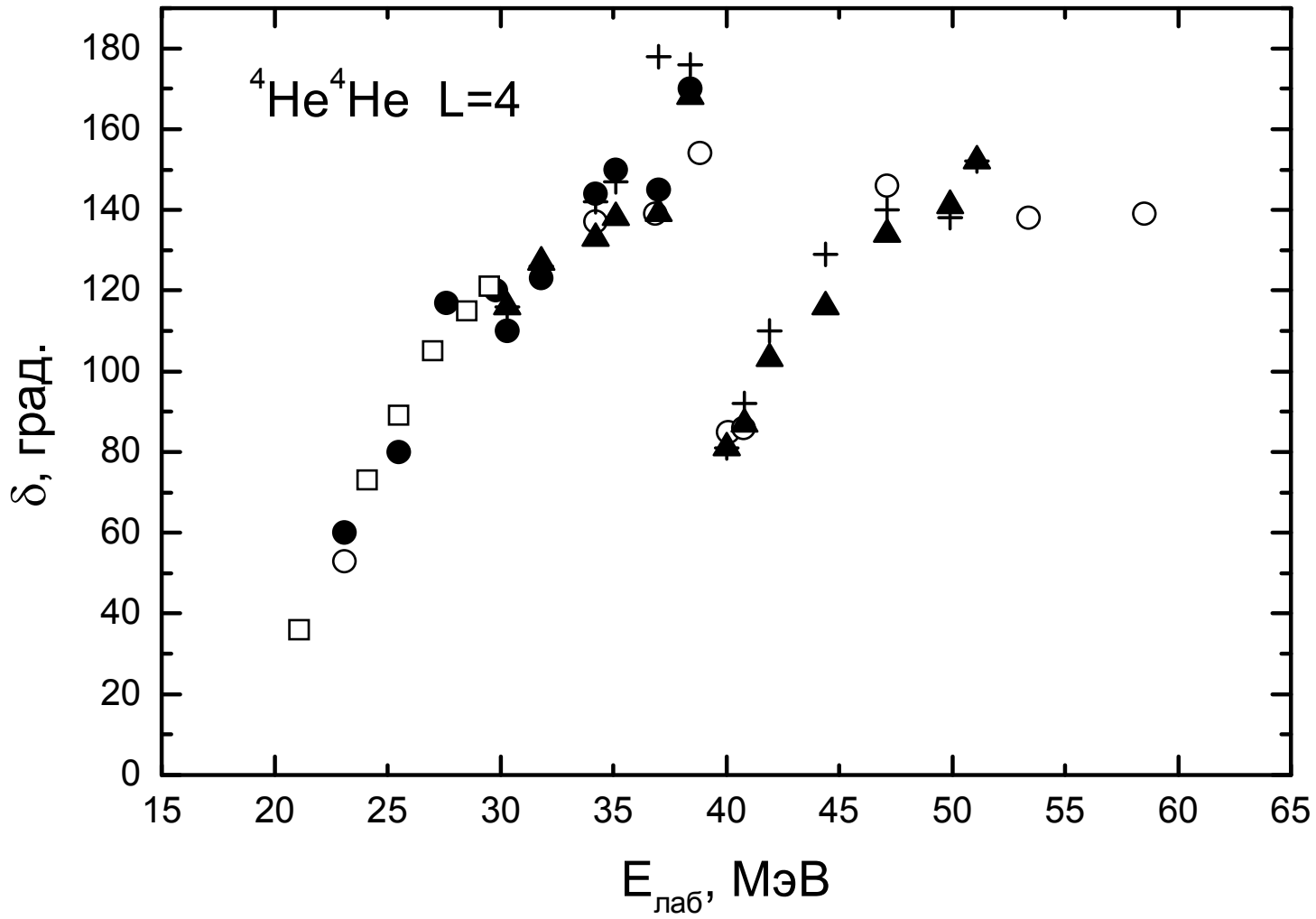

Рис.2.1.7. Фазы упругого $^4He^4He$ рассеяния при $L = 4$.
Точки: ○ – данные работ [138,141] при 51.1 МэВ и [136] при энергии больше 53 МэВ; • – результаты работы [135] при 23÷38 МэВ, которые определялись из рисунков; □ – данные [134] в области энергий 20÷30 МэВ; + и ▲– наши результаты.

При энергии 22.2 МэВ [145] в спектре уровней $^8$Be имеется состояние с моментом $2^+$ и шириной 0.8 МэВ при нулевом изоспине. Действительно, *D* фаза при энергии 44.41 МэВ испытывает заметный скачок вверх. К сожалению, в этой области энергий имеется только одно измерение дифференциальных сечений и нельзя точно воспроизвести поведение *D* фазы рассеяния.

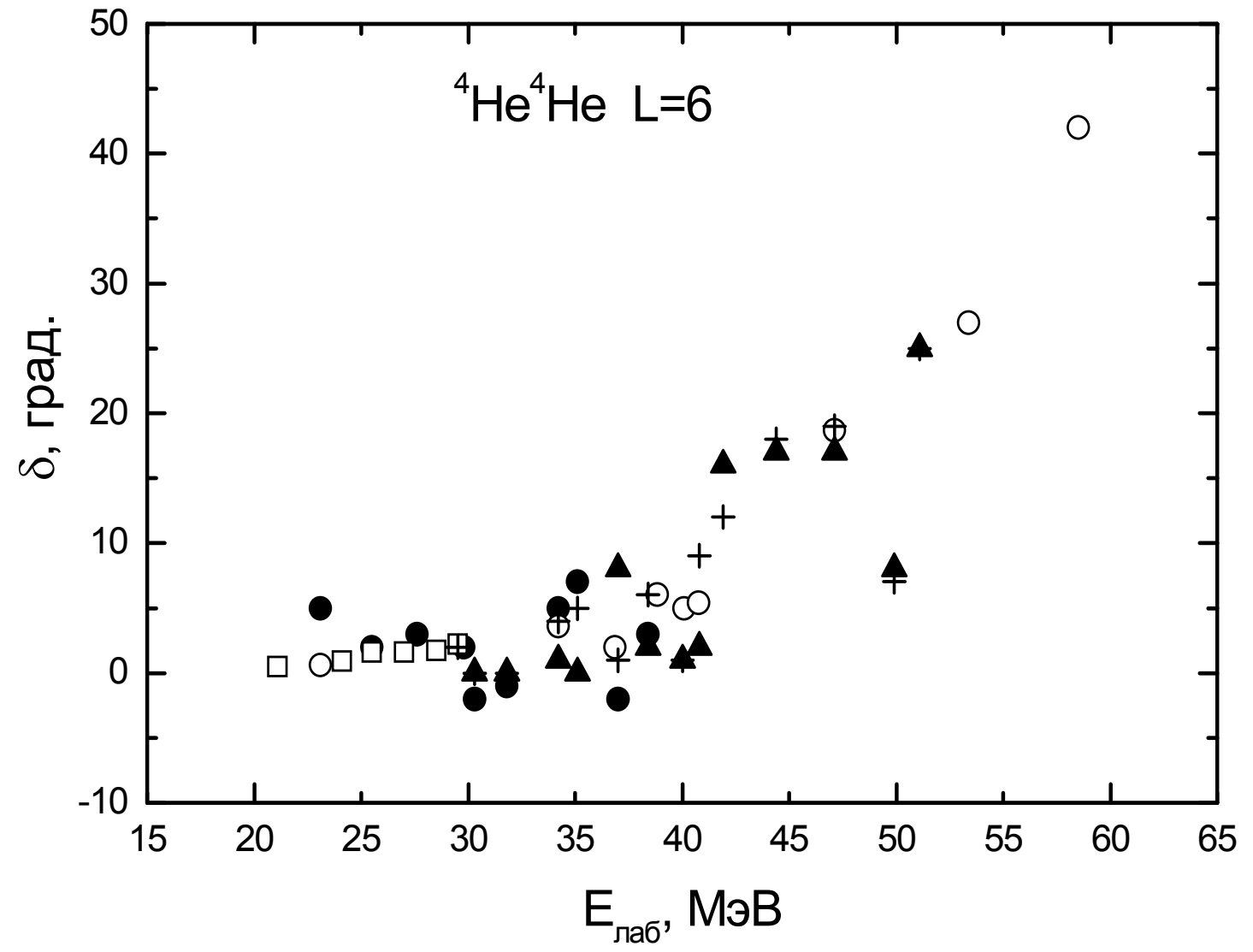

Рис.2.1.8. Фазы упругого $^4$He$^4$He рассеяния при $L = 6$.
Точки: ○ – данные работ [138,141] при 51.1 МэВ и [136] при энергии больше 53 МэВ; • – результаты работы [135] при 23÷38 МэВ, которые определялись из рисунков; □ – данные [134] в области энергий 20÷30 МэВ. + и ▲– наши результаты.

В спектрах ядра $^8$Be при 25.2 и 25.5 МэВ присутствуют уровни с моментом $2^+$ и $4^+$ и изоспином, равным нулю. Рассматриваемая нами энергия 49.9 МэВ попадает в область первого из них и в *D* волне рассеяния наблюдается резкий спад фазы. Такой же скачок имеется и в *S* волне, однако, его нельзя сопоставить какому-либо уровню ядра $^8$Be. То же самое относится к парциальной волне с $L = 6$, где фаза рассея-ния также претерпевает резкий скачок. Энергия 51.1 МэВ попадает на резонанс при 25.5 МэВ в *G* волне, но ее влияние явно не просматривается – возможно, это очень узкий уровень ядра $^8$Be.

Таким образом, при разных энергиях в области 30÷40 МэВ выполнено уточнение известных значений фаз упругого $^4$He$^4$He рассеяния. Фазовый анализ при энергиях 40.77, 41.9, 44.41, 47.1, 49.9 и 51.1 МэВ приводит нас к вполне разумным результатам и, в целом, согласуется с данными других работ и спектрами ядра $^8$Be [92].

Для более детального изучения поведения фаз рассеяния в различных парциальных волнах желательно иметь более подробные экспериментальные измерения дифференциальных сечений в области энергий 39÷41 МэВ, 43÷45 МэВ и 49÷52 МэВ с шагом порядка 0.1÷0.3 МэВ [92].

### 2.1.5 Программа для $^4$He$^4$He и $^4$He$^{12}$C фазового анализа
### The program for $^4$He$^4$He and $^4$He$^{12}$C phase shifts analysis

Приведем распечатку компьютерной программы на языке Fortran, предназначенной для $^4$He$^4$He и $^4$He$^{12}$C фазового анализа. Какой именно анализ будет выполняться, зависит от значения параметра NYS и LH. В случае $^4$He$^4$He они должны быть равны 1 и 2 соответственно, для выполнения $^4$He$^{12}$C фазового анализа их значения равны 0 и 1. Кроме того, для больших энергий, порядка 30÷40 МэВ, параметр NP должен быть равен

NP = 2*LMA+LH ,

а не LMA, что позволяет учитывать комплексную часть фаз рассеяния.

Обозначения некоторых начальных параметров программы приведены ниже

```
Z1=2.0D-000; Z2=2.0D-000 ! Заряды
AM1=4.0D-000; AM2=4.0D-000 ! Массы
LMI=0 ! Начальный орбитальный момент
LH=1 ! Шаг по моменту
LMA=2 ! Максимальный момент
EP=1.0D-010 ! Точность поиска минимума
NI=10 ! Число итераций
NT=22 ! Число экспериментальных точек
EL=12.30D-000 ! Лабораторная энергия
```

Приведем текст, т.е. распечатку самой компьютерной программы.

```
PROGRAM FAZ_AL_AL
! ПРОГРАММА ФАЗОВОГО АНАЛИЗА ДЛЯ AL-AL AND
!AL-12C
IMPLICIT REAL(8) (A - Z)
INTEGER
L,I,NT,LMI,LMA,LH,NYS,NP,NTT,NV,NI,LMI1,LH1,NPP
DIMENSION ST(0:50),FR(0:50),FM(0:50),ET(0:50),
XP(0:50),ETA(0:50)
COMMON /A/ PI,NT,TT(0:50),GG,SS,LMI,LMA,LH,NYS,NP
COMMON /B/ SE(0:50),DS(0:50),DE(0:50),NTT
COMMON /C/ LH1,LMI1,P1,NPP
! ************* НАЧАЛЬНЫЕ ЗНАЧЕНИЯ **************
PI=4.0D-000*DATAN(1.0D-000)
P1=PI
Z1=2.0D-000; Z2=2.0D-000
AM1=4.0D-000; AM2=4.0D-000
AM=AM1+AM2
A1=41.46860D-000
PM=AM1*AM2/(AM1+AM2)
B1=2.0D-000*PM/A1
LMI=0; LMI1=LMI; LH=2; LH1=LH
LMA=4; LMA1=LMA
NYS=1; ! IF =1 THEN 4HE4HE, IF = 0 THEN 4HE12C
EP=1.0D-010; NV=1
FH=0.010D-000; NI=10
```

```
!NP=2*LMA+LH
NP=LMA;  NPP=NP
! **************** CROSS SECTIONS ****************
SE(1)=1357.0D-000; SE(2)=1203.0D-000; SE(3)=1074.0D-000;
SE(4)=870.0D-000; SE(5)=759.0D-000
SE(6)=688.0D-000; SE(7)=467.0D-000; SE(8)=271.0D-000;
SE(9)=196.0D-000; SE(10)=130.0D-000
SE(11)=93.90D-000;SE(12)=57.0D-000; SE(13)=32.50D-000;
SE(14)=12.30D-000; SE(15)=2.280D-000
SE(16)=24.7; SE(17)=86.5; SE(18)=157; SE(19)=270
SE(20)=337.0D-000; SE(21)=408.0D-000; SE(22)=418.0D-000
DE(1)=39.0D-000; DE(2)=40.0D-000; DE(3)=24.0D-000;
DE(4)=20.0D-000; DE(5)=16.0D-000;
DE(6)=17.0D-000
DE(7)=12.0D-000; DE(8)=7.0D-000; DE(9)=4.10D-000;
DE(10)=3.60D-000; DE(11)=2.20D-000;
DE(12)=1.50D-000
DE(13)=1.1; DE(14)=1.0; DE(15)=0.4; DE(16)=0.7;
DE(17)=2.0; DE(18)=3.6
DE(19)=6.50D-000; DE(20)=7.40D-000; DE(21)=8.20D-000;
DE(22)=8.30D-000
TT(1)=22.0D-000;  TT(2)=24.0D-000;  TT(3)=26.0D-000;
TT(4)=28.0D-000; TT(5)=30.0D-000
TT(6)=32.0D-000;  TT(7)=35.0D-000;  TT(8)=40.0D-000;
TT(9)=42.0D-000
TT(10)=45.0D-000; TT(11)=46.0D-000; TT(12)=48.0D-000;
TT(13)=50.0D-000
TT(14)=52.0D-000; TT(15)=55.0D-000; TT(16)=60.0D-000;
TT(17)=65.0D-000
TT(18)=70.0D-000; TT(19)=75.0D-000; TT(20)=80.0D-000;
TT(21)=85.0D-000; TT(22)=90.0D-000
! ************** FOR AL-AL ON E=12.3 **************
NT=22;  NTT=NT; EL=12.30D-000
FR(0)=29.0D-000;  FR(2)=103.0D-000;  FR(4)=3.0D-000
FM(0)= 0.0D-000;  FM(2)=  0.0D-000;  FM(4)=0.0D-000
OPEN (4,FILE='FAZ.DAT')
DO L=LMI,LMA,LH
READ(4,*) L,FR(L),FM(L)
```

```
ENDDO
CLOSE(4)
 ! ************* ENERGY IN LAB. SYSTEM *************
 DO L=LMI,LMA,LH
 FM(L)=FM(L)*PI/180.0D-000
 FR(L)=FR(L)*PI/180.0D-000
 ET(L)=DEXP(-2.0D-000*FM(L))
 ENDDO
 FH=FH*PI/180.0D-000
 DO I=LMI,LMA,LH
 XP(I)=FR(I)
 XP(I+LMA+LH)=FM(I)
 ENDDO
 ! *********** TRANSFORM TO C.M. *******************
 EC=EL*PM/AM1
 SK=EC*B1
 SS=DSQRT(SK)
 GG=3.44476D-002*Z1*Z2*PM/SS
 ! ******* DIFFERENTIAL CROSSS SECTION ************
 CALL VAR(ST,FH,NI,XP,EP,XI,NV)
 PRINT*, "XI-KV=; NI=; EL=",XI,NI,EL
! ********** TOTAL CROSSS SECTION *****************
 SIGMAR=0.0D-000; SIGMAS=0.0D-000
 DO L=LMI,LMA,LH
 FR(L)=XP(L)
 FM(L)=XP(L+LMA+LH)
 A=FR(L)
 ETA(L)=1
!ETA(L)=DEXP(-2.0D-000*FM(L))
 SIGMAR=SIGMAR+(2*L+1)*(1-(ETA(L))**2)
 SIGMAS=SIGMAS+(2*L+1)*(ETA(L))**2*(DSIN(A))**2
 ENDDO
 SIGMAR=10.0D-000*4.0D-000*PI*SIGMAR/SK
 SIGMAS=10.0D-000*4.0D-000*PI*SIGMAS/SK
 PRINT*, "SIGMR-TOT=",SIGMAR
 PRINT*, "SIGMS-TOT=",SIGMAS
 PRINT*, "    T          SE          ST          XI"
! ****************** RESULTS ************************
```

```
 DO I=1,NT
 WRITE(*,2) TT(I),SE(I),ST(I),DS(I)
 ENDDO
 PRINT*
 PRINT*, "    L     FR(L)      FM(L)"
 DO L=LMI,LMA,LH
 FM(L)=FM(L)*180.0D-000/PI
 FR(L)=FR(L)*180.0D-000/PI
 WRITE(*,1) L,FR(L),FM(L)
 ENDDO
OPEN (4,FILE='SEC-AL-AL.DAT')
WRITE(4,*)  "             AL-AL LAB E=; XI=",EL,XI
WRITE(4,*) "       T             SE              ST             XI"
DO I=1,NT
WRITE(4,2) TT(I),SE(I),ST(I),DS(I)
ENDDO
WRITE(4,*)
WRITE(4,*) "    L     FR(L)      FM(L)"
DO L=LMI,LMA,LH
WRITE(4,1) L,FR(L),FM(L)
ENDDO
CLOSE(4)
OPEN (4,FILE='FAZ.DAT')
DO L=LMI,LMA,LH
WRITE(4,1) L,FR(L),FM(L)
ENDDO
CLOSE(4)
1 FORMAT(1X,I5,E15.6,2X,E15.6)
2 FORMAT(1X,4(E10.3,2X))
3 FORMAT(1X,E15.5,2X,I5)
 END
 SUBROUTINE VAR(ST,PHN,NI,XP,EP,AMIN,NV)
 IMPLICIT REAL(8) (A - Z)
 INTEGER I,NP,LMI,LH,NT,NV,NI,IIN,NN,IN
 DIMENSION XPN(0:50),XP(0:50),ST(0:50)
 COMMON /C/ LH,LMI,PI,NP
 COMMON /B/ SE(0:50),DS(0:50),DE(0:50),NT
 !SHARED LH,LMI,NT,PI,DS(),NP
```

```
! ************* ПОИСК МИНИМУМА ***************
 DO I=LMI,NP,LH
 XPN(I)=XP(I)
 ENDDO
 NN=LMI
 PH=PHN
 CALL DET(XPN,ST,ALA)
 B=ALA
 IF (NV==0) GOTO 3012
 DO IIN=1,NI
 NN=-LH
 GOTO 1119
1159 XPN(NN)=XPN(NN)-PH*XP(NN)
1119 NN=NN+LH
 IF (NN>NP) GOTO 3012
 IN=0
2229 A=B
 XPN(NN)=XPN(NN)+PH*XP(NN)
 IF (XPN(NN)<0.0D-000) GOTO 1159
 IN=IN+1
 CALL DET(XPN,ST,ALA)
 B=ALA
IF (B<A) GOTO 2229
 C=A
 XPN(NN)=XPN(NN)-PH*XP(NN)
 IF (IN>1) GOTO 3339
 PH=-PH
 GOTO 5559
3339 IF (ABS((C-B)/(B))<EP) GOTO 4449
  PH=PH/2.0D-000
5559 B=C
 GOTO 2229
4449 PH=PHN
 B=C
 IF (NN<NP) GOTO 1119
 AMIN=B
 PH=PH/NI
 ENDDO
```

```
3012  AMIN=B
 DO I=LMI,NP,LH
 XP(I)=XPN(I)
 ENDDO
 END
 SUBROUTINE DET(XP,ST,XI)
 IMPLICIT REAL(8) (A - Z)
 INTEGER I,N
 DIMENSION XP(0:50),ST(0:50)
 COMMON /B/ SE(0:50),DS(0:50),DE(0:50),N
! **************** ДЕТЕРМИНАНТ*******************
 S=0.0D-000
 CALL SEC(XP,ST)
 DO I=1,N
 S=S+((ST(I)-SE(I))/DE(I))**2
 DS(I)=((ST(I)-SE(I))/DE(I))**2
 ENDDO
 XI=S/N
 END
 SUBROUTINE SEC(XP,S)
 IMPLICIT REAL(8) (A - Z)
 INTEGER I,NP,LH,LMI,LMA,NT,NYS,L
 DIMENSION                S0(0:50),P(0:50),FR(0:50),ET(0:50),
S(0:50),XP(0:50)
 COMMON /A/ PI,NT,TT(0:50),GG,SS,LMI,LMA,LH,NYS,NP
! ************** РАСЧЕТ СЕЧЕНИЙ**** ***************
 DO I=LMI,LMA,LH
 FR(I)=XP(I)
 ET(I)=1.0D-000
! IF NP=LMA GOTO 1234
! ET(I)=EXP(-2*XP(I+LMA+LH))
 ENDDO
 RECUL1=0.0D-000; AIMCUL1=0.0D-000
 CALL CULFAZ(GG,S0)
 DO I=1,NT
 T=TT(I)*PI/180.0D-000
 X=DCOS(T)
 A=2.0D-000/(1-X)
```

```
 S00=2.0D-000*S0(0)
 BB=-GG*A
 ALO=GG*DLOG(A)+S00
 RECUL=BB*DCOS(ALO)
 AIMCUL=BB*DSIN(ALO)
 IF (NYS==0) GOTO 555
 X1=DCOS(T)
 A1=2.0D-000/(1.0D-000+X1)
 BB1=-GG*A1
 ALO1=GG*DLOG(A1)+S00
 RECUL1=BB1*COS(ALO1)
 AIMCUL1=BB1*SIN(ALO1)
555 RENUC=0.0D-000; AIMNUC=0.0D-000
 DO L=LMI,LMA,LH
 AL=ET(L)*DCOS(2.0D-000*FR(L))-1.0D-000
 BE=ET(L)*DSIN(2.0D-000*FR(L))
 LL=2.0D-000*L+1.0D-000
 SL=2.0D-000*S0(L)
 CALL POLLEG(X,L,P)
 RENUC=RENUC+LL*(BE*DCOS(SL)+AL*DSIN(SL))*P(L)
 AIMNUC=AIMNUC+LL*(BE*DSIN(SL)-
AL*DCOS(SL))*P(L)
 ENDDO
 IF (NYS==0) GOTO 556
 AIMNUC=2.0D-000*AIMNUC
 RENUC=2.0D-000*RENUC
556 RE=RECUL+RECUL1+RENUC
 AIM=AIMCUL+AIMCUL1+AIMNUC
 S(I)=10.0D-000*(RE**2+AIM**2)/4.0D-000/SS**2
 ENDDO
 END
 SUBROUTINE POLLEG(X,L,P)
 IMPLICIT REAL(8) (A - Z)
 INTEGER I,L
 DIMENSION P(0:50)
 ! ************* ПОЛИНОМЫ ЛЕЖАНДРА************
 P(0)=1.0D-000
 P(1)=X
```

```
DO I=2,L
A=I*1.0D-000
P(I)=(2.0D-000*A-1)*X/A*P(I-1)-(A-1.0D-000)/A*P(I-2)
ENDDO
END
SUBROUTINE CULFAZ(G,F)
! ************* КУЛОНОВСКИЕ ФАЗЫ *************
IMPLICIT REAL(8) (A - Z)
INTEGER I
DIMENSION F(0:50)
C=0.577215665D-000
S=0.0D-000; N=50
A1=1.202056903D-000/3.0D-000
A2=1.0369277550D-000/5.0D-000
DO I=1,N
AA=I*1.0D-000
A=G/AA-DATAN(G/AA)-(G/AA)**3/3.0D-
000+(G/AA)**5/5.0D-000
S=S+A
ENDDO
FAZ=-C*G+A1*G**3-A2*G**5+S
F(0)=FAZ
DO I=1,20
A=I*1.0D-000
F(I)=F(I-1)+DATAN(G/A)
ENDDO
END
```

Далее приведен контрольный счет поиска фаз упругого $^4He^4He$ рассеяния для энергии 12.3 МэВ, при которой в работе [132] были получены фазы $\delta_0 = 29° \pm 4°$, $\delta_2 = 103° \pm 8°$, $\delta_4 = 3° \pm 1.5°$. Эти фазы принимаются в качестве начальных, а при дальнейшем их варьировании и по предыдущей программе на языке Turbo Basic [17] и на основе приведенной выше, находим $\delta_0 = 28.37°$, $\delta_2 = 105.03°$, $\delta_4 = 2.62°$ при $\chi^2 = 3.43$, что видно из приведенной далее распечатки работы программы на языке Fortran-90:

| $\chi^2$ | NI |
|---|---|
| 3.943880354539536 | 1 |
| 3.437578175376997 | 2 |
| 3.432394633290976 | 3 |
| 3.432107724455588 | 4 |
| 3.432044190539325 | 5 |
| 3.432020692528544 | 6 |
| 3.432007556422437 | 7 |
| 3.432005733725609 | 8 |
| 3.432004938728721 | 9 |
| 3.432004743298262 | 10 |

$\chi^2$ = ; NI = ; EL = 3.43200 10 12.300

| θ | $\sigma_e$ | $\sigma_t$ | $\chi^2_i$ |
|---|---|---|---|
| .22000E+02 | .13570E+04 | .12796E+04 | .39401E+01 |
| .24000E+02 | .12030E+04 | .11346E+04 | .29238E+01 |
| .26000E+02 | .10740E+04 | .10008E+04 | .93096E+01 |
| .28000E+02 | .87000E+03 | .87464E+03 | .53776E-01 |
| .30000E+02 | .75900E+03 | .75523E+03 | .55540E-01 |
| .32000E+02 | .68800E+03 | .64266E+03 | .71117E+01 |
| .35000E+02 | .46700E+03 | .48793E+03 | .30429E+01 |
| .40000E+02 | .27100E+03 | .27340E+03 | .11779E+00 |
| .42000E+02 | .19600E+03 | .20443E+03 | .42244E+01 |
| .45000E+02 | .13000E+03 | .12007E+03 | .76028E+01 |
| .46000E+02 | .93900E+02 | .97112E+02 | .21321E+01 |
| .48000E+02 | .57000E+02 | .58891E+02 | .15896E+01 |
| .50000E+02 | .32500E+02 | .30767E+02 | .24826E+01 |
| .52000E+02 | .12300E+02 | .12398E+02 | .95282E-02 |
| .55000E+02 | .22800E+01 | .20301E+01 | .39027E+00 |
| .60000E+02 | .24700E+02 | .24774E+02 | .11250E-01 |
| .65000E+02 | .86500E+02 | .85514E+02 | .24313E+00 |
| .70000E+02 | .15700E+03 | .16800E+03 | .93323E+01 |
| .75000E+02 | .27000E+03 | .25509E+03 | .52640E+01 |
| .80000E+02 | .33700E+03 | .33073E+03 | .71799E+00 |
| .85000E+02 | .40800E+03 | .38184E+03 | .10177E+02 |
| .90000E+02 | .41800E+03 | .39987E+03 | .47720E+01 |

| L | $\delta_r$ | $\delta_m$ |
|---|---|---|
| 0 | .283717E+02 | .000000E+00 |
| 2 | .105030E+03 | .000000E+00 |
| 4 | .261561E+01 | .000000E+00 |

Здесь в распечатке результатов приняты следующие обозначения: $\theta$ – угол рассеяния, $\sigma_e$ – экспериментальные сечения, $\sigma_t$ – вычисленные сечения, $\chi^2_i$ – парциальные $\chi^2$ для *i*-го угла, $\delta_r$ – реальная часть фазы, $\delta_m$ – мнимая часть фазы, $\chi^2$ – среднее значение по всем точкам, EL – энергия в лабораторной системе. В первой строчке при распечатке фаз указан орбитальный момент $L = 0$, во второй $L = 1$ и в третьей $L = 2$ и т.д. Для поиска минимума, как и в работе [17], используется 10 итераций NI. Только в данном случае точность EP задавалась равной $10^{-10}$, а в предыдущем варианте счета $10^{-5}$ [17]. Первые столбцы $\chi^2$ и NI приведенной распечатки показывают сходимость $\chi^2$ от номера итерации NI.

Далее приведен контрольный счет для энергии 29.5 МэВ, для которой в [17] были получены фазы $\delta_0 = 150.76$, $\delta_2 = 86.61$, $\delta_4 = 121.00$, $\delta_6 = 2.16$, $\delta_8 = 0.09$, приведенные также в табл.2.1.1, при $\chi^2 = 0.602$. С такими фазами по приведенной выше программе получается $\chi^2 = 0.600$, как это видно из приведенной ниже распечатки с числом итераций NI = 0:

$\chi^2$ = ; NI = ; EL = 0.600328 0 29.50

| $\theta$ | $\sigma_e$ | $\sigma_t$ | $\chi^2_i$ |
|---|---|---|---|
| .22040E+02 | .15230E+04 | .15134E+04 | .64891E+00 |
| .24050E+02 | .11640E+04 | .11664E+04 | .55570E-01 |
| .26050E+02 | .88590E+03 | .86843E+03 | .38892E+01 |
| .28050E+02 | .61610E+03 | .61957E+03 | .23236E+00 |
| .30060E+02 | .42260E+03 | .41922E+03 | .35643E+00 |
| .32060E+02 | .27000E+03 | .26762E+03 | .34211E+00 |
| .34060E+02 | .16020E+03 | .16006E+03 | .24306E-02 |
| .36070E+02 | .91500E+02 | .91163E+02 | .44395E-01 |
| .38070E+02 | .55530E+02 | .55230E+02 | .16734E+00 |
| .40070E+02 | .44680E+02 | .44765E+02 | .12421E+00 |
| .42080E+02 | .52960E+02 | .52530E+02 | .76193E+00 |
| .44080E+02 | .71740E+02 | .71346E+02 | .28489E+00 |
| .46080E+02 | .95440E+02 | .94808E+02 | .57926E+00 |
| .48080E+02 | .11846E+03 | .11754E+03 | .12657E+01 |
| .50090E+02 | .13558E+03 | .13552E+03 | .72584E-02 |
| .52090E+02 | .14562E+03 | .14590E+03 | .26152E+00 |
| .54090E+02 | .14760E+03 | .14758E+03 | .53350E-03 |
| .56090E+02 | .13986E+03 | .14072E+03 | .14564E+01 |
| .58100E+02 | .12710E+03 | .12655E+03 | .43671E+00 |
| .60100E+02 | .10783E+03 | .10745E+03 | .23062E+00 |
| .62100E+02 | .86660E+02 | .86114E+02 | .46841E+00 |
| .64100E+02 | .66120E+02 | .65522E+02 | .75651E+00 |
| .66100E+02 | .48430E+02 | .48481E+02 | .10080E-01 |
| .68110E+02 | .37430E+02 | .37301E+02 | .17753E+00 |
| .70110E+02 | .33770E+02 | .33732E+02 | .43056E-01 |
| .72110E+02 | .38340E+02 | .38461E+02 | .13210E+00 |
| .74110E+02 | .50740E+02 | .51308E+02 | .10928E+01 |
| .76110E+02 | .70820E+02 | .71187E+02 | .24362E+00 |
| .78110E+02 | .95550E+02 | .96239E+02 | .60125E+00 |
| .80110E+02 | .12420E+03 | .12403E+03 | .27579E-01 |
| .82110E+02 | .15340E+03 | .15182E+03 | .22344E+01 |
| .84110E+02 | .17750E+03 | .17683E+03 | .42129E+00 |
| .86110E+02 | .19700E+03 | .19657E+03 | .17685E+00 |
| .88110E+02 | .20974E+03 | .20905E+03 | .51222E+00 |
| .90110E+02 | .21120E+03 | .21302E+03 | .29660E+01 |

| L | $\delta_r$ | $\delta_m$ |
|---|---|---|
| 0 | .150760E+03 | .000000E+00 |
| 2 | .866100E+02 | .000000E+00 |
| 4 | .121000E+03 | .000000E+00 |
| 6 | .216000E+01 | .000000E+00 |
| 8 | .900000E-01 | .000000E+00 |

Отличие величины $\chi^2$ в 0.02 связано с ошибками округления фаз при записи их значений в файл из программы на Turbo Basic с дальнейшим использованием в программе на языке Fortran-90. Здесь и далее в этом параграфе мнимая часть фазы FM(L) равна нулю для всех значений орбитального момента.

Используя эти фазы в качестве начальных, при дополнительном варьировании по новой программе на Fortran-90, приведенной выше, с 10 итерациями и максимально возможной точностью, получаем приведенный далее результат для фаз рассеяния с наименьшим значением $\chi^2$

$\chi^2$ = ; NI = ; EL = 0.571434      10    29.500

| $\theta$ | $\sigma_e$ | $\sigma_t$ | $\chi^2_i$ |
|---|---|---|---|
| .22040E+02 | .15230E+04 | .15144E+04 | .52763E+00 |
| .24050E+02 | .11640E+04 | .11671E+04 | .90939E-01 |
| .26050E+02 | .88590E+03 | .86891E+03 | .36792E+01 |
| .28050E+02 | .61610E+03 | .61990E+03 | .27763E+00 |
| .30060E+02 | .42260E+03 | .41943E+03 | .31337E+00 |
| .32060E+02 | .27000E+03 | .26775E+03 | .30469E+00 |
| .34060E+02 | .16020E+03 | .16015E+03 | .33034E-03 |
| .36070E+02 | .91500E+02 | .91219E+02 | .30734E-01 |
| .38070E+02 | .55530E+02 | .55273E+02 | .12333E+00 |
| .40070E+02 | .44680E+02 | .44803E+02 | .25848E+00 |
| .42080E+02 | .52960E+02 | .52569E+02 | .62951E+00 |
| .44080E+02 | .71740E+02 | .71391E+02 | .22342E+00 |
| .46080E+02 | .95440E+02 | .94863E+02 | .48325E+00 |
| .48080E+02 | .11846E+03 | .11761E+03 | .10869E+01 |
| .50090E+02 | .13558E+03 | .13560E+03 | .13302E-02 |
| .52090E+02 | .14562E+03 | .14600E+03 | .47518E+00 |
| .54090E+02 | .14760E+03 | .14770E+03 | .17885E-01 |
| .56090E+02 | .13986E+03 | .14085E+03 | .19094E+01 |
| .58100E+02 | .12710E+03 | .12668E+03 | .25354E+00 |
| .60100E+02 | .10783E+03 | .10757E+03 | .10341E+00 |
| .62100E+02 | .86660E+02 | .86228E+02 | .29377E+00 |
| .64100E+02 | .66120E+02 | .65613E+02 | .54400E+00 |
| .66100E+02 | .48430E+02 | .48542E+02 | .48101E-01 |
| .68110E+02 | .37430E+02 | .37326E+02 | .11519E+00 |
| .70110E+02 | .33770E+02 | .33722E+02 | .70632E-01 |
| .72110E+02 | .38340E+02 | .38419E+02 | .55430E-01 |
| .74110E+02 | .50740E+02 | .51240E+02 | .84900E+00 |
| .76110E+02 | .70820E+02 | .71105E+02 | .14712E+00 |
| .78110E+02 | .95550E+02 | .96154E+02 | .46141E+00 |
| .80110E+02 | .12420E+03 | .12395E+03 | .59440E-01 |
| .82110E+02 | .15340E+03 | .15175E+03 | .24240E+01 |
| .84110E+02 | .17750E+03 | .17678E+03 | .48540E+00 |
| .86110E+02 | .19700E+03 | .19654E+03 | .20525E+00 |
| .88110E+02 | .20974E+03 | .20903E+03 | .54586E+00 |
| .90110E+02 | .21120E+03 | .21300E+03 | .29054E+01 |

| L | $\delta_r$ | $\delta_m$ |
|---|---|---|
| 0 | .150729E+03 | .000000E+00 |
| 2 | .865625E+02 | .000000E+00 |
| 4 | .120914E+03 | .000000E+00 |
| 6 | .211725E+01 | .000000E+00 |
| 8 | .662343E-01 | .000000E+00 |

Как видно, удается улучшить $\chi^2$ до 0.571, что оказывается возможным благодаря существенно повышенной, по сравнению с [17], точностью поиска минимума такой вариационной задачи.

## *2.2 Фазовый анализ упругого $^4He^{12}C$ рассеяния Phase shifts analysis of $^4He^{12}C$ scattering*

Рассмотрим теперь методы фазового анализа для нетождественных частиц со спином 0 + 0 и результаты фазового анализа для упругого рассеяния в $^4He^{12}C$ системе при астрофизических энергиях. Для получения этих результатов использовалась та же компьютерная программа, что и для упругого $^4He^4He$ рассеяния, но с другим программным параметром (см. параграф 2.1.5) [146].

### 2.2.1 Дифференциальные сечения Differential cross sections

В случае упругого рассеяния нетождественных частиц с нулевым спином выражение для сечения принимает наиболее простой вид [88]

$$\frac{d\sigma(\theta)}{d\Omega} = |f(\theta)|^2 ,$$

где полная амплитуда рассеяния $f(\theta)$ представляется в виде суммы кулоновской $f_c(\theta)$ и ядерной $f_N(\theta)$ амплитуд

$$f(\theta) = f_c(\theta) + f_N(\theta) \; ,$$

которые выражаются через ядерные $\delta_L \to \delta_L + i\Delta_L$ и кулоновские $\sigma_L$ фазы рассеяния, а вид самих амплитуд приведен в начале этого раздела (см. параграф 2.1.2).

Для полного сечения упругого рассеяния при $f_c = 0$ будем иметь

$$\sigma_s = \frac{\pi}{k^2}\sum_L \left[(2L+1)\left(|1-S_L|^2\right)\right] = \frac{4\pi}{k^2}\sum_L (2L+1)\eta_L^2 \sin^2\delta_L \quad .$$

Суммирование в этом выражении, в отличие от $^4He^4He$

рассеяния, выполняется по всем возможным $L$.

### 2.2.2 Фазовый анализ
### Phase shifts analysis

Приведем результаты фазового анализа, полученные для $^4He^{12}C$ упругого рассеяния в области энергий от 1.5 МэВ до 6.5 МэВ. Ранее фазовый анализ дифференциальных сечений при энергиях 2.5÷5.0 МэВ был проведен в работе [147], а потенциальное описание таких фаз рассеяния на основе потенциалов с запрещенными состояниями было выполнено нами в работе [52]. Здесь, используя экспериментальные данные по функциям возбуждения при семи углах из [147], мы повторили фазовый анализ [147] в области энергий 2.5÷4.5 МэВ, который можно рассматривать, как контрольный тест для нашей новой компьютерной программы и использованных нами методов расчета.

Заметим, что измерение функций возбуждения при таком небольшом количестве точек по углам рассеяния недостаточно, чтобы с хорошей точностью воспроизвести форму угловых распределений $^4He^{12}C$ упругого рассеяния даже при низких энергиях. Возможно, поэтому, фазовый анализ не позволяет получить полностью однозначные значения фаз рассеяния, особенно для $S$ волны, несмотря на то, что в качестве начальных фаз [17] были использованы известные результаты фазового анализа [147].

Результаты нашего анализа приведены далее на рис.2.2.1 ÷рис.2.2.5. На рис.2.2.1 вместе с $S$ фазой рассеяния показаны значения среднего $\chi^2$, которые получены для различных энергий рассеяния. Как видно из этих рисунков, $S$ фаза несколько отличается от данных работы [147], а фазы во всех остальных парциальных волнах вполне согласуются с этими результатами [147]. Однако, поскольку, данные по функциям возбуждения брались из рисунков работы [147], возможная ошибка в настоящем фазовом анализе, зависящая от точности данных по сечениям, может составлять примерно 2°÷3°, что в целом позволяет объяснить расхождение в результатах этих двух фазовых анализов для *S* фазы упругого рассеяния.

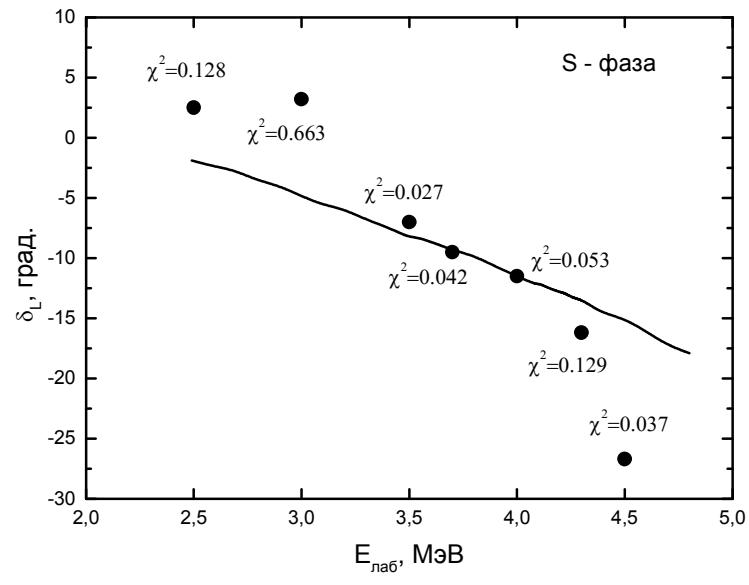

Рис.2.2.1. *S* фаза упругого $^4He^{12}C$ рассеяния. Кривая – данные работы [147]. Точки – наши результаты, полученные на основе данных [147].

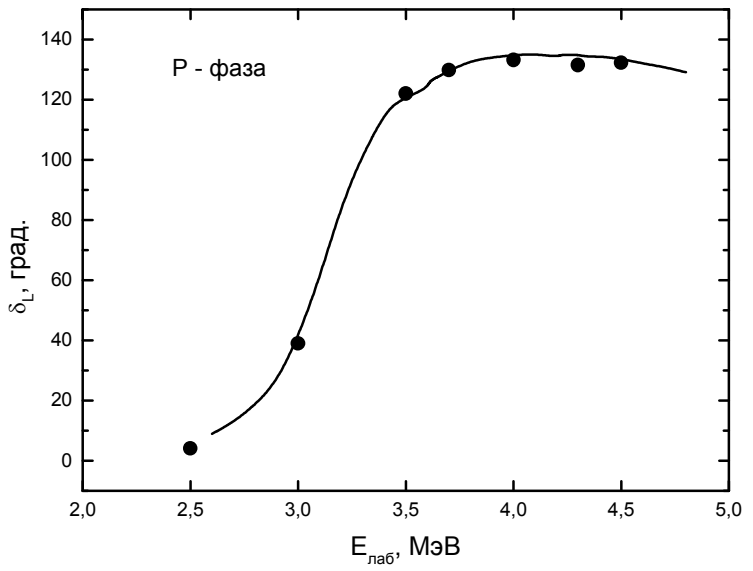

Рис.2.2.2. *P* фаза упругого $^4He^{12}C$ рассеяния. Обозначения, как на рис.2.2.1.

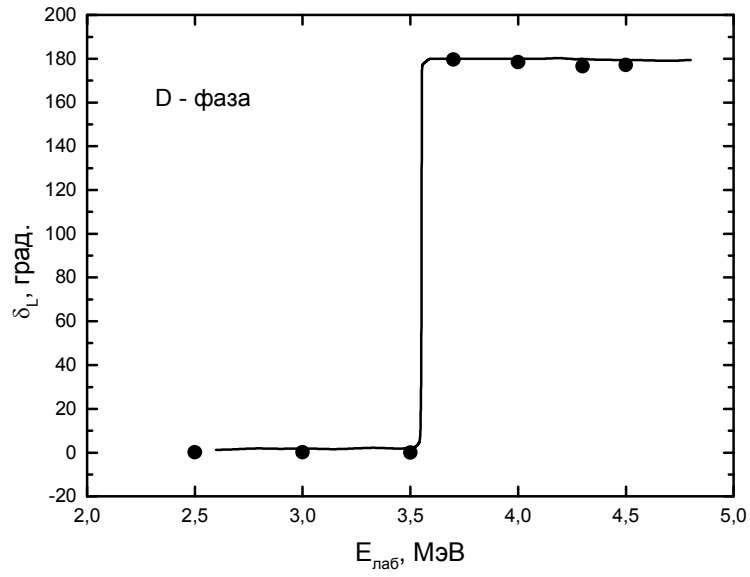

Рис.2.2.3. *D* фаза упругого $^4He^{12}C$ рассеяния. Обозначения, как на рис.2.2.1.

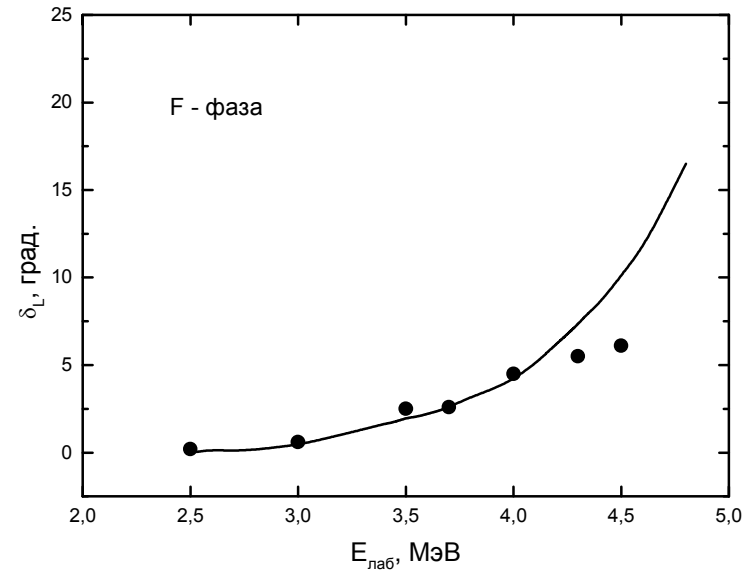

Рис.2.2.4. *F* фаза упругого $^4He^{12}C$ рассеяния. Обозначения, как на рис.2.2.1.

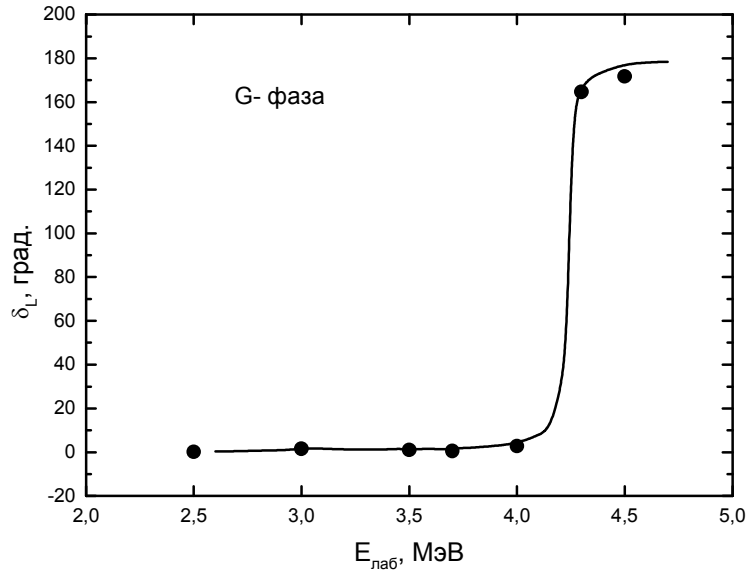

Рис.2.2.5. *G* фаза упругого $^4He^{12}C$ рассеяния. Обозначения, как на рис.2.2.1.

Далее, в работе [148], был выполнен очень аккуратный фазовый анализ экспериментальных данных при 49 энергиях в области от 1.5 до 6.5 МэВ. Используя эти данные, мы провели новый фазовый анализ при энергиях 1.466, 1.973, 2.073, 2.870, 3.371, 4.851, 5.799 и 6.458 МэВ. Результаты, полученные в нашем анализе, представлены в табл.2.2.1÷табл.2.2.8 вместе со средними значениями $\chi^2$ в сравнении с табличными данными работы [148]. В табл.2.2.9 показан спектр резонансных уровней, наблюдаемых в упругом $^4$He$^{12}$C рассеянии [149].

Табл.2.2.1. Результаты фазового анализа $^4$He$^{12}$C упругого рассеяния и их сравнение с данными работы [148] при энергии 1.466 МэВ.

| $E_{лаб}$ = 1.466 МэВ ($\chi^2$ = 0.055) | | |
|---|---|---|
| $L$ | $\delta^0$ (Наш) | $\delta^0$ [148] |
| 0 | -0.2 | 0.5±1.0 |
| 1 | -0.4 | -0.1±1.0 |
| 2 | -1.1 | -0.8±1.0 |

Табл.2.2.2. Результаты фазового анализа $^4$He$^{12}$C упругого рассеяния и их сравнение с данными работы [148] при энергии 1.973 МэВ.

| $E_{лаб}$ = 1.973 МэВ ($\chi^2$ = 0.077) | | |
|---|---|---|
| $L$ | $\delta^0$ (Наш) | $\delta^0$ [148] |
| 0 | -2.6 | -0.5±1.0 |
| 1 | 0.0 | 0.9±1.7 |
| 2 | -1.2 | -0.1±1.3 |

Табл.2.2.3. Результаты фазового анализа $^4$He$^{12}$C упругого рассеяния и их сравнение с данными работы [148] при энергии 2.073 МэВ.

| $E_{лаб}$ = 2.073 МэВ ($\chi^2$ = 0.029) | | |
|---|---|---|
| $L$ | $\delta^0$ (Наш) | $\delta^0$ [148] |
| 0 | -1.2 | 0±0.8 |
| 1 | -0.1 | 0.1±1.2 |
| 2 | -1.1 | -0.6±0.9 |

Табл.2.2.4. Результаты фазового анализа $^4$He$^{12}$C упругого рассеяния и их сравнение с данными работы [148] при энергии 2.870 МэВ.

| $E_{лаб}$ = 2.870 МэВ ($\chi^2$ = 0.038) | | |
|---|---|---|
| $L$ | $\delta^0$ (Наш) | $\delta^0$ [148] |
| 0 | -3.1 | -2.1±1.1 |
| 1 | 21.3 | 22.0±2.1 |
| 2 | 0.0 | 0.4±0.9 |
| 3 | 0.5 | 1.0±0.5 |

Табл.2.2.5. Результаты фазового анализа $^4He^{12}C$ упругого рассеяния и их сравнение с данными работы [148] при энергии 3.371 МэВ.

| $E_{лаб}$ = 3.371 МэВ ($\chi^2$ = 0.31) | | |
|---|---|---|
| $L$ | $\delta^0$ (Наш) | $\delta^0$ [148] |
| 0 | 169.4 | - |
| 1 | 103.4 | 103.7±1.7 |
| 2 | -1.7 | 0.0±0.7 |
| 3 | 0.2 | 0.8±0.6 |

Табл.2.2.6. Результаты фазового анализа $^4He^{12}C$ упругого рассеяния и их сравнение с данными работы [148] при энергии 4.851 МэВ.

| $E_{лаб}$ = 4.851 МэВ ($\chi^2$ = 0.26) | | |
|---|---|---|
| $L$ | $\delta^0$ (Наш) | $\delta^0$ [148] |
| 0 | 164.2 | 164±1.1 |
| 1 | 128.4 | 129.5±0.9 |
| 2 | 177.1 | 178.8±0.9 |
| 3 | 15.5 | 16.4±0.8 |
| 4 | 176.9 | 177.2±0.8 |
| 5 | -0.3 | 0.5±0.5 |

Табл.2.2.7. Результаты фазового анализа $^4He^{12}C$ упругого рассеяния и их сравнение с данными работы [148] при энергии 5.799 МэВ.

| $E_{лаб}$ = 5.799 МэВ ($\chi^2$ = 0.37) | | |
|---|---|---|
| $L$ | Re$\delta^0$ (Наш) | Re$\delta^0$ [148] |
| 0 | 162.2 | - |
| 1 | 128.2 | - |
| 2 | 83.2 | 82.3±0.6 |
| 3 | 86.0 | - |
| 4 | 173.8 | 175.3±0.7 |
| 5 | -1.0 | 0.2±0.4 |

Табл.2.2.8. Результаты фазового анализа $^4He^{12}C$ упругого рассеяния и их сравнение с данными работы [148] при энергии 6.458 МэВ.

| $E_{лаб}$ = 6.458 МэВ ($\chi^2$ = 0.41) | | |
|---|---|---|
| $L$ | $\delta^0$ (Наш) | $\delta^0$ [148] |
| 0 | 151.2 | 153±2.5 |
| 1 | 115.8 | 119.4±2.1 |
| 2 | 172.2 | 172.2±1.9 |
| 3 | 120.8 | 122.0±2.4 |
| 4 | 176.4 | 179.1±1.2 |
| 5 | 0.8 | 2.2±0.8 |
| 6 | 0.1 | 0.4±0.4 |

Из приведенных таблиц видно, что энергия $^4He^{12}C$ рассеяния 3.371 МэВ приходится на уровень 3.324 МэВ с шириной 480±20 кэВ, приведенный в табл.2.2.9 и обзоре [149]. Хотя в таблицах работы [148] не приводится фаза для $S$ волны (прочерки в табл.2.2.5) при этой энергии, наш фазовый анализ на основе только действительных фаз рассеяния позволяет определить все фазы рассеяния. Величина $S$ фазы приведена в табл.2.2.5 с $\chi^2 = 0.31$ при 10% ошибках определения экспериментальных дифференциальных сечений из рисунков работы [148].

Энергия 5.799 МэВ приходится на уровень 5.809±18 МэВ с шириной 73(3) кэВ [149] (см. ниже табл.2.2.9) и в таблицах работы [148] не приводятся значения фаз рассеяния для некоторых парциальных волн (прочерки в табл.2.2.7). В проведенном здесь фазовом анализе вполне удается описать эти дифференциальные сечения упругого рассеяния со средним значением $\chi^2 = 0.37$ и найти все парциальные фазы, как это показано в табл.2.2.7.

Табл.2.2.9. Спектр уровней ядра $^{16}O$ в упругом $^4He^{12}C$ рассеянии при изоспине $T = 0$ [149].
Здесь $J^\pi$ – полный момент и четность, $E_{лаб}$ – энергия налетающей $\alpha$ - частицы в лабораторной системе, $\Gamma_{цм}$ – ширина уровня в системе центра масс. Курсивом в таблице выделены уровни с шириной меньше 1 кэВ, а жирным показаны состояния, которые рассматриваются в тексте.

| $E_{лаб}$, МэВ | $J^\pi$ | $\Gamma_{цм}$, кэВ |
|---|---|---|
| **3.324** | **$1^-$** | **480±20** |
| *3.5770±0.5* | *$2^+$* | *0.625±0.1* |
| 4.259 | $4^+$ | 27±3 |
| *5.245±8* | *$4^+$* | *0.28±0.05* |
| 5.47 | $0^+$ | 2500 |
| **5.809±18** | **$2^+$** | **73±5** |
| 5.92±20 | $3^-$ | 800±100 |
| 6.518±10 | $0^+$ | 1.5±0.5 |
| 7.043±4 | $1^-$ | 99±7 |

Нерезонансные энергии при 2.870, 4.851 и 6.458 МэВ (см. табл.2.2.4, 2.2.6, 2.2.8) описываются фазами, которые в пределах, приведенных в работе [148] погрешностей, и с учетом возможных 10% ошибок нашего извлечения экспериментальных данных из рисунков [148] совпадают с результатами работы [148]. Последние три энергии 1.466, 1.973 и 2.073 МэВ (см. табл.2.2.1, 2.2.2, .2.2.3) совместимы с нулевыми значениями ядерных фаз, т.е. имеют значения ±1°±1°, и соответствуют чисто кулоновскому, т.е. резерфордовскому рассеянию.

На рис.2.2.6÷2.2.13 показано качество описания экспериментальных дифференциальных сечений упругого $^4He^{12}C$ рассеяния [148] с приведенными в табл.2.2.1÷2.2.8 фазами рассеяния при всех рассмотренных энергиях. В процессе нашего фазового анализа пришлось перевести дифференциальные сечения и углы, приведенные в работе [148] в лабораторной системе, в систему центра масс, которая наиболее удобна для расчетов.

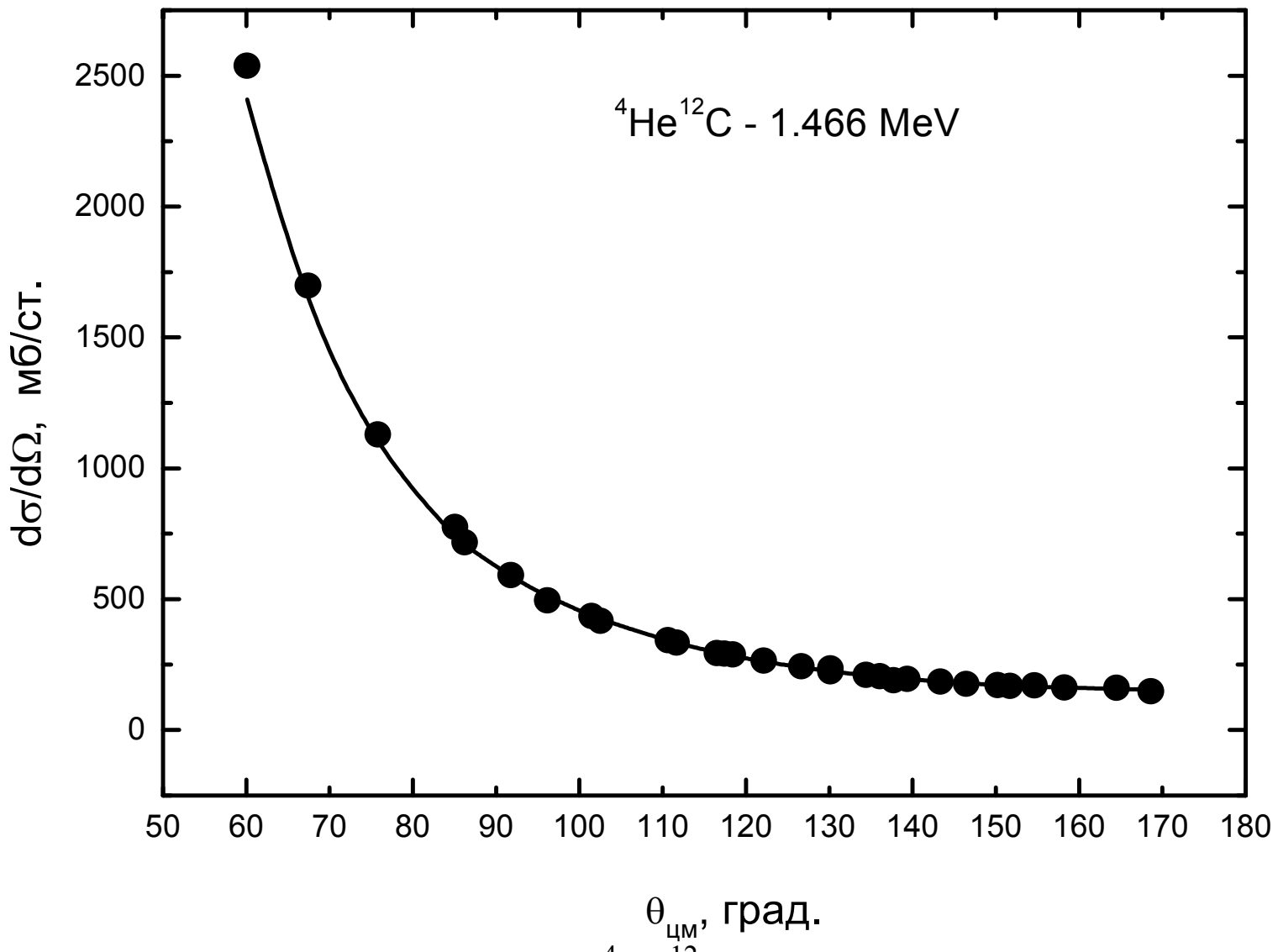

Рис.2.2.6. Сечения упругого $^4He^{12}C$ рассеяния при 1.466 МэВ. Точки – экспериментальные данные работы [148], кривая – результаты нашего фазового анализа.

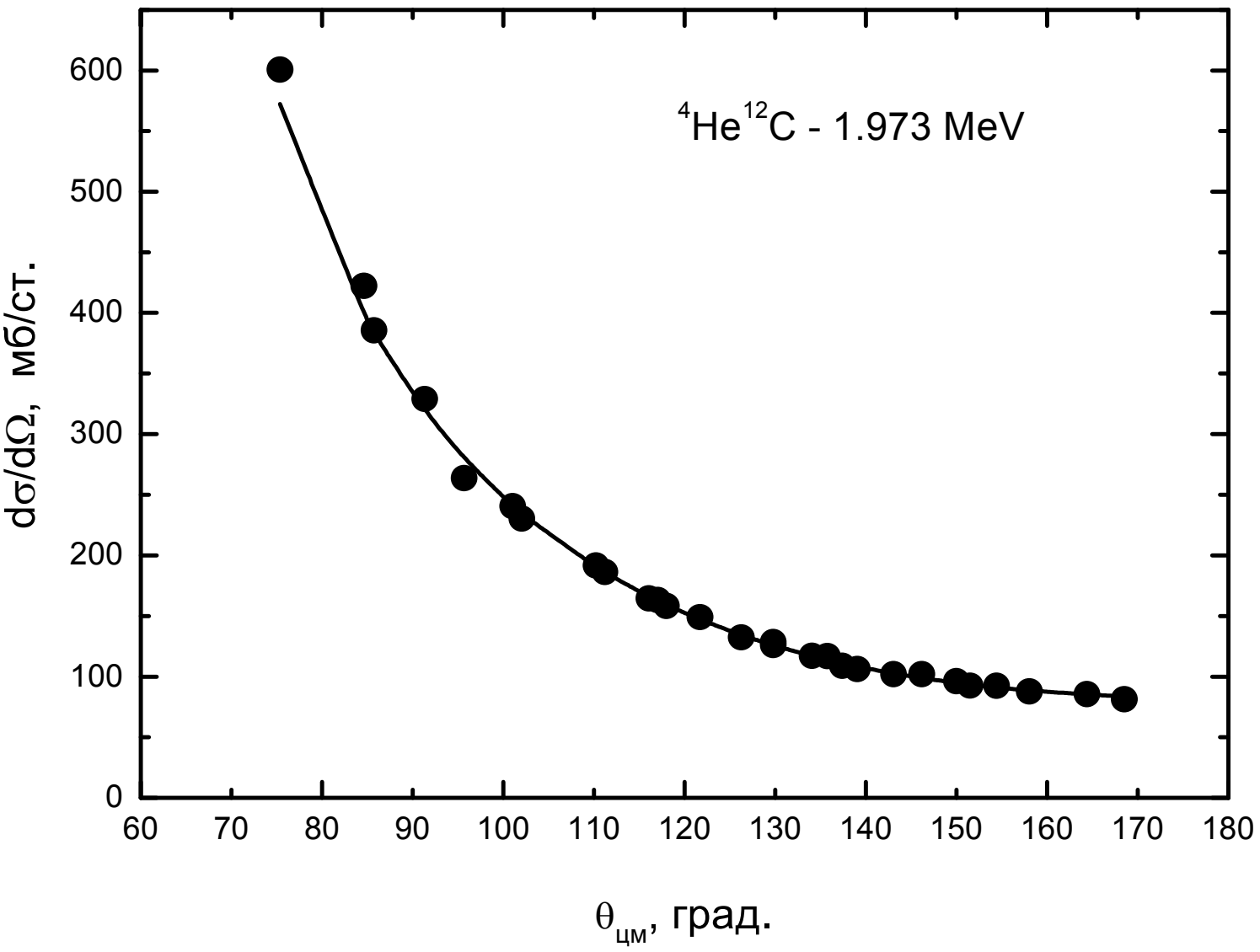

Рис.2.2.7. То же, что на рис.2.2.6, но при энергии 1.973 МэВ.

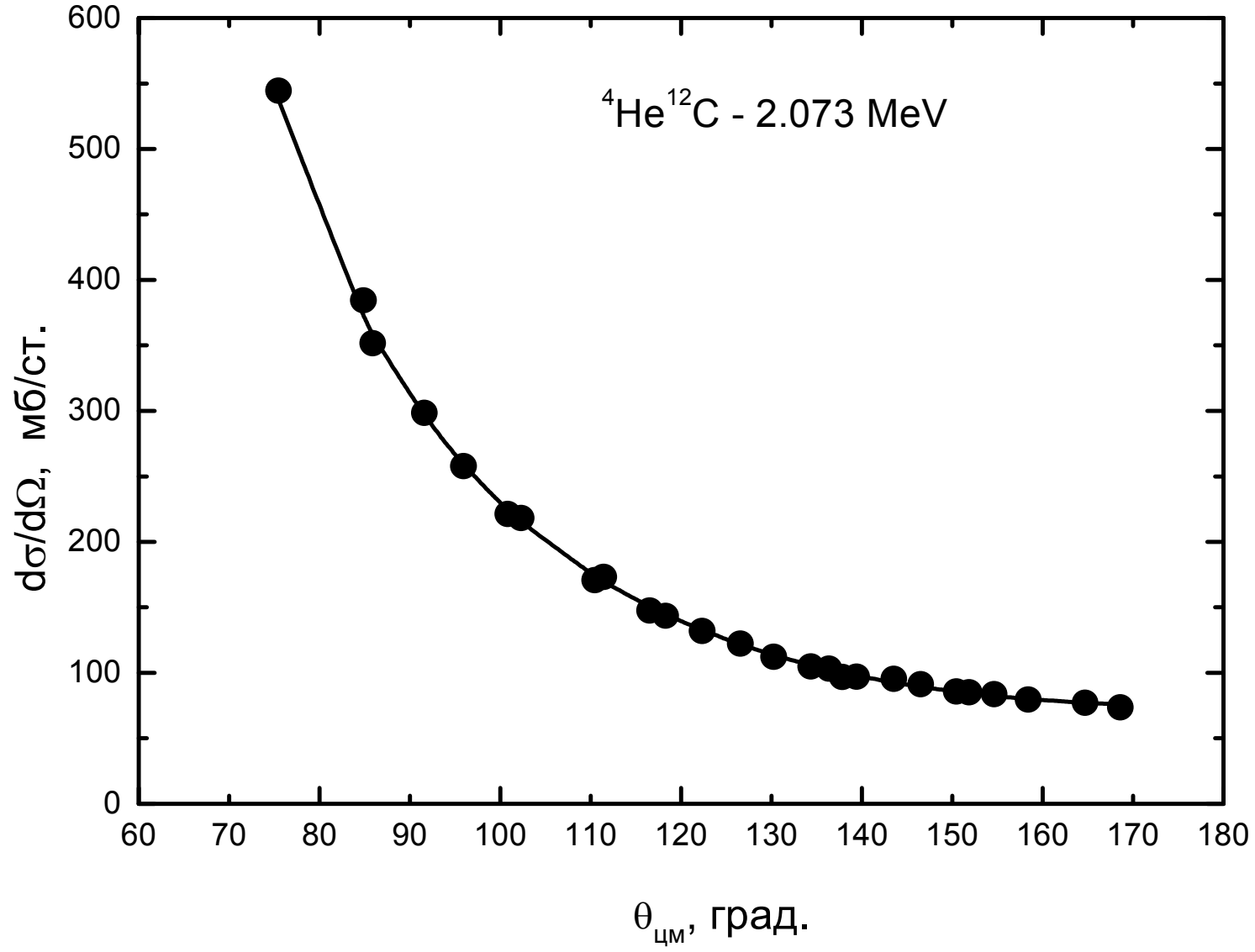

Рис.2.2.8. То же, что на рис.2.2.6, но при энергии 2.073 МэВ.

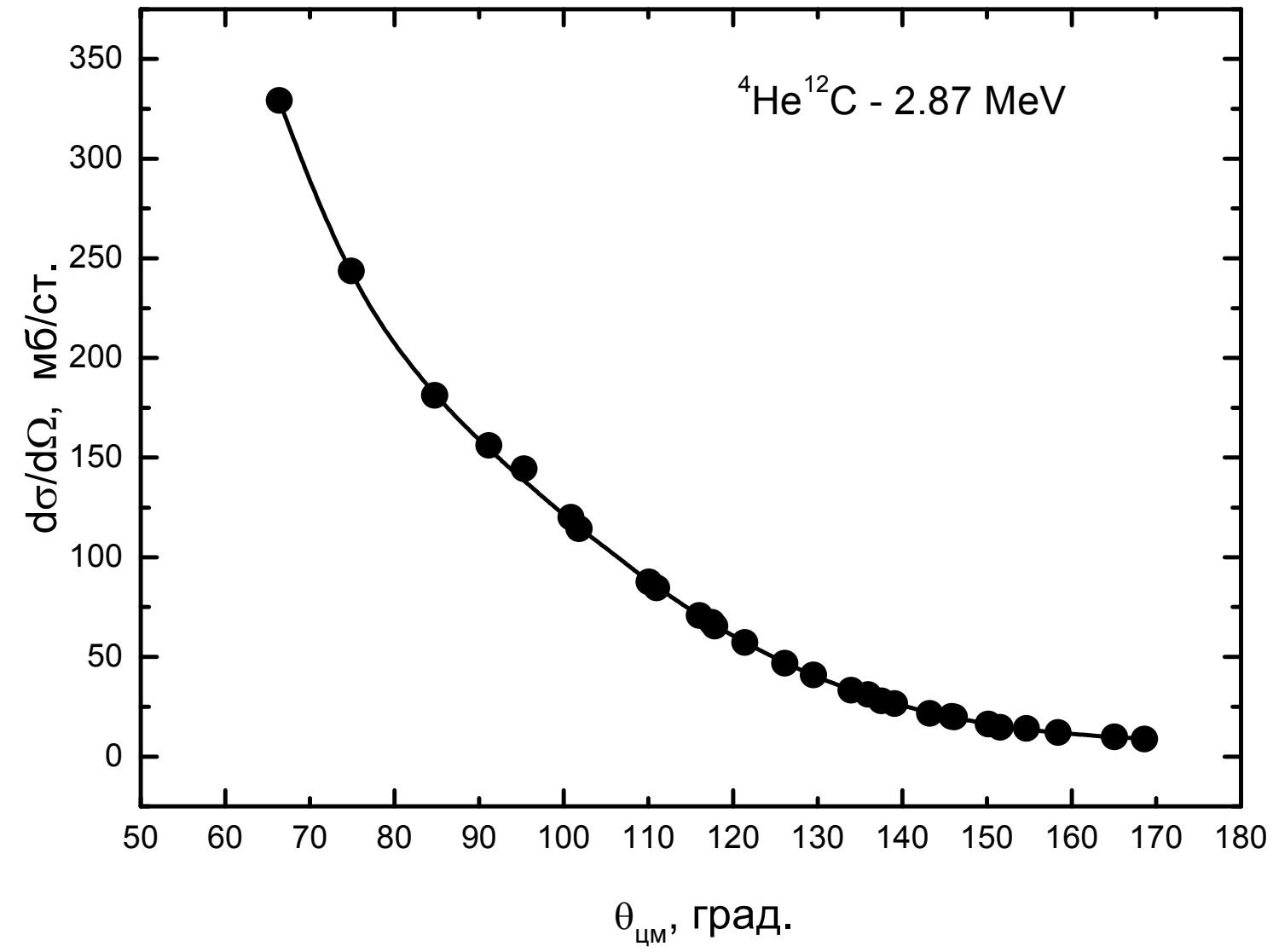

Рис.2.2.9. То же, что на рис.2.2.6, но при энергии 2.87 МэВ.

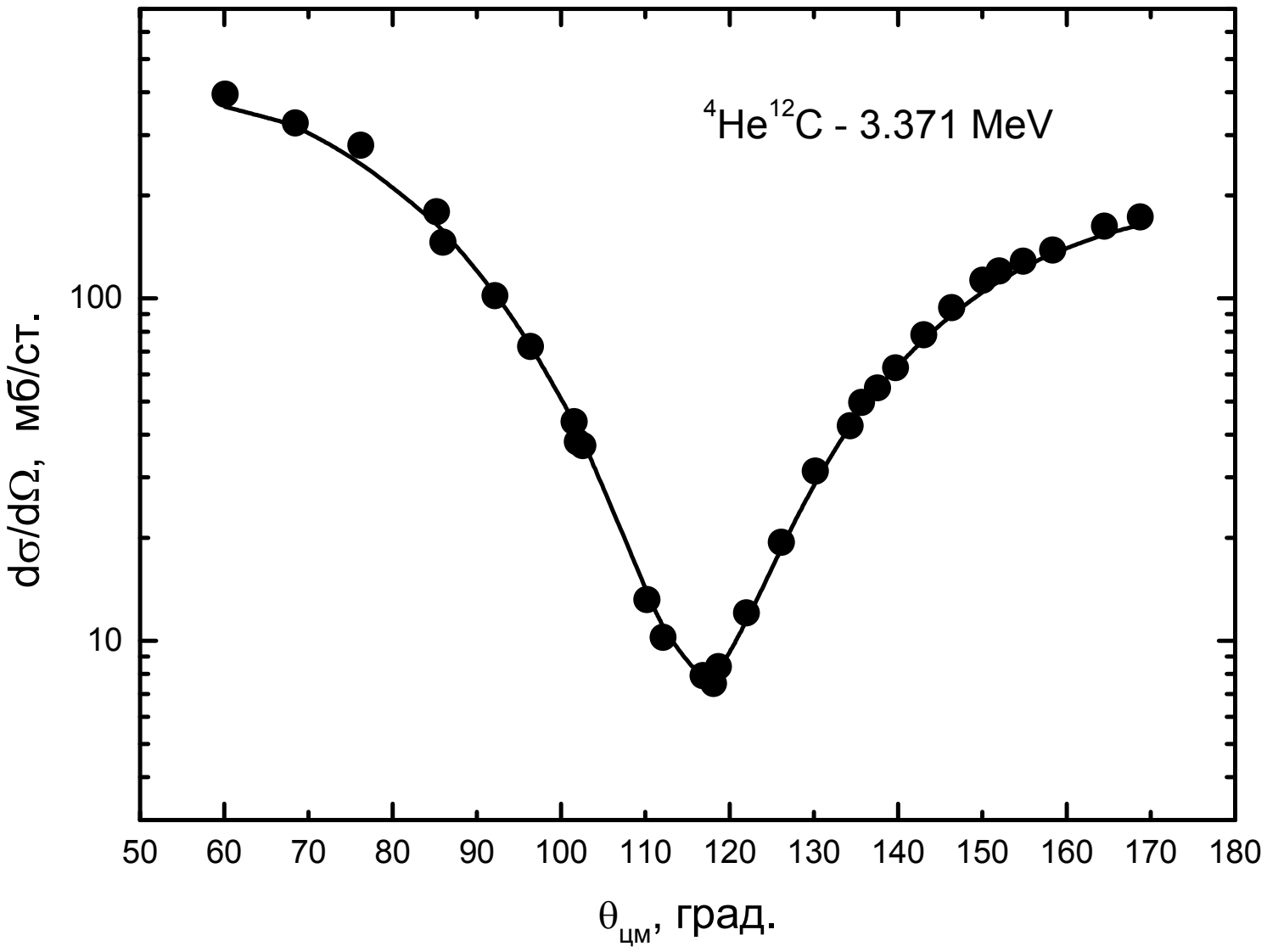

Рис.2.2.10. То же, что на рис.2.2.6, но при энергии 3.371 МэВ.

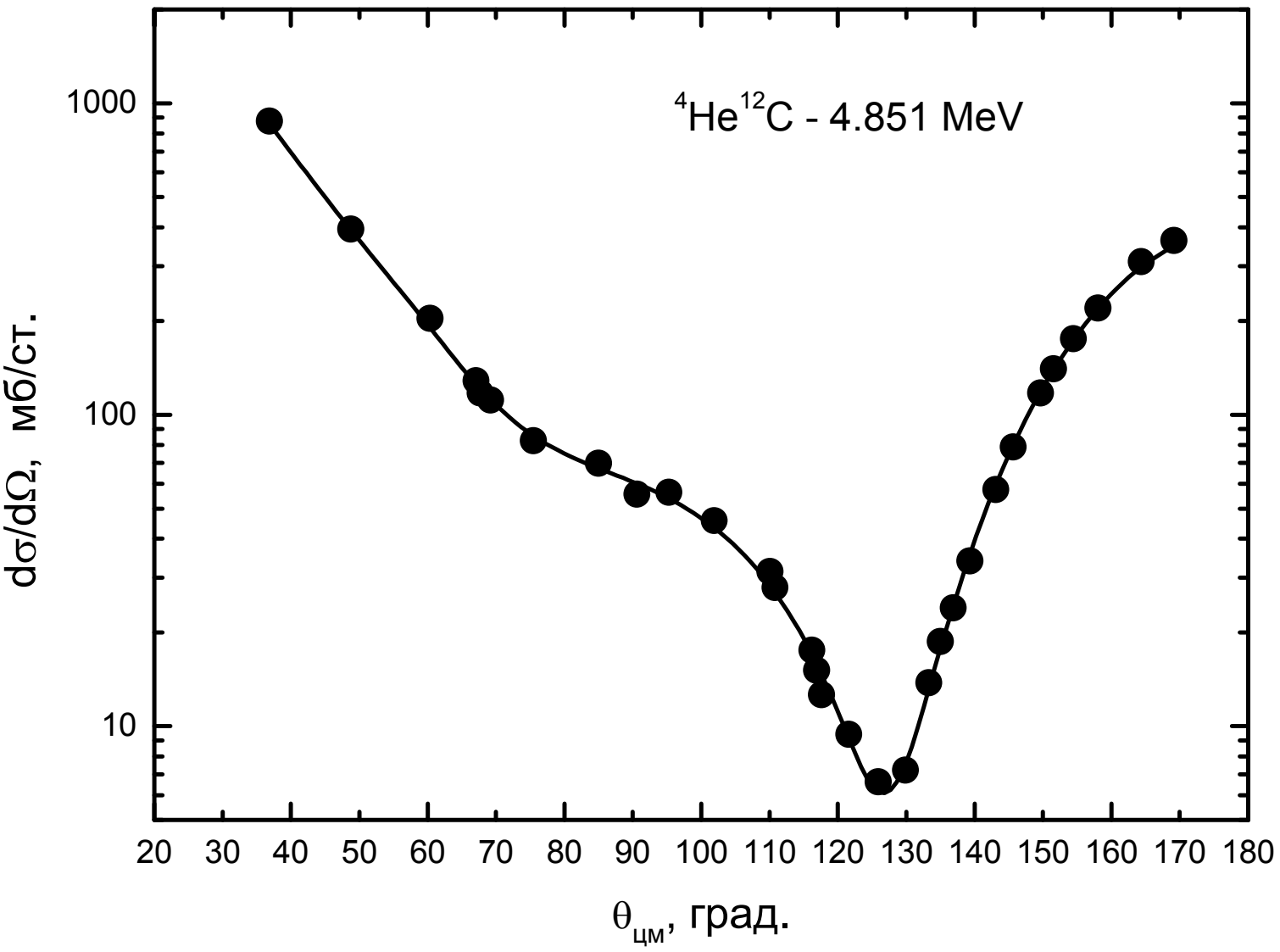

Рис.2.2.11. То же, что на рис.2.2.6, но при энергии 4.851 МэВ.

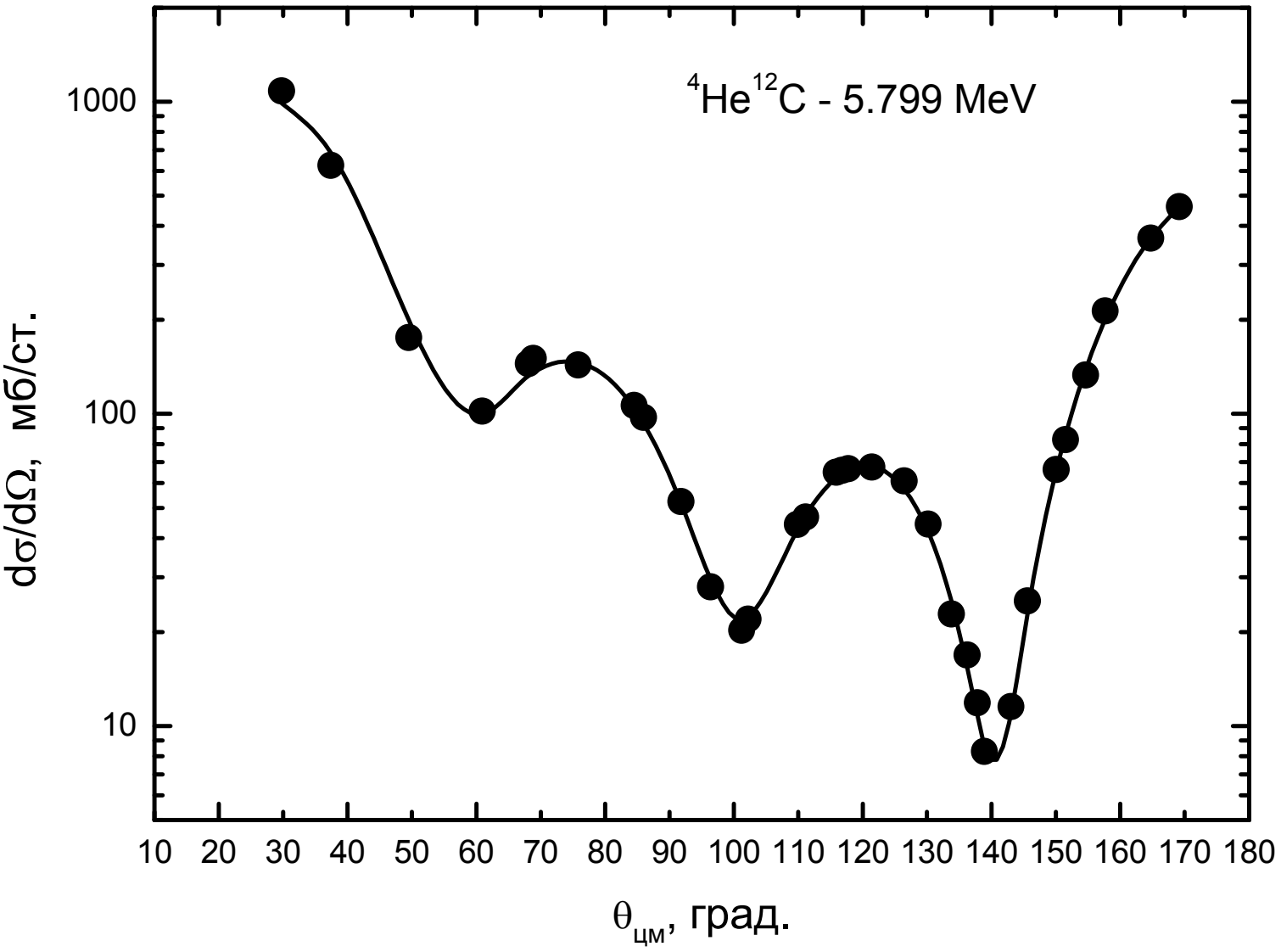

Рис.2.2.12. То же, что на рис.2.2.6, но при энергии 5.799 МэВ.

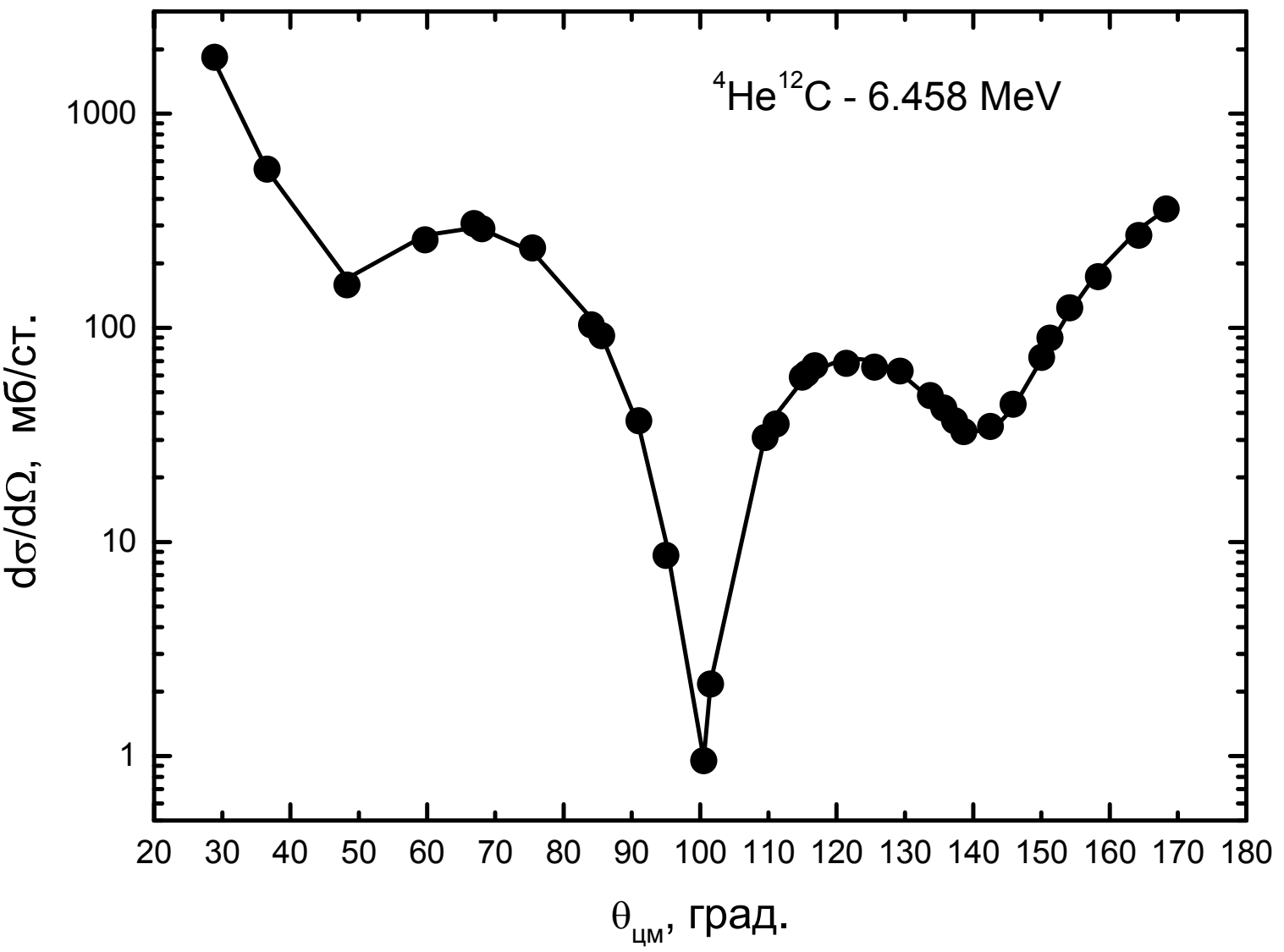

Рис.2.2.13. То же, что на рис.2.2.6, но при энергии 6.458 МэВ.

Небольшие отличия в фазах рассеяния могут быть обусловлены, кроме этого, различными значениями констант или масс частиц, которые используются в таком анализе. Например, можно использовать точные значения масс частиц [120] или же их целые величины, а константа $\hbar^2 / m$ может быть принята равной, например, 41.47 или более точно 41.4686 МэВ·Фм$^2$ (хотя возможны и другие ее значения) Кроме того, точность определения фаз рассеяния, в проведенном фазовом анализе, на основе данных [148] оценивается нами на уровне $1°\div2°$.

На рис.2.2.14÷рис.2.2.18 приведено сравнение результатов последнего фазового анализа [150], полученного на основе экспериментальных данных работы [148] (точки) и фазового анализа работы [147] (квадраты). Как видно из приведенных в табл.2.2.1÷табл.2.2.8 и рис.2.2.14÷рис.2.2.18 результатов, полученные фазы практически совпадают с выводами работы [148], но несколько отличаются от данных [147], особенно в *S* волне рассеяния.

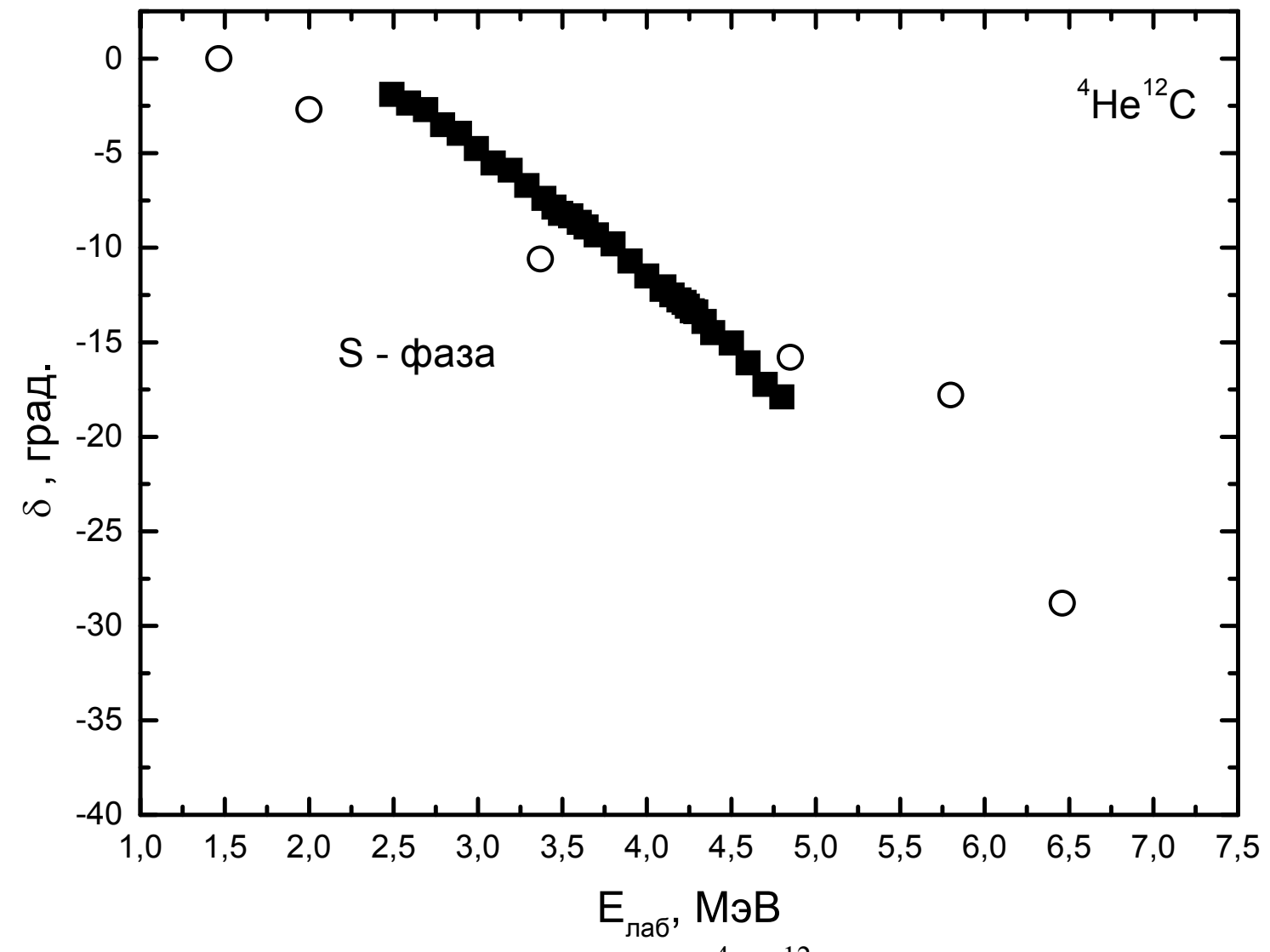

Рис.2.2.14 *S* фаза упругого $^4He^{12}C$ рассеяния.
Квадраты – данные работы [147]. Кружки – наши результаты [150], полученные на основе данных [148].

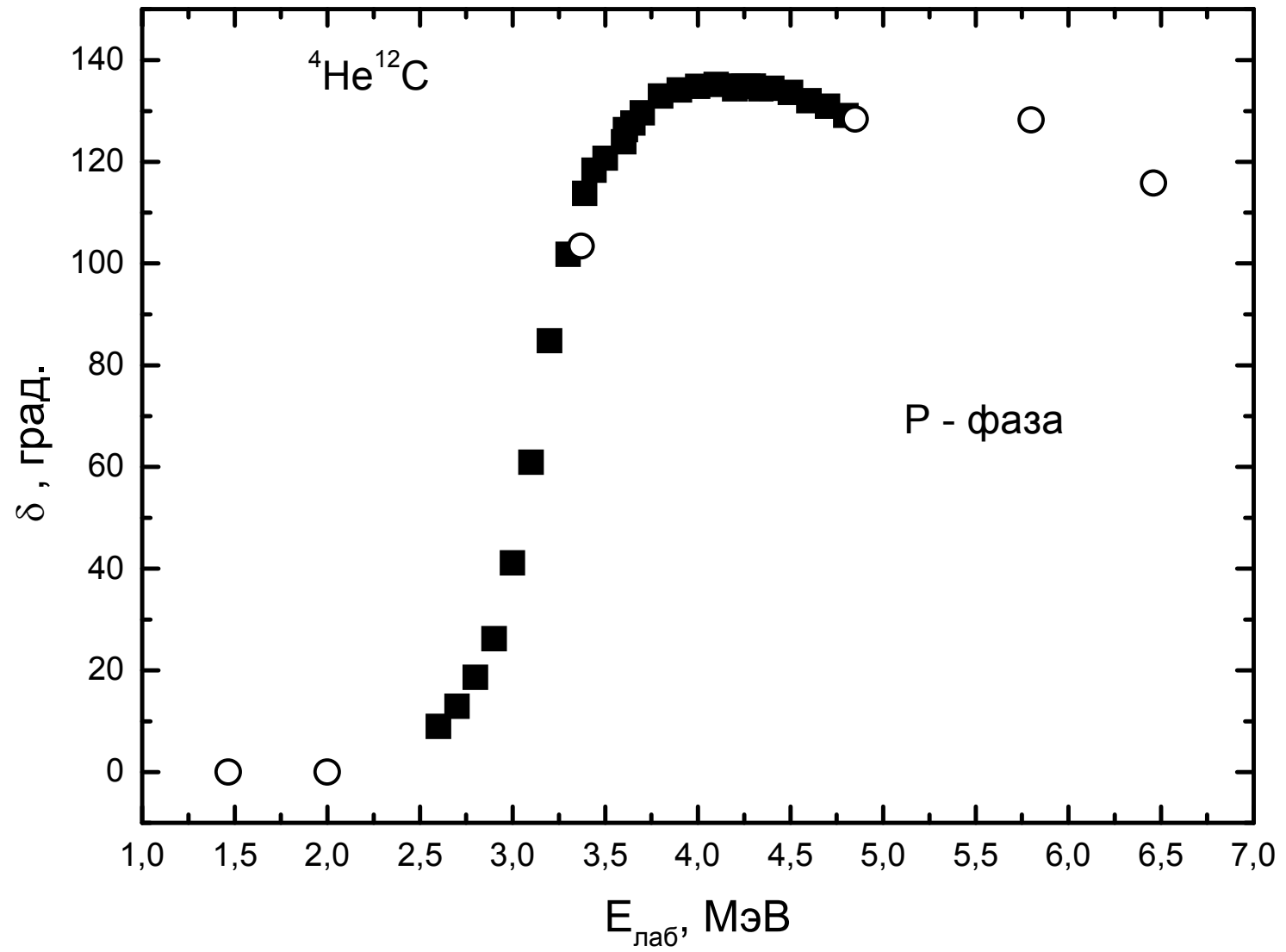

Рис.2.2.15. *P* фаза упругого $^4He^{12}C$ рассеяния.
Обозначения, как на рис.2.2.14.

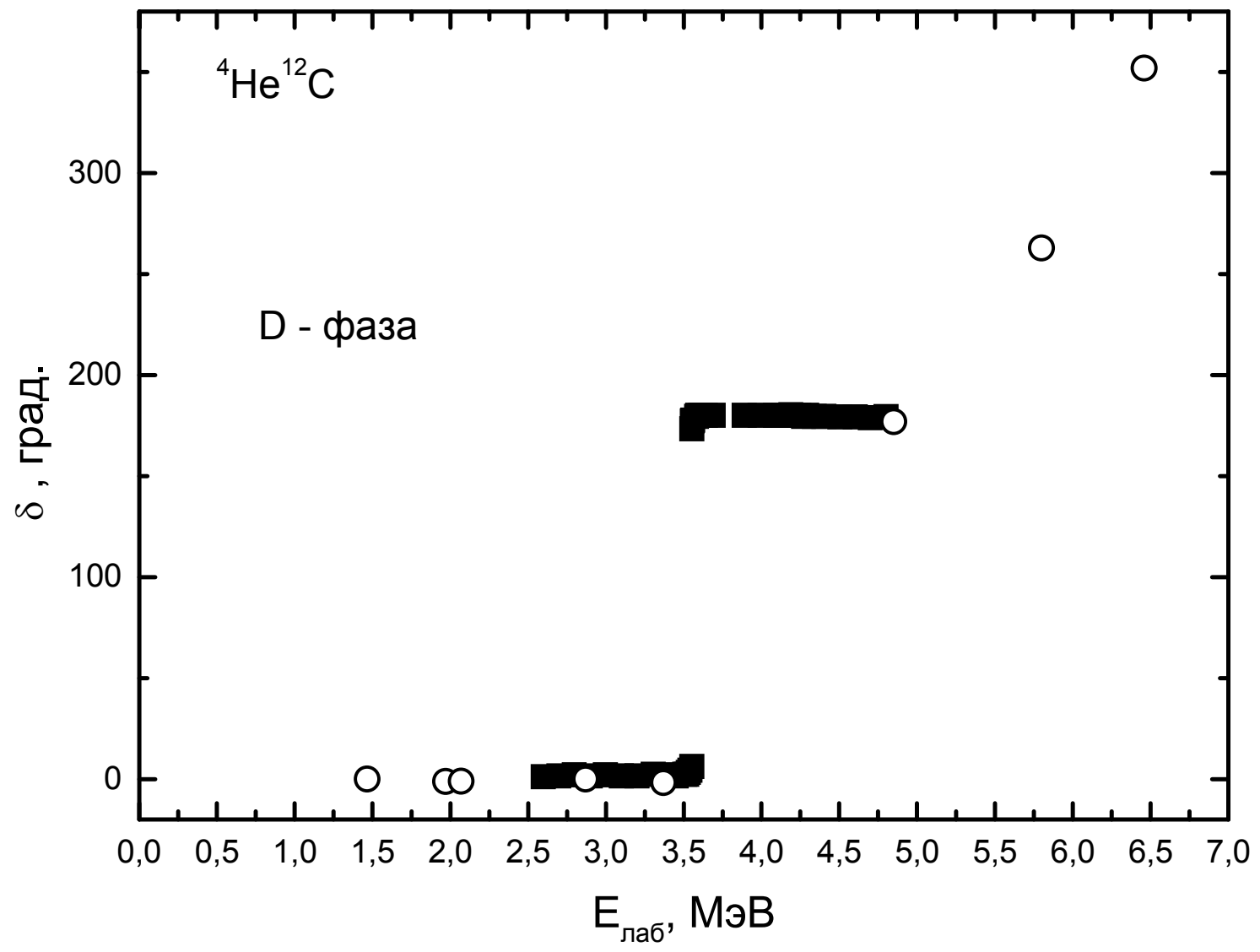

Рис.2.2.16. *D* фаза упругого $^4He^{12}C$ рассеяния.
Обозначения, как на рис.2.2.14.

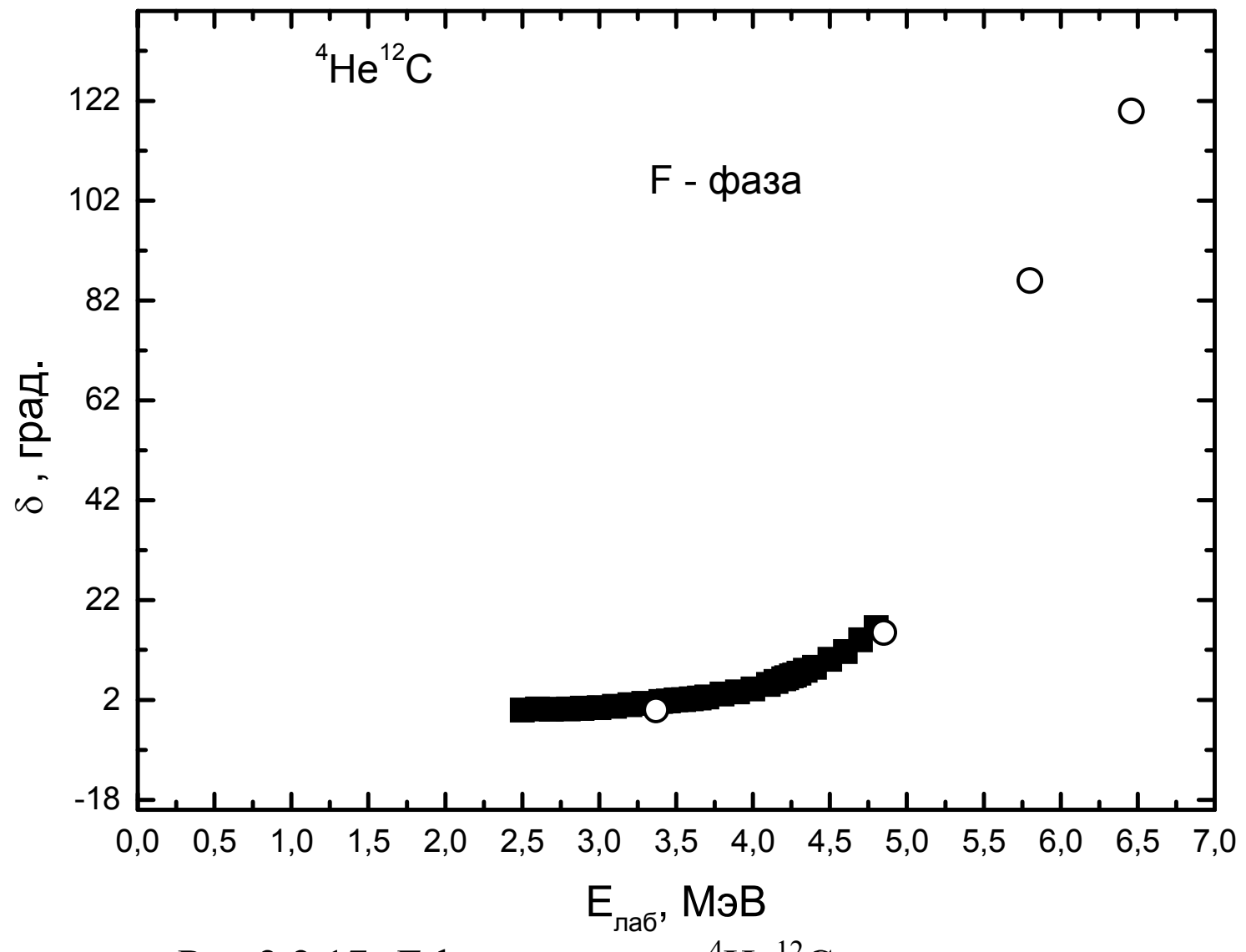

Рис.2.2.17. *F* фаза упругого $^4He^{12}C$ рассеяния.
Обозначения, как на рис.2.2.14.

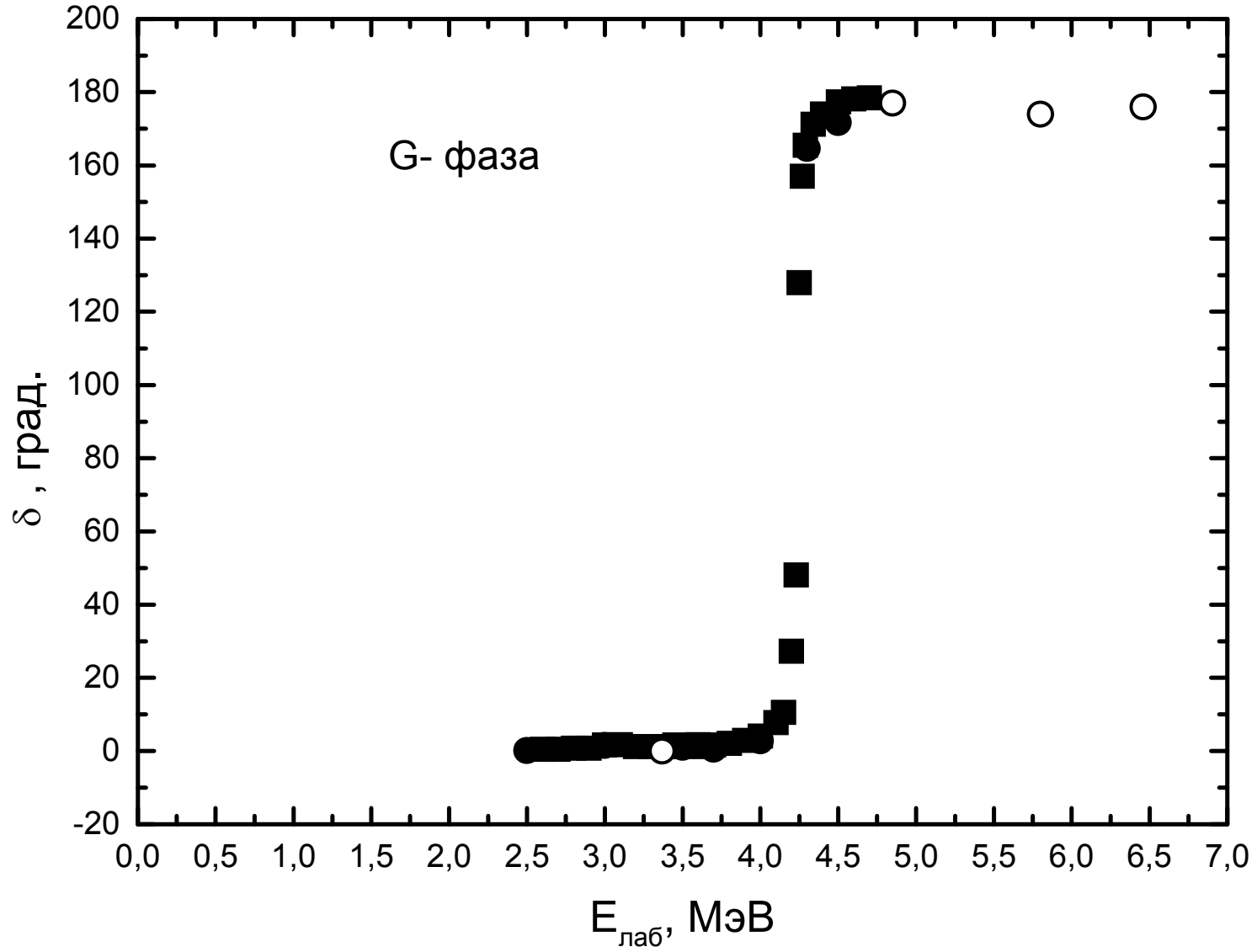

Рис.2.2.18. *G* фаза упругого $^4He^{12}C$ рассеяния. Обозначения, как на рис.2.2.14.

В целом это понятно, поскольку анализ [147] выполнялся в начале 60-х годов, когда вычислительной техники еще практически не существовало, а программные и вычислительные средства только начинали развиваться. В тоже время, работа [148] была проведена в конце 80-х при развитых компьютерных и программных системах и средствах.

В результате выполненного здесь фазового анализа можно сделать следующие выводы:

1. Энергия 3.371 МэВ приходится на уровень 3.324 МэВ с довольно большой шириной 0.48 МэВ и в *P* волне хорошо просматривается плавный резонанс.

2. По данным работ [147,148] в *D* волне наблюдается резонанс, который соответствует уровню при 3.577 МэВ с очень маленькой шириной 0.625 кэВ. При этой энергии на рис.2.2.16 наблюдается переход фазы через 90°, приведенный также в табл.2.2.5 и 2.2.6. Кроме того, в результатах анализа [148] появляется еще один резонанс при 5.799 МэВ, посколь-ку при этой энергии в *D* волне имеется уровень 5.809 МэВ с относительно большой шириной 73 кэВ. При этой энергии величина фазы (см. табл.2.2.7) почти равна 90°. В результате *D* фаза рассеяния, приведенная на рис.2.2.16, дважды переходит через 90°, т.е. соответствует двум резонансам, присутствующим в этой парциальной волне.

3. При энергии 4.259 МэВ с шириной 27 кэВ в *G* волне имеется резонансный уровень и в результатах работ [147] и [148] виден скачек фазы, показанный на рис.2.2.18.

4. Энергия 5.799 МэВ находится около резонанса при 5.47 МэВ с очень большой шириной 2.5 МэВ и моментом $0^+$, и *S* фаза испытывает некоторый подъем.

5. При энергии 5.92 МэВ имеется резонанс с шириной 0.8 МэВ и $J^\pi = 3^-$ для *F* волны и *F* фаза, показанная на рис.2.2.17, при этой энергии имеет величину около $90^o$ (см. также результаты, приведенные в табл.2.2.7 и 2.2.8 для *F* фазы рассеяния).

6. При трех энергиях в области 1.47-2.07 МэВ наблюдается чисто резерфордовское рассеяние и все фазы практически равны нулю.

В дальнейшем, полученные здесь фазы упругого рассеяния [151], используются для построения межкластерных потенциалов и расчетов астрофизического *S*-фактора реакции радиационного $^{12}C(^4He,\gamma)^{16}O$ захвата [152]. Этот процесс, наряду с тройным гелиевым циклом, присутствует в цепочке термоядерных реакций на горячей стадии развития звезд, когда температура внутри звезды достигает сотен миллионов градусов Кельвина [153].

При столь высокой температуре взаимодействующие частицы имеют достаточную энергию для существенного увеличения вероятности прохождения через кулоновский барьер. В таком случае они попадают в область сильного взаимодействия, что ведет к увеличению вклада такой реакции, т.е. ее энергетического выхода в полный энергетический баланс звезды.

# 2.3 Фазовый анализ упругого $p^{12}C$ рассеяния Phase shifts analysis of $p^{12}C$ scattering

В этом параграфе будут рассмотрены методы фазового анализа для частиц со спинами 0 + 1/2 и результаты фазового анализа для $p^{12}C$ системы. По-видимому, самые последние экспериментальные данные, полученные при низких, астрофизических энергиях приведены в работах Института Ядерной Физики (ИЯФ) Национального Ядерного Центра (НЯЦ) Республики Казахстан (РК). Именно они и будут использованы далее в нашем фазовом анализе.

## 2.3.1 Дифференциальные сечения Differential cross sections

При рассмотрении упругого рассеяния в системе частиц со спинами 0 и 1/2 учтем спин-орбитальное расщепление фаз, которое имеет место в ядерных системах типа $N^4He$, $^3H^4He$, $p^{12}C$. В этом случае упругое рассеяние ядерных частиц полностью описывается двумя независимыми спиновыми амплитудами ($A$ и $B$), а сечение представляется в следующем виде [88]:

$$\frac{d\sigma(\theta)}{d\Omega} = |A(\theta)|^2 + |B(\theta)|^2 ,$$

где

$$A(\theta) = f_c(\theta) + \frac{1}{2ik}\sum_{L=0}^{\infty}\{(L+1)S_L^+ + LS_L^- - (2L+1)\}\exp(2i\sigma_L)P_L(\cos\theta) ,$$

$$B(\theta) = \frac{1}{2ik}\sum_{L=0}^{\infty}(S_L^+ - S_L^-)\exp(2i\sigma_L)P_L^1(\cos\theta) .$$

Здесь $S_L^{\pm} = \eta_L^{\pm}\exp(2i\delta_L^{\pm})$ – матрица рассеяния, $\eta^{\pm}_L$ – параметры неупругости, а знаки “±“ соответствуют полному моменту системы $J = L \pm 1/2$, $f_c$ – кулоновская амплитуда, представляется в виде

$$f_c(\theta) = -\left(\frac{\eta}{2k\sin^2(\theta/2)}\right)\exp\{i\eta\ln[\sin^{-2}(\theta/2)] + 2i\sigma_0\} \quad ,$$

где $P_n^m(x)$ – присоединенные полиномы Лежандра, $\eta$ – кулоновский параметр, $\mu$ – приведенная масса частиц, $k$ – волновое число относительного движения частиц во входном канале.

Присоединенные полиномы или функция Лежандра $P_n^m(x)$ представляются в виде

$$P_n^m(x) = (1 - x^2)^{m/2} \frac{d^m P_n(x)}{dx^m} \quad ,$$

и для $m = 1$ их можно вычислять по рекуррентным формулам вида

$$P_{L+1}^1(x) = \frac{(2L+1)x}{L} P_L^1(x) - \frac{L+1}{L} P_{L-1}^1(x)$$

с начальными значениями

$$P_0^1(x) = 0 \quad , \qquad P_1^1(x) = (1 - x^2)^{1/2} \quad , \qquad P_2^1(x) = 3xP_1^1(x) \quad .$$

Величины $P_n^2(x)$, которые встретятся нам далее, вычисляются по другим рекуррентным формулам

$$P_{L+1}^2(x) = \frac{(2L+1)x}{L-1} P_L^2(x) - \frac{L+2}{L-1} P_{L-1}^2(x)$$

с начальными значениями

$$P_0^2(x) = P_1^2(x) = 0 \ , \quad P_2^2 = 3(1-x^2) \ , \quad P_3^2(x) = 5xP_2^2(x).$$

Через приведенные амплитуды $A$ и $B$ можно выразить и векторную поляризацию в упругом рассеянии таких частиц [88]

$$P(\theta) = \frac{2\,\mathrm{Im}(AB^*)}{|A|^2 + |B|^2} \quad .$$

Расписывая выражение для амплитуды $B(\Theta)$, получим

$$\mathrm{Re}\,B = \frac{1}{2k}\sum_{\mathrm{L}=0}^{\infty}[a\cdot\sin(2\sigma_{\mathrm{L}}) + b\cdot\cos(2\sigma_{\mathrm{L}})]P_{\mathrm{L}}^1(x) \quad ,$$

$$\mathrm{Im}\,B = \frac{1}{2k}\sum_{\mathrm{L}=0}^{\infty}[b\cdot\sin(2\sigma_{\mathrm{L}}) - a\cdot\cos(2\sigma_{\mathrm{L}})]P_{\mathrm{L}}^1(x) \quad ,$$

где

$$a = \eta_{\mathrm{L}}^+\cos(2\delta_{\mathrm{L}}^+) - \eta_{\mathrm{L}}^-\cos(2\delta_{\mathrm{L}}^-) \quad ,$$

$$b = \eta_{\mathrm{L}}^+\sin(2\delta_{\mathrm{L}}^+) - \eta_{\mathrm{L}}^-\sin(2\delta_{\mathrm{L}}^-) \quad .$$

Аналогичным способом, для амплитуды $A(\Theta)$ можно найти следующую форму записи [154]

$$\mathrm{Re}\,A = \mathrm{Re}\,f_{\mathrm{c}} + \frac{1}{2k}\sum_{\mathrm{L}=0}^{\infty}[c\cdot\sin(2\sigma_{\mathrm{L}}) + d\cdot\cos(2\sigma_{\mathrm{L}})]P_{\mathrm{L}}(x) \quad ,$$

$$\mathrm{Im}\,A = \mathrm{Im}\,f_{\mathrm{c}} + \frac{1}{2k}\sum_{\mathrm{L}=0}^{\infty}[d\cdot\sin(2\sigma_{\mathrm{L}}) - c\cdot\cos(2\sigma_{\mathrm{L}})]P_{\mathrm{L}}(x) \quad ,$$

где

$$c = (L+1)\eta_L^+\cos(2\delta_L^+) + L\eta_L^-\cos(2\delta_L^-) - (2L+1) \quad ,$$

$$d = (L+1)\eta_L^+\sin(2\delta_L^+) + L\eta_L^-\sin(2\delta_L^-) \quad .$$

Для полного сечения упругого рассеяния можно получить [88]

$$\sigma_s = \frac{\pi}{k^2}\sum_L \left[ (L+1)\left|1-S_L^+\right|^2 + L\left|1-S_L^-\right|^2 \right]$$

или

$$\sigma_s = \frac{4\pi}{k^2}\sum_L \left\{ (L+1)[\eta_L^+\sin\delta_L^+]^2 + L[\eta_L^-\sin\delta_L^-]^2 \right\} \quad .$$

Все эти выражения использовались далее для выполнения фазового анализа $p^{12}C$ упругого рассеяния при энергии до 1.1 МэВ [155].

### 2.3.2 Контроль компьютерной программы Computer program check

Текст нашей компьютерной программы для расчета полных и дифференциальных сечений упругого рассеяния частиц с полуцелым спином, которая использовалась для выполнения соответствующего фазового анализа, полностью приведен далее и протестирован на упругом рассеяние в $p^4He$ системе. Здесь приведен только один вариант контрольного счета по этой программе для $p^4He$ рассеяния, в сравнении с данными из работы [156], где выполнен фазовый анализ для энергии 9.89 МэВ, получены положительные $D$ фазы и среднее по всем точкам значение $\chi^2 = 0.60$.

В анализе [156] использованы 22 точки по сечениям из работы [157] при энергии 9.954 МэВ (в [156] не указано, какие именно 22 точки были взяты из 24-х, приведенных в работе [157]) и несколько точек по поляризациям из работ [156,158]. В последнем случае, по-видимому, использовано 10 данных при 8-ми углах $46.5^0$, $55.9^0$, $56.2^0$, $73.5^0$, $89.7^0$, $99.8^0$, $114.3^0$, $128.3^0$ и энергиях 9.89, 9.84, 9.82 МэВ.

Фазы из работы [156] приведены в табл.2.3.1, а среднее $\chi^2_\sigma$ только для дифференциальных сечений по нашей программе с учетом 24 точек из [157] (энергия задавалась равной 9.954 МэВ) и этими фазами получается равным 0.586. Результаты этих расчетов показаны в табл.2.3.2.

Табл.2.3.1. Фазы (в град.) упругого $p^4He$ рассеяния из работы [156].

| $E$, МэВ | $S_0$ | $P_{3/2}$ | $P_{1/2}$ | $D_{5/2}$ | $D_{3/2}$ |
|---|---|---|---|---|---|
| 9.954 | $119,3^{+2.0}_{-1.8}$ | $112,4^{+3.5}_{-5.2}$ | $65,7^{+2.7}_{-3.2}$ | $5,3^{+1.6}_{-2.5}$ | $3,7^{+1.6}_{-2.8}$ |

Табл.2.3.2. Дифференциальные сечения упругого $p^4He$ рассеяния. Здесь $\theta^\circ$ – угол рассеяния, $\sigma_e$ – экспериментальные и $\sigma_t$ – расчетные сечения.

| $\theta^\circ$ | $\sigma_e$, мбн/ст | $\sigma_t$, мбн/ст | $\chi^2_i$ | $\theta^0$ | $\sigma_e$, мбн/ст | $\sigma_t$, мбн/ст | $\chi^2_i$ |
|---|---|---|---|---|---|---|---|
| 25.10 | 371.00 | 366.85 | 0.31 | 109.90 | 21.00 | 20.70 | 0.51 |
| 30.89 | 339.00 | 331.54 | 1.21 | 120.60 | 23.00 | 22.59 | 0.79 |
| 35.07 | 305.00 | 308.40 | 0.31 | 122.80 | 24.50 | 24.19 | 0.40 |
| 49.03 | 232.00 | 230.61 | 0.09 | 130.13 | 31.90 | 31.91 | 0.00 |
| 54.70 | 205.00 | 199.10 | 2.07 | 130.90 | 33.20 | 32.90 | 0.21 |
| 60.00 | 176.00 | 170.56 | 2.39 | 134.87 | 37.80 | 38.44 | 0.71 |
| 70.10 | 124.00 | 120.59 | 1.89 | 140.80 | 47.30 | 47.69 | 0.17 |
| 80.00 | 82.00 | 79.77 | 1.85 | 145.00 | 54.00 | 54.62 | 0.33 |
| 90.00 | 49.20 | 48.82 | 0.15 | 149.40 | 61.60 | 61.88 | 0.05 |
| 94.07 | 39.10 | 39.44 | 0.19 | 154.90 | 70.60 | 70.54 | 0.00 |
| 102.17 | 26.20 | 26.39 | 0.13 | 160.00 | 78.40 | 77.71 | 0.19 |
| 106.90 | 22.00 | 22.15 | 0.12 | 164.40 | 83.00 | 82.94 | 0.00 |

Для 10-ти экспериментальных данных из работ [156,158] по поляризациям при энергиях 9.82÷9.89 МэВ при восьми углах рассеяния с фазами из [156], можно получить $\chi^2_P$ = 0.589 (энергия, по-прежнему, задается равной 9.954 МэВ). Эти результаты приведены в табл.2.3.3.

Табл.2.3.3. Поляризации в упругом p$^4$He рассеянии. Здесь $\theta^\circ$ – угол рассеяния, $P_e$ – экспериментальные поляризации, $\Delta P_e$ – экспериментальные ошибки для поляризаций и $P_t$ – расчетные поляризации.

| $\theta^\circ$ | $P_e$, % | $\Delta P_e$ | $P_t$, % | $\chi^2_i$ |
|---|---|---|---|---|
| 46.50 | -32.30 | 2.10 | -33.11 | 0.15 |
| 55.90 | -41.30 | 2.20 | -42.50 | 0.30 |
| 56.20 | -44.40 | 0.90 | -42.81 | 3.11 |
| 73.50 | -62.60 | 3.00 | -62.84 | 0.01 |
| 73.50 | -64.80 | 1.90 | -62.84 | 1.06 |
| 89.70 | -76.10 | 3.60 | -76.33 | 0.01 |
| 89.70 | -75.50 | 2.40 | -76.33 | 0.12 |
| 99.80 | -59.30 | 2.50 | -58.55 | 0.09 |
| 114.30 | 48.20 | 3.20 | 51.03 | 0.78 |
| 128.30 | 99.40 | 3.30 | 97.66 | 0.28 |

Если усреднить $\chi^2$ по всем точкам (24 + 10 = 34), т.е. использовать более общее выражение для $\chi^2$

$$\chi^2 = \frac{1}{(N_\sigma + N_P)}\left\{\sum_{i=1}^{N}\left[\frac{\sigma_i^t - \sigma_i^e}{\Delta\sigma_i^e}\right]^2 + \sum_{i=1}^{N}\left[\frac{P_i^t - P_i^e}{\Delta P_i^e}\right]^2\right\} = ,$$
$$= \frac{1}{(N_\sigma + N_P)}\{\chi^2_\sigma + \chi^2_P\}$$

то получается величина $\chi^2 = 0.5875 \approx 0.59$ в хорошем согласии с результатами работы [156]. Здесь $N_\sigma$ и $N_P$ – число данных по сечениям (24 точки) и поляризациям (10 точек), $\sigma^e$, $P^e$, $\sigma^t$, $P^t$ – экспериментальные и теоретические значения сечений и поляризаций, $\Delta\sigma$ и $\Delta P$ – их ошибки.

Если выполнить дополнительную минимизацию $\chi^2$ по нашей программе, то для $\chi^2_\sigma$ по сечениям получим 0.576, для поляризаций $\chi^2_p = 0.561$ и среднее $\chi^2 = 0.572 \approx 0.57$ при следующих значениях фаз

$S_0 = 119.01^\circ$ , $P_{3/2} = 112.25^\circ$ , $P_{1/2} = 65.39^\circ$ , $D_{5/2} = 5.24^\circ$ , $D_{3/2} = 3.63^\circ$ ,

которые полностью находятся в полосе ошибок, приведенных в [156] и показаны в табл.2.3.1.

Таким образом, представленная далее программа позволяет получить результаты, хорошо совпадающие с ранее выполненным анализом. Далее она тестировалась по фазовому анализу, проведенному в других работах при низких энергиях, но уже непосредственно для упругого рассеяния в $p^{12}C$ системе.

Ранее фазовый анализ функций возбуждения для упругого $p^{12}C$ рассеяния, измеренных в [159], при энергиях в области 400÷1300 кэВ (л.с.) и углах 106°÷169°, был выполнен в работе [160], где получено, что, например, при $E_{лаб} = 900$ кэВ $S$ фаза должна лежать в области 153°÷154°. С теми же экспериментальными данными нами получено значение 152.7°. Для получения этого результата использовались сечения рассеяния из функций возбуждения работы [159] при энергиях 866÷900 кэВ. Результаты наших расчетов $\sigma_t$ в сравнении с экспериментальными данными $\sigma_e$ приведены в табл.2.3.4.

Табл.2.3.4. Сравнение теоретических и экспериментальных сечений $p^{12}C$ упругого рассеяния при энергии 900 кэВ.

| $\theta^\circ$ | $\sigma_e$, мбн/ст | $\sigma_t$, мбн/ст | $\chi^2_i$ |
|---|---|---|---|
| 106 | 341 | 341.5 | 1.90E-04 |
| 127 | 280 | 282.1 | 5.76E-03 |
| 148 | 241 | 251.2 | 1.80E-01 |
| 169 | 250 | 237.5 | 2.50E-01 |

В последнем столбце таблицы даны парциальные значения $\chi^2_i$ на каждую точку при 10% ошибках в экспериментальных сечениях, а для среднего по всем экспериментальным точкам $\chi^2$ была получена величина 0.11.

При энергии 751 кэВ (л.с.) в работе [160] для *S* фазы были найдены значения в интервале 155°÷157°. Результаты, полученные нами для этой энергии, приведены в табл.2.3.5. Использовались данные по сечениям из функций возбуждения в диапазоне энергий 749÷754 кэВ и для *S* фазы найдено 156.8° при среднем $\chi^2 = 0.30$.

Табл.2.3.5. Сравнение теоретических и экспериментальных сечений $p^{12}C$ упругого рассеяния при энергии 750 кэВ.

| $\theta^0$ | $\sigma_e$, мбн/ст | $\sigma_t$, мбн/ст | $\chi^2_i$ |
|---|---|---|---|
| 106 | 428 | 428.3 | 3.44E-05 |
| 127 | 334 | 342.8 | 6.91E-02 |
| 148 | 282 | 299.1 | 3.66E-01 |
| 169 | 307 | 279.9 | 7.82E-01 |

Таким образом, по нашей программе, при двух энергиях упругого $p^{12}C$ рассеяния получены фазы, совпадающие с результатами анализа, выполненного на основе функций возбуждения в работе [160].

## 2.3.3 Фазовый анализ упругого $p^{12}C$ рассеяния Phase shifts analysis of $p^{12}C$ scattering

Приведенные выше контрольные счеты хорошо согласуются с более ранними результатами, поэтому по нашей программе, был выполнен фазовый анализ [161] новых экспериментальных данных по дифференциальным сечениям $p^{12}C$ рассеяния в диапазоне энергий 230÷1200 кэВ (л.с.) [162]. Результаты этого анализа приведены в табл.2.3.6 и представлены точками на рис.2.3.1 в сравнении с данными работы [160], которые показаны штриховой линией.

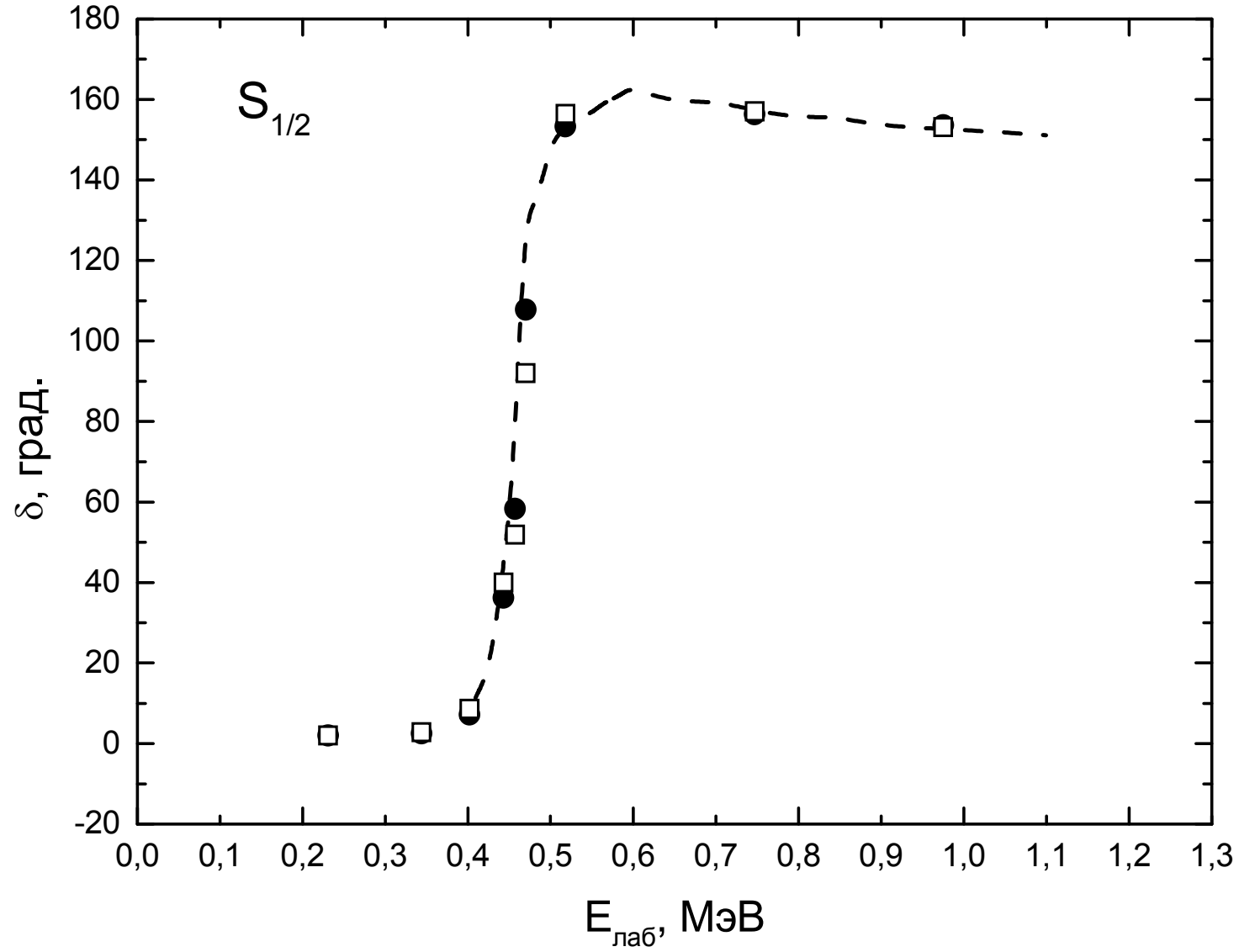

Рис.2.3.1. $^2S$ фаза $p^{12}C$ рассеяния при низких энергиях. Точки – результаты фазового анализа для *S* фазы с учетом в фазовом анализе только *S* волны, открытые квадраты – результаты фазового анализа для *S* фазы с учетом *S* и *P* волн, штриховая кривая – результаты [160].

Табл.2.3.6. Результаты фазового анализа $p^{12}C$ упругого рассеяния при низких энергиях с учетом только *S* фазы.

| $E_{цм}$, кэВ | $S_{1/2}$, град. | $\chi^2$ |
|---|---|---|
| 213 | 2.0 | 1.35 |
| 317 | 2.5 | 0.31 |
| 371 | 7.2 | 0.51 |
| 409 | 36.2 | 0.98 |
| 422 | 58.2 | 3.69 |
| 434 | 107.8 | 0.78 |
| 478 | 153.3 | 2.56 |
| 689 | 156.3 | 2.79 |
| 900 | 153.6 | 2.55 |
| 1110 | 149.9 | 1.77 |

На рис.2.3.2а,б,в точками представлены экспериментальные дифференциальные сечения в области резонанса при 457 кэВ (л.с.). Результаты расчета этих сечений на основе формулы Резерфорда (точечная кривая), а также сечения, полученные из нашего фазового анализа (непрерывная линия), который учитывает только *S* фазу и анализа при учете *S* и *P* фаз рассеяния (пунктирная кривая). Из рисунков видно, что в области резонанса не удается хорошо описать сечение только на основе одной *S* фазы.

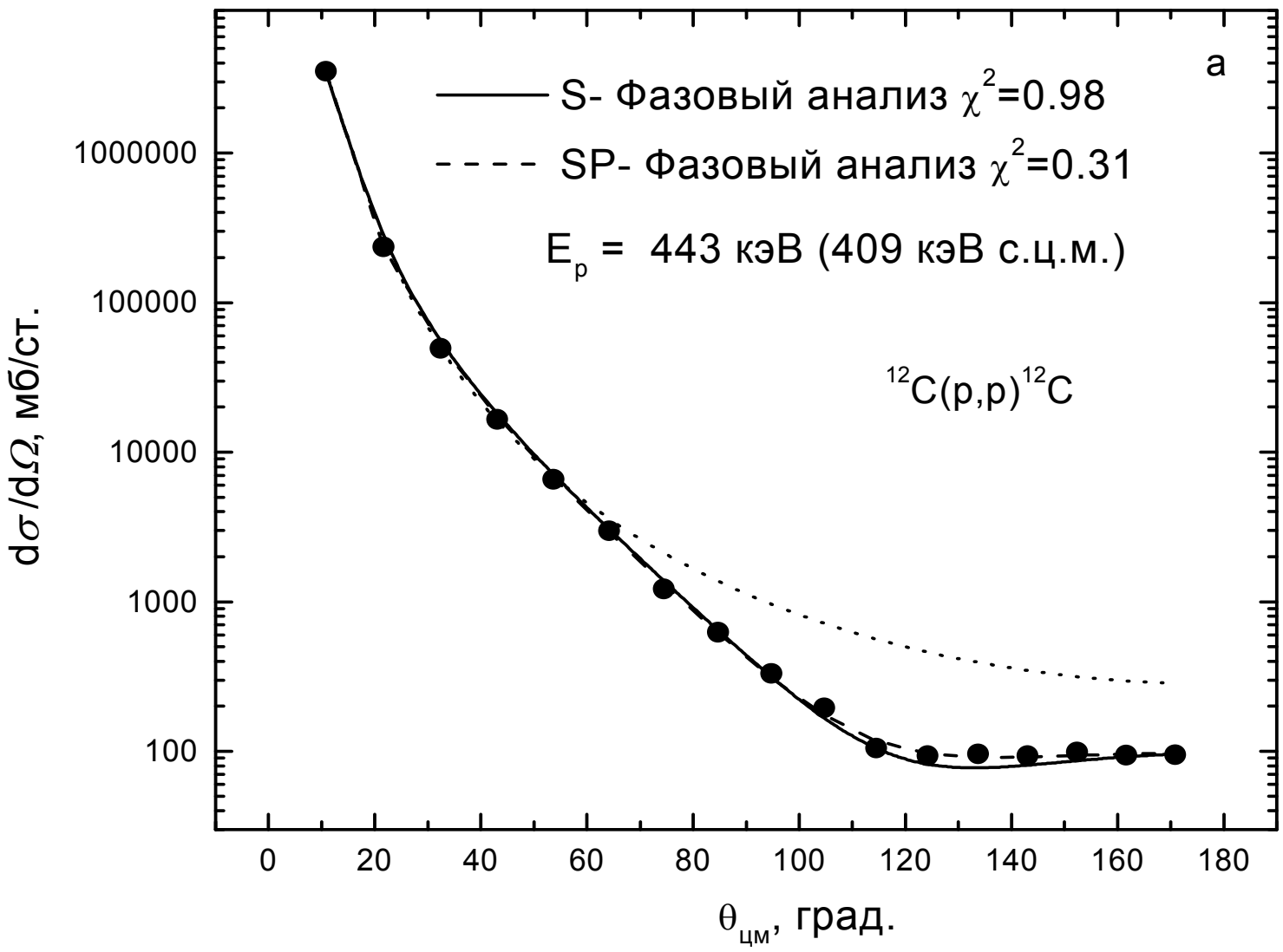

Рис.2.3.2а. Дифференциальные сечения упругого $p^{12}C$ рассеяния. Непрерывная кривая получена на основе фазового анализа с учетом только *S* волны, точечная кривая – резерфордовское рассеяние, пунктир – результаты фазового анализа с учетом *S* и *P* волн, точки – эксперимент [162].

Заметную роль начинает играть *P* волна, представленная на рис.2.3.3, учет которой заметно улучшает описание экспериментальных данных. При резонансной энергии 457 кэВ (л.с.), сечения для которой показаны на рис.2.3.2б, учет *P* волны уменьшает величину $\chi^2$ с 3.69 до 0.79.

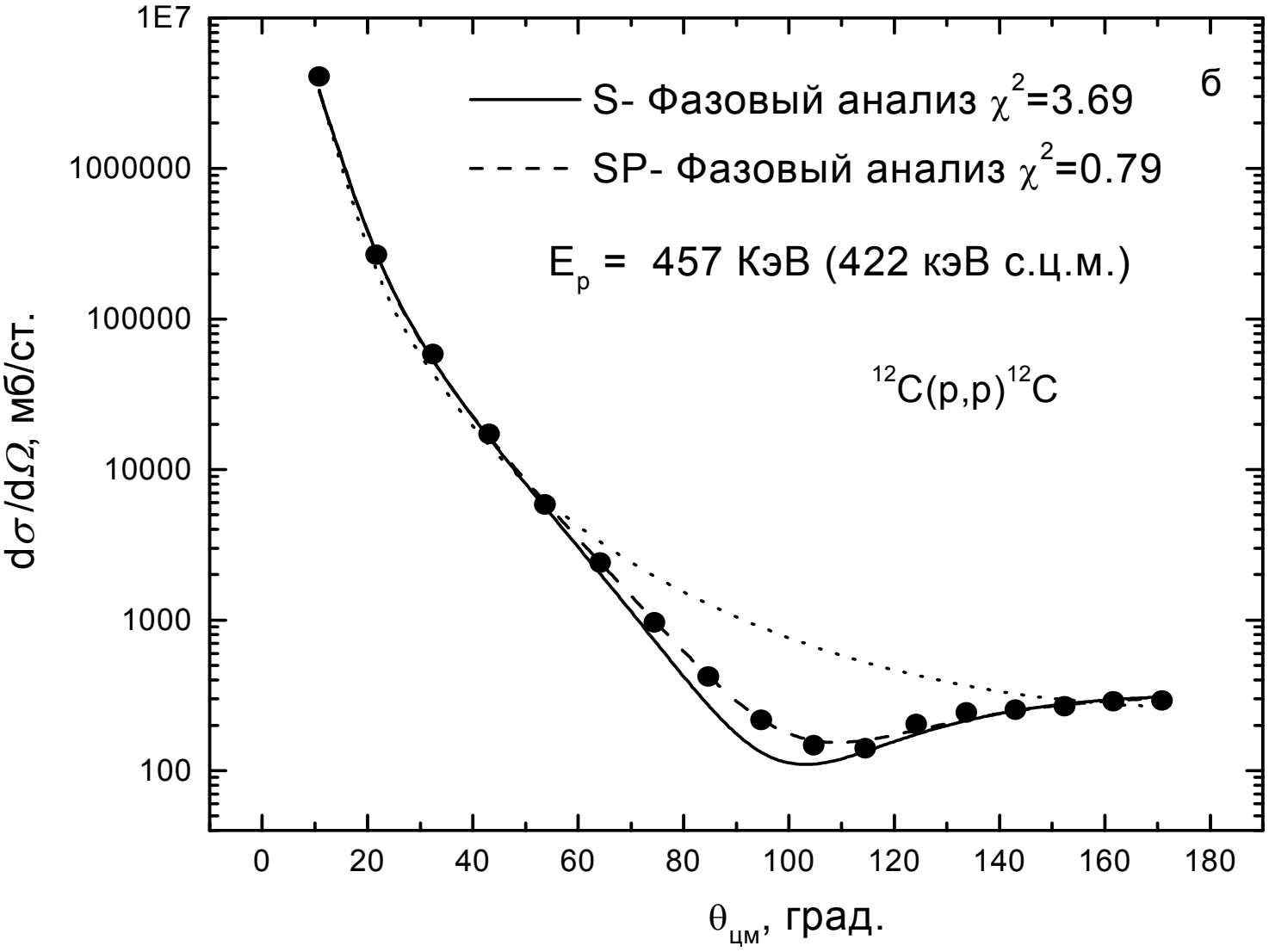

Рис.2.3.2б. Дифференциальные сечения $p^{12}C$ рассеяния. Подписи, как на рис.2.3.2а.

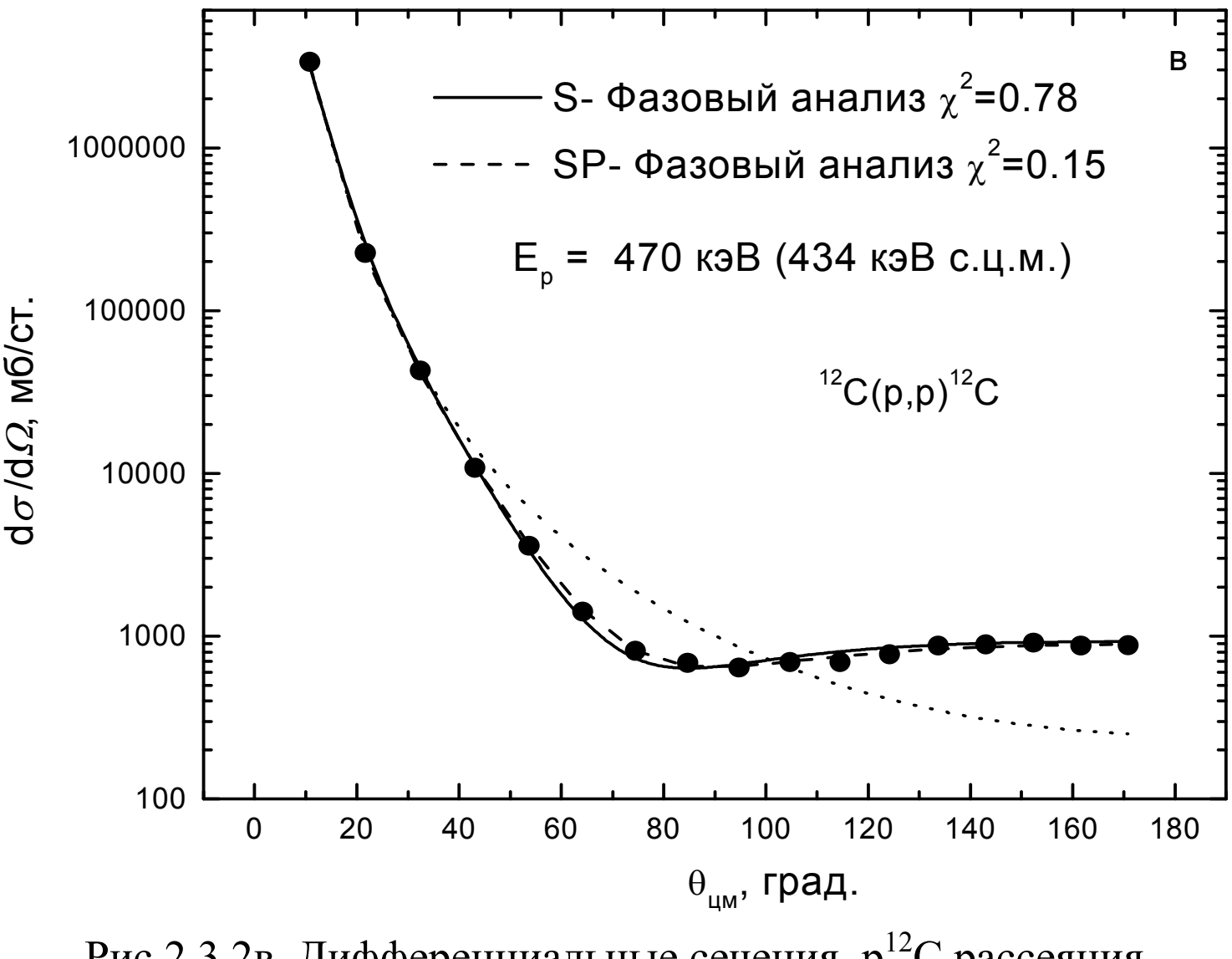

Рис.2.3.2в. Дифференциальные сечения $p^{12}C$ рассеяния. Подписи, как на рис.2.3.2а.

На рис.2.3.3 видно, что при низких энергиях $P_{1/2}$ фаза идет выше, чем $P_{3/2}$, но при энергии порядка 1.2 МэВ они пересекаются и далее $P_{3/2}$ идет выше в отрицательной области углов [163,164]. Величина $S$ фазы при учете $P$ волны практически не меняется и ее форма показана на рис.2.3.1 открытыми квадратами. Учет $D$ волны в фазовом анализе приводит к ее величине порядка одного градуса в области резонанса и практически не влияет на поведение расчетных дифференциальных сечений.

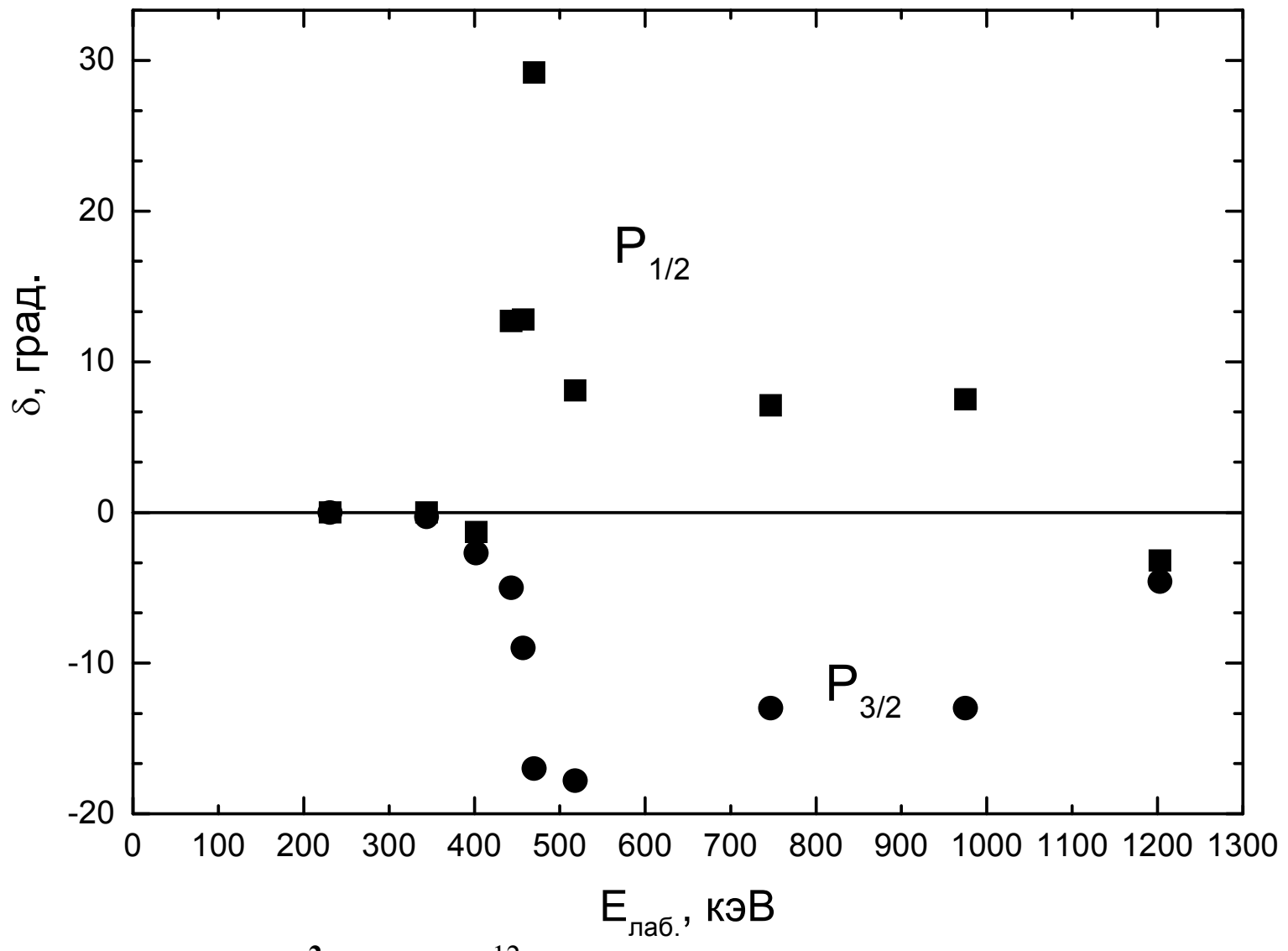

Рис.2.3.3. $^2P$ фазы $p^{12}C$ рассеяния при низких энергиях. Точки – $P_{3/2}$ и квадраты – $P_{1/2}$ фазы, полученные в результате фазового анализа с учетом $S$ и $P$ волн.

Полные сечения упругого рассеяния представлены точками на рис.2.3.4. Они рассчитаны на основе извлеченных из экспериментальных дифференциальных сечений [162] $S$ фаз рассеяния. Кружками представляют полные сечения, полученные из фазового анализа работы [160]. На рис.2.3.4 в области энергий 200÷300 кэВ (л.с.) наблюдается некоторое плато.

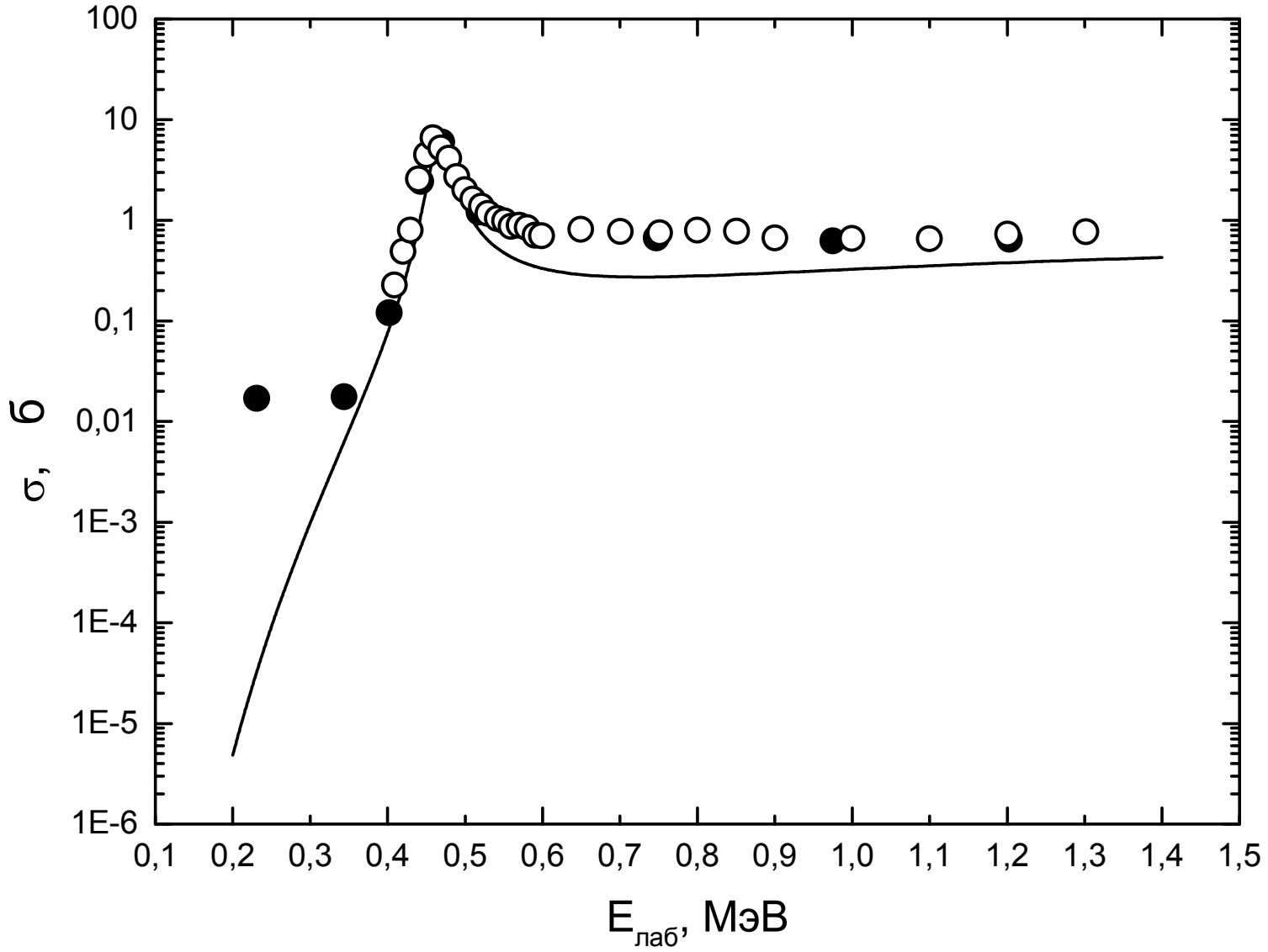

Рис.2.3.4. Полные сечения упругого $p^{12}C$ рассеяния.
Точки – полные сечения, найденные из фаз рассеяния работ [162], кружки – тоже для работ [159,160].

В данный момент не понятно, вызвано оно экспериментальными неточностями, непредвиденными ошибками фазового анализа или действительно существует при этих энергиях. Для выяснения этого вопроса требуются новые измерения угловых распределений упругого $p^{12}C$ рассеяния в области энергий от 100÷150 до 300÷400 кэВ с шагом по энергии порядка 50 кэВ или функций возбуждения при разных углах [155].

В заключение этого параграфа заметим, что в данном случае использовалось несколько другое, чем обычно, значение $\hbar^2/m_0 = 41.80159$ МэВ·Фм$^2$, которое было получено с несколько другими значениями фундаментальных констант $\hbar^2$ и $m_0$, где последняя обозначает атомную единицу массы. Однако, как будет видно далее в п.п.2.3.5, это различие практически не сказывается на полученных результатах по фазам рассеяния.

## 2.3.4 Фазовый анализ упругого $n^{12}C$ рассеяния Phase shifts analysis of $n^{12}C$ scattering

В данном параграфе продолжим рассмотрение фазового анализа упругого рассеяния на легких ядрах и на основе измерений [165] выполним фазовый анализ $n^{12}C$ рассеяния. Нам не удалось найти результаты ранее проведенного фазового анализа для $n^{12}C$ системы даже при низких энергиях, хотя они должны заметно отличаться от аналогичного анализа $p^{12}C$ рассеяния [161]. Причиной таких различий является существенное отличие структуры спектров уровней ядер $^{13}C$ и $^{13}N$ при энергиях до 1.5÷2 МэВ выше порогов $n^{12}C$ и $p^{12}C$ кластерных каналов [166].

В частности, в упругом $p^{12}C$ рассеянии имеется надпороговый резонансный уровень ядра $^{13}N$ при энергии 0.42 МэВ с $J^{\pi} = 1/2^{+}$, который приводит к резонансу $^{2}S_{1/2}$ фазы рассеяния, что и было показано в нашей работе [161] в сравнении с более ранними результатами фазового анализа [159,160]. В случае $n^{12}C$ рассеяния никаких резонансов в спектрах ядра $^{13}C$ вплоть до энергии 1.9 МэВ не наблюдается. Именно поэтому представляется интересным выяснить форму фаз упругого $n^{12}C$ рассеяния, извлеченных из экспериментальных данных при низких и, в первую очередь, астрофизических энергиях.

Тем более что реакция радиационного захвата в $n^{12}C$ канале, по-видимому, при любых энергиях входит в основную цепочку термоядерных реакций первичного нуклеосинтеза [167]

$$^{7}\mathrm{Li}(n,\gamma)^{8}\mathrm{Li}(^{4}\mathrm{He},n)^{11}\mathrm{B}(n,\gamma)^{12}\mathrm{B}(\beta^{-})^{12}\mathrm{C}(n,\gamma)^{13}\mathrm{C}\ldots ,$$

который и привел, в конечном итоге, к начальному формированию Солнца, звезд и всей нашей Вселенной [3,9,15,168, 169].

Извлекаемые из экспериментальных дифференциальных сечений фазы упругого рассеяния позволяют построить потенциалы взаимодействия двух частиц в непрерывном спектре и выполнить расчеты некоторых характеристик их взаи-модействия в процессах рассеяния и реакций. Например, это могут быть астрофизические *S*-факторы [6] или полные сечения реакций [170], в том числе, радиационного $n^{12}C$ захвата в астрофизической области энергий, которые обычно рассматривается в наших предыдущих работах [24,37,168].

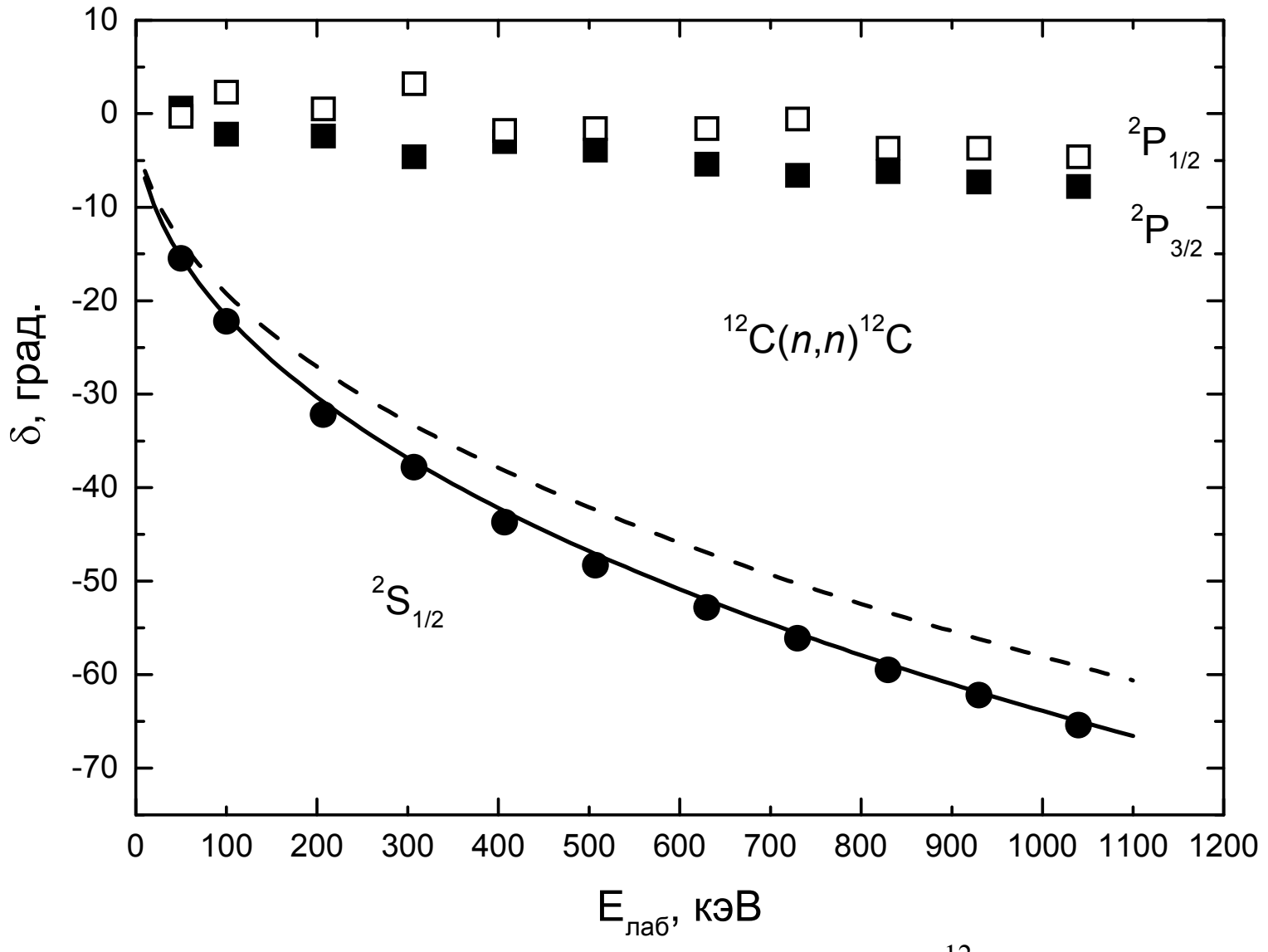

Рис.2.3.5. Результаты фазового анализа упругого $n^{12}C$ рассеяния для варианта с отрицательными значениями $^2S$ фаз.
Точки – результаты фазового анализа для $^2S_{1/2}$ фазы, открытые квадраты – результаты фазового анализа для $^2P_{1/2}$ фазы, черные квадраты – результаты фазового анализа для $^2P_{3/2}$ фазы. Кривые – результаты расчета с разными потенциалами, которые объяснены в тексте.

На основе описанных выше методов был выполнен фазовый анализ известных экспериментальных данных по дифференциальным сечениям упругого $n^{12}C$ рассеяния в диапазоне энергий 50÷1040 кэВ (л.с.) [165]. Результаты нашего фазового анализа для *S* и *P* фаз рассеяния представлены на рис.2.3.5 и приведены в табл.2.3.7. При выполнении анализа величина $\chi^2$ вычислялась при 10% экспериментальных ошиб-ках дифференциальных сечений [165]. В последнем столбце табл.2.3.7 приведено значение $\chi^2$ для каждой рассмотренной энергии.

Табл.2.3.7. Результаты фазового анализа (в град.) упругого $n^{12}C$ рассеяния при низких энергиях для отрицательных значений $^2S$ фаз.

| $E_{лаб}$, кэВ | $^2S_{1/2}$ | $^2P_{3/2}$ | $^2P_{1/2}$ | $\chi^2$ |
|---|---|---|---|---|
| 50 | -15.5 | 0.6 | -0.3 | 0.023 |
| 100 | -22.2 | -2.2 | 2.3 | 0.016 |
| 157 | -28.3 | -1.0 | -1.0 | 0.007 |
| 207 | -32.2 | -2.4 | 0.5 | 0.014 |
| 257 | -35.6 | -2.5 | -0.4 | 0.009 |
| 307 | -37.8 | -4.6 | 3.3 | 0.033 |
| 357 | -41.3 | -2.7 | -1.3 | 0.007 |
| 407 | -43.7 | -3.0 | -1.8 | 0.027 |
| 457 | -45.3 | -5.1 | 2.5 | 0.029 |
| 507 | -48.3 | -3.9 | -1.6 | 0.019 |
| 530 | -49.1 | -3.6 | -3.4 | 0.020 |
| 630 | -52.8 | -5.4 | -1.6 | 0.015 |
| 730 | -56.0 | -6.6 | -0.6 | 0.031 |
| 830 | -59.5 | -6.2 | -3.7 | 0.044 |
| 930 | -62.1 | -7.3 | -3.7 | 0.042 |
| 1040 | -65.4 | -7.7 | -4.6 | 0.096 |

Из рис.2.3.5 видно, что все $^2P$ фазы имеют значения по модулю не более 10°, но их влияние уменьшает величину $\chi^2$ примерно на порядок. Величина $^2S$ фазы плавно спадает и, как будет видно далее, хорошо описывается гауссовым потенциалом во всей области низких энергий, которые были включены в анализ.

На рис.2.3.6 точками представлены дифференциальные сечения для некоторых рассмотренных энергий из работы [165] и сечения, полученные из нашего фазового анализа (непрерывная линия), который учитывает $S$ и $P$ фазы, приведенные в табл.2.3.7. Учет полученных небольших значений $^2P$ фаз позволил заметно улучшить качество описания дифференциальных сечений даже при самых низких энергиях.

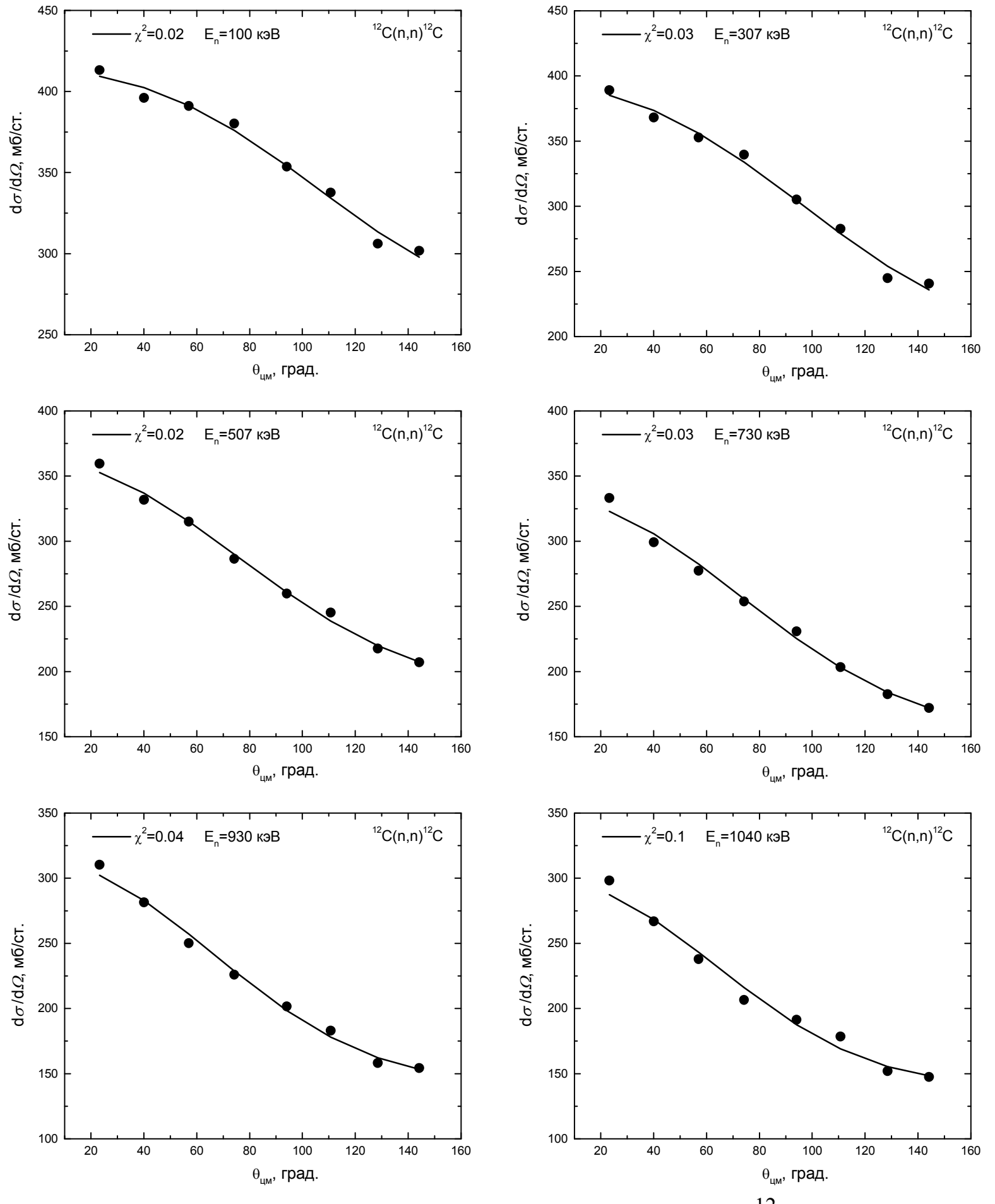

Рис.2.3.6. Дифференциальные сечения упругого $n^{12}C$ рассеяния. Непрерывная кривая – результаты расчета сечений с найденными фазами, точки – экспериментальные дифференциальные сечения рассеяния [165].

На рис.2.3.7 и в табл.2.3.8 приведен еще один вариант для набора фаз, в котором $^2S$ фаза имеет положительные зна-чения, также как и величины обеих $^2P$ фаз. Значения $\chi^2$ для этого варианта фаз практически не отличаются от первого набора из табл.2.3.7, но $^2S$ фаза имеет форму близкую к резонансной, хотя в спектрах ядра $^{13}$C резонансов не наблюдается [166]. Поэтому второй набор фаз рассеяния, по-видимому, не соответствует реальной ситуации и приводится здесь лишь для демонстрации имеющейся неоднозначности фазовых наборов.

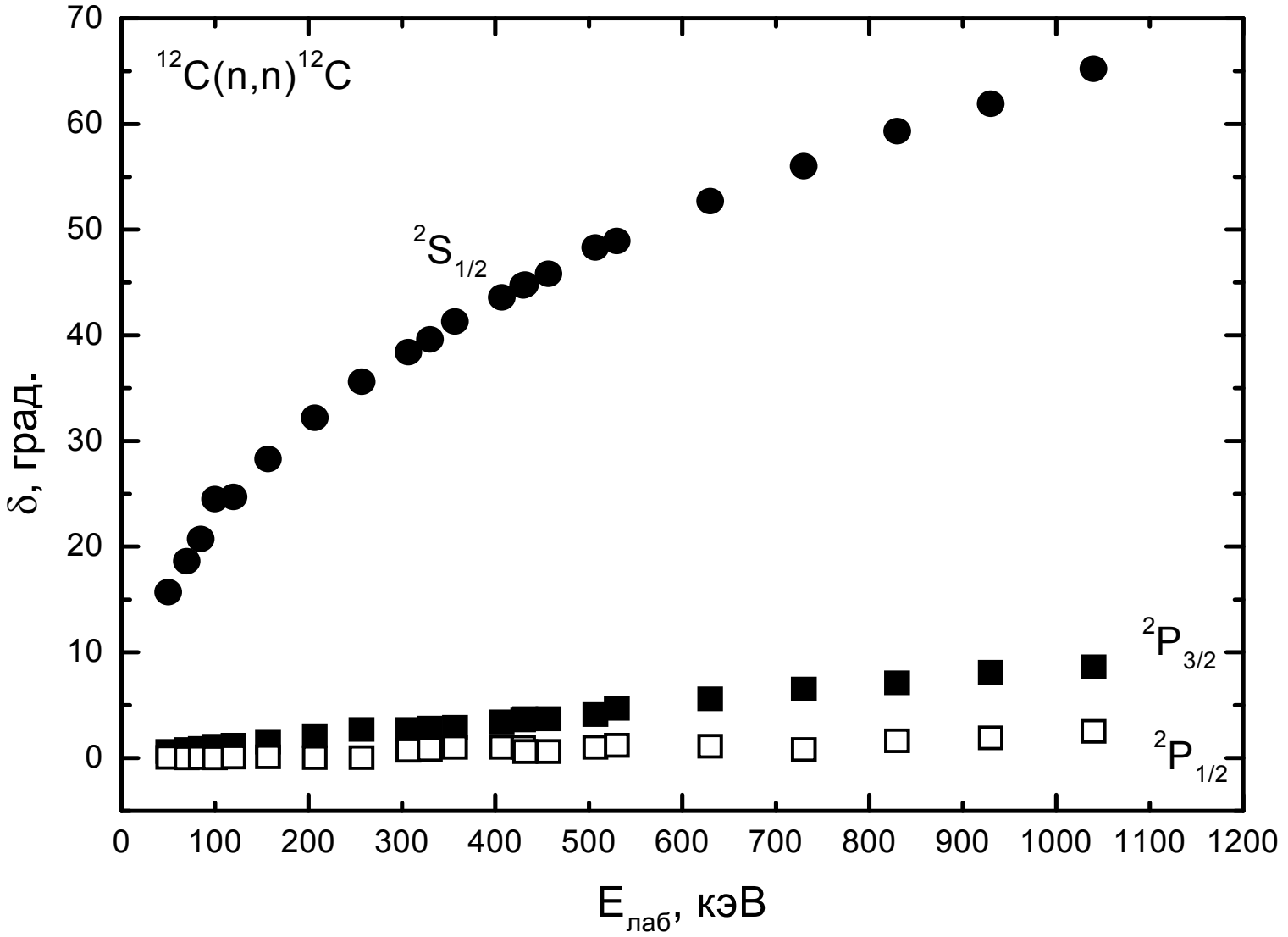

Рис.2.3.7. Результаты фазового анализа упругого n$^{12}$C рассеяния для варианта с положительными значениями $^2S$ фаз.
Точки – результаты фазового анализа для $^2S_{1/2}$ фазы, открытые квадраты – результаты фазового анализа для $^2P_{1/2}$ фазы, черные квадраты – результаты фазового анализа для $^2P_{3/2}$ фазы.

Для исключения имеющейся неоднозначности требуется использовать определенные физические принципы, например, информацию о структуре спектров уровней ядра [166]. Такая информация позволяет окончательно выбрать реальные наборы фаз упругого рассеяния, варианты которых могут быть получены в таком анализе. Кроме того, можно исполь-зовать потенциальное описание процесса рассеяния, которое, как это будет показано далее, позволяет оценить разницу фаз для рассеяния нейтронов и протонов на одном ядре в одном потенциале, т.е. при изменении только кулоновского взаимодействия.

Табл.2.3.8. Результаты фазового анализа (в град.) упругого $n^{12}C$ рассеяния при низких энергиях для положительных значений $^2S$ фаз.

| $E_{лаб}$, кэВ | $^2S_{1/2}$ | $^2P_{3/2}$ | $^2P_{1/2}$ | $\chi^2$ |
|---|---|---|---|---|
| 50 | 15.7 | 0.6 | 0.06 | 0.036 |
| 100 | 24.5 | 1.1 | 0.0 | 0.029 |
| 157 | 28.3 | 1.5 | 0.1 | 0.007 |
| 207 | 32.3 | 2.1 | 0.0 | 0.014 |
| 257 | 35.6 | 2.7 | 0.0 | 0.009 |
| 307 | 38.4 | 2.7 | 0.1 | 0.05 |
| 357 | 41.3 | 2.9 | 1.0 | 0.007 |
| 407 | 43.6 | 3.4 | 1.0 | 0.028 |
| 457 | 45.8 | 3.8 | 0.7 | 0.037 |
| 507 | 48.3 | 4.2 | 1.0 | 0.019 |
| 530 | 50.0 | 4.7 | 1.2 | 0.021 |
| 630 | 52.8 | 5.6 | 1.1 | 0.015 |
| 730 | 56.1 | 6.5 | 0.8 | 0.031 |
| 830 | 59.3 | 7.1 | 1.6 | 0.045 |
| 930 | 61.9 | 8.1 | 1.9 | 0.044 |
| 1040 | 65.2 | 8.6 | 2.5 | 0.098 |

Например, на рис.2.3.5 пунктирной кривой показана $^2S$ фаза потенциала с точечным кулоновским членом, имеющего простой гауссов вид (1.2.2) и параметры

$V_0$ = -102.05 МэВ, $\alpha$ = 0.195 Фм$^{-2}$.

Он был получен ранее для $p^{12}C$ рассеяния при описании астрофизического $S$-фактора радиационного $p^{12}C$ захвата в нашей работе [171]. Такой потенциал правильно описывал резонансную форму $^2S$ фазы $p^{12}C$ рассеяния и приводил к приемлемому согласию с экспериментальными данными при расчетах астрофизического $S$-фактора.

На рис.2.3.5 расчеты $^2S$ фазы проводились для приведенного выше $p^{12}C$ потенциала с выключенным кулоновским взаимодействием, т.е. для $n^{12}C$ рассеяния. Из рис. 2.3.5 видно, что $^2S$ фаза процесса $n^{12}C$ рассеяния не содержит резонанса. Это находится в полном согласии с наблюдаемыми спектрами ядра $^{13}C$ [166]. Кроме того видно, что результаты расчета $^2S$ фазы вполне приемлемо передают результаты фазового $n^{12}C$ анализа, особенно при самых низких энергиях.

Для более точного описания полученных данных по фазам рассеяния требуется несколько изменить глубину потенциала при той же геометрии и задать -97.0 МэВ, что примерно на 5% отличается от первоначальных параметров. Результаты расчета $^2S$ фазы с таким потенциалом приведены на рис. 2.3.5 непрерывной кривой, которая точно передает положение точек извлеченной из эксперимента $^2S$ фазы упругого $n^{12}C$ рассеяния.

Такой потенциал, как и предыдущий, по-прежнему содержит одно связанное запрещенное состояние, наличие которого следует из анализа структуры запрещенных и разрешенных связанных состояний в $N^{12}C$ системе, проведенного в [168,171]. Несколько округляя ширину потенциала $\alpha$, получим

$$V_0 = -99.0 \text{ МэВ}, \alpha = 0.2 \text{ Фм}^{-2} .$$

Фаза для такого варианта потенциала не отличается от приведенной на рис.2.3.5 непрерывной кривой.

Таким образом, в результате проведенного фазового анализа экспериментальных дифференциальных сечений получен набор фаз упругого $n^{12}C$ рассеяния при энергиях до 1.0 МэВ, который согласуется со спектрами уровней ядра $^{13}C$ [166] и потенциальными расчетами фаз процесса рассеяния, выполненными на основе предложенного ранее $p^{12}C$ потенциала взаимодействия. Полученный набор фаз рассеяния по-зволяет приемлемо описать величину и форму дифференциальных сечений угловых распределений упругого $n^{12}C$ рассеяния при низких энергиях, которые могут представлять интерес в некоторых задачах ядерной астрофизики [168].

Еще раз подчеркнем, что результаты выполненного фазового анализа, т.е. фазы упругого рассеяния системы частиц, в данном случае $n^{12}C$, позволяют параметризовать, как это было сделано выше, межкластерные парциальные потенциалы взаимодействия процессов рассеяния в этой системе. Такие потенциалы, в свою очередь, могут использоваться далее для выполнения определенных расчетов в области различных астрофизических приложений, частично рассмотренных, например, в наших работах [9,15,168].

## 2.3.5 Программа для $n^{12}C$ и $p^{12}C$ фазового анализа
## The program for $n^{12}C$ and $p^{12}C$ phase shifts analysis

Приведем текст компьютерной программы для выполнения фазового анализа упругого $p^{12}C$ рассеяния на языке Fortran-90. Программа ищет фазы упругого рассеяния двух частиц по экспериментальным дифференциальным сечениям методами, описанными в предыдущих параграфах.

Описание основных параметров, переменных, параметров потенциалов взаимодействия, блоков программы и подпрограмм дано далее в распечатке самой программы и практически не отличается от обозначений в предыдущих программах.

```
PROGRAM FAZ_ANAL_p12C
!  * Программа фазового анализа упругого p12C рассеяния *
IMPLICIT REAL(8) (A-Z)
INTEGER   I,L,LMA,NI,NV,LMI,LH,LMII,LMAA,LHH,NTTT,
NTT,NPP,NT,NTP
CHARACTER(34) AA
CHARACTER(33) BB
CHARACTER(25) AC
CHARACTER(24) BC
CHARACTER(3) NOM
```

```
CHARACTER(6) EX,EX1
COMMON /A1/
SE(0:50),DS(0:50),DE(0:50),NT,POLE(0:50),POLED(0:50),DS1
(0:50),NTP,XIS,XIP,XI1
COMMON /A2/ NTTT,GG,SS,LMII,LMAA,LHH,NP
COMMON /A3/ POL(0:50),TT(0:100),REZ(0:50)
COMMON /A4/ LH,LMI,NTT,NPP
COMMON/A5/ PI
DIMENSION ST(0:50),FP(0:50),FM(0:50),XP(0:50)
! *************** Начальные значения ******************
PI=4.0D-000*DATAN(1.0D-000)
Z1=1.0D-000 ! Заряд P
Z2=6.0D-000 ! Заряд 12C
AM1=1.00727646577D-000; ! Масса P
AM2=12.0D-000; ! Масса 12C
AM=AM1+AM2
A1=41.80159D-000
PM=AM1*AM2/AM
B1=2.0D-000*PM/A1
LMI=0; LH=1; LMA=0
! Минимальный LMI и максимальный LMA орбитальный
!момент
EP=1.0D-05; LMII=LMI; LHH=LH; LMAA=LMA
! EP - точность поиска минимума хи квадрат
NV=1; ! 1 - Проводить минимизацию, 0 - без минимизации
FH=0.0123D-000 ! Начальный шаг
NI=10 ! Число итераций
NP=2*LMA; NPP=NP
! *************** Задание энергии в ц.м. ****************
ECM=0.422D-000 ! Энергия в ц.м.
NT=17; NTT=NT; NTTT=NT ! Число точек по углам
NOM='422'
EX='-1.TXT'; EX1='-R.DAT'
AC='G:\BASICA\FAZ-ANAL\p12C\c'
BC='G:\BASICA\FAZ-ANAL\p12C\'
AA=AC//NOM//EX
BB=BC//NOM//EX1
OPEN (1,FILE=AA)
```

```
DO L=1,NT
READ(1,*) TT(L),SE(L),DE(L)
SE(L)=SE(L)*1000.0D-000
DE(L)=SE(L)*0.10D-000
ENDDO
CLOSE(1)
OPEN (1,FILE="G:\BASICA\FAZ-ANAL\p12C\FAZ.DAT")
DO L=LMI,LMA,LH
READ(1,*) FP(L),FM(L)
ENDDO
CLOSE(1)
! ******* Перевод начальных фаз в радианы **************
DO L=LMI,LMA,LH
FM(L)=FM(L)*PI/180.0D-000
FP(L)=FP(L)*PI/180.0D-000
ENDDO
FH=FH*PI/180.0D-000
DO I=LMI,LMA,LH
XP(I)=FP(I)
IF (I==LMA) GOTO 112
XP(I+LMA+LH)=FM(I+1)
112 ENDDO
! *********** Поиск минимума хи квадрат ***************
EL=ECM*AM1/PM
SK=ECM*B1
SS=DSQRT(SK)
GG=3.4495312D-002*Z1*Z2*PM/SS
CALL VAR(ST,FH,NI,XP,EP,XI,NV)
FM(0)=XP(0)
DO I=LMI,LMA,LH
FP(I)=XP(I)
IF (I==LMA) GOTO 111
FM(I+1)=XP(I+LMA+LH)
111 ENDDO
! ********** Печать результатов ************************
PRINT *,"   EL, ECM, SK, SS=",EL,ECM,SK,SS
PRINT *
PRINT *,"              T                    SE                    ST
```

```
XI "
DO I=1,NT
PRINT *,TT(I),SE(I),ST(I),DS(I)
ENDDO
PRINT *
PRINT *,"   XI=(XIS+XIP),XIS,XIP=",XI,XIS,XIP
PRINT *
PRINT *,"           L           FP           FM"
DO L=LMI,LMA,LH
FM(L)=FM(L)*180.0D-000/PI
FP(L)=FP(L)*180.0D-000/PI
PRINT *,L,FP(L),FM(L)
ENDDO
OPEN (1,FILE=BB)
WRITE(1,*) " EL,ECM=",EL,ECM
WRITE(1,*) "XI=(XIS+XIP),XIS,XIP=",XI,XIS,XIP
WRITE(1,*) "            T                    SE                    ST
XI"
DO I=1,NT
WRITE(1,*) TT(I),SE(I),ST(I),DS(I)
ENDDO
WRITE(1,*) "
WRITE(1,*) "          L           FP            FM"
DO I=LMI,LMA,LH
WRITE(1,*) I,FP(I),FM(I)
ENDDO
OPEN (1,FILE="G:\BASICA\FAZ-ANAL\p12C\FAZ.DAT")
DO I=LMI,LMA,LH
WRITE(1,*) FP(I),FM(I)
ENDDO
CLOSE(1)
END
SUBROUTINE VAR(ST,PHN,NI,XP,EP,AMIN,NV)
! Вариационная подпрограмма для минимизации хи квадрат
IMPLICIT REAL(8) (A-Z)
INTEGER I,NI,NT,NV,NP,LMI,LH,NN,IN,NTT,NTP
COMMON                                                /A1/
SE(0:50),DS(0:50),DE(0:50),NT,POLE(0:50),POLED(0:50),DS1
```

```
(0:50),NTP,XIS,XIP,XI1
COMMON /A3/ POL(0:50),TT(0:100),REZ(0:50)
COMMON /A4/ LH,LMI,NTT,NP
COMMON/A5/ PI
DIMENSION XPN(0:50),XP(0:50),ST(0:50)
DO I=LMI,NP,LH
XPN(I)=XP(I)
ENDDO
NN=LMI
PH=PHN
CALL DET(XPN,ST,ALA)
B=ALA
IF (NV==0) GOTO 3012
DO IIN=1,NI
NN=-LH
PRINT *,'FF=',ALA,IIN
1119  NN=NN+LH
IN=0
2229 A=B
XPN(NN)=XPN(NN)+PH*XP(NN)
IN=IN+1
CALL DET(XPN,ST,ALA)
B=ALA
IF (B<A) GOTO 2229
C=A
XPN(NN)=XPN(NN)-PH*XP(NN)
IF (IN>1) GOTO 3339
PH=-PH
GOTO 5559
3339 IF (ABS((C-B)/(B))<EP) GOTO 4449
PH=PH/2.0D-000
5559 B=C
GOTO 2229
4449 PH=PHN
B=C
IF (NN<NP) GOTO 1119
AMIN=B
PH=PHN
```

```
ENDDO
3012 AMIN=B
DO I=LMI,NP,LH
XP(I)=XPN(I)
ENDDO
END
SUBROUTINE DET(XP,ST,XI)
! ******* Подпрограмма вычисления хи квадрат **********
IMPLICIT REAL(8) (A-Z)
INTEGER I,NT,NTP
COMMON   /A1/   SE(0:50),DS(0:50),DE(0:50),NT,POLE(0:50),
POLED(0:50),DS1(0:50),NTP,XIS,XIP,XI1
COMMON /A3/ POL(0:50),TT(0:100),REZ(0:50)
DIMENSION XP(0:50),ST(0:50)
S=0.0D-000
CALL SEC(XP,ST)
S1=0.0D-000
DO I=1,NT
DS(I)=((ST(I)-SE(I))/DE(I))**2
S=S+DS(I)
ENDDO
XI=S/NT
END
SUBROUTINE SEC(XP,S)
! *** Подпрограмма вычисления сечения рассеяния ********
IMPLICIT REAL(8) (A-Z)
INTEGER I,NT,LMI,LMA,LH,L
COMMON /A2/ NT,GG,SS,LMI,LMA,LH,NP
COMMON /A3/ POL(0:50),TT(0:100),REZ(0:50)
COMMON/A5/ PI
DIMENSION
S0(0:50),P(0:50),PP(0:50),FP(0:50),FM(0:50),XP(0:50),S(0:50)
DO I=LMI,LMA,LH
FP(I)=XP(I)
IF (I==LMA) GOTO 111
FM(I+1)=XP(I+LMA+LH)
111 ENDDO
FM(0)=FP(0)
```

```
CALL CULFAZ(GG,S0)
DO I=1,NT
T=TT(I)*PI/180.0D-000
X=DCOS(T)
A=2.0D-000/(1.0D-000-X)
S00=2.0D-000*S0(0)
BB=-GG*A
ALO=GG*DLOG(A)+S00
REC=BB*DCOS(ALO)
AMC=BB*DSIN(ALO)
REZ1=REC**2+AMC**2
REA=0.0D-000
AMA=0.0D-000
REB=0.0D-000
AMB=0.0D-000
DO L=LMI,LMA,LH
FPP=2.0D-000*FP(L)
FMP=2.0D-000*FM(L)
AA=DCOS(FPP)-DCOS(FMP)
BB=DSIN(FPP)-DSIN(FMP)
SL=2.0D-000*S0(L)
CALL FUNLEG(X,L,PP)
REB=REB+(BB*DCOS(SL)+AA*DSIN(SL))*PP(L)
AMB=AMB+(BB*DSIN(SL)-AA*DCOS(SL))*PP(L)
LL=2*L+1
JJ=L+1
AA=JJ*DCOS(FPP)+L*DCOS(FMP)-LL
BB=JJ*DSIN(FPP)+L*DSIN(FMP)
CALL POLLEG(X,L,P)
REA=REA+(BB*DCOS(SL)+AA*DSIN(SL))*P(L)
AMA=AMA+(BB*DSIN(SL)-AA*DCOS(SL))*P(L)
ENDDO
REA=REC+REA
AMA=AMC+AMA
RE=REA**2+AMA**2
AM=REB**2+AMB**2
S(I)=10.0D-000*(RE+AM)/4.0D-000/SS**2
REZ(I)=REZ1*10.0D-000/4.0D-000/SS**2
```

```
POL(I)=2.0D-000*(REB*AMA-REA*AMB)/(RE+AM)
ENDDO
END
SUBROUTINE POLLEG(X,L,P)
! ***** Подпрограмма вычисления полиномов Лежандра ***
IMPLICIT REAL(8) (A-Z)
INTEGER I,L
DIMENSION P(0:50)
P(0)=1.0D-000; P(1)=X
DO I=2,L
P(I)=(2*I-1)*X/I*P(I-1)-(I-1)/I*P(I-2)
ENDDO
END
SUBROUTINE FUNLEG(X,L,P)
! *** Подпрограмма вычисления функций Лежандра *******
IMPLICIT REAL(8) (A-Z)
INTEGER I,L
DIMENSION P(0:50)
P(0)=0.0D-000;      P(1)=DSQRT(ABS(1.-X**2));      P(2)=3.0D-
000*X*P(1)
IF (L>=3) THEN
DO I=2,L
P(I+1)=(2*I+1)*X/I*P(I)-(I+1)/I*P(I-1)
ENDDO
ENDIF
END
SUBROUTINE CULFAZ(G,F)
! *** Подпрограмма вычисления кулоновских фаз *********
IMPLICIT REAL(8) (A-Z)
INTEGER I,N
DIMENSION F(0:50)
C=0.5772156650D-000; S=0.0D-000; N=50
A1=1.202056903D-000/3.0D-000; A2=1.036927755D-000/5.0D-
000
DO I=1,N
A=G/I-DATAN(G/I)-(G/I)**3/3.0D-000+(G/I)**5/5.0D-000
S=S+A
ENDDO
```

```
FAZ=-C*G+A1*G**3-A2*G**5+S
F(0)=FAZ
DO I=1,20
F(I)=F(I-1)+DATAN(G/I)
ENDDO
END
```

Следующий контрольный счет по этой программе выполнен при резонансной энергии 422 кэВ (ц.м.) в $p^{12}C$ упругом рассеянии с учетом в фазовом анализе только одной *S* волны:

EL, ECM, SK, SS = 4.574E-1 4.22E-1 1.87E-2 1.37E-1

| $\theta$ | $\sigma_e$ | $\sigma_t$ | $\chi^2_i$ |
|---|---|---|---|
| 10.83 | 409561 | 3256822.3428083 | 4.194371298676935 |
| 21.63 | 266969 | 227766.5412064077 | 2.156278956304461 |
| 32.39 | 58583.1 | 50674.784355610 | 1.8223143351212 |
| 43.07 | 17267.6 | 15925.09271621951 | 6.044624704055184E-1 |
| 53.66 | 5839.94 | 5640.62956376928 | 1.164778401875084E-1 |
| 64.14 | 2414.18 | 2060.5610732571380 | 2.145515754644307 |
| 74.5 | 965.24 | 740.3435278409627 | 5.428685268597148 |
| 84.71 | 422.29 | 265.4550687576415 | 13.793167214162040 |
| 94.79 | 217.01 | 120.1919175449456 | 19.9046247867550 |
| 104.7 | 147.11 | 103.6791667064438 | 8.715893867706498 |
| 114.5 | 140.23 | 133.5569838776437 | 2.264448667637590E-1 |
| 124.15 | 204.09 | 175.8612452117759 | 1.913110216888019 |
| 133.67 | 243.52 | 216.8648984979974 | 1.198095428652019 |
| 143.08 | 253.77 | 251.4932553653157 | 8.049114108914179E-3 |
| 152.39 | 267.64 | 278.2186995788807 | 1.562293138659170E-1 |
| 161.64 | 288.94 | 296.9745596276764 | 7.732302205565983E-2 |
| 170.84 | 292.72 | 308.0225110939944 | 2.732880319244108E-1 |

$\chi^2 = 3.69$

| L | $\delta_p$ | $\delta_m$ |
|---|---|---|
| 0 | 58.15 | 58.15 |

В результате получаем значение *S* фазы $58.15^\circ$ при сравнительно большом значении $\chi^2$ = 3.69. Величины фаз и неко-торые другие характеристики показаны в округленном, до второго знака после запятой, виде.

Если в анализе учитывать и *P* волну, то для фаз рассеяния получим следующий контрольный счет:

EL, ECM, SK, SS = 4.57E-1 4.22E-1 1.87E-2 1.37E-1

| $\theta$ | $\sigma_e$ | $\sigma_t$ | $\chi^2_i$ |
|---|---|---|---|
| 10.83 | 4095610 | 3264781.541334357 | 4.115148756111661 |
| 21.63 | 266969 | 223190.0387772885 | 2.689114135211925 |
| 32.39 | 58583.1 | 49592.26685532391 | 2.355348513821907 |
| 43.07 | 17267.6 | 16111.41279780082 | 4.483243796400418E-1 |
| 53.66 | 5839.94 | 6059.579098437353 | 1.414498640631696E-1 |
| 64.14 | 2414.18 | 2415.589085838059 | 3.406713446204295E-5 |
| 74.5 | 965.24 | 982.7590318182862 | 3.294197102373393E-2 |
| 84.71 | 422.29 | 414.5062045985828 | 3.397513798561286E-2 |
| 94.79 | 217.01 | 205.28584114367 | 2.918797023030877E-1 |
| 104.71 | 147.11 | 149.1168415589131 | 1.860980323212324E-2 |
| 114.5 | 140.23 | 155.8558179281346 | 1.241662746301032 |
| 124.15 | 204.09 | 185.2840079884762 | 8.490808445689716E-1 |
| 133.67 | 243.52 | 219.3107960699812 | 9.883067178217969E-1 |
| 143.08 | 253.77 | 250.1596154521572 | 2.024073897229597E-2 |
| 152.39 | 267.64 | 274.7849530022851 | 7.126835104038670E-2 |
| 161.64 | 288.94 | 292.3439981186766 | 1.387916330759821E-2 |
| 170.84 | 292.72 | 302.7569167687375 | 1.175698198093965E-1 |

$\chi^2$ = 7.90E-1

| L | $\delta_p$ | $\delta_m$ |
|---|---|---|
| 0 | 52.13 | 52.13 |
| 1 | -8.97 | 12.82 |

Видно, что при учете в фазовом анализе *S* и *P* волн рассеяния, величина $\chi^2$ уменьшается с 3.69 до 0.79.

В распечатках использованы следующие обозначения: EL – энергия частиц в лабораторной системе, ECM – энергия частиц в системе центра масс, SK – квадрат волнового числа $k^2$, SS – волновое число $k$, L – орбитальный момент, $\theta$ – угол рассеяния, $\sigma_e$ – экспериментальные сечения, $\sigma_t$ – вычисленные сечения, $\chi^2_i$ – парциальные $\chi^2$ для $i$-го угла, $\delta_p$ – фаза с $J$ $= L + 1/2$, $\delta_m$ – фаза при $J = L - 1/2$, $\chi^2$ – среднее значение по всем точкам.

Если использовать для величины константы $\hbar^2 / m_0$ обычное значение 41.4686 МэВ·Фм$^2$, то, как видно из приведенной ниже распечатки, величина $\chi^2$ изменяется примерно на 10%, а некоторые фазы на 0.5$^\circ$, что практически не сказывается на результатах даже при резонансной энергии:

$\chi^2$ = 8.85E-1

| L | $\delta_p$ | $\delta_m$ |
|---|---|---|
| 0 | 51.63 | 51.63 |
| 1 | -9.40 | 12.87 |

Из приведенных результатов видно, что при 422 кэВ (ц.м.) *S* фаза почти достигает своего резонансного значения в 90$^\circ$. При энергиях в области резонанса, поскольку его ширина менее 32 кэВ [166], наблюдается настолько резкий подъем фазы, что изменение энергии примерно на 1 кэВ может привести к изменению фазы на 40$^\circ$÷50$^\circ$. Заметим, что точность определения энергии в рассматриваемом эксперименте составляет около 1 кэВ, а точность определения энергии резонансного уровня имеет величину 0.6 кэВ [166].

Найденные в результате фазового анализа фазы рассеяния использованы далее для построения межкластерных $p^{12}C$ потенциалов, которые, в свою очередь, применялись для расчетов астрофизических S-факторов радиационного $p^{12}C$ захвата [171]. Данный процесс является первой термоядерной реакцией CNO-цикла, который присутствует на более поздней стадии развития звезд, когда происходит частичное выгорание водорода. По мере его выгорания, ядро звезды начинает заметно сжиматься, приводя в результате к увеличению давления и температуры внутри звезды, и наряду с протон-протонным циклом вступает в действие следующая цепочка термоядерных процессов, называемая, CNO или углеродным циклом [1-3,168].

## 2.4 Фазовый анализ упругого $p^6Li$ рассеяния Phase shifts analysis of $p^6Li$ scattering

Рассмотрим процессы упругого $p^6Li$ рассеяния при астрофизических энергиях и выполним фазовый анализ экспериментальных данных. Результаты такого фазового анализа потребуются для построения в дальнейшем потенциалов межкластерного взаимодействия при расчетах, связанных с решением астрофизических задач [172].

### 2.4.1 Дифференциальные сечения Differential cross sections

При рассмотрении процессов рассеяния в системе частиц со спинами 1/2 и 1 без учета спин-орбитального расщепления фаз сечение упругого рассеяния представляется в наиболее простом виде [88]

$$\frac{d\sigma(\theta)}{d\Omega} = \frac{2}{6}\frac{d\sigma_d(\theta)}{d\Omega} + \frac{4}{6}\frac{d\sigma_q(\theta)}{d\Omega} ,$$

где индексы d и q относятся к дублетному (со спином 1/2) и квартетному (со спином 3/2) состоянию $p^6Li$ системы, а сами сечения выражаются через амплитуды рассеяния, которые записываются подобно выражениям для фазового анализа в $^4He^4He$ системе

$$\frac{d\sigma_d(\theta)}{d\Omega} = |f_d(\theta)|^2 , \qquad \frac{d\sigma_q(\theta)}{d\Omega} = |f_q(\theta)|^2 ,$$

где

$$f_{d,q}(\theta) = f_c(\theta) + f^N_{d,q}(\theta) ,$$

и

$$f_{\mathrm{c}}(\theta)=-\left(\frac{\eta}{2k\sin^2(\theta/2)}\right)\exp\{i\eta\ln[\sin^{-2}(\theta/2)]+2i\sigma_0\} \;,$$

$$f_{\mathrm{d}}^{\mathrm{N}}(\theta)=\frac{1}{2ik}\sum_{\mathrm{L}}(2L+1)\exp(2i\sigma_{\mathrm{L}})[S_{\mathrm{L}}^{\mathrm{d}}-1]P_{\mathrm{L}}(\cos\theta) \;,$$

$$f_{\mathrm{q}}^{\mathrm{N}}(\theta)=\frac{1}{2ik}\sum_{\mathrm{L}}(2L+1)\exp(2i\sigma_{\mathrm{L}})[S_{\mathrm{L}}^{\mathrm{q}}-1]P_{\mathrm{L}}(\cos\theta) \;,$$

где $S_{\mathrm{L}}^{\mathrm{d,q}}=\eta_{\mathrm{L}}^{\mathrm{d,q}}\exp[2i\delta_{\mathrm{L}}^{\mathrm{d,q}}(k)]$ - матрица рассеяния в дублетном или квартетном спиновом состоянии [88].

Возможность использования простых выражений для расчетов сечений упругого рассеяния обусловлена тем, что в области низких энергий спин-орбитальное расщепление фаз оказывается сравнительно мало, так как в упругом рассеянии отсутствуют резонансы, что подтверждается и результатами фазового анализа, выполненного ранее в работе [173], в которой учитывалось спин-орбитальное расщепление фаз рассеяния.

### 2.4.2 Фазовый анализ
### Phase shifts analysis

Ранее фазовый анализ дифференциальных сечений и функций возбуждения для упругого $p^6Li$ рассеяния без явного учета дублетной $^2P$ волны был выполнен в [173]. Наш фазовый анализ проводится при более низких энергиях, имеющих значение для ядерной астрофизики, учитывает все низшие парциальные волны, в том числе дублетную $^2P$ волну и основан на дифференциальных сечениях, приведенных в работах [174,175] и [176].

При энергии 500 кэВ, на основе данных [176], находим $^2S$ и $^4S$ фазы рассеяния, которые даны в табл.2.4.1 под №1. Полученные результаты расчета сечений вполне согласуются с экспериментальными данными при среднем по всем точкам $\chi^2$ = 0.15. Ошибка дифференциальных сечений этих данных принималась равной 10%. Учет дублетной $^2P$ и квартетной $^4P$ фаз показал, что их численные значения меньше 0.1°.

Табл.2.4.1. Результаты фазового анализа (в град.) упругого p$^6$Li рассеяния.

| № | $E_p$, кэВ | $^2S$ | $^4S$ | $^2P$ | $^4P$ | $\chi^2$ |
|---|---|---|---|---|---|---|
| 1 | 500 | 176.2 | 178.7 | – | – | 0.15 |
| 2 | 593.0 | 174.2 | 178.8 | – | – | 0.15 |
| 3-1 | 746.4 | 170.1 | 180.0 | – | – | 0.24 |
| 3-2 | 746.4 | 172.5 | 179.9 | 1.7 | – | 0.16 |
| 4-1 | 866.8 | 157.8 | 180.0 | – | – | 0.39 |
| 4-2 | 866.8 | 170.2 | 174.9 | 3.9 | – | 0.22 |
| 4-3 | 866.8 | 169.6 | 175.0 | 3.5 | 0.1 | 0.23 |
| 5-1 | 976.5 | 160.0 | 178.5 | – | – | 0.12 |
| 5-2 | 976.5 | 167.0 | 174.5 | 1.1 | – | 0.12 |
| 6-1 | 1136.3 | 144.9 | 180.0 | – | – | 0.58 |
| 6-2 | 1136.3 | 164.7 | 171.1 | 5.8 | – | 0.32 |
| 6-3 | 1136.3 | 166.4 | 169.9 | 5.5 | 0.1 | 0.32 |

Следующие пять энергий относятся к новым результатам измерений дифференциальных сечений, выполненных в работах [174,175] в самое последнее время. Первая из них, 593.0 кэВ, дает возможность найти $^{2,4}S$ фазы, которые мало отличаются от фаз для предыдущей энергии, имеют такой же $\chi^2$ и показаны в табл.2.4.1 под №2, а фазы для $^{2,4}P$ волн также стремятся к нулю.

При энергии 746.7 кэВ находим $^{2,4}S$ фазы (табл.2.4.1 №3-1), которые позволяют описать сечения с точностью $\chi^2 = 0.24$. Несмотря на малость величины $\chi^2$, была предпринята попытка учесть $^{2,4}P$ фазы. Вначале полагалось, что квартетная $^4P$ фаза пренебрежимо мала, что следует из результатов [173], в которой их учет начинался только с 1.0÷1.5 МэВ. Результаты нашего анализа с учетом только $^2P$ фазы представлены на рис.2.4.1а и в табл.2.4.1 под №3-2. Видно, что учет неболь-шой дублетной $^2P$ фазы несколько изменяет величину $^2S$ волны, увеличивая ее значение, и уменьшает $\chi^2 = 0.16$. Учет квартетной $^4P$ фазы дал для ее численного значения пренебрежимо малую величину, меньше 0.1°, что соответствует результатам [173].

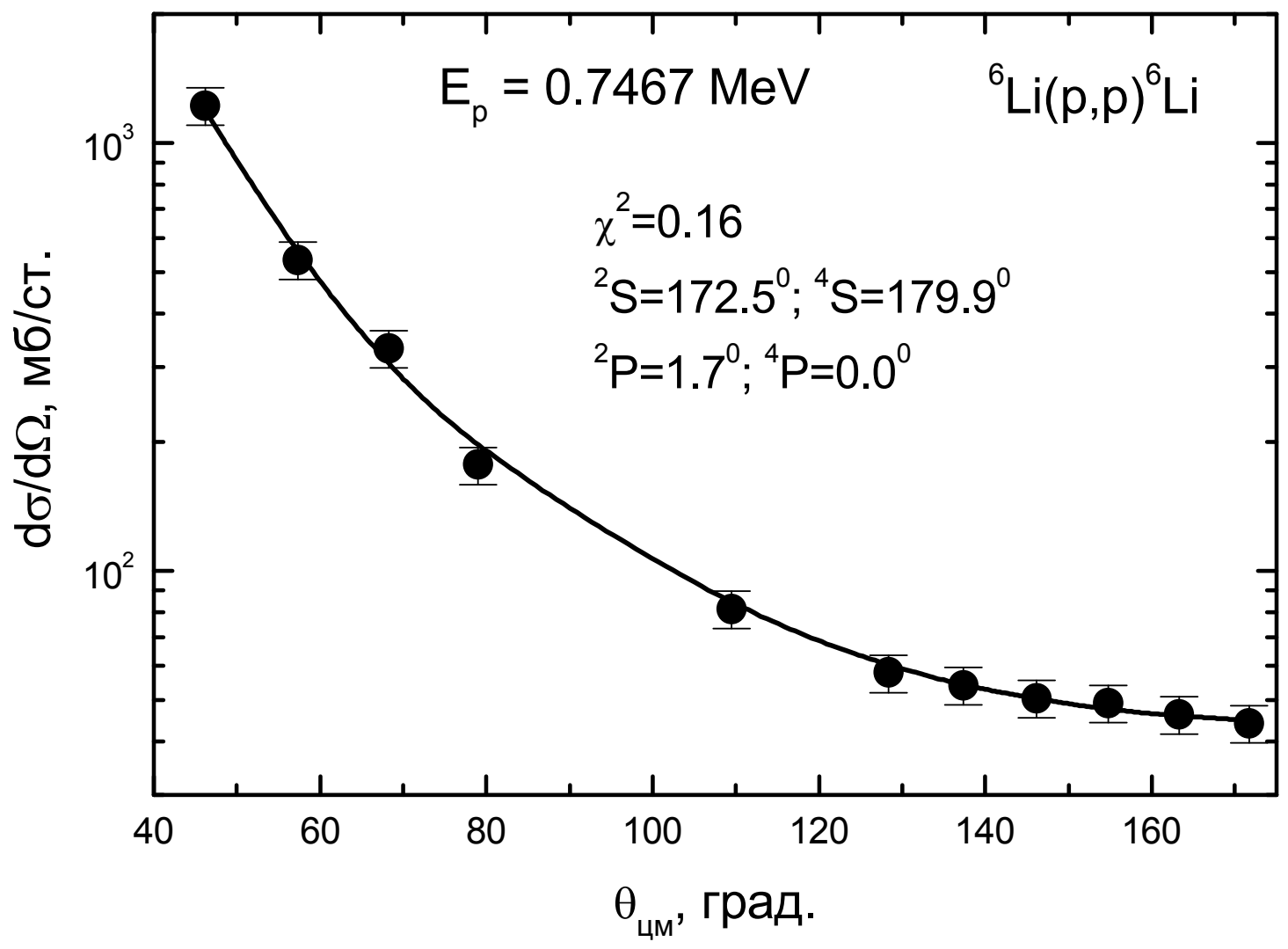

Рис.2.4.1а. Сечения упругого $p^6Li$ рассеяния при 746.7 кэВ. Непрерывная кривая – расчет сечений с найденными фазами.

Результат поиска фаз для энергии 866.8 кэВ с учетом только $^{2,4}S$ волн приведен в табл.2.4.1 под №4-1 при $\chi^2 = 0.39$. Как видно, величина $^2S$ фазы резко спадает по сравнению с предыдущей энергией. Учет же $^2P$ волны заметно увеличивает ее значение (рис.2.4.1б и табл.2.4.1 №4-2) и почти в два раза уменьшает величину $\chi^2$. Попытка учесть квартетную $^4P$ фазу привела к значению около 0.1° (табл.2.4.1 №4-3).

Любое изменение $^4P$ волны в бóльшую сторону, в том числе, при других значениях остальных фаз, приводило к увеличению $\chi^2$. При этой энергии, как и всех других рассмотренных энергиях из работ [174,175], не удается найти какой-либо вариант для ненулевой квартетной фазы при стремле-нии величины $\chi^2$ к минимуму.

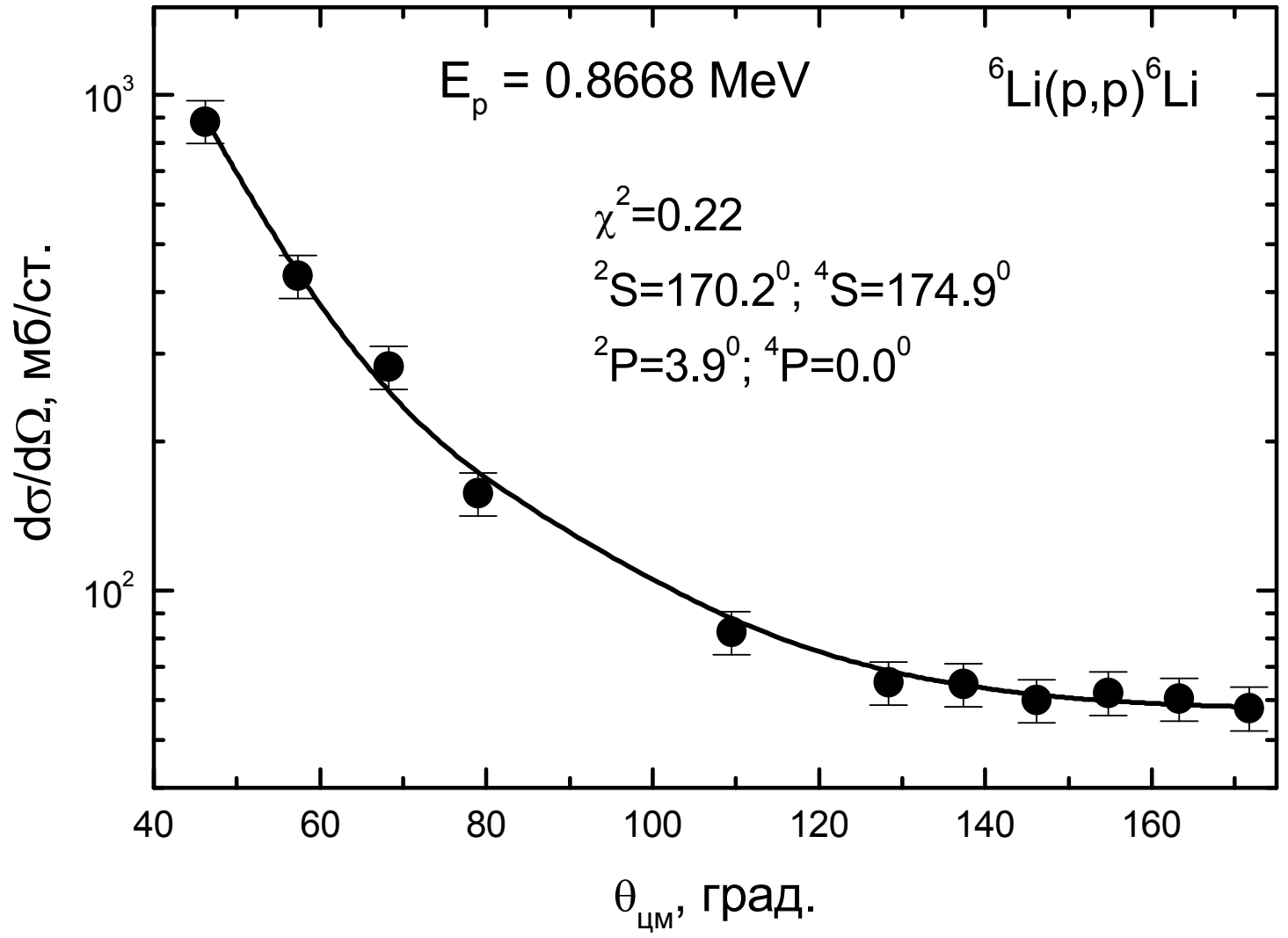

Рис.2.4.1б. Сечения упругого p$^6$Li рассеяния при 866.8 кэВ. Непрерывная кривая – расчет сечений с найденными фазами.

Для следующей энергии, 976.5 кэВ, без учета $^{2,4}P$ волн найдены значения $^2S$ и $^4S$ фаз, приведенные в табл.2.4.1 с номером 5-1. Последующий учет $^2P$ волны заметно увеличивает значения $^2S$ фазы, если пренебречь $^4P$ волной, как это видно на рис.2.4.1в и табл.2.4.1 №5-2 при $\chi^2$ = 0.12. Если включить в анализ квартетную $^4P$ волну, то она стремится к нулю при уменьшении величины $\chi^2$.

Последняя из рассмотренных энергий 1.1363 МэВ из работ [174,175] даже при учете только $^{2,4}S$ волн приводит к сравнительно малому $\chi^2$, равному 0.58, как это видно в табл.2.4.1 №6-1. И в этом случае, учет $^2P$ волны приводит к заметному увеличению значения $^2S$ фазы – соответствующие результаты расчета сечений показаны на рис.2.4.1г и табл.2.4.1 под №6-2. Учет квартетной $^4P$ волны и при этой энергии приводит к ее значению порядка 0.1°, как показано в табл.2.4.1 под №6-3.

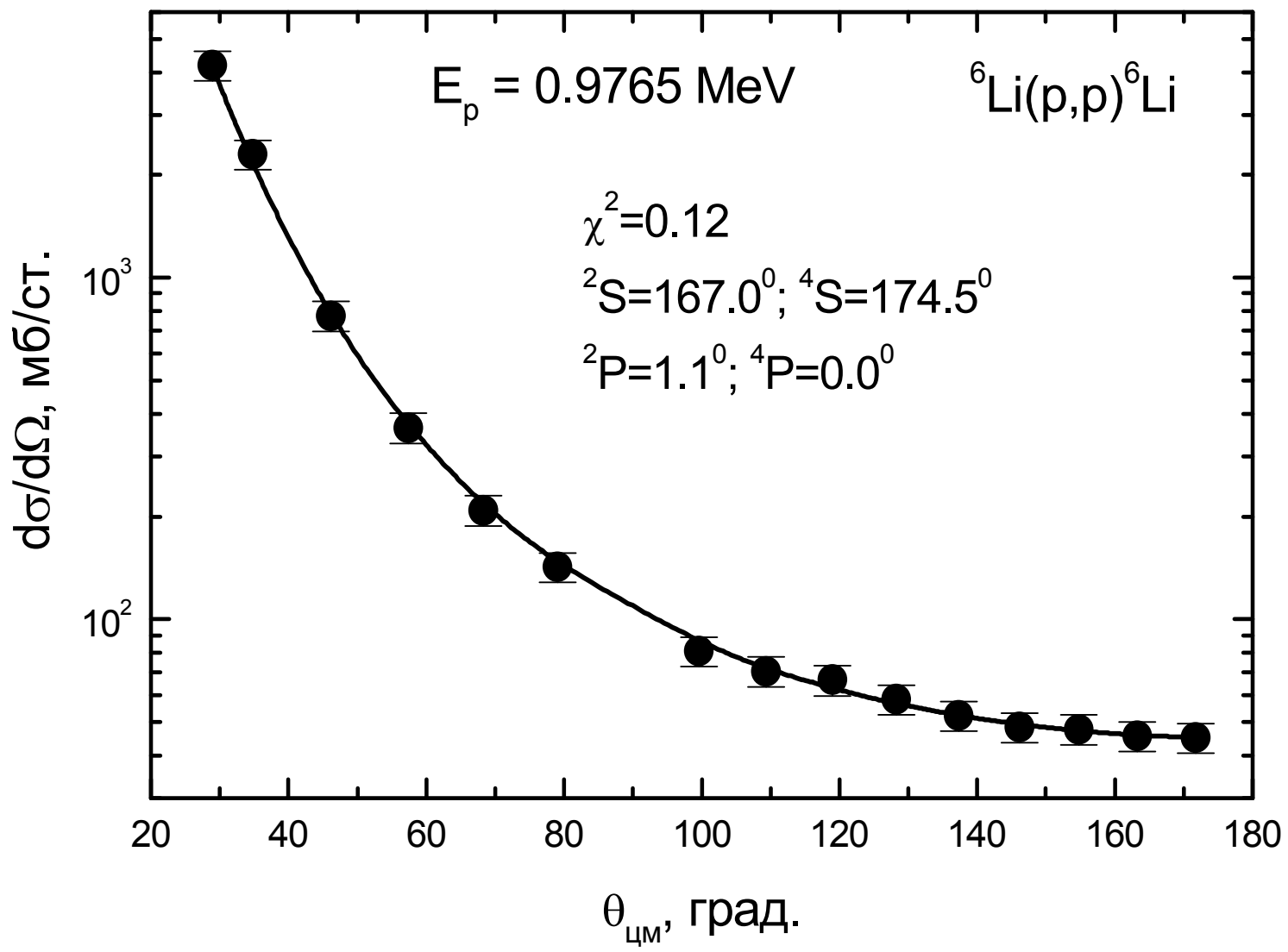

Рис.2.4.1в. Сечения упругого $p^6Li$ рассеяния при 976.5 кэВ. Непрерывная кривая – расчет сечений с найденными фазами.

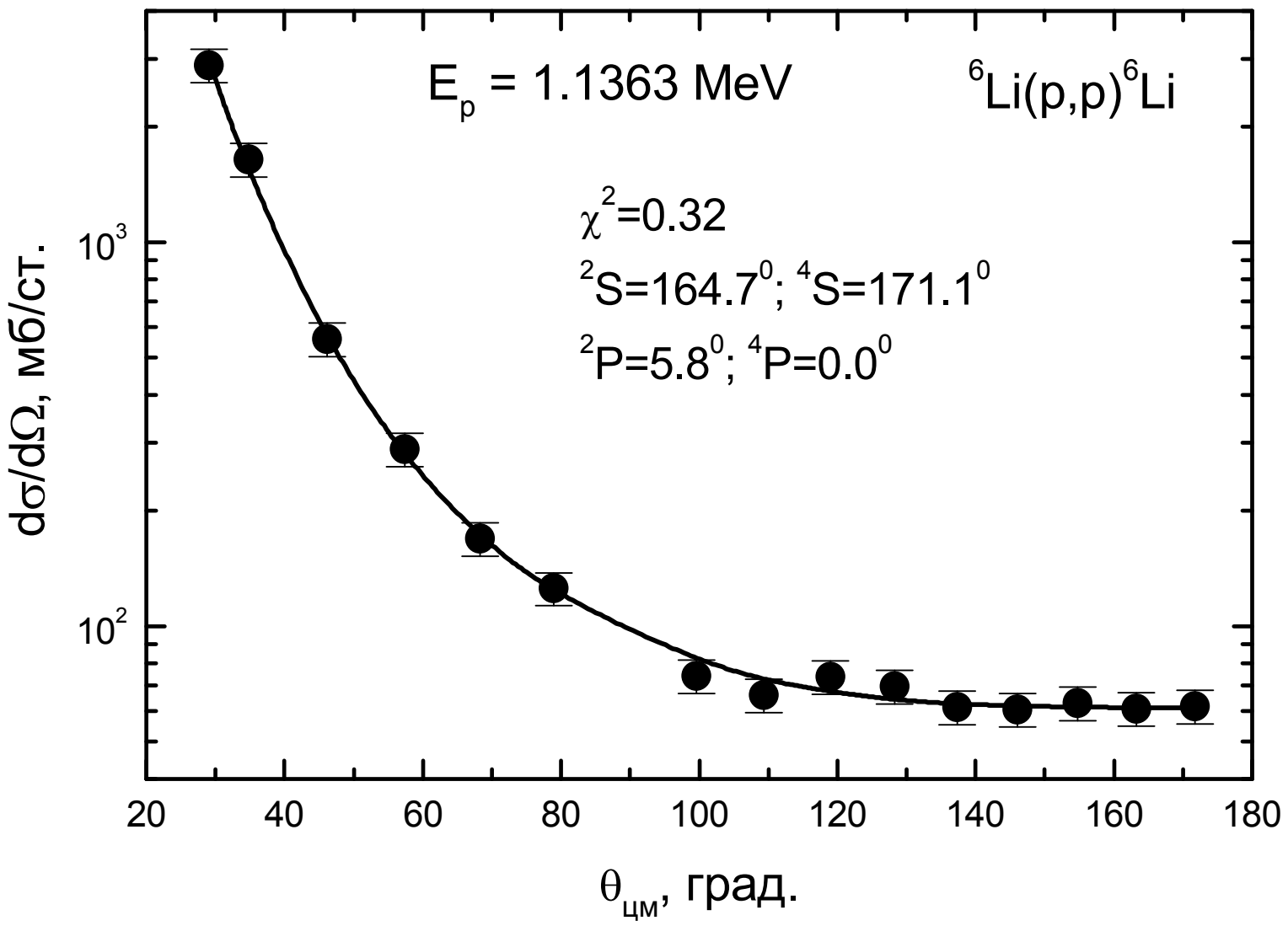

Рис.2.4.1г. Сечения упругого $p^6Li$ рассеяния при 1136.3 кэВ. Непрерывная кривая – расчет сечений с найденными фазами.

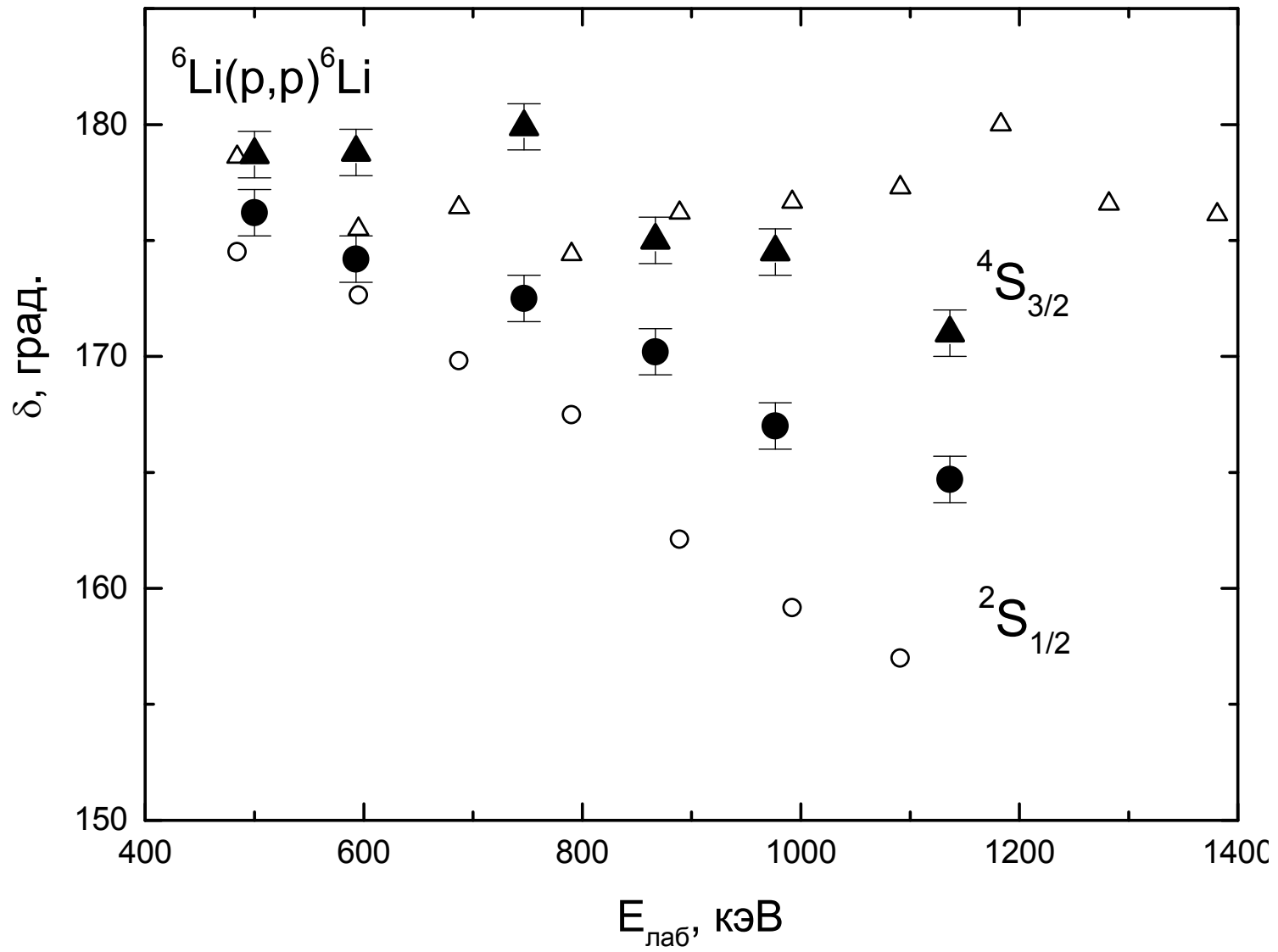

Рис.2.4.2а. Дублетные и квартетные $S$ фазы упругого $p^6Li$ рассеяния при низких энергиях.
Приведены дублетные и квартетные $S$ фазы при наличии $^2P$ волны, когда $^4P$ фаза принималась равной нулю. Точки $^2S$ и треугольники $^4S$ фазы, полученные по данным работ [174,175,176]. Для сравнения открытыми треугольниками и кружками приведены результаты фазового анализа [173].

Таким образом, при описании всех экспериментальных данных из работ [174,175] не требуется учета квартетных $^4P$ волн в этой области энергии, т.е. их величина равна или меньше 0.1°. Это согласуется с результатами [173], однако, дублетная $^2P$ фаза доходит до 5.5°÷6° и ее значением нельзя пренебречь.

Общий вид $^2S$ и $^4S$ фаз рассеяния показан на рис.2.4.2а, а дублетные $^2P$ фазы приведены на рис.2.4.2б. Несмотря на довольно большой разброс результатов для $^4S$ фаз, дублетная $^2S$ фаза имеет определенную тенденцию к убыванию, но происходит это заметно медленнее, чем следует из результатов анализа [173]. Если в нашем анализе не учитывать дублетную $^2P$ волну, то для $^2S$ фазы получаются результаты очень близ-кие к результатам фазового анализа работы [173].

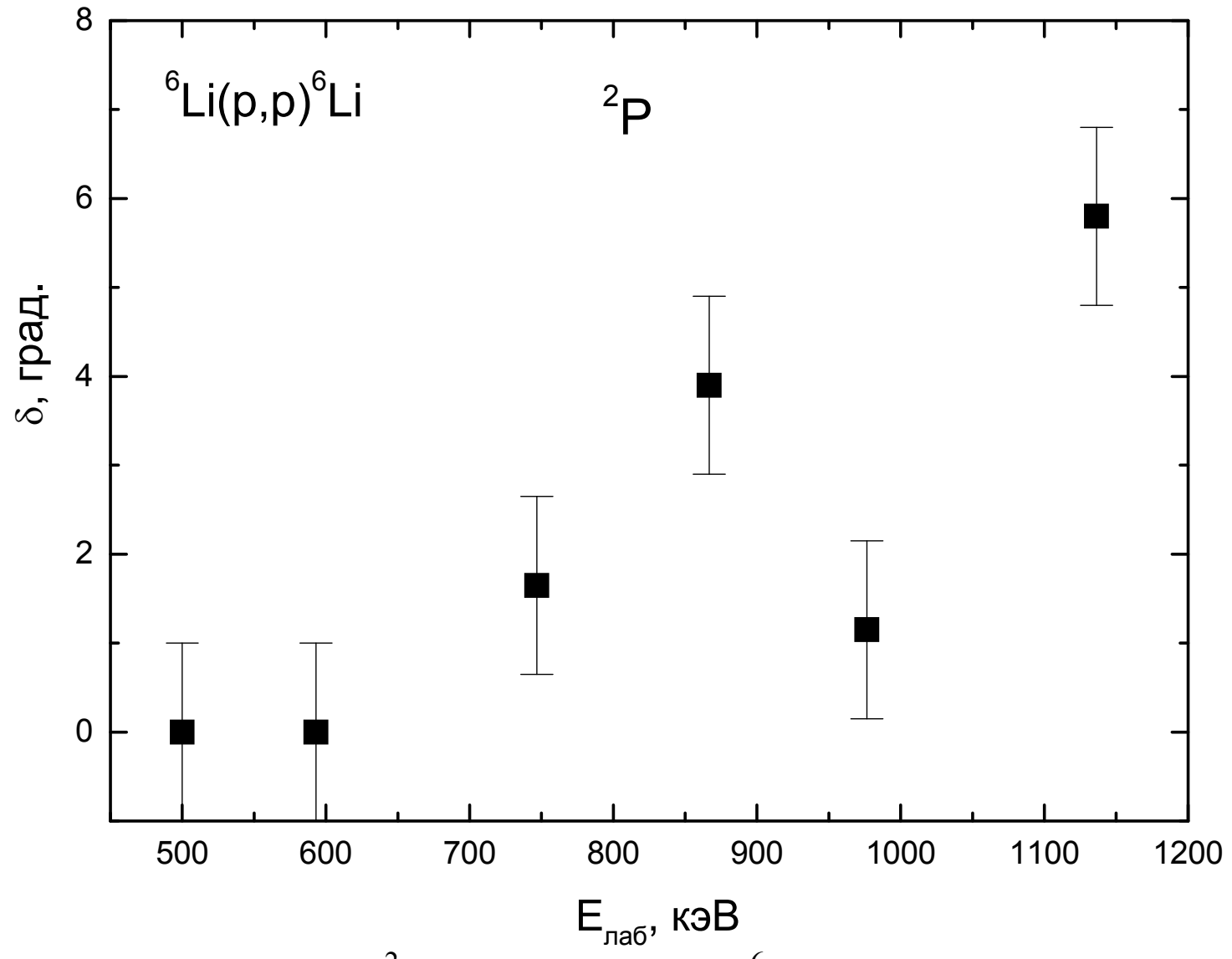

Рис.2.4.2б. Дублетные $^2P$ фазы упругого p$^6$Li рассеяния при низких энергиях.
Квадраты – результаты нашего фазового анализа для $^2P$ фазы при $^4P = 0$.

Ошибки фаз упругого рассеяния, приведенных на рис.2.4.2, определяются неоднозначностью фазового анализа, а именно, при практически одном и том же значении $\chi^2$, которое может отличаться на 5÷10%, оказывается возможным получить несколько разные значения самих фаз рассеяния. Эта неоднозначность для $^{2,4}S$ и $^2P$ фаз оценивается на уровне 1°÷1.5°.

## 2.4.3 Программа для фазового анализа Program for phase shifts analysis

Приведем текст компьютерной программы для выполнения p$^6$Li фазового анализа при низких энергиях без спин-орбитального расщепления фаз, которая использовалась для получения описанных выше результатов. Обозначения основных параметров программы совпадают с предыдущими распечатками.

**PROGRAM FAZ_ANAL_P6LI**

```
! CALCULATE OF CROSS SECTION WITH COMPLEX
! PHASE SHIFTS FOR SYSTEM SPIN 1/2+1 WITHOUT LS
IMPLICIT REAL(8) (A-Z)
INTEGER
L,LL,TMA,T,LN,LV,NV,Z1,Z2,LN1,NT,NP,NPP,NI,NYS,NNN
REAL(8)
SECT(0:200),FK(0:200),FD(0:200),XP(0:200),FK1(0:200),FD1(
0:200),S0(0:200)
COMMON REAL(8) /A/
PI,ET(0:200),ES(0:200),ST(0:200),SSE(0:200),TE(0:200) /R/
LN1,NT,P1,NP,NPP,LV1,NNN,BB /T/
SECE(0:200),DS(0:200),DSEC(0:200),TMA /W/
GG,SS,LN,LV,NYS
CHARACTER(7) BB
CHARACTER(12) AA
CHARACTER(13) CC
! *************** INPUT PARAMETERS ***************
!AA='SEC11363.DAT'
!AA='SEC9765.DAT'
!AA='SEC8668.DAT'
!AA='SEC7467.DAT'
!AA='SEC495.DAT'
!AA='SEC989.DAT'
AA='SEC640.DAT'
!AA='SEC1585.DAT'
BB='FAZ.DAT'
!CC='SECT11363.DAT'
!CC='SECT9765.DAT'
!CC='SECT8668.DAT'
!CC='SECT7467.DAT'
!CC='SECT495.DAT'
!CC='SECT989.DAT'
CC='SECT640.DAT'
```

```
!CC='SECT1585.DAT'
!EL=1.1363D-000
!EL=0.9765D-000
!EL=0.8668D-000
!EL=0.7467D-000
EL=0.7464D-000
!EL=0.989D-000
!EL=1.585D-000
PI=4.0D-000*DATAN(1.0D-000)
P1=PI
NYS=0; ! =0 - P-6LI; =1 - 3HE-3HE
NNN=1;  ! ЧИСЛО НА КОТОРОЕ УМЕНЬШАЕТСЯ ШАГ
ПРИ КАЖДОЙ ИТЕРАЦИИ
NV=1 ! =0 БЕЗ ВАРИАЦИИ = 1 С ВАРЬИРОВАНИЕМ
FH=0.0123D-000; ! НАЧАЛЬНЫЙ ШАГ
NI=10; ! ЧИСЛО ИТЕРАЦИЙ
EPP=1.0D-010; ! ТОЧНОСТЬ
LN=0; LN1=LN
LV=0
LV1=LV; NPP=2*LV
TMA=11
NT=TMA
AM1=1.0D-000; AM2=6.0D-000; Z1=1; Z2=3
A1=41.4686D-000
PM=AM1*AM2/(AM1+AM2)
B1=2.0D-000*PM/A1
NP=NPP+2
! ************** PHASE SHIFKS FOR P- 6LI  ************
FD(0)=160.0D-000
FD(1)=5.D-000
FD(2)=1.D0-000
FK(0)=179.D-000
FK(1)=3.0D-000
FK(2)=1.1D-000
OPEN (1,FILE=BB)
DO I=LN,LV
READ(1,*) FD(I),FK(I)
ENDDO
```

```
CLOSE(1)
OPEN (1,FILE=AA)
DO L=1,NT
READ(1,*) TE(L),SECE(L)
!,DSEC(L)
PRINT *, TE(L),SECE(L),DSEC(L)
ENDDO
CLOSE(1)
DO L=LN,LV
FD(L)=FD(L)*PI/180.0D-000
FK(L)=FK(L)*PI/180.0D-000
ET(L)=1.0D-000
ES(L)=1.0D-000
ENDDO
FH=FH*PI/180.0D-000
!DO L=LN,LV
!XP(L)=FD(L)
!XP(L+LV+1)=FK(L)
!ENDDO
DO I=LN,LV
XP(2*I)=FD(I)
XP(2*I+1)=FK(I)
ENDDO
!   *********** TRANSFORM TO C.M. ******************
ECM=EL*PM/AM1
! ********  TOTAL CROSS SECTION  ******************
SK=ECM*B1
SS=DSQRT(SK)
GG=3.44476D-002*Z1*Z2*PM/SS
CALL CULFAZ(GG,S0)
DO I=1,NT
TT=TE(I)*PI/180.0D-000
S00=2.0D-000*S0(0)
X=DCOS(TT)
A=2.0D-000/(1.0D-000-X)
BBB=-GG*A
ALO=GG*DLOG(A)+S00
RECUL=BBB*DCOS(ALO)
```

```
AIMCUL=BBB*DSIN(ALO)
CUL=RECUL**2+AIMCUL**2
SCUL=CUL*10.0D-000/4.0D-000/SS**2
!SECE(I)=SECE(I)*SCUL
DSEC(I)=SECE(I)*0.1
!DSEC(I)=DSEC(I)*SCUL
ENDDO
SIGS=0.0D-000; SIGT=0.0D-000
DO LL=LN,LV
FK11=FK(LL)
FD11=FD(LL)
SIGS=SIGS+(2.0D-000*LL+1.0D-000)*DSIN(FK11)**2
SIGT=SIGT+(2.0D-000*LL+1.0D-000)*DSIN(FD11)**2
ENDDO
SIGS=10.0D-000*4.0D-000*PI*SIGS/SK
SIGT=10.0D-000*4.0D-000*PI*SIGT/SK
SIG=1.0D-000/4.0D-000*SIGS+3.0D-000/4.0D-000*SIGT
PRINT *,"                    SIGMS-TOT=",SIG
! ********** DIFFERENTIAL CROSS SECTION **********
CALL VAR(SECT,FH,NI,XP,EPP,XI,NV)
PRINT *,'    T       SE        ST        XI'
DO T=1,TMA
WRITE(*,1) TE(T),SECE(T),SECT(T),DS(T)
ENDDO
1 FORMAT(1X,F8.3,3E12.5)
PRINT *,'  FD        FK'
DO L=LN,LV
!FD(L)=XP(L)*180.0D-000/PI
!FK(L)=XP(L+LV+1)*180.0D-000/PI
FD(L)=XP(2*L)*180.0D-000/PI
FK(L)=XP(2*L+1)*180.0D-000/PI
IF (FD(L)<+0.0D-000) THEN
FD1(L)=FD(L)+180.0D-000
ELSE
FD1(L)=FD(L)
ENDIF
IF (FK(L)<+0.0D-000) THEN
FK1(L)=FK(L)+180.0D-000
```

```
ELSE
FK1(L)=FK(L)
ENDIF
WRITE(*,3) FD1(L),FK1(L)
ENDDO
WRITE(*,5) XI
5 FORMAT(F12.5)
OPEN (1,FILE=CC)
WRITE(1,*)"      EL       ECM        XI"
WRITE(1,4) EL,ECM,XI
WRITE(1,*) "    T            SE             DS             ST
XI"
DO I=1,NT
WRITE(1,6)TE(I),SECE(I),DSEC(I),SECT(I),DS(I)
ENDDO
WRITE(1,*) "     FD           FK"
DO I=LN,LV
WRITE(1,3)FD1(I),FK1(I)
ENDDO
CLOSE(1)
OPEN (1,FILE=BB)
DO I=LN,LV
WRITE(1,3)FD(I),FK(I)
ENDDO
CLOSE(1)
6 FORMAT(1X,F8.3,4E12.5)
4 FORMAT(3F10.5)
2 FORMAT(4F10.5)
3 FORMAT(2F10.5)
END
SUBROUTINE VAR(ST,PHN,NI,XP,EP,AMIN,NV)
IMPLICIT REAL(8) (A-Z)
INTEGER I,LMI,NV,NI,NPP,NT,IIN,NN,NP,IN,KK,NNN
REAL(8) XPN(0:50),XP(0:50),ST(0:50),FD(0:50),FK(0:50)
COMMON REAL(8) /R/ LMI,NT,PI,NP,NPP,LV,NNN,BB
CHARACTER(7) BB
! ************************************************
!BB='FAZ.DAT'
```

```
DO I=LMI,NP
XPN(I)=XP(I)
ENDDO
CALL DET(XPN,ST,B)
IF (NV==0) GOTO 3013
! ------------------------------------------------------------------
KK=1
DO IIN=1,NI
PH=PHN/KK
NN=-1
1119  NN=NN+1
IN=0
2229 A=B
XPN(NN)=XPN(NN)+PH*XPN(NN)
IF (XPN(NN)>PI)GOTO 1118
IF (NN>NPP+2) THEN
IF (XPN(NN)<0.0D-000) GOTO 1118
ELSE
IN=IN+1
ENDIF
! ------------------------------------------------------------------
CALL DET(XPN,ST,B)
IF (B<A) GOTO 2229
1118 XPN(NN)=XPN(NN)-PH*XPN(NN)
IF (XPN(NN)>PI .AND. NN/=0 .AND. NN/=1)THEN
XPN(NN)=XPN(NN)-PI
ENDIF
PRINT *,'1=',B,NN,XPN(NN)*180.0/PI
IF (IN>1) THEN
IF (ABS(B-A)<EP .OR. ABS(PH*XPN(NN))<EP) GOTO 4449
PH=PH/2.0D-000
ELSE
PH=-PH/2.0D-000
ENDIF
GOTO 2229
! ------------------------------------------------------------------
4449 IF (NN+1<NP) THEN
PH=PHN
```

```
!/KK
GOTO 1119
ENDIF
PRINT *,'-----------------------------------------------------------'
PRINT*, 'B=',B,IIN,PHN/KK/PI*180
PRINT *,'-----------------------------------------------------------'
DO L=LMI,LV
FD(L)=XPN(2*L)*180.0D-000/PI
FK(L)=XPN(2*L+1)*180.0D-000/PI
ENDDO
OPEN (5,FILE=BB)
DO I=LMI,LV
WRITE(5,3)FD(I),FK(I)
ENDDO
CLOSE(5)
KK=NNN*KK
ENDDO
3013 AMIN=B
DO I=LMI,NP
XP(I)=XPN(I)
3 FORMAT(2F10.5)
ENDDO
END
SUBROUTINE DET(XP,ST,XI)
IMPLICIT REAL(8) (A-Z)
INTEGER I,NT
REAL(8) XP(0:50),ST(0:50)
COMMON REAL(8) /T/
SECE(0:200),DS(0:200),DSEC(0:200),NT
S=0.0D-000
CALL SEC(XP,ST)
DO I=1,NT
DS(I)=((ST(I)-SECE(I))/DSEC(I))**2
S=S+DS(I)
ENDDO
XI=S/NT
END
SUBROUTINE SEC(XP,ST)
```

```
IMPLICIT REAL(8) (A-Z)
INTEGER TT,L,TMA,LMA,LMI,NYS
REAL(8)
S0(0:20),P(0:20),ST(0:200),FK(0:200),FD(0:200),XP(0:200)
COMMON REAL(8) /A/
PI,ET(0:200),ES(0:200),SKS(0:200),SDS(0:200),TE(0:200) /T/
SE(0:200),DS(0:200),DE(0:200),TMA /W/
GG,SS,LMI,LMA,NYS
RECUL1=0.0D-000; AIMCUL1=0.0D-000
DO L=LMI,LMA
!FD(L)=XP(L)
!FK(L)=XP(L+LMA+1)
FD(L)=XP(2*L)
FK(L)=XP(2*L+1)
ENDDO
CALL CULFAZ(GG,S0)
DO TT=1,TMA
T=TE(TT)*PI/180.0D-000
S00=2.0D-000*S0(0)
X=DCOS(T)
A=2.0D-000/(1.0D-000-X)
BB=-GG*A
ALO=GG*DLOG(A)+S00
RECUL=BB*DCOS(ALO)
AIMCUL=BB*DSIN(ALO)
IF (NYS==0) GOTO 555
X1=DCOS(T)
A1=2.0D-000/(1.0D-000+X1)
BB1=-GG*A1
ALO1=GG*DLOG(A1)+S00
RECUL1=BB1*DCOS(ALO1)
AIMCUL1=BB1*DSIN(ALO1)
555 RET=0.0D-000; AIT=0.0D-000; RES=0.0D-000; AIS=0.0D-
000
DO L=LMI,LMA
AL=ET(L)*DCOS(2.0D-000*FK(L))-1.0D-000
BE=ET(L)*DSIN(2.0D-000*FK(L))
LL=2.0D-000*L+1.0D-000
```

```
SL=2.0D-000*S0(L)
CALL POLLEG(X,L,P)
RET=RET+LL*(BE*DCOS(SL)+AL*DSIN(SL))*P(L)
AIT=AIT+LL*(BE*DSIN(SL)-AL*DCOS(SL))*P(L)
AL=ES(L)*DCOS(2.0D-000*FD(L))-1.0D-000
BE=ES(L)*DSIN(2.0D-000*FD(L))
RES=RES+LL*(BE*DCOS(SL)+AL*DSIN(SL))*P(L)
AIS=AIS+LL*(BE*DSIN(SL)-AL*DCOS(SL))*P(L)
ENDDO
IF (NYS==0) GOTO 556
AIT=2.0D-000*AIT
RET=2.0D-000*RET
AIS=2.0D-000*AIS
RES=2.0D-000*RES
556 RETR=RECUL+RECUL1+RET
AITR=AIMCUL+AIMCUL1+AIT
RESI=RECUL+RECUL1+RES
AISI=AIMCUL+AIMCUL1+AIS
CUL=RECUL**2+AIMCUL**2
SCUL=CUL*10.0D-000/4.0D-000/SS**2
SKS(TT)=10.0D-000*(RETR**2+AITR**2)/4.0D-000/SS**2
SDS(TT)=10.0D-000*(RESI**2+AISI**2)/4.0D-000/SS**2
ST(TT)=(2.0D-000/6.0D-000*SDS(TT)+4.0D-000/6.0D-
000*SKS(TT))
!/SCUL
ENDDO
END
SUBROUTINE POLLEG(X,L,P)
IMPLICIT REAL(8) (A-Z)
INTEGER L,I
REAL(8) P(0:20)
P(0)=1.0D-000; P(1)=X
DO I=2,L
P(I)=(2*I-1)*X/I*P(I-1)-(I-1)/I*P(I-2)
!P(I)=(2.0D-000*I-1.0D-000)*X/(1.0D-000*I)*P(I-1)-(1.0D-
000*I-1.0D-000)/(1.0D-000*I)*P(I-2)
ENDDO
END
```

```
SUBROUTINE CULFAZ(G,F)
IMPLICIT REAL(8) (A-Z)
INTEGER I,N,J
REAL(8) F(0:20)
C=0.577215665D-000; S=0.0D-000; N=1000
A1=1.2020569030D-000/3.0D-000; A2=1.0369277550D-
000/5.0D-000
DO I=1,N
A=G/(1.0D-000*I)-DATAN(G/(1.0D-000*I))-(G/(1.0D-
000*I))**3/3.0D-000+(G/(1.0D-000*I))**5/5.0D-000
S=S+A
ENDDO
FAZ=-C*G+A1*G**3-A2*G**5+S; F(0)=FAZ
DO J=1,10
F(J)=F(J-1)+DATAN(G/(J*1.0D-000))
ENDDO
END
```

Теперь выполним некоторый контрольный счет по этой программе для поиска фаз рассеяния при энергии 640 кэВ в с.ц.м. (в табл.2.4.1 это вариант 3) с учетом, вначале *S*, а затем, *S* и *P* парциальных волн. Первый из контрольных счетов, проведенный только для нулевой парциальной волны, приводит к следующим результатам:

| $\theta$ | $\sigma_e$ | $\sigma_t$ | $\chi^2_i$ |
|---|---|---|---|
| 46.150 | .12221E+04 | .12206E+04 | .14291E-03 |
| 57.340 | .53259E+03 | .56204E+03 | .30591E+00 |
| 68.300 | .33140E+03 | .31229E+03 | .33259E+00 |
| 79.020 | .17700E+03 | .19794E+03 | .13992E+01 |
| 109.450 | .81443E+02 | .83291E+02 | .51460E-01 |
| 128.310 | .57902E+02 | .60517E+02 | .20406E+00 |
| 137.350 | .54128E+02 | .54215E+02 | .25420E-03 |
| 146.160 | .50402E+02 | .49847E+02 | .12139E-01 |
| 154.800 | .49110E+02 | .46852E+02 | .21136E+00 |
| 163.290 | .46213E+02 | .44907E+02 | .79861E-01 |
| 171.680 | .43990E+02 | .43811E+02 | .16535E-02 |

| $\delta_d$ | $\delta_k$ |
|---|---|
| 170.08 | 179.99 |

$\chi^2 = 0.236$

В этом контрольном счете получаются фазы и $\chi^2$, совпадающие с результатами, приведенными выше в табл.2.4.1. под номером 3-1. В приведенных здесь распечатках приняты следующие обозначения: $\theta$ – угол рассеяния, $\sigma_e$ – экспериментальные сечения, $\sigma_t$ – вычисленные сечения, $\chi^2_i$ – парциальные $\chi^2$ для *i*-го угла, $\delta_d$ – дублетная фаза, $\delta_k$ – квартетная фаза, $\chi^2$ – среднее значение по всем экспериментальным точкам. При учете в фазовом анализе двух парциальных волн имеем

| $\theta$ | $\sigma_e$ | $\sigma_t$ | $\chi^2_i$ |
|---|---|---|---|
| 46.150 | .12221E+04 | .11875E+04 | .80049E-01 |
| 57.340 | .53259E+03 | .54118E+03 | .26020E-01 |
| 68.300 | .33140E+03 | .29882E+03 | .96658E+00 |
| 79.020 | .17700E+03 | .18906E+03 | .46396E+00 |
| 109.450 | .81443E+02 | .80736E+02 | .75339E-02 |
| 128.310 | .57902E+02 | .59753E+02 | .10226E+00 |
| 137.350 | .54128E+02 | .54042E+02 | .25495E-03 |
| 146.160 | .50402E+02 | .50126E+02 | .29924E-02 |
| 154.800 | .49110E+02 | .47469E+02 | .11164E+00 |
| 163.290 | .46213E+02 | .45758E+02 | .97024E-02 |
| 171.680 | .43990E+02 | .44800E+02 | .33889E-01 |

| $\delta_d$ | $\delta_k$ |
|---|---|
| 172.49 | 179.93 |
| 1.67139 | .00000 |

$\chi^2 = 0.164$

В этом варианте счета результаты совпадают с данными, приведенными выше в табл.2.4.1 под номером 3-2 [177].

Полученные в этом параграфе фазы упругого рассеяния использовались далее для построения межкластерных потенциалов [177] и расчетов астрофизического *S*-фактора радиационного $p^6Li$ захвата [172,175,178].

## *2.5 Фазовый анализ и компьютерные программы для рассеяния частиц со спином 1/2+1/2*
## *Phase shifts analysis and computer program for scattering particles with spin 1/2+1/2*

Рассмотрим теперь методы и компьютерные программы для фазового анализа процессов упругого рассеяния нетождественных частиц с полуцелым спином, например, это может быть упругое рассеяние $p^3He$, $p^3H$ или $p^{13}C$ и т.д.

### 2.5.1 Система со спин-орбитальным взаимодействием
### System with spin-orbit interaction

Использованные здесь выражения для дифференциальных сечений, на основе которых написана программа, приведены в нашей работе [17] вместе с аналогичными программами на языке Turbo Basic и для триплетного спинового состояния имеют вид, который учитывает только спин-орбитальное расщепление фаз рассеяния

$$\frac{d\sigma_t(\theta)}{d\Omega} = \frac{1}{3}\left[|A|^2 + 2\left(|B|^2 + |C|^2 + |D|^2 + |E|^2\right)\right] \quad .$$

где

$$A = f_c(\theta) + \frac{1}{2ik}\sum_{L=0}\{(L+1)\alpha_L^+ + L\alpha_L^-\}\exp(2i\sigma_L)P_L(\cos\theta) \quad ,$$

$$B = f_c(\theta) + \frac{1}{4ik}\sum_{L=0}\{(L+2)\alpha_L^+ + (2L+1)\alpha_L^0 + (L-1)\alpha_L^-\}\cdot \quad ,$$
$$\cdot\exp(2i\sigma_L)P_L(\cos\theta)$$

$$C = \frac{1}{2ik\sqrt{2}} \sum_{\mathrm{L}=1} \{\alpha_{\mathrm{L}}^{+} - \alpha_{\mathrm{L}}^{-}\} \exp(2i\sigma_{\mathrm{L}}) P_{\mathrm{L}}^{1}(\cos\theta) \quad ,$$

$$D = \frac{1}{2ik\sqrt{2}} \sum_{\mathrm{L}=1} \frac{1}{L(L+1)} \{L(L+2)\alpha_{\mathrm{L}}^{+} - (2L+1)\alpha_{\mathrm{L}}^{0} - (L-1)(L+1)\alpha_{\mathrm{L}}^{-}\} \cdot$$

$$\cdot \exp(2i\sigma_{\mathrm{L}}) P_{\mathrm{L}}^{1}(\cos\theta) \quad ,$$

$$E = \frac{1}{4ik} \sum_{\mathrm{L}=2} \frac{1}{L(L+1)} \{L\alpha_{\mathrm{L}}^{+} - (2L+1)\alpha_{\mathrm{L}}^{0} + (L+1)\alpha_{\mathrm{L}}^{-}\} \cdot$$

$$\cdot \exp(2i\sigma_{\mathrm{L}}) P_{\mathrm{L}}^{2}(\cos\theta) \quad .$$

Здесь определена величина $\alpha$ для каждого состояния с полным моментом $J = L \pm 1$ ($\alpha^{+}$ и $\alpha^{-}$) и $J = L$ ($\alpha^{0}$). Отметим, что существует и другая форма записи выражений для сечений, представленная через производные полиномов Лежандра [179].

Для рассеяния в синглетом спиновом состоянии имеем

$$\frac{d\sigma_{\mathrm{s}}(\theta)}{d\Omega} = |f_{\mathrm{s}}(\theta)|^{2} \quad ,$$

где

$$f_{\mathrm{s}}(\theta) = f_{\mathrm{c}}(\theta) + f^{\mathrm{N}}{}_{\mathrm{s}}(\theta) \, ,$$

$$f_{\mathrm{c}}(\theta) = -\left(\frac{\eta}{2k\sin^{2}(\theta/2)}\right) \exp\{i\eta \ln[\sin^{-2}(\theta/2)] + 2i\sigma_{0}\} \quad ,$$

$$f_{\mathrm{s}}^{\mathrm{N}}(\theta) = \frac{1}{2ik} \sum_{\mathrm{L}} (2L+1) \exp(2i\sigma_{\mathrm{L}}) [S_{\mathrm{L}}^{\mathrm{s}} - 1] P_{\mathrm{L}}(\cos\theta) \quad .$$

Здесь приняты обозначения переменных такими же, как для $^{4}$He$^{4}$He рассеяния. Суммарное сечение упругого рассея-ния определяется теперь в виде

$$\frac{d\sigma(\theta)}{d\Omega} = \frac{1}{4}\frac{d\sigma_s(\theta)}{d\Omega} + \frac{3}{4}\frac{d\sigma_t(\theta)}{d\Omega} \quad .$$

Приведем текст компьютерной программы для поиска фаз рассеяния в системе нетождественных частиц с полуцелым спином, а именно, $p^3He$. Данная программа учитывает спин-орбитальное расщепление фаз рассеяния, но не учитывает триплет-синглетное смешивание, которое будет включено в следующем параграфе.

```
PROGRAM FAZOVIY_ANALIZ_p3He_WITH_LS
IMPLICIT REAL(8) (A-Z)
INTEGER
I,L,Z1,Z2,LMI,LH,LMA,LN,LV,NV,NI,NPP,NT,NTT,NP
DIMENSION ST(0:50),FT(0:50),XP(0:50)
COMMON /A/ LH,LMI,NT,PI,NP,NPP
COMMON /B/ SE(0:50),DS(0:50),DE(0:50),NTT
COMMON /C/ SS,GG,LN,LV,POL(0:50),TT(0:50)
COMMON /D/  FP(0:50),FPI(0:50),EP(0:50),F0(0:50),F0I(0:50),
E0(0:50),M(0:50),FMI(0:50),EM(0:50),FS(0:50),FSI(0:50),
ES(0:50)
CHARACTER(9) BB
CHARACTER(7) AA
CHARACTER(15) CC
!  ********* INPUT PARAMETERS ********************
AA='SEC.DAT'
BB='FAZLS.DAT'
PI=4.0D-000*DATAN(1.0D-000)
P1=PI
Z1=1 ! Заряд p
Z2=2 ! Заряд 3He
AM1=1.0D-000 ! Масса p
AM2=3.0D-000 ! Масса 3He
AM=AM1+AM2
A1=41.46860D-000
```

```
PM=AM1*AM2/(AM1+AM2)
B1=2.0D-000*PM/A1
LMI=0 ! Начальный орбитальный момент
LH=1 ! Шаг по моменту
LMA=2 ! Максимальный момент
LN=LMI
LV=LMA
EPP=1.0D-005
NV=1 ! Если 1 то варьировать фазы, если 0 – без варьирова-
ния
FH=0.01D-000
NI=5 ! Число итераций
NPP=2*LMA
! ****** ECSPERIMENTAL CROSS SECTION 11.48 *******
NT=17 ! Число экспериментальных точек
NTT=NT
! ************ FOR P-3HE ON E=11.48 *****************
OPEN (1,FILE=AA)
DO L=1,NT
READ(1,*) TT(L),SE(L)
ENDDO
CLOSE(1)
OPEN (1,FILE=BB)
DO I=LN,LV
READ(1,*) FP(I),F0(I),FM(I),FS(I)
ENDDO
CLOSE(1)
! ******* TRANSFORM TO RADIANS ******************
DO L=LN,LV,LH
FM(L)=FM(L)*PI/180.0D-000
FP(L)=FP(L)*PI/180.0D-000
F0(L)=F0(L)*PI/180.0D-000
FMI(L)=FMI(L)*PI/180.0D-000
FPI(L)=FPI(L)*PI/180.0D-000
F0I(L)=F0I(L)*PI/180.0D-000
FT(L)=FT(L)*PI/180.0D-000
FS(L)=FS(L)*PI/180.0D-000
FSI(L)=FSI(L)*PI/180.0D-000
```

```
EP(L)=DEXP(-2.0D-000*FPI(L))
EM(L)=DEXP(-2.0D-000*FMI(L))
E0(L)=DEXP(-2.0D-000*F0I(L))
ES(L)=DEXP(-2.0D-000*FSI(L))
ENDDO
! ****************************************************
FH=FH*PI/180.0D-000
NP=2*NPP+1
DO I=LMI,LMA,LH
XP(I)=FP(I)
ENDDO
DO I=LMI,LMA-1,LH
XP(I+LMA+1)=F0(I+1)
ENDDO
DO I=LMI,LMA-1,LH
XP(I+2*LMA+1)=FM(I+1)
ENDDO
DO I=LMI,LMA,LH
XP(I+3*LMA+1)=FS(I)
ENDDO
DO I=LMI,LMA,LH
XP(I+4*LMA+2)=FPI(I)
ENDDO
DO I=LMI,LMA-1,LH
XP(I+5*LMA+3)=F0I(I+1)
ENDDO
DO I=LMI,LMA-1,LH
XP(I+6*LMA+3)=FMI(I+1)
ENDDO
DO I=LMI,LMA,LH
XP(I+7*LMA+3)=FSI(I)
ENDDO
!   ********** TRANSFORM TO C.M. ******************
EL=11.48D-000
CC='SECTLS.DAT'
EC=EL*PM/AM1
SK=EC*B1
SS=DSQRT(SK)
```

```
GG=3.44476D-002*Z1*Z2*PM/SS
CALL VAR(ST,FH,NI,XP,EPP,XI,NV)
PRINT *,"                    XI-KV=",XI
! ********* TOTAL CROSSS SECTION *****************
DO I=LMI,LMA,LH
FP(I)=XP(I)
ENDDO
DO I=LMI,LMA-1,LH
F0(I+1)=XP(I+LMA+1)
ENDDO
DO I=LMI,LMA-1,LH
FM(I+1)=XP(I+2*LMA+1)
ENDDO
DO I=LMI,LMA,LH
FS(I)=XP(I+3*LMA+1)
ENDDO
F0(0)=FP(0); FM(0)=FP(0)
DO I=LMI,LMA,LH
FPI(I)=XP(I+4*LMA+2)
ENDDO
DO I=LMI,LMA-1,LH
F0I(I+1)=XP(I+5*LMA+3)
ENDDO
DO I=LMI,LMA-1,LH
FMI(I+1)=XP(I+6*LMA+3)
ENDDO
DO I=LMI,LMA,LH
FSI(I)=XP(I+7*LMA+3)
ENDDO
F0I(0)=FPI(0); FMI(0)=FPI(0)
DO L=LN,LV,LH
EP(L)=DEXP(-2.0D-000*FPI(L))
EM(L)=DEXP(-2.0D-000*FMI(L))
E0(L)=DEXP(-2.0D-000*F0I(L))
ES(L)=DEXP(-2.0D-000*FSI(L))
ENDDO
SRT=0.0D-000;  SRS=0.0D-000;     SST=0.0D-000;    SSS=0.0D-
000
```

```
DO L=LN,LV,LH
AP=FP(L)
AM=FM(L)
A0=F0(L)
ASS=FS(L)
L1=2*L+3
L2=2*L+1
L3=2*L-1
SRT=SRT+L1*(1.0D-000-EP(L)**2)+L2*(1.0D-000-
E0(L)**2)+L3*(1.0D-000-EM(L)**2)
SRS=SRS+L2*(1.0D-000-ES(L)**2)
SST=SST+L1*EP(L)**2*DSIN(AP)**2+L2*E0(L)**2*DSIN(A
0)**2+L3*EM(L)**2*DSIN(AM)**2
SSS=SSS+L2*ES(L)**2*DSIN(ASS)**2
ENDDO
SRT=10.0D-000*PI*SRT/SK/3.0D-000
SRS=10.0D-000*PI*SRS/SK
SIGR=1.0D-000/4.0D-000*SRS+3.0D-000/4.0D-000*SRT
SST=10.0D-000*4.0D-000*PI*SST/SK/3.0D-000
SSS=10.0D-000*4.0D-000*PI*SSS/SK
SIGS=1.0D-000/4.0D-000*SSS+3.0D-000/4.0D-000*SST
!PRINT *,"                    SIGMS-TOT=",SIGS
PRINT *,"    T     SE        ST          XI"
DO I=1,NT
WRITE(*,2)TT(I),SE(I),ST(I),DS(I)
ENDDO
PRINT *,'    FP      F0      FM      FS'
DO L=LMI,LMA,LH
FM(L)=FM(L)*180.0D-000/PI
FP(L)=FP(L)*180.0D-000/PI
FMI(L)=FMI(L)*180.0D-000/PI
FPI(L)=FPI(L)*180.0D-000/PI
F0(L)=F0(L)*180.0D-000/PI
F0I(L)=F0I(L)*180.0D-000/PI
FS(L)=FS(L)*180.0D-000/PI
FSI(L)=FSI(L)*180.0D-000/PI
WRITE(*,2) FP(L),F0(L),FM(L),FS(L)
ENDDO
```

```
OPEN (1,FILE=CC)
WRITE(1,*)"      EL       ECM       XI"
WRITE(1,4) EL,EC,XI
WRITE(1,*) "    T            SE             ST             XI"
DO I=1,NT
WRITE(1,2) TT(I),SE(I),ST(I),DS(I)
ENDDO
WRITE(1,*) "     FP(L)    F0(L)    FM(L)    FS(L)"
DO L=LN,LV
WRITE(1,2) FP(L),F0(L),FM(L),FS(L)
ENDDO
CLOSE(1)
OPEN (1,FILE=BB)
DO L=LN,LV
WRITE(1,3) FP(L),F0(L),FM(L),FS(L)
ENDDO
CLOSE(1)
4 FORMAT(1x,3F10.3)
2 FORMAT(1x,4F10.3)
3 FORMAT(1x,4F14.7)
END
SUBROUTINE VAR(ST,PHN,NI,XP,EP,AMIN,NV)
IMPLICIT REAL(8) (A-Z)
INTEGER I,LMI,LH,NV,NI,NPP,NT,IIN,NN,NP
DIMENSION XPN(0:50),XP(0:50),ST(0:50)
COMMON /A/ LH,LMI,NT,PI,NP,NPP
! **************** ПОИСК ФАЗ ***********************
DO I=LMI,NP,LH
XPN(I)=XP(I)
ENDDO
NN=LMI
PRINT *,NN,XPN(NN)*180.0D-000/PI
PH=PHN
CALL DET(XPN,ST,ALA)
B=ALA
IF (NV==0) GOTO 3013
PRINT *,ALA
DO IIN=1,NI
```

```
NN=-LH
PRINT *,ALA,IIN
GOTO 1119
1159 XPN(NN)=XPN(NN)-PH*XP(NN)
1119  NN=NN+LH
IN=0
2229 A=B
XPN(NN)=XPN(NN)+PH*XP(NN)
IF (NP==2*NPP+1) GOTO 7777
IF (NN<(NP/2)) GOTO 7777
IF (XPN(NN)<0) GOTO 1159
7777 IN=IN+1
CALL DET(XPN,ST,ALA)
B=ALA
GOTO 5678
5678 CONTINUE
IF (B<A) GOTO 2229
C=A
XPN(NN)=XPN(NN)-PH*XP(NN)
IF (IN>1) GOTO 3339
PH=-PH
GOTO 5559
3339 IF (ABS((C-B)/(B))<EP) GOTO 4449
PH=PH/2
5559 B=C
GOTO 2229
4449 PH=PHN
B=C
IF (NN<NP) GOTO 1119
AMIN=B
PH=PHN
ENDDO
3013 AMIN=B
DO I=LMI,NP,LH
XP(I)=XPN(I)
ENDDO
END
SUBROUTINE DET(XP,ST,XI)
```

```
! ***************** ДЕТЕРМИНАНТ ****************
IMPLICIT REAL(8) (A-Z)
INTEGER I,NT
DIMENSION XP(0:50),ST(0:50)
COMMON /B/ SE(0:50),DS(0:50),DE(0:50),NT
S=0.0D-000
CALL SEC(XP,ST)
DO I=1,NT
DS(I)=((ST(I)-SE(I)))**2/(0.025*SE(I))**2
S=S+DS(I)
ENDDO
XI=S/NT
END
SUBROUTINE SEC(XP,ST)
! ***************** СЕЧЕНИЕ **********************
IMPLICIT REAL(8) (A-Z)
INTEGER I,L,II,LH,LN,LV,NP,NPP,LMI,NT
COMMON /D/
FP(0:50),FPI(0:50),EP(0:50),F0(0:50),F0I(0:50),E0(0:50),FM(0:5
0),FMI(0:50),EM(0:50),FS(0:50),FSI(0:50),ES(0:50)
COMMON /C/ SS,GG,LN,LV,POL(0:50),TT(0:50)
COMMON /A/ LH,LMI,NT,PI,NP,NPP
DIMENSION
S0(0:50),P(0:50),P1(0:50),P2(0:50),ST(0:50),XP(0:50)
DO I=LN,LV,LH
FP(I)=XP(I)
ENDDO
DO I=LN,LV-1,LH
II=I+LV+1
F0(I+1)=XP(II)
ENDDO
DO I=LN,LV-1,LH
II=I+2*LV+1
FM(I+1)=XP(II)
ENDDO
DO I=LN,LV,LH
II=I+3*LV+1
FS(I)=XP(II)
```

```
ENDDO
F0(0)=FP(0); FM(0)=FP(0)
DO I=LN,LV,LH
II=I+4*LV+2
FPI(I)=XP(II)
ENDDO
DO I=LN,LV-1,LH
II=I+5*LV+3
F0I(I+1)=XP(II)
ENDDO
DO I=LN,LV-1,LH
II=I+6*LV+3
FMI(I+1)=XP(II)
ENDDO
DO I=LN,LV,LH
II=I+7*LV+3
FSI(I)=XP(II)
ENDDO
F0I(0)=FPI(0); FMI(0)=FPI(0)
DO L=LN,LV,LH
EP(L)=DEXP(-2.0D-000*FPI(L))
EM(L)=DEXP(-2.0D-000*FMI(L))
E0(L)=DEXP(-2.0D-000*F0I(L))
ES(L)=DEXP(-2.0D-000*FSI(L))
ENDDO
CALL CULFAZ(GG,S0)
DO I=1,NT
T=TT(I)*PI/180.0D-000
X=DCOS(T)
CALL CULAMP(X,GG,S0,RECUL,AMCUL)
CALL POLLEG(X,LV,P)
CALL FUNLEG1(X,LV,P1)
CALL FUNLEG2(X,LV,P2)
RES=0.0D-000;  AMS=0.0D-000; REA=0.0D-000; AMA=0.0D-
000; REB=0.0D-000; AMB=0.0D-000
REC=0.0D-000; AMC=0.0D-000; RED=0.0D-000; AMD=0.0D-
000; REE=0.0D-000; AME=0.0D-000
DO L=LN,LV,LH
```

```
FP1=2.0D-000*FP(L);      FM1=2.0D-000*FM(L);      F01=2.0D-000*F0(L)
SL=2.0D-000*S0(L);  C=DCOS(SL);  S=DSIN(SL);  FS1=2.0D-000*FS(L)
AL1P=EP(L)*DCOS(FP1)-1.0D-000
AL2P=EP(L)*DSIN(FP1)
AL1M=EM(L)*DCOS(FM1)-1.0D-000
AL2M=EM(L)*DSIN(FM1)
AL10=E0(L)*DCOS(F01)-1.0D-000
AL20=E0(L)*DSIN(F01)
A1=(L+1)*AL1P+L*AL1M
A2=(L+1)*AL2P+L*AL2M
REA=REA+(A2*C+A1*S)*P(L)
AMA=AMA+(A2*S-A1*C)*P(L)
B1=(L+2)*AL1P+(2*L+1)*AL10+(L-1)*AL1M
B2=(L+2)*AL2P+(2*L+1)*AL20+(L-1)*AL2M
REB=REB+(B2*C+B1*S)*P(L)/2.0D-000
AMB=AMB+(B2*S-B1*C)*P(L)/2.0D-000
IF (L<1) GOTO 2111
C1=AL1P-AL1M
C2=AL2P-AL2M
CC1=1.0D-000/(DSQRT(2.0D-000))
REC=REC+(C2*C+C1*S)*P1(L)*CC1
AMC=AMC+(C2*S-C1*C)*P1(L)*CC1
DD1=1.0D-000/(DSQRT(2.0D-000)*L*(L+1))
D1=L*(L+2)*AL1P-(2*L+1)*AL10-(L**2-1)*AL1M
D2=L*(L+2)*AL2P-(2*L+1)*AL20-(L**2-1)*AL2M
RED=RED+(D2*C+D1*S)*P1(L)*DD1
AMD=AMD+(D2*S-D1*C)*P1(L)*DD1
2111 IF (L<2) GOTO 2222
EE1=1.0D-000/(2*L*(L+1))
E1=L*AL1P-(2*L+1)*AL10+(L+1)*AL1M
E2=L*AL2P-(2*L+1)*AL20+(L+1)*AL2M
REE=REE+(E2*C+E1*S)*P2(L)*EE1
AME=AME+(E2*S-E1*C)*P2(L)*EE1
2222 ENDDO
RES=0.0D-000;  AMS=0.0D-000
DO L=LN,LV,LH
```

```
SL=2.0D-000*S0(L)
C=DCOS(SL)
S=DSIN(SL)
FS1=2.0D-000*FS(L)
ALS=ES(L)*DCOS(FS1)-1.0D-000
BS=ES(L)*DSIN(FS1)
RES=RES+(2*L+1)*(BS*C+ALS*S)*P(L)
AMS=AMS+(2*L+1)*(BS*S-ALS*C)*P(L)
ENDDO
9191 CONTINUE
RES=RECUL+RES
AMS=AMCUL+AMS
SES=10.0D-000*(RES**2+AMS**2)/4.0D-000/SS**2
REA=RECUL+REA
AMA=AMCUL+AMA
REB=RECUL+REB
AMB=AMCUL+AMB
AA=REA**2+AMA**2
BB=REB**2+AMB**2
CC=REC**2+AMC**2
DD=RED**2+AMD**2
EE=REE**2+AME**2
SET=10.0D-000*(AA+2*(BB+CC+DD+EE))/4.0D-
000/SS**2/3.0D-000
S=3.0D-000/4.0D-000*SET+1.0D-000/4.0D-000*SES
ST(I)=S
ENDDO
END
SUBROUTINE CULAMP(X,GG,S0,RECUL,AMCUL)
! ********* КУЛОНОВСКАЯ АМПЛИТУДА  ***********
IMPLICIT REAL(8) (A-Z)
DIMENSION S0(0:20)
A=2.0D-000/(1.0D-000-X)
S00=2.0D-000*S0(0)
BB=-GG*A
AL=GG*DLOG(A)+S00
RECUL=BB*DCOS(AL)
AMCUL=BB*DSIN(AL)
```

```
END
SUBROUTINE POLLEG(X,L,P)
! ****************** ПОЛИНОМ ЛЕЖАНДРА **********
IMPLICIT REAL(8) (A-Z)
INTEGER I,L
DIMENSION P(0:20)
P(0)=1.0D-000
P(1)=X
DO I=2,L
P(I)=(2.0D-000*I-1.0D-000)*X/I*P(I-1)-(I-1.0D-000)/I*P(I-2)
ENDDO
END
SUBROUTINE FUNLEG1(X,L,P)
! ****************** ФУНКЦИЯ ЛЕЖАНДРА **********
IMPLICIT REAL(8) (A-Z)
INTEGER I,L
DIMENSION P(0:20)
P(0)=0.0D-000
P(1)=DSQRT(ABS(1.0D-000-X**2))
DO I=2,L
P(I)=(2.0D-000*I-1.0D-000)*X/(I-1.0D-000)*P(I-1)-I/(I-1.0D-
000)*P(I-2)
ENDDO
END
SUBROUTINE FUNLEG2(X,L,P)
! ****************** ФУНКЦИЯ ЛЕЖАНДРА **********
IMPLICIT REAL(8) (A-Z)
INTEGER L
DIMENSION P(0:20)
P(0)=0.0D-000
P(1)=0.0D-000
P(2)=3.0D-000*ABS(1.0D-000-X**2)
DO I=3,L
P(I)=(2.0D-000*I-1.0D-000)*X/(I-2.0D-000)*P(I-1)-(I+1.0D-
000)/(I-2.0D-000)*P(I-2)
ENDDO
END
SUBROUTINE CULFAZ(G,F)
```

```
! ***************** КУЛОНОВСКИЕ ФАЗЫ **********
IMPLICIT REAL(8) (A-Z)
INTEGER I,N
DIMENSION F(0:20)
C=0.577215665D-000
S=0.D-000
N=50
A1=1.202056903D-000/3.D-000
A2=1.036927755D-000/5.D-000
DO I=1,N
A=G/I-DATAN(G/I)-(G/I)**3/3.D-000+(G/I)**5/5.D-000
S=S+A
ENDDO
FAZ=-C*G+A1*G**3-A2*G**5+S
F(0)=FAZ
DO I=1,20
F(I)=F(I-1)+DATAN(G/I)
ENDDO
END
```

Приведем теперь результаты контрольного счета по этой программе для случая упругого $p^3$He рассеяния при энергии 11.48 МэВ, где с учетом синглет-триплетного смешивания для $\chi^2$ было получено 0.45 [180,181]. В этих работах в табличном виде приведены дифференциальные сечения, их ошибки и фазы рассеяния, полученные при фазовом анализе, которые мы используем для контрольных расчетов, т.е. считаем сечения и $\chi^2$ с полученными в [180,181] фазами.

Далее в распечатке результатов приняты следующие обозначения: $\theta$ – угол рассеяния, $\sigma_e$ – экспериментальные сечения, $\sigma_t$ – вычисленные сечения, $\chi^2_i$ – парциальные $\chi^2$ для $i$-го угла, $\chi^2$ – среднее значение $\chi^2$, $\delta_m$ – триплетная фаза с $J = L - 1$, $\delta_0$ – фаза при $J = L$ , $\delta_p$ – фаза при $J = L + 1$, $\delta_s$ – синглетная фаза рассеяния.

В первой строчке распечатки фазы соответствуют орбитальному моменту $L = 0$, во второй $L = 1$ и в третьей $L = 2$

$\chi^2$ = 7.375E-001

| $\theta$ | $\sigma_e$ | $\sigma_t$ | $\chi^2_i$ |
|---|---|---|---|
| 27.640 | 223.100 | 229.159 | 1.179 |
| 31.970 | 222.000 | 222.687 | 0.015 |
| 36.710 | 211.900 | 211.146 | 0 .020 |
| 82.530 | 54.270 | 53.522 | 0.302 |
| 90.000 | 36.760 | 36.249 | 0.309 |
| 96.030 | 25.700 | 25.467 | 0.133 |
| 103.800 | 16.780 | 16.162 | 2.165 |
| 110.550 | 13.210 | 12.598 | 3.444 |
| 116.570 | 13.210 | 13.120 | 0.075 |
| 125.270 | 20.260 | 19.962 | 0.341 |
| 133.480 | 32.210 | 32.333 | 0.023 |
| 140.790 | 45.950 | 46.975 | 0.794 |
| 147.210 | 58.820 | 61.388 | 3.052 |
| 153.900 | 75.460 | 76.517 | 0.313 |
| 162.140 | 92.720 | 93.062 | 0.022 |
| 165.670 | 97.700 | 98.823 | 0.212 |
| 166.590 | 101.100 | 100.157 | 0.139 |

| $\delta_p$ | $\delta_0$ | $\delta_m$ | $\delta_s$ |
|---|---|---|---|
| -88.800 | -88.800 | -88.800 | -84.600 |
| 66.700 | 49.400 | 44.300 | 21.400 |
| 2.500 | 2.500 | 2.500 | -18.600 |

Поскольку эта программа не учитывает синглет-триплетного смешивания, то и результаты для $\chi^2$ получаются несколько больше, чем в работах [180,181].

Теперь для сравнения приведем результаты с теми же фазами, при выполнении контрольных расчетов по программам на языке Turbo Basic в нашей предыдущей работе [17], где была получена величина $\chi^2$ = 0.74:

| $\theta$ | $\sigma_e$ | $\sigma_t$ |
|---|---|---|
| 27.64 | 223.10 | 229.16 |
| 31.97 | 222.00 | 222.69 |
| 36.71 | 211.90 | 211.15 |
| 82.53 | 54.27 | 53.52 |
| 90.00 | 36.76 | 36.25 |
| 96.03 | 25.70 | 25.47 |
| 103.80 | 16.78 | 16.16 |
| 110.55 | 13.21 | 12.60 |
| 116.57 | 13.21 | 13.12 |
| 125.27 | 20.26 | 19.96 |
| 133.48 | 32.21 | 32.33 |
| 140.79 | 45.95 | 46.98 |
| 147.21 | 58.82 | 61.39 |
| 153.90 | 75.46 | 76.52 |
| 162.14 | 92.72 | 93.06 |
| 165.67 | 97.70 | 98.82 |
| 166.59 | 101.10 | 100.16 |

Видно, что эти результаты совпадают с точностью до ошибок округления, т.е. не зависят от языка, на котором написана программа. Однако переход на язык Fortran-90 позволил существенно повысить скорость работы всех компьютерных программ и возможность использования более высокой точности поиска минимума $\chi^2$.

Если теперь выполнить полное варьирование фаз по нашей новой компьютерной программе, с поиском минимального значения $\chi^2$, то после, примерно, 3 000 итераций получим следующий результат:

$$\chi^2 = 2.27E\text{-}001$$

| $\theta$ | $\sigma_e$ | $\sigma_t$ | $\chi^2_i$ |
|---|---|---|---|
| 27.640 | 223.100 | 224.854 | 0.099 |
| 31.970 | 222.000 | 221.150 | 0.023 |
| 36.710 | 211.900 | 211.107 | 0.022 |
| 82.530 | 54.270 | 53.832 | 0.104 |
| 90.000 | 36.760 | 36.715 | 0.002 |
| 96.030 | 25.700 | 26.026 | 0.257 |
| 103.800 | 16.780 | 16.737 | 0.011 |
| 110.550 | 13.210 | 13.072 | 0.175 |
| 116.570 | 13.210 | 13.423 | 0.415 |
| 125.270 | 20.260 | 19.928 | 0.430 |
| 133.480 | 32.210 | 31.955 | 0.100 |
| 140.790 | 45.950 | 46.330 | 0.109 |
| 147.210 | 58.820 | 60.563 | 1.405 |
| 153.900 | 75.460 | 75.569 | 0.003 |
| 162.140 | 92.720 | 92.038 | 0.087 |
| 165.670 | 97.700 | 97.785 | 0.001 |
| 166.590 | 101.100 | 99.116 | 0.616 |

| $\delta_p$ | $\delta_0$ | $\delta_m$ | $\delta_s$ |
|---|---|---|---|
| -84.530 | -84.530 | -84.530 | -93.630 |
| 56.144 | 46.185 | 42.904 | 44.694 |
| 3.404 | 3.584 | 4.982 | -19.086 |

Полученные новые фазы рассеяния несколько отличаются от предыдущих результатов, а $\chi^2$ уменьшился более чем в три раза.

## 2.5.2 Система со спин-орбитой и синглет-триплетным смешиванием System with spin-orbit and singlet-triplet mixing

Рассмотрим далее рассеяние нетождественных частиц с полуцелым спином, с учетом спин-орбитальных взаимодействий, смешивания различных орбитальных состояний за счет тензорных сил и смешивания синглет-триплетных состояний. Дифференциальное сечение рассеяния имеет более сложный вид, в формулы для сечений входят, как фазы рассеяния, так и параметры смешивания состояний с разным спином и орбитальным моментом, а наиболее полные выражения для таких сечений приведены в работе [182]

$$\frac{d\sigma(\theta)}{d\Omega}=\frac{1}{2k^2}\{|A|^2+|B|^2+|C|^2+|D|^2+|E|^2+|F|^2+|G|^2+|H|^2\} \ .$$

Амплитуды рассеяния записываются в виде

$$A = f_{\rm c}^{'} + \frac{1}{4}\sum_{\rm L=0}^{\infty} P_{\rm L}(x)\Big\{-\sqrt{L(L-1)}U_{\rm L,1;L\text{-}2,1}^{\rm L\text{-}1} + (L+2)U_{\rm L,1;L,1}^{\rm L+1} + (2L+1)U_{\rm L,1;L,1}^{\rm L} +$$
$$+(L-1)U_{\rm L,1;L,1}^{\rm L\text{-}1} - \sqrt{(L+1)(L+2)}\,U_{\rm L,1;L+2,1}^{\rm L+1}\Big\} \quad ,$$

$$B = f_{\rm c}^{'} + \frac{1}{4}\sum_{\rm L=0}^{\infty} P_{\rm L}(x)\Big\{\sqrt{L(L-1)}U_{\rm L,1;L\text{-}2,1}^{\rm L\text{-}1} + (L+1)U_{\rm L,1;L,1}^{\rm L+1} + (2L+1)U_{\rm L,0;L,0}^{\rm L} +$$
$$+LU_{\rm L,1;L,1}^{\rm L\text{-}1} + \sqrt{(L+1)(L+2)}\,U_{\rm L,1;L+2,1}^{\rm L+1}\Big\} \quad ,$$

$$C = \frac{1}{4}\sum_{\rm L=0}^{\infty} P_{\rm L}(x)\Big\{\sqrt{L(L-1)}U_{\rm L,1;L\text{-}2,1}^{\rm L\text{-}1} + (L+1)U_{\rm L,1;L,1}^{\rm L+1} - (2L+1)U_{\rm L,0;L,0}^{\rm L} +$$
$$+LU_{\rm L,1;L,1}^{\rm L\text{-}1} + \sqrt{(L+1)(L+2)}\,U_{\rm L,1;L+2,1}^{\rm L+1}\Big\} \quad ,$$

$$D = -\frac{1}{4}i\sin\theta\sum_{\rm L=1}^{\infty} P_{\rm L}^{'}(x)/\sqrt{L(L+1)}\Big\{-\sqrt{(L+1)(L-1)}U_{\rm L,1;L\text{-}2,1}^{\rm L\text{-}1} + \sqrt{L(L+1)}U_{\rm L,1;L,1}^{\rm L+1} -$$
$$-\sqrt{L(L+1)}U_{\rm L,1;L,1}^{\rm L\text{-}1} + \sqrt{L(L+2)}\,U_{\rm L,1;L+2,1}^{\rm L+1} - (2L+1)U_{\rm L,1;L,0}^{\rm L}\Big\},$$

$$E = -\frac{1}{4}i\sin\theta\sum_{L=1}^{\infty} P_{\rm L}^{'}(x)/\sqrt{L(L+1)}\Big\{-\sqrt{(L+1)(L-1)}U_{\rm L,1;L\text{-}2,1}^{\rm L\text{-}1} + \sqrt{L(L+1)}U_{\rm L,1;L,1}^{\rm L+1} -$$
$$-\sqrt{L(L+1)}U_{\rm L,1;L,1}^{\rm L\text{-}1} + \sqrt{L(L+2)}U_{\rm L,1;L+2,1}^{\rm L+1} + (2L+1)U_{\rm L,1;L,0}^{\rm L}\Big\} \quad ,$$

$$F = -\frac{1}{4}i\sin^2\theta\sum_{\rm L=2}^{\infty} P_{\rm L}^{''}(x)/\sqrt{(L-1)L(L+1)(L+2)}\Big\{-\sqrt{(L+1)(L+2)}U_{\rm L,1;L\text{-}2,1}^{\rm L\text{-}1} +$$
$$+\sqrt{\frac{L(L-1)(L+2)}{L+1}}U_{\rm L,1;L,1}^{\rm L+1} - (2L+1)\sqrt{\frac{(L-1)(L+2)}{L(L+1)}}U_{\rm L,1;L,1}^{\rm L} +$$
$$+\sqrt{\frac{(L-1)(L+1)(L+2)}{L}}U_{\rm L,1;L,1}^{\rm L\text{-}1} - \sqrt{L(L-1)}\,U_{\rm L,1;L+2,1}^{\rm L+1}\Big\},$$

$$G = -\frac{1}{4}i\sin\theta\sum_{\rm L=1}^{\infty} P_{\rm L}^{'}(x)/\sqrt{L(L+1)}\left\{\sqrt{(L-1)(L+1)}U_{\rm L,1;L\text{-}2,1}^{\rm L\text{-}1} + (L+2)\sqrt{\frac{L}{L+1}}U_{\rm L,1;L,1}^{\rm L+1} -\right.$$
$$\left.-\frac{(2L+1)}{\sqrt{L(L+1)}}U_{\rm L,1;L,1}^{\rm L} - (L-1)\sqrt{\frac{(L+1)}{L}}U_{\rm L,1;L,1}^{\rm L\text{-}1} - \sqrt{L(L+2)}U_{\rm L,1;L+2,1}^{\rm L+1} - (2L+1)U_{\rm L,0;L,1}^{\rm L}\right\} \quad ,$$

$$H = -\frac{1}{4} i \sin\theta \sum_{L=1}^{\infty} P_L^{'}(x) / \sqrt{L(L+1)} \left\{ \sqrt{(L-1)(L+1)} U^{L-1}_{L,1;L-2,1} + (L+2)\sqrt{\frac{L}{L+1}} U^{L+1}_{L,1;L,1} - \right.$$

$$\left. - \frac{(2L+1)}{\sqrt{L(L+1)}} U^{L}_{L,1;L,1} - (L-1)\sqrt{\frac{(L+1)}{L}} U^{L-1}_{L,1;L,1} - \sqrt{L(L+2)} U^{L+1}_{L,1;L+2,1} + (2L+1)\, U^{L}_{L,0;L,1} \right\} .$$

Матрица рассеяния представляется в виде

$$U^{J}_{L,S;L',S'} = U^{J}_{L',S';L,S} = \exp[i(\alpha_L + \alpha_{L'})](S^{J}_{L',S';L,S} - \delta_{L,L'}\delta_{S,S'})$$

и, например, при $L = 1$ и $J = 1$ с учетом смешивания $\varepsilon^{1}_{1,0} = \varepsilon^{J}_{S,S'}$ синглетного и триплетного состояний записывается

$$S^{1}_{1,0;1,0} = \cos^2\varepsilon^{1}_{1,0} \exp(2i\delta^{1}_{0,1}) + \sin^2\varepsilon_{1,1} \exp(2i\delta^{1}_{1,1}) \quad ,$$

$$S^{1}_{1,1;1,1} = \sin^2\varepsilon^{1}_{1,0} \exp(2i\delta^{1}_{0,1}) + \cos^2\varepsilon_{1,1} \exp(2i\delta^{1}_{1,1}) \quad ,$$

$$S^{1}_{1,0;1,1} = S^{1}_{1,1;1,0} = \frac{1}{2}\sin(2\varepsilon^{1}_{1,0})\Big(\exp(2i\delta^{1}_{0,1}) - \exp(2i\delta^{1}_{1,1})\Big) \quad ,$$

где $\delta_{k,k'}$ – символ Кронекера, $x = \cos(\theta)$, величины без штриха обозначают начальное состояние, а со штрихом – конечное при таком же полном моменте $J$, кулоновские фазы $\alpha_L$ определены в параграфе 2.1, а ядерные фазы $\delta^{J}_{S,L}$ считаются комплексными, чтобы учесть неупругие каналы.

Смешивание $\varepsilon_1 = \varepsilon^{J}_{S,S'}$ триплетных (спины $S$ и $S'$, и полный момент равны 1) $S$ и $D$ состояний определяется следующими выражениями для матрицы рассеяния

$$S^{1}_{0,0} = \cos^2\varepsilon_1 \exp(2i\delta^{1}_{0}) + \sin^2\varepsilon_1 \exp(2i\delta^{1}_{2}) \ ,$$

$$S^{1}_{2,2} = \sin^2\varepsilon_1 \exp(2i\delta^{1}_{0}) + \cos^2\varepsilon_1 \exp(2i\delta^{1}_{2}) \ ,$$

$$S^1_{0,2} = S^1_{2,0} = \frac{1}{2}\sin(2\varepsilon_1)\Big[\exp(2i\delta^1_0) - \exp(2i\delta^1_2)\Big] .$$

Штрихи у полиномов Лежандра обозначают производные, а кулоновская амплитуда рассеяния записана в форме

$$f_c^{'}(\theta) = -\left(\frac{i\eta}{2\sin^2(\theta/2)}\right)\exp\left\{i\eta\ln[\sin^{-2}(\theta/2)]\right\} .$$

Производные полиномов Лежандра связаны с функциями Лежандра, о которых говорилось ранее, следующим образом

$$P^{\mathrm{m}}_{\mathrm{n}}(x) = (1-x^2)^{\mathrm{m}/2}\frac{d^{\mathrm{m}}P_{\mathrm{n}}(x)}{dx^{\mathrm{m}}} = (1-x^2)^{\mathrm{m}/2}\frac{d^{\mathrm{m+n}}(x^2-1)^{\mathrm{n}}}{dx^{\mathrm{m+n}}} =$$
$$= sin^{\mathrm{m}}\theta\frac{d^{\mathrm{m}}P_{\mathrm{n}}(\cos\theta)}{(d\cos\theta)^{\mathrm{m}}} .$$

Если в приведенных выше выражениях пренебречь тензорным взаимодействием и синглет-триплетным смешиванием, то матрица рассеяния примет обычный вид $\exp(2i\delta_{\mathrm{S,L}})$ с комплексной фазой.

Следующая программа основана на всех этих выражениях и включает синглет-триплетное смешивание EPS [182], которое учитывается только для $L = 1$. В программе явно заданы значения экспериментальных сечений, их ошибок и фаз рассеяния. Обозначения переменных, параметров и блоков, как в предыдущей программе.

Операторы

```
READ(1,*) TT(L),SE(L),DE(L)
```

и

```
READ(1,*) FP(I),F0(I),FM(I),FS(I),EPS(I)
```

вначале программы считывают из файлов

```
AA='SEC.DAT'
BB='FAZTENZ.DAT'
```

те же данные, которые явно заданы в самой программе, т.е. исходные данные можно использовать из программы или из этих файлов.

**PROGRAM FAZ_ANAL_P_3He_WIRH_LS_AND_TS**

```
IMPLICIT REAL(8) (A-Z)
INTEGER
I,L,LH,LN,LV,LMA,LMI,NT,NP,NPP,Z1,Z2,LMI1,LH1,NT2,N
T1,NI
DIMENSION
ST(0:50),FP(0:50),F0(0:50),FM(0:50),FS(0:50),ES(0:50)
DIMENSION
XP(0:50),FPI(0:50),F0I(0:50),FMI(0:50),FSI(0:50),EPS(0:50),EP
(0:50),E0(0:50),EM(0:50)
COMMON /B/ SE(0:50),DS(0:50),DE(0:50),NT
COMMON /C/ LH,LMI,NT1,PI,NP,NPP
COMMON /A/ SS,GG,P1,LMI1,LMA,LH1,NT2,POL(0:50),
TT(0:50)
CHARACTER(11) BB
CHARACTER(7) AA
CHARACTER(12) CC
! *************** INPUT PARAMETERS ****************
AA='SEC.DAT'
BB='FAZTENZ.DAT'
CC='SECTTENZ.DAT'
PI=4.0D-000*DATAN(1.0D-000)
P1=PI
Z1=1
Z2=2
AM1=1.0D-000
AM2=3.0D-000
AM=AM1+AM2
A1=41.4686D-000
```

```
PM=AM1*AM2/(AM1+AM2)
B1=2.0D-000*PM/A1
LMI=0
LH=1
LMA=2
LN=LMI
LMI1=LMI
LV=LMA
LH1=LH
EPP=1.0D-05
NV=1
FH=0.01D-000
NI=500
NPP=2*LMA
! **** ECSPERIMENTAL CROSS SECTION 11.48 *********
SE(1)=223.1D-000; SE(2)=222.0D-000; SE(3)=211.9D-000;
SE(4)=54.27D-000; SE(5)=36.76D-000
SE(6)=25.7D-000; SE(7)=16.78D-000; SE(8)=13.21D-000;
SE(9)=13.21D-000; SE(10)=20.26D-000
SE(11)=32.21D-000; SE(12)=45.95D-000; SE(13)=58.82D-000;
SE(14)=75.46D-000
SE(15)=92.72D-000; SE(16)=97.7D-000; SE(17)=101.1D-000
DE(1)=5.58D-000; DE(2)=5.55D-000; DE(3)=5.3D-000;
DE(4)=1.36D-000; DE(5)=0.92D-000
DE(6)=0.64D-000; DE(7)=0.42D-000; DE(8)=0.33D-000;
DE(9)=0.33D-000; DE(10)=0.51D-000
DE(11)=0.81D-000; DE(12)=1.15D-000; DE(13)=1.47D-000;
DE(14)=1.89D-000
DE(15)=2.32D-000; DE(16)=2.44D-000; DE(17)=2.53D-000
TT(1)=27.64D-000; TT(2)=31.97D-000; TT(3)=36.71D-000;
TT(4)=82.53D-000
TT(5)=90.0D-000; TT(6)=96.03D-000; TT(7)=103.8D-000;
TT(8)=110.55D-000
TT(9)=116.57D-000; TT(10)=125.27D-000; TT(11)=133.48D-
000
TT(12)=140.79D-000; TT(13)=147.21D-000; TT(14)=153.9D-
000
TT(15)=162.14D-000; TT(16)=165.67D-000; TT(17)=166.59D-
```

```
000
NT=17
NT1=NT
NT2=NT
! *********** DO P-3HE ON E=11.48 *******************
FP(0)=-88.8D-000;    FPI(0)=0.D-000
FP(1)=66.7D-000;     FPI(1)=0.D-000
FP(2)=2.5D-000;      FPI(2)=0.D-000
FP(3)=1.D-000;       FPI(3)=0.D-000
F0(0)=FP(0);         F0I(0)=0.D-000
F0(1)=49.4D-000;     F0I(1)=0.D-000
F0(2)=2.5D-000;      F0I(2)=0.D-000
F0(3)=1.D-000;       F0I(3)=0.D-000
FM(0)=FP(0);         FMI(0)=0.D-000
FM(1)=44.3D-000;     FMI(1)=0.D-000
FM(2)=2.5D-000;      FMI(2)=0.D-000
FM(3)=1.D-000;       FMI(3)=0.D-000
FS(0)=-84.6D-000;    FSI(0)=0.D-000
FS(1)=21.4D-000;     FSI(1)=0.D-000
FS(2)=-18.6D-000;    FSI(2)=0.D-000
FS(3)=1.D-000;       FSI(3)=0.D-000
EPS(1)=11.2D-000
EPS(3)=0.D-000
OPEN (1,FILE=AA)
DO L=1,NT
READ(1,*) TT(L),SE(L),DE(L)
ENDDO
CLOSE(1)
OPEN (1,FILE=BB)
DO I=LN,LV
READ(1,*) FP(I),F0(I),FM(I),FS(I),EPS(I)
ENDDO
CLOSE(1)
! ************ TRANSFORM TO RADIANS **************
DO L=LN,LV,LH
FM(L)=FM(L)*PI/180.0D-000
FP(L)=FP(L)*PI/180.0D-000
F0(L)=F0(L)*PI/180.0D-000
```

```
EPS(L)=EPS(L)*PI/180.0D-000
FMI(L)=FMI(L)*PI/180.0D-000
FPI(L)=FPI(L)*PI/180.0D-000
F0I(L)=F0I(L)*PI/180.0D-000
FS(L)=FS(L)*PI/180.0D-000
FSI(L)=FSI(L)*PI/180.0D-000
EP(L)=EXP(-2.0D-000*FPI(L))
EM(L)=EXP(-2.0D-000*FMI(L))
E0(L)=EXP(-2.0D-000*F0I(L))
ES(L)=EXP(-2.0D-000*FSI(L))
ENDDO
! ************************************************
FH=FH*PI/180.0D-000
NP=5*LMA+2
IF (NP>(5*LMA+2)) GOTO 9988
DO L=LN,LV,LH
FMI(L)=0
FPI(L)=0
F0I(L)=0
FSI(L)=0
ENDDO
9988 DO I=LMI,LMA,LH
XP(I)=FP(I)
ENDDO
DO I=LMI,LMA-1,LH
XP(I+LMA+1)=F0(I+1)
ENDDO
DO I=LMI,LMA-1,LH
XP(I+2*LMA+1)=FM(I+1)
ENDDO
DO I=LMI,LMA,LH
XP(I+3*LMA+1)=FS(I)
ENDDO
DO I=LMI,LMA,LH
XP(I+4*LMA+2)=EPS(I)
ENDDO
DO I=LMI,LMA,LH
XP(I+5*LMA+3)=FPI(I)
```

```
ENDDO
DO I=LMI,LMA-1,LH
XP(I+6*LMA+4)=F0I(I+1)
ENDDO
DO I=LMI,LMA-1,LH
XP(I+7*LMA+4)=FMI(I+1)
ENDDO
DO I=LMI,LMA,LH
XP(I+8*LMA+4)=FSI(I)
ENDDO
 ! ************ TRANSFOM TO C.M. *******************
EL=11.480D-000
EC=EL*PM/AM1
SK=EC*B1
SS=DSQRT(SK)
GG=3.44476D-002*Z1*Z2*PM/SS
CALL VAR(ST,FH,LMA,NI,XP,EPP,XI,NV)
PRINT *,"                    XI-KV=",XI
 ! ************* TOTAL CROSSS SECTION *************
DO I=LMI,LMA,LH
FP(I)=XP(I)
ENDDO
DO I=LMI,LMA-1,LH
F0(I+1)=XP(I+LMA+1)
ENDDO
DO I=LMI,LMA-1,LH
FM(I+1)=XP(I+2*LMA+1)
ENDDO
DO I=LMI,LMA,LH
FS(I)=XP(I+3*LMA+1)
ENDDO
DO I=LMI,LMA,LH
EPS(I)=XP(I+4*LMA+2)
ENDDO
F0(0)=FP(0); FM(0)=FP(0)
DO I=LMI,LMA,LH
FPI(I)=XP(I+5*LMA+3)
ENDDO
```

```
DO I=LMI,LMA-1,LH
F0I(I+1)=XP(I+6*LMA+4)
ENDDO
DO I=LMI,LMA-1,LH
FMI(I+1)=XP(I+7*LMA+4)
ENDDO
DO I=LMI,LMA,LH
FSI(I)=XP(I+8*LMA+4)
ENDDO
F0I(0)=FPI(0); FMI(0)=FPI(0)
DO L=LN,LV,LH
EP(L)=EXP(-2.0D-000*FPI(L))
EM(L)=EXP(-2.0D-000*FMI(L))
E0(L)=EXP(-2.0D-000*F0I(L))
ES(L)=EXP(-2.0D-000*FSI(L))
ENDDO
SRT=0.0D-000;  SRS=0.0D-000;     SST=0.0D-000;   SSS=0.0D-
000
DO L=LN,LV,LH
AP=FP(L)
AM=FM(L)
A0=F0(L)
ASS=FS(L)
L1=2*L+3
L2=2*L+1
L3=2*L-1
SRT=SRT+L1*(1.0D-000-EP(L)**2)+L2*(1.0D-000-
E0(L)**2)+L3*(1.0D-000-EM(L)**2)
SRS=SRS+L2*(1.0D-000-ES(L)**2)
SST=SST+L1*EP(L)**2*SIN(AP)**2+L2*E0(L)**2*SIN(A0)*
*2+L3*EM(L)**2*SIN(AM)**2
SSS=SSS+L2*ES(L)**2*SIN(ASS)**2
ENDDO
SRT=10.0D-000*PI*SRT/SK/3.0D-000
SRS=10.0D-000*PI*SRS/SK
SIGR=1.0D-000/4.0D-000*SRS+3.0D-000/4.0D-000*SRT
SST=10*4*PI*SST/SK/3
SSS=10.0D-000*4*PI*SSS/SK
```

```
SIGS=1.0D-000/4.0D-000*SSS+3.0D-000/4.0D-000*SST
PRINT *,"                    SIGMS-TOT=",SIGS
PRINT *,"     T      SE        ST       XI"
DO I=1,NT
WRITE(*,2) TT(I),SE(I),ST(I),DS(I)
ENDDO
PRINT *,"     FP       F0       FM       FS       EPS"
DO L=LMI,LMA,LH
FM(L)=FM(L)*180.0D-000/PI
FP(L)=FP(L)*180.0D-000/PI
FMI(L)=FMI(L)*180.0D-000/PI
FPI(L)=FPI(L)*180.0D-000/PI
F0(L)=F0(L)*180.0D-000/PI
F0I(L)=F0I(L)*180.0D-000/PI
FS(L)=FS(L)*180.0D-000/PI
FSI(L)=FSI(L)*180.0D-000/PI
EPS(L)=EPS(L)*180.0D-000/PI
WRITE(*,2) FP(L),F0(L),FM(L),FS(L),EPS(L)
ENDDO
 !READ *, A
 !IF (A==0) GOTO 1111
 !PRINT
 !PRINT *,"      T       POL "
 !DO I=1,NT
 !PRINT *,TT(I);POL(I)
 !ENDDO
OPEN (1,FILE=CC)
WRITE(1,*)"      EL       ECM       XI"
WRITE(1,4) EL,EC,XI
WRITE(1,*) "     T               SE               DE               ST
XI"
DO I=1,NT
WRITE(1,3) TT(I),SE(I),DE(I),ST(I),DS(I)
ENDDO
WRITE(1,*) "     FP(L)    F0(L)    FM(L)    FS(L)"
DO L=LN,LV
WRITE(1,2) FP(L),F0(L),FM(L),FS(L),EPS(L)
ENDDO
```

```
CLOSE(1)
OPEN (1,FILE=BB)
DO L=LN,LV
WRITE(1,3) FP(L),F0(L),FM(L),FS(L),EPS(L)
ENDDO
CLOSE(1)
!OPEN (1,FILE=AA)
!DO L=1,NT
!WRITE(1,3) TT(L),SE(L),DE(L)
!ENDDO
!CLOSE(1)
4 FORMAT(3F10.5)
2 FORMAT(5F10.5)
3 FORMAT(4F14.7)
END
SUBROUTINE VAR(ST,PHN,LMA,NI,XP,EP,AMIN,NV)
! **************** ПОИСК ФАЗ ***********************
IMPLICIT REAL(8) (A-Z)
INTEGER I,LH,NT,NP,NPP,NN,NI,LMA,LMI
DIMENSION XPN(0:50),XP(0:50),ST(0:50)
COMMON /C/ LH,LMI,NT,PI,NP,NPP
DO I=LMI,NP,LH
XPN(I)=XP(I)
ENDDO
NN=LMI
PH=PHN
CALL DET(XPN,ST,ALA)
B=ALA
IF (NV==0) GOTO 3013
DO IIN=1,NI
NN=-LH
PRINT *,ALA,IIN
GOTO 1119
1159 XPN(NN)=XPN(NN)-PH*XP(NN)
1119  NN=NN+LH
IN=0
2229 A=B
XPN(NN)=XPN(NN)+PH*XP(NN)
```

```
IF (NN<(5*LMA+3)) GOTO 7777
IF (XPN(NN)<0) GOTO 1159
7777 IN=IN+1
CALL DET(XPN,ST,ALA)
B=ALA
IF (B<A) GOTO 2229
C=A
XPN(NN)=XPN(NN)-PH*XP(NN)
IF (IN>1) GOTO 3339
PH=-PH
GOTO 5559
3339 IF (ABS((C-B)/ABS(B))<EP) GOTO 4449
PH=PH/2
5559 B=C
GOTO 2229
4449 PH=PHN
B=C
IF (NN<NP) GOTO 1119
AMIN=B
PH=PHN
ENDDO
3013 AMIN=B
DO I=LMI,NP,LH
XP(I)=XPN(I)
ENDDO
END
SUBROUTINE DET(XP,ST,XI)
! **************** ДЕТЕРМИНАНТ ********************
IMPLICIT REAL(8) (A-Z)
INTEGER I,NT
DIMENSION XP(0:50),ST(0:50)
COMMON /B/ SE(0:50),DS(0:50),DE(0:50),NT
S=0.0D-000
CALL SEC(XP,ST)
DO I=1,NT
DS(I)=( (ST(I)-SE(I) )/DE(I) )**2
S=S+DS(I)
ENDDO
```

```
XI=S/NT
END
SUBROUTINE SEC(XP,ST)
! **************** СЕЧЕНИЕ ***********************
IMPLICIT REAL(8) (A-Z)
INTEGER I,II,L,LN,LV,LH,NT
COMMON /A/ SS,GG,PI,LN,LV,LH,NT,POL(0:50),TT(0:50)
DIMENSION
S0(0:50),P(0:50),P1(0:50),P2(0:50),XP(0:50),ST(0:50),F0(0:50),
FP(0:50),FM(0:50),FS(0:50),EPS(0:50)
DIMENSION
F0I(0:50),FPI(0:50),FMI(0:50),FSI(0:50),EP(0:50),EM(0:50),ES(
0:50),E0(0:50)
DO I=LN,LV,LH
FP(I)=XP(I)
ENDDO
DO I=LN,LV-1,LH
II=I+LV+1
F0(I+1)=XP(II)
ENDDO
DO I=LN,LV-1,LH
II=I+2*LV+1
FM(I+1)=XP(II)
ENDDO
DO I=LN,LV,LH
II=I+3*LV+1
FS(I)=XP(II)
ENDDO
DO I=LN,LV,LH
II=I+4*LV+2
EPS(I)=XP(II)
ENDDO
F0(0)=FP(0); FM(0)=FP(0)
DO I=LN,LV,LH
II=I+5*LV+3
FPI(I)=XP(II)
ENDDO
DO I=LN,LV-1,LH
```

```
II=I+6*LV+4
F0I(I+1)=XP(II)
ENDDO
DO I=LN,LV-1,LH
II=I+7*LV+4
FMI(I+1)=XP(II)
ENDDO
DO I=LN,LV,LH
II=I+8*LV+4
FSI(I)=XP(II)
ENDDO
F0I(0)=FPI(0); FMI(0)=FPI(0)
DO L=LN,LV,LH
EP(L)=DEXP(-2.0D-000*FPI(L))
EM(L)=DEXP(-2.0D-000*FMI(L))
E0(L)=DEXP(-2.0D-000*F0I(L))
ES(L)=DEXP(-2.0D-000*FSI(L))
ENDDO
CALL CULFAZ(GG,S0)
DO I=1,NT
T=TT(I)*PI/180.0D-000
X=DCOS(T)
CALL CULAMP(X,GG,S0,RECUL,AMCUL)
CALL POLLEG(X,LV,P)
CALL FUNLEG1(X,LV,P1)
CALL FUNLEG2(X,LV,P2)
REA=0.0D-000; AMA=0.0D-000; REB=0.0D-000; AMB=0.0D-
000; REC=0.0D-000; AMC=0.0D-000; RED=0.0D-000;
AMD=0.0D-000
REE=0.0D-000; AME=0.0D-000; RRG=0.0D-000; AAG=0.0D-
000; REH=0.0D-000; AMH=0.0D-000; REF=0.0D-000;
AMF=0.0D-000
DO L=LN,LV,LH
FP1=2.0D-000*FP(L)
FM1=2.0D-000*FM(L)
F01=2.0D-000*F0(L)
SL=2.0D-000*S0(L)
C=DCOS(SL)
```

```
S=DSIN(SL)
FS1=2.0D-000*FS(L)
SO=DSIN(EPS(L))**2
CO=DCOS(EPS(L))**2
AL1P=EP(L)*DCOS(FP1)-1.0D-000
AL2P=EP(L)*DSIN(FP1)
AL1M=EM(L)*DCOS(FM1)-1.0D-000
AL2M=EM(L)*DSIN(FM1)
AL10=SO*ES(L)*DCOS(FS1)+CO*E0(L)*DCOS(F01)-1.0D-
000
AL20=SO*ES(L)*DSIN(FS1)+CO*E0(L)*DSIN(F01)
A1=(L+2.0D-000)*AL1P+(2.0D-000*L+1.0D-000)*AL10+(L-
1.0D-000)*AL1M
A2=(L+2.0D-000)*AL2P+(2.0D-000*L+1.0D-000)*AL20+(L-
1.0D-000)*AL2M
REA=REA+(A1*C-A2*S)*P(L)/2.0D-000
AMA=AMA+(A1*S+A2*C)*P(L)/2.0D-000
ALS=CO*ES(L)*DCOS(FS1)+SO*E0(L)*DCOS(F01)-1.0D-000
BS=CO*ES(L)*DSIN(FS1)+SO*E0(L)*DSIN(F01)
RES=(2.0D-000*L+1.0D-000)*(ALS*C-BS*S)
AMS=(2.0D-000*L+1.0D-000)*(ALS*S+BS*C)
B1=(L+1.0D-000)*AL1P+L*AL1M
B2=(L+1.0D-000)*AL2P+L*AL2M
REB=REB+(B1*C-B2*S+RES)*P(L)/2.0D-000
AMB=AMB+(B1*S+B2*C+AMS)*P(L)/2.0D-000
REC=REC+(B1*C-B2*S-RES)*P(L)/2.0D-000
AMC=AMC+(B1*S+B2*C-AMS)*P(L)/2.0D-000
IF (L<1) GOTO 1211
SI2=1.0D-000/2.0D-000*DSIN(2.0D-000*EPS(L))
AL1=SI2*(ES(L)*DCOS(FS1)-E0(L)*DCOS(F01))
AL2=SI2*(ES(L)*DSIN(FS1)-E0(L)*DSIN(F01))
RE1=(2.0D-000*L+1.0D-
000)*(AL2*C+AL1*S)/DSQRT(L*(L+1.0D-000))
AM1=(2.0D-000*L+1.0D-000)*(AL2*S-
AL1*C)/DSQRT(L*(L+1.0D-000))
C1=AL1P-AL1M
C2=AL2P-AL2M
RED=RED+(C2*C+C1*S-RE1)*P1(L)/2.0D-000
```

```
AMD=AMD+(C2*S-C1*C-AM1)*P1(L)/2.0D-000
REE=REE+(C2*C+C1*S+RE1)*P1(L)/2.0D-000
AME=AME+(C2*S-C1*C+AM1)*P1(L)/2.0D-000
D1=(L+2.0D-000)/(L+1.0D-000)*AL1P-(2.0D-000*L+1.0D-
000)/(L*(L+1.0D-000))*AL10-(L-1.0D-000)/L*AL1M
D2=(L+2.0D-000)/(L+1.0D-000)*AL2P-(2.0D-000*L+1.0D-
000)/(L*(L+1.0D-000))*AL20-(L-1.0D-000)/L*AL2M
RRG=RRG+(D2*C+D1*S-RE1)*P1(L)/2.0D-000
AAG=AAG+(D2*S-D1*C-AM1)*P1(L)/2.0D-000
REH=REH+(D2*C+D1*S+RE1)*P1(L)/2.0D-000
AMH=AMH+(D2*S-D1*C+AM1)*P1(L)/2.0D-000
1211 IF (L<2) GOTO 2122
F1=1.0D-000/(L+1.0D-000)*AL1P-(2.0D-000*L+1.0D-
000)/(L*(L+1.0D-000))*AL10+AL1M/L
F2=1.0D-000/(L+1.0D-000)*AL2P-(2.0D-000*L+1.0D-
000)/(L*(L+1.0D-000))*AL20+AL2M/L
REF=REF+(F2*C+F1*S)*P2(L)/2.0D-000
AMF=AMF+(F2*S-F1*C)*P2(L)/2.0D-000
2122 ENDDO
RET=0.0D-000; AMT=0.0D-000; RES=0.0D-000;  AMS=0.0D-
000
DO L=LN,LV,LH
SL=2.0D-000*S0(L)
C=DCOS(SL)
S=DSIN(SL)
FS1=2.0D-000*FS(L)
ALS=ES(L)*DCOS(FS1)-1.0D-000
BS=ES(L)*DSIN(FS1)
RES=RES+(2.0D-000*L+1.0D-000)*(BS*C+ALS*S)*P(L)
AMS=AMS+(2.0D-000*L+1.0D-000)*(BS*S-ALS*C)*P(L)
ENDDO
RES=RECUL+RES
AMS=AMCUL+AMS
REA=RECUL+REA
AMA=AMCUL+AMA
REB=RECUL+REB
AMB=AMCUL+AMB
AA=REA**2+AMA**2
```

```
BB=REB**2+AMB**2
CC=REC**2+AMC**2
DD=RED**2+AMD**2
EE=REE**2+AME**2
FF=REF**2+AMF**2
HH=REH**2+AMH**2
GGG=RRG**2+AAG**2
SUM=AA+BB+CC+DD+EE+GGG+HH+FF
S=10.0D-000*SUM/2.0D-000/SS**2/4.0D-000
ST(I)=S
! POL(I) = - 2*(REA*REE+AMA*AME+REB*REH
+AMB*AMH+REC*RRG+AMC*AAG+RED*REF+AMD*AM
F)/SUM
ENDDO
END
SUBROUTINE CULAMP(X,GG,S0,RECUL,AMCUL)
! ************ КУЛОНОВСКАЯ АМПЛИТУДА **********
IMPLICIT REAL(8) (A-Z)
DIMENSION S0(0:50)
A=2.0D-000/(1.0D-000-X)
S00=2.0D-000*S0(0)
BB=-GG*A
AL=GG*DLOG(A)+S00
RECUL=-BB*DSIN(AL)
AMCUL=BB*DCOS(AL)
END
SUBROUTINE POLLEG(X,L,P)
! **************** ПОЛИНОМЫ ЛЕЖАНДРА ***********
IMPLICIT REAL(8) (A-Z)
INTEGER I,L
DIMENSION P(0:50)
P(0)=1.0D-000; P(1)=X
DO I=2,L
P(I)=(2.0D-000*I-1.0D-000)*X/I*P(I-1)-(I-1.0D-000)/I*P(I-2)
ENDDO
END
SUBROUTINE FUNLEG1(X,L,P)
! **************** ФУНКЦИЯ ЛЕЖАНДРА ***********
```

```
IMPLICIT REAL(8) (A-Z)
INTEGER I,L
DIMENSION P(0:50)
P(0)=0.0D-000; P(1)=DSQRT(DABS(1.0D-000-X**2))
DO I=2,L
P(I)=(2.0D-000*I-1.0D-000)*X/(I-1.0D-000)*P(I-1)-I/(I-1.0D-
000)*P(I-2)
ENDDO
END
SUBROUTINE FUNLEG2(X,L,P)
! **************** ФУНКЦИЯ ЛЕЖАНДРА ***********
IMPLICIT REAL(8) (A-Z)
INTEGER I,L
DIMENSION P(0:50)
P(0)=0.0D-000;   P(1)=0.0D-000;   P(2)=3.0D-000*DABS(1.0D-
000-X**2)
DO I=3,L
P(I)=(2.0D-000*I-1.0D-000)*X/(I-2.0D-000)*P(I-1)-(I+1.0D-
000)/(I-2.0D-000)*P(I-2)
ENDDO
END
SUBROUTINE CULFAZ(G,F)
! **************** КУЛОНОВСКИЕ ФАЗЫ ***********
IMPLICIT REAL(8) (A-Z)
INTEGER I,N
DIMENSION F(0:50)
C=0.577215665D-000; S=0.0D-000; N=50
A1=1.202056903D-000/3.0D-000;  A2=1.036927755D-000/5.0D-
000
DO I=1,N
A=G/I-DATAN(G/I)-(G/I)**3/3.0D-000+(G/I)**5/5.0D-000
S=S+A
ENDDO
FAZ=-C*G+A1*G**3-A2*G**5+S
F(0)=FAZ
DO I=1,20
F(I)=F(I-1)+DATAN(G/(I))
ENDDO
```

**END**

С такими же фазами, как в предыдущем случае [180,181] и без учета смешивания получаем по этой программе тот же результат, который приведен первым в качестве контрольного счета для предыдущей программы с $\chi^2$ = 7.375E-001:

$\chi^2$ = 7.375E-001

| $\theta$ | $\sigma_e$ | $\sigma_t$ | $\chi^2_i$ |
|---|---|---|---|
| 27.64000 | 223.10000 | 229.15944 | 1.17923 |
| 31.97000 | 222.00000 | 222.68747 | .01534 |
| 36.71000 | 211.90000 | 211.14558 | .02026 |
| 82.53000 | 54.27000 | 53.52241 | .30217 |
| 90.00000 | 36.76000 | 36.24893 | .30859 |
| 96.03000 | 25.70000 | 25.46703 | .13250 |
| 103.80000 | 16.78000 | 16.16202 | 2.16494 |
| 110.55000 | 13.21000 | 12.59755 | 3.44437 |
| 116.57000 | 13.21000 | 13.11973 | .07482 |
| 125.27000 | 20.26000 | 19.96234 | .34064 |
| 133.48000 | 32.21000 | 32.33282 | .02299 |
| 140.79000 | 45.95000 | 46.97500 | .79443 |
| 147.21000 | 58.82000 | 61.38791 | 3.05158 |
| 153.90000 | 75.46000 | 76.51701 | .31277 |
| 162.14000 | 92.72000 | 93.06232 | .02177 |
| 165.67000 | 97.70000 | 98.82336 | .21196 |
| 166.59000 | 101.10000 | 100.15680 | .13898 |

| $\delta_p$ | $\delta_0$ | $\delta_m$ | $\delta_s$ | $\varepsilon$ |
|---|---|---|---|---|
| -88.800 | -88.800 | -88.800 | -84.600 | 0.000 |
| 66.700 | 49.400 | 44.300 | 21.400 | 0.000 |
| 2.500 | 2.500 | 2.500 | -18.600 | 0.000 |

Введение смешивания в *P* волне, как это было сделано в работах [180,181], позволяет лучше описать сечения упругого рассеяния при тех же фазах [180,181]

$\chi^2$ = 2.93E-001

| θ | $\sigma_e$ | $\sigma_t$ | $\chi^2_i$ |
|---|---|---|---|
| 27.640 | 223.100 | 228.038 | 0.783 |
| 31.970 | 222.000 | 221.732 | 0.002 |
| 36.710 | 211.900 | 210.383 | 0.082 |
| 82.530 | 54.270 | 54.215 | 0.002 |
| 90.000 | 36.760 | 36.966 | 0.050 |
| 96.030 | 25.700 | 26.151 | 0.497 |
| 103.800 | 16.780 | 16.739 | 0.010 |
| 110.550 | 13.210 | 13.031 | 0.295 |
| 116.570 | 13.210 | 13.394 | 0.310 |
| 125.270 | 20.260 | 19.972 | 0.400 |
| 133.480 | 32.210 | 32.075 | 0.028 |
| 140.790 | 45.950 | 46.482 | 0.214 |
| 147.210 | 58.820 | 60.704 | 1.642 |
| 153.900 | 75.460 | 75.656 | 0.011 |
| 162.140 | 92.720 | 92.029 | 0.089 |
| 165.670 | 97.700 | 97.733 | 0.000 |
| 166.590 | 101.100 | 99.054 | 0 .654 |

| $\delta_p$ | $\delta_0$ | $\delta_m$ | $\delta_s$ | ε |
|---|---|---|---|---|
| -88.800 | -88.800 | -88.800 | -84.600 | 0 .000 |
| 66.700 | 49.400 | 44.300 | 21.400 | 11.200 |
| 2.500 | 2.500 | 2.500 | -18.600 | 0.000 |

Полученное среднее значение $\chi^2$ несколько меньше, приведенной в работах [180,181] величины 0.45. Вычисления [180,181] выполнялись в начале-середине 60-х годов прошлого века на сравнительно простых вычислительных машинах и, по-видимому, при сравнительно не высокой точности, что вполне может привести к такому небольшому различию результатов для $\chi^2$. Кроме того, в настоящих расчетах, возможно, использовались несколько другие значения констант, например, $\hbar^2 / m_0$ и масс частиц, а округление приводимых в статьях [180,181] фаз до второго знака после запятой также может оказать влияние на величину $\chi^2$. Поэтому вполне можно считать, что в приведенных контрольных счетах получено приемлемое согласие величины $\chi^2$ с результатами работ [180,

181].

Для сравнения приведем результаты при тех же фазах и параметре смешивания, которые были получены в нашей предыдущей работе [17] для программ на языке Turbo Basic при среднем $\chi^2 = 0.29$

| $\theta$ | $\sigma_e$ | $\sigma_t$ | $\chi^2_i$ |
|---|---|---|---|
| 27.64 | 223.10 | 228.04 | 0.78 |
| 31.97 | 222.00 | 221.73 | 0.00 |
| 36.71 | 211.90 | 210.38 | 0.08 |
| 82.53 | 54.27 | 54.22 | 0.00 |
| 90.00 | 36.76 | 36.97 | 0.05 |
| 96.03 | 25.70 | 26.15 | 0.49 |
| 103.80 | 16.78 | 16.74 | 0.01 |
| 110.55 | 13.21 | 13.03 | 0.29 |
| 116.57 | 13.21 | 13.39 | 0.31 |
| 125.27 | 20.26 | 19.97 | 0.32 |
| 133.48 | 32.21 | 32.07 | 0.03 |
| 140.79 | 45.95 | 46.48 | 0.21 |
| 147.21 | 58.82 | 60.70 | 1.64 |
| 153.90 | 75.46 | 75.66 | 0.01 |
| 162.14 | 92.72 | 92.03 | 0.09 |
| 165.67 | 97.70 | 97.73 | 0.00 |
| 166.59 | 101.10 | 99.05 | 0.66 |

И здесь результаты с точностью до ошибок округления совпадают между собой и не зависят от языка программирования.

Если выполнить детальное варьирование фаз рассеяния и параметра смешивания ε, то в результате примерно 10 000 итераций получим, что для лучшего описания имеющихся экспериментальных данных требуется практически нулевое смешивание

$\chi^2 =$ 2.25E-001

| $\theta$ | $\sigma_e$ | $\sigma_t$ | $\chi^2_i$ |
|---|---|---|---|
| 27.640 | 223.100 | 224.736 | 0.086 |
| 31.970 | 222.000 | 221.181 | 0.022 |
| 36.710 | 211.900 | 211.217 | 0.017 |
| 82.530 | 54.270 | 53.839 | 0.100 |
| 90.000 | 36.760 | 36.717 | 0.002 |
| 96.030 | 25.700 | 26.025 | 0.259 |
| 103.800 | 16.780 | 16.736 | 0.011 |
| 110.550 | 13.210 | 13.072 | 0.175 |
| 116.570 | 13.210 | 13.423 | 0.415 |
| 125.270 | 20.260 | 19.927 | 0.427 |
| 133.480 | 32.210 | 31.953 | 0.101 |
| 140.790 | 45.950 | 46.328 | 0.108 |
| 147.210 | 58.820 | 60.562 | 1.405 |
| 153.900 | 75.460 | 75.570 | 0.003 |
| 162.140 | 92.720 | 92.043 | 0.085 |
| 165.670 | 97.700 | 97.791 | 0.001 |
| 166.590 | 101.100 | 99.123 | 0.611 |

| $\delta_p$ | $\delta_0$ | $\delta_m$ | $\delta_s$ | $\varepsilon$ |
|---|---|---|---|---|
| -83.342 | -83.342 | -83.342 | -97.861 | 0 .000 |
| 52.520 | 50.021 | 50.444 | 44.905 | -0.00017 |
| 3.222 | 0.122 | 2.667 | -19.114 | 0.000 |

По-видимому, можно считать, что вводить смешивание для наилучшего описания имеющихся экспериментальных данных не требуется, хотя точка при 147°, как и в случае расчетов по предыдущей программе с $\chi^2$ = 2.27E-001, описывается сравнительно плохо, хотя средние $\chi^2$ практически совпадают. Заметим, что изменение параметра смешивания $\varepsilon$ в любую сторону приводит к заметному подъему среднего $\chi^2$, заметно ухудшая описание используемых экспериментальных данных.

Заметим, что приведенные в этом параграфе результаты непосредственно относятся к проверке работоспособности компьютерных программ, написанных на основе известных методов расчета дифференциальных сечений, а не к получению новых результатов по фазовому анализу упругого $p^3$He рассеяния [17]. Новые результаты для фазового анализа в системе частиц со спином 1/2 + 1/2 приведены в следующем параграфе для $p^{13}C$ системы.

### 2.5.3 Система частиц $p^{13}C$ с полуцелым спином и спин-орбитой The $p^{13}C$ system with spin 1/2 and spin-orbit interaction

Рассмотрим теперь систему $p^{13}C$ частиц при низких энергиях, которой соответствуют спины 1/2 + 1/2. При энергиях ниже 0.5÷1 МэВ отклонение измеренного дифференциального сечения от резерфордовской величины вполне может характеризоваться только одной точкой при определенном значении угла рассеяния и заданной энергии. Такое отклонение, как это будет показано далее, вполне однозначно описывается одной триплетной *S* фазой рассеяния даже в области резонанса в системе $p^{13}C$.

На основе этих представлений в [183,184] было выполнено измерение дифференциальных сечений, а именно, функций возбуждения для упругого $p^{13}C$ рассеяния при энергиях от 0.25 до 0.75 МэВ. Измерения сечений при каждой энергии проводились только при одном угле рассеяния, и для разных энергий использовалось четыре значения углов. Эти данные мы используем далее для выполнения фазового анализа упругого $p^{13}C$ рассеяния при низких энергиях и определения резонансной формы триплетной $^3S_1$ фазы при энергии 0.55 МэВ.

На основе проверенных в предыдущем параграфе компьютерных программ со спин-орбитальным смешиванием, был выполнен фазовый анализ экспериментальных данных по дифференциальным сечениям упругого $p^{13}C$ рассеяния (функциям возбуждения) в диапазоне энергий 300÷750 кэВ (л.с.), приведенных в [183,184]. В результате фазового анализа было показано, что синглетная $^1S_0$ фаза упругого $p^{13}C$ рассеяния в рассматриваемой области энергий оказалась близка к нулю – ее величина не превышала 2÷3 градуса. Вид триплетной $^3S_1$ фазы показан ниже на рис.2.5.1, а ее значения приведены в табл.2.5.1 и 2.5.2. Эта фаза имеет явно выраженный резонанс, который соответствует уровню ядра $^{14}$N в кластерном p$^{13}$C канале с $J^{\pi}T = 1^{-}1$ при энергии 0.55 МэВ (л.с.) [166].

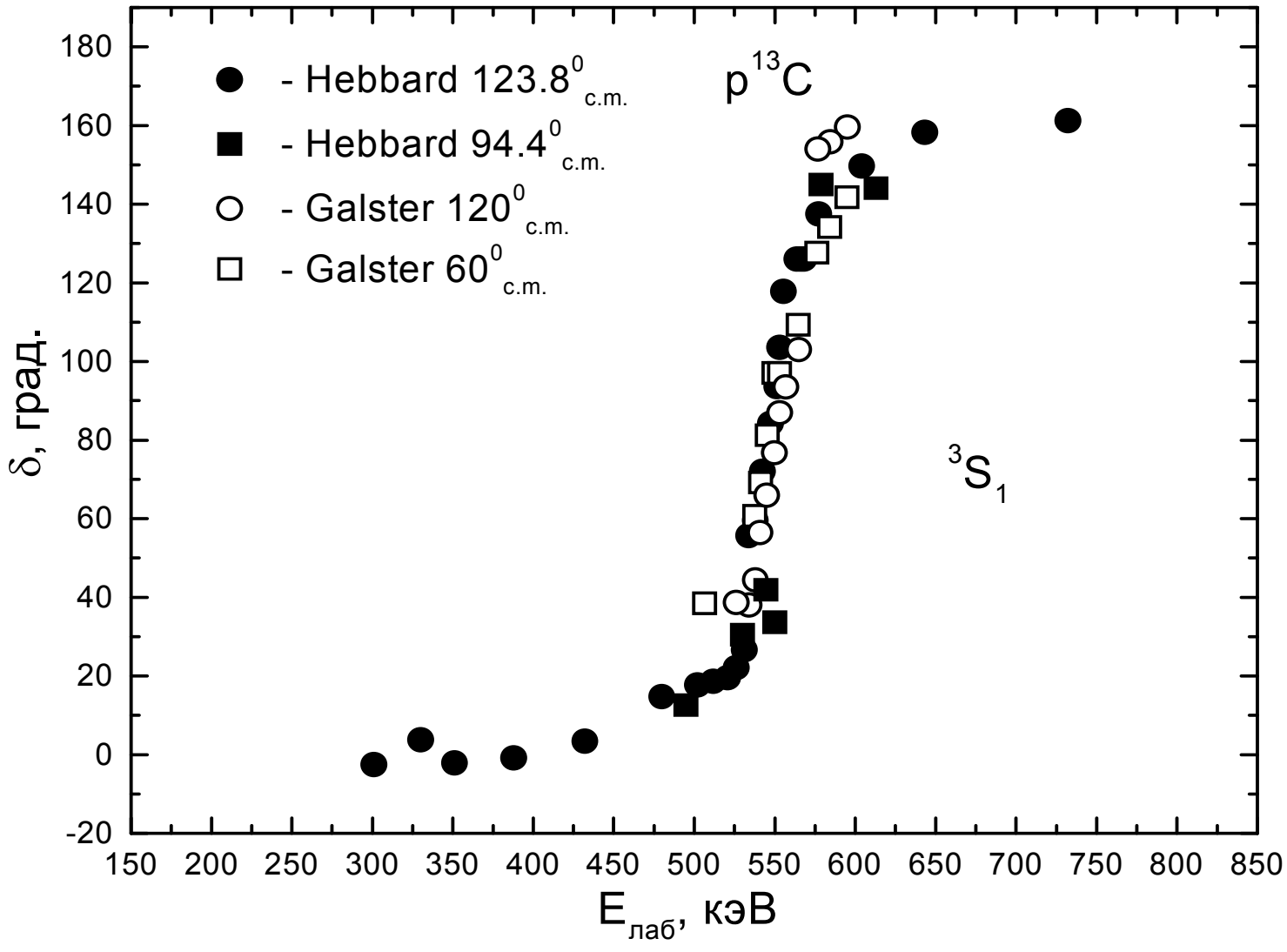

Рис.2.5.1. Триплетная $^3S_1$ фаза p$^{13}$C рассеяния при низких энергиях. Точки и квадраты – результаты нашего фазового анализа по данным [183] при 94.4° и 123.8° в с.ц.м., кружки и открытые квадраты – результаты нашего фазового анализа по данным [184] при 60° и 120° в с.ц.м.

Далее в табл.2.5.1 приведены результаты фазового анализа экспериментальных данных из работы [183] в области энергий 301-732 кэВ и 495-613 кэВ при двух значениях углов, которые представлены на рис.2.5.1 черными точками и квадратами соответственно.

Табл.2.5.2 содержит результаты фазового анализа экспериментальных данных из [184], которые представлены на рис.2.5.1 кружками (при энергиях 526-595 кэВ) и открытыми квадратами (для энергии 506-595 кэВ). По данным работы [183] извлеченная фаза упругого рассеяния переходит 90° при энергии 551 кэВ, а для данных [184] это происходит при 557 кэВ.

Табл.2.5.1. Результаты фазового анализа $p^{13}C$ упругого рассеяния при низких энергиях с учетом $^3S_1$ фазы по экспериментальным данным [183].

| $^3S_1$ фаза получена по данным [183] при 123.8° (ц.м.). Точки на рис.2.5.1. | |
|---|---|
| $E_{лаб}$, кэВ | $^3S_1$, град. |
| 301 | -2,6 |
| 330 | 3,7 |
| 351 | -2,1 |
| 388 | -0,8 |
| 432 | 3,4 |
| 479,8 | 14,7 |
| 502 | 17,7 |
| 512 | 18,6 |
| 520,8 | 19,5 |
| 526 | 22,1 |
| 531,2 | 26,7 |
| 533,7 | 55,6 |
| 538 | 59,6 |
| 542,4 | 72,1 |
| 547,5 | 84,2 |
| 551,2 | 93,6 |
| 553,1 | 103,6 |
| 555,4 | 117,8 |
| 563,9 | 126 |
| 566 | 126 |
| 568,3 | 126 |
| 577,3 | 137,5 |
| 604 | 149,7 |
| 643,3 | 158,3 |
| 732,1 | 161,2 |
| $^3S_1$ фаза получена по данным [183] при 94.4° (ц.м.). Квадраты на рис.2.5.1. | |
| $E_{лаб}$, кэВ | $^3S_1$, град. |
| 495 | 12,6 |
| 530 | 30,4 |
| 544,6 | 41,9 |
| 550 | 33,7 |
| 579 | 145 |
| 613 | 144 |

Табл.2.5.2. Результаты фазового анализа $p^{13}C$ упругого рассеяния при низких энергиях с учетом $^3S_1$ фазы по экспериментальным данным [184].

| $^3S_1$ фаза получена по данным [184] при 120° (ц.м.). Кружки на рис.2.51. | |
|---|---|
| $E_{лаб}$, кэВ | $^3S_1$, град. |
| 526 | 38,7 |
| 534 | 38 |
| 538 | 44,4 |
| 541 | 56,5 |
| 545 | 66 |
| 549,9 | 76,7 |
| 553 | 87 |
| 557 | 93,5 |
| 565,1 | 103 |
| 576,8 | 154 |
| 584,4 | 155,8 |
| 595,3 | 159,6 |
| $^3S_1$ фаза получена по данным [184] при 60° (ц.м.). Открытые квадраты на рис.2.5.1. | |
| $E_{лаб}$, кэВ | $^3S_1$, град. |
| 506,4 | 38,5 |
| 537,8 | 60,7 |
| 541,1 | 69,2 |
| 545,4 | 81,2 |
| 549,5 | 97 |
| 553,1 | 97 |
| 564,7 | 109,3 |
| 576,3 | 127,6 |
| 584,3 | 134,2 |
| 594,9 | 141,7 |

Ширина резонанса находится в диапазоне 20÷25 кэВ, что вполне согласуется с результатами работы [166], где приводится величина 23(1) кэВ при энергии резонанса 551 кэВ. Ширина резонанса, извлеченная из экспериментальных данных для триплетной $^3S_1$ фазы, имеет меньшее, чем в $p^{12}C$ рассеянии [161], значение и далее для ее описания потребуется очень узкий потенциал, что должно привести нас к параметру ширины порядка $\beta = 2{\div}3$ Фм$^{-2}$ [185].

Однако сравнительно точное извлечение на основе выполненного фазового анализа триплетной $S$ фазы из экспериментальных данных позволяет надеяться на вполне однозначное построение потенциала $p^{13}C$ взаимодействия. Резонансная форма фазы при столь малой ширине резонанса позволяет, при построении потенциалов, избежать непрерывной неоднозначности, свойственной оптической модели [88], по-тому что только вполне определенная ширина потенциала способна правильно описать ее резонансный вид.

В дальнейшем полученные фазы рассеяния используются для построения межкластерных потенциалов взаимодействия [185] и расчетов астрофизического *S*-фактора радиационного $p^{13}C$ захвата [186]. Этот процесс входит в CNO-термоядерный цикл, является его второй реакцией и дает, по-видимому, существенный вклад в энергетический выход термоядерных реакций [1,3,170,186,187], приводящих к горению Солнца и звезд нашей Вселенной [168].

# *Заключение*
# *Conclusion*

Таким образом, использованные алгоритмы применения численных методов для нахождения частных решений общей многопараметрической вариационной задачи для функционала $\chi^2$, который определяет точность описания экспериментальных данных на основе выбранного теоретического представления, позволяет вполне однозначно определять фазы рассеяния ядерных частиц [17].

Использованные методы и алгоритмы позволили получить новые результаты для фазового анализа в упругом $p^3He$, $p^6Li$, $p^{12}C$, $n^{12}C$, $p^{13}C$ и $^4He^4He$, $^4He^{12}C$ рассеянии при любых, в том числе, астрофизических энергиях. Найденные, таким образом, фазы рассеяния хорошо описывают экспериментальные сечения упругого рассеяния, определяя, тем самым, положение некоторых низколежащих резонансных уровней рассматриваемых в этих каналах атомных ядер.

Полученные здесь фазы рассеяния применяются в дальнейшем для построения межкластерных потенциалов взаимодействия, которые могут быть использованы для расчетов полных сечений фотопроцессов и астрофизических *S*-факторов при сверхнизких или тепловых энергиях [9,14,15, 168,188].

# III. ТРЕХТЕЛЬНАЯ МОДЕЛЬ
## Three-body model

*В настоящее время нет точной теории, которая объясняла бы все свойства атомных ядер. Поэтому для описания структуры ядер используются различные модели, каждая из которых базируется на тех или иных экспериментальных фактах и позволяет объяснить некоторые выделенные свойства ядра [2].*

### *Введение*
### *Introduction*

Ранее в работах [58-63] были подробно рассмотрены возможности трехтельной модели $^6$Li и показана ее способность описывать почти все наблюдаемые характеристики этого ядра, и продемонстрирована большая степень его кластеризации в $^2$H$^4$He канал. Такой результат вполне может объяснить определенные успехи простых двухкластерных моделей легких ядер с запрещенными состояниями, в частности, $^2$H$^4$He и $^3$H$^4$He моделей $^6$Li и $^7$Li [9,12,13,37], в которых получается хорошее описание многих экспериментальных характеристик этих ядер.

В этом разделе будут рассмотрены некоторые наши результаты, полученные для трехчастичной вариационной задачи дискретного спектра с разложением волновой функции по неортогональному гауссовому базису при независимом варьировании всех параметров разложения. Методы решения этой задачи, описанные в первой главе, применяются к рассмотрению некоторых характеристик связанных состояний легких атомных ядер в трехчастичных моделях.

Трехтельная модель, рассмотренных далее легких атомных ядер $^7$Li, $^9$Be и $^{11}$B, позволяет провести определенную проверку и оценку полученных по фазам рассеяния парных межкластерных потенциалов. Это дает возможность убедиться в целесообразности их дальнейшего использования для расчетов, связанных с рассмотрением некоторых, в первую очередь, астрофизических характеристик ядерных систем и процессов при низких и сверхнизких энергиях, которые протекают на Солнце, звездах и всей нашей Вселенной [10,12,13,168,189].

## 3.1 Трехтельные конфигурации ядра $^7$Li и трехтельная программа Three-body configuration of $^7$Li and three-body program

Рассмотрим теперь трехтельную одноканальную кластерную $^4$He$^2$Hn модель ядра $^7$L. В любой одноканальной модели возможны три варианта расположения частиц в вершинах треугольника – три возможных канала. Причем, один из таких каналов может иметь наибольшую вероятность существования. Как говорилось в первом разделе, некоторые ядерные характеристики отдельных ядер могут быть обусловлены преимущественно одной кластерной конфигурацией трехкластерной системы при малом вкладе других возможных кластерных структур. В этом случае используемая одноканальная трехкластерная модель позволяет идентифицировать доминирующий кластерный канал и выделить некоторые основные свойства такой ядерной системы, которые им обусловлены.

Кроме того, одноканальная модель заметно проще в численном и программном исполнении, по сравнению с ее многоканальным вариантом. И наконец, для выделения преобладающего кластерного канала и тестирования межкластерных потенциалов требуется рассмотрение именно одноканального варианта трехтельной модели. Далее такие потенциалы могут быть использованы для разнообразных астрофизических приложений, в том числе, при рассмотрении термоядерных процессов на Солнце и Вселенной [168].

### 3.1.1 Потенциалы и фазы Potentials and phase shifts

В рассматриваемом варианте расположения частиц будем считать, что в основании треугольника находятся n$^2$H кластеры (частицы 23) с радиус-вектором относительного расстояния $\boldsymbol{r} = \boldsymbol{r}_{23}$ и орбитальным моментом относительного движения λ, который может принимать значения 1/2 и 3/2.

Ядро $^4$He (частица 1) находится в вершине треугольника и его положение относительно центра масс двухкластерной системы определяется вектором $\boldsymbol{R} = \boldsymbol{R}_{(23),1}$ и моментом $l$. Причем $\boldsymbol{l} = \boldsymbol{l}_{12} + \boldsymbol{l}_{13}$ , где $\boldsymbol{l}_{1i}$ – орбитальные моменты между частицами 1 и $i$, принимающего значения 2 и 3 (см. рис.3.1.1).

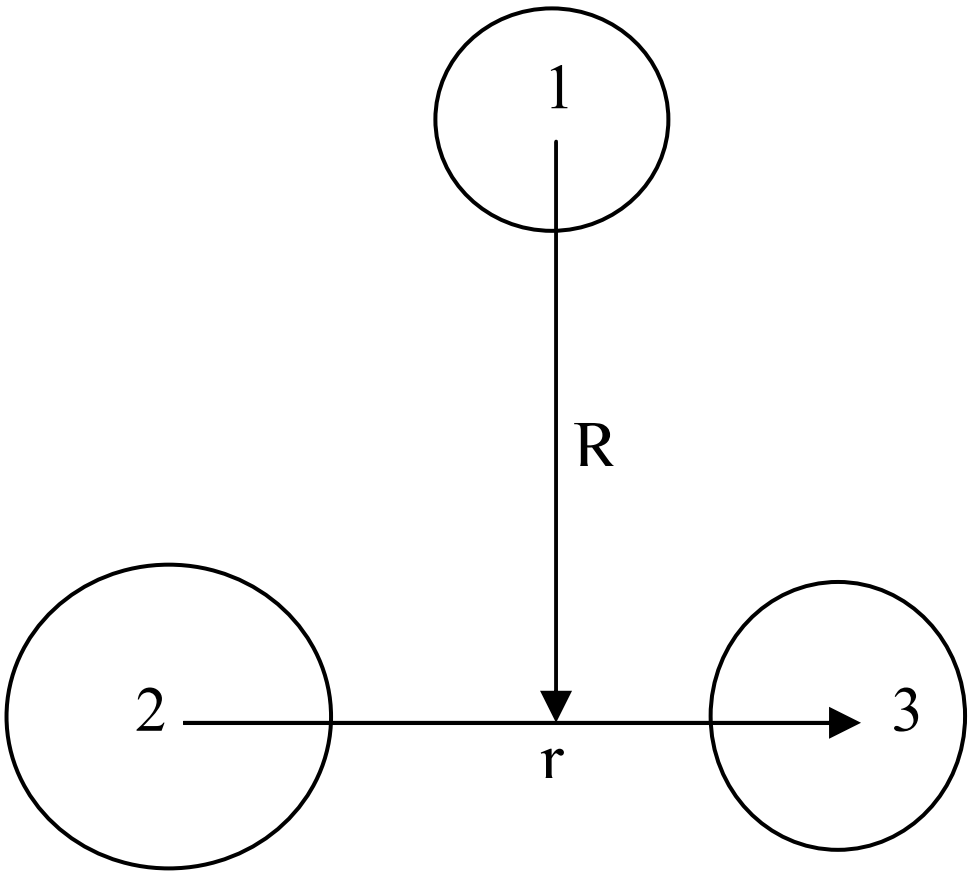

Рис.3.1.1. Векторная схема трехкластерной модели.

Полный спин системы $\boldsymbol{S} = \boldsymbol{S}_3 + \boldsymbol{S}_2$ при $S_1 = 0$ может иметь значения 1/2 и 3/2, т.е. система n$^2$H может находиться в дублетном и квартетном спиновом состояниях. Первое из них соответствует основному состоянию ядра трития при $\lambda = 0$ – именно его будем рассматривать далее в используемом здесь одноканальном варианте трехчастичной модели ядра $^7$Li. Полный орбитальный момент системы $\boldsymbol{L} = \boldsymbol{l} + \lambda$, равный единице, может быть получен, например, из комбинации $l = 1$ и $\lambda = 0$, которая позволяет рассматривать систему n$^2$H, как ядро трития [13]. Полный момент $\boldsymbol{J} = \boldsymbol{S} + \boldsymbol{L}$ основного состояния $^7$Li, равный $3/2^-$, может быть получен из комбинации $L = 1$ и $S = 1/2$, которая приводит к $J = 1/2$ и 3/2 с отрицательной четностью. Основному состоянию ядра соответствует момент $J^\pi = 3/2^-$, а первому возбужденному состоянию при энергии 0.478 МэВ момент $1/2^-$.

В качестве парных межкластерных потенциалов выбира-лись взаимодействия гауссовой формы с отталкивающим кором (1.2.1), позволяющие правильно передавать соответствующие фазы упругого рассеяния. В паре частиц (13) используется чистый по схемам Юнга $n^4He$ потенциал для *S* волны ($l_{13}$ = 0) с параметрами, описывающими экспериментальную фазу [190], как показано на рис.3.1.2.

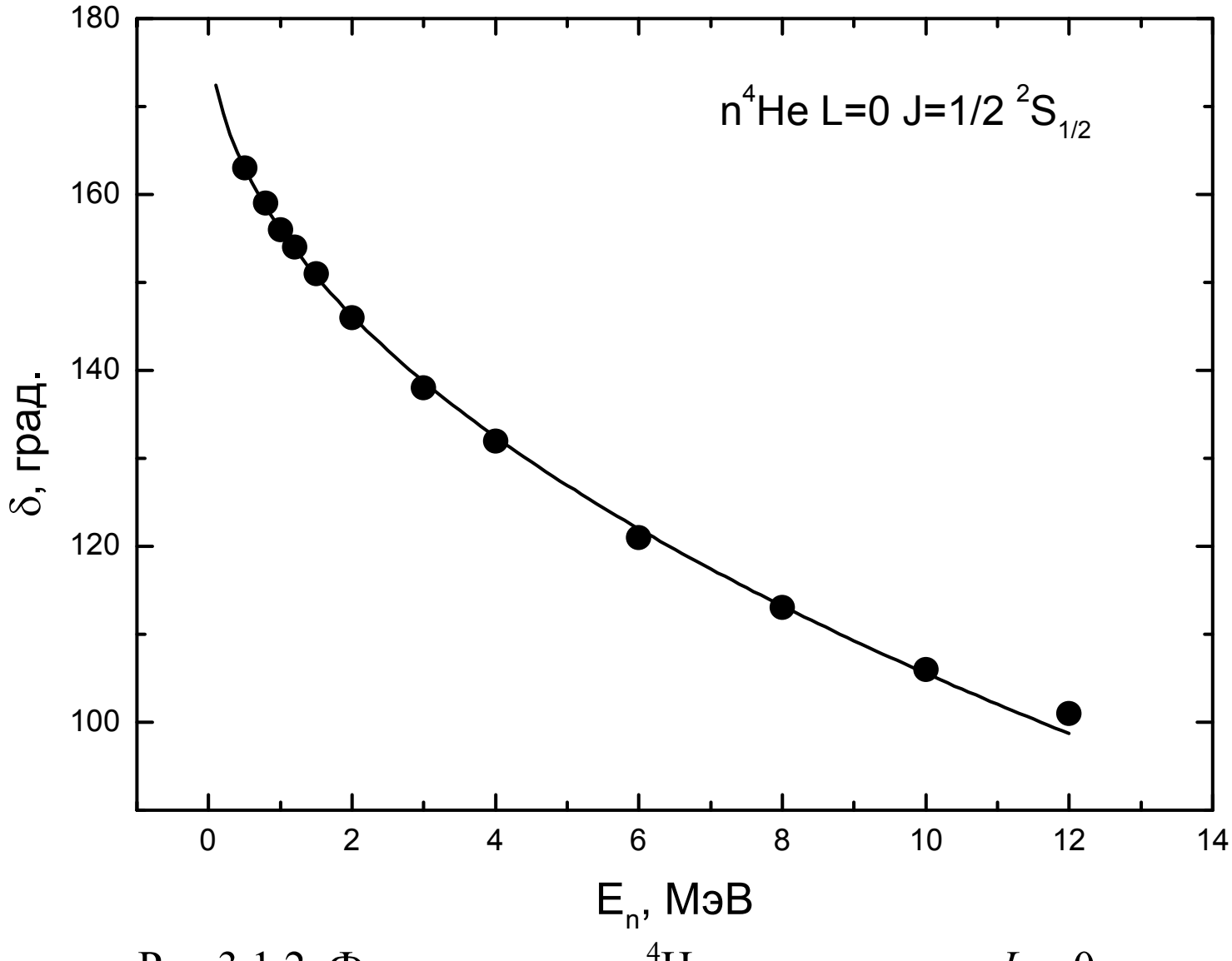

Рис.3.1.2. Фазы упругого $n^4He$ рассеяния для $L$ = 0.
Экспериментальные данные из [190].

В паре (12) использован триплетный $P_0$ потенциал $^4He^2H$ взаимодействия ($l_{12}$ = 1), параметры которого в целом уточнялись по трехтельной энергии связи $^7Li$, поскольку $^3P_0$ фазы, показанные на рис.3.1.3 и полученные в разных работах [191,192,193,194,195], имеют большую неоднозначность. В паре частиц (23) взято чистое по орбитальным симметриям $^2Hn$ дублетное *S* взаимодействие ($l_{23} = \lambda = 0$) с отталкиванием, параметры которого фиксированы по характеристикам связанного состояния ядра трития, а фазы показаны на рис.3.1.4 непрерывной линией. Параметры этих потенциалов приведены в табл.3.1.1.

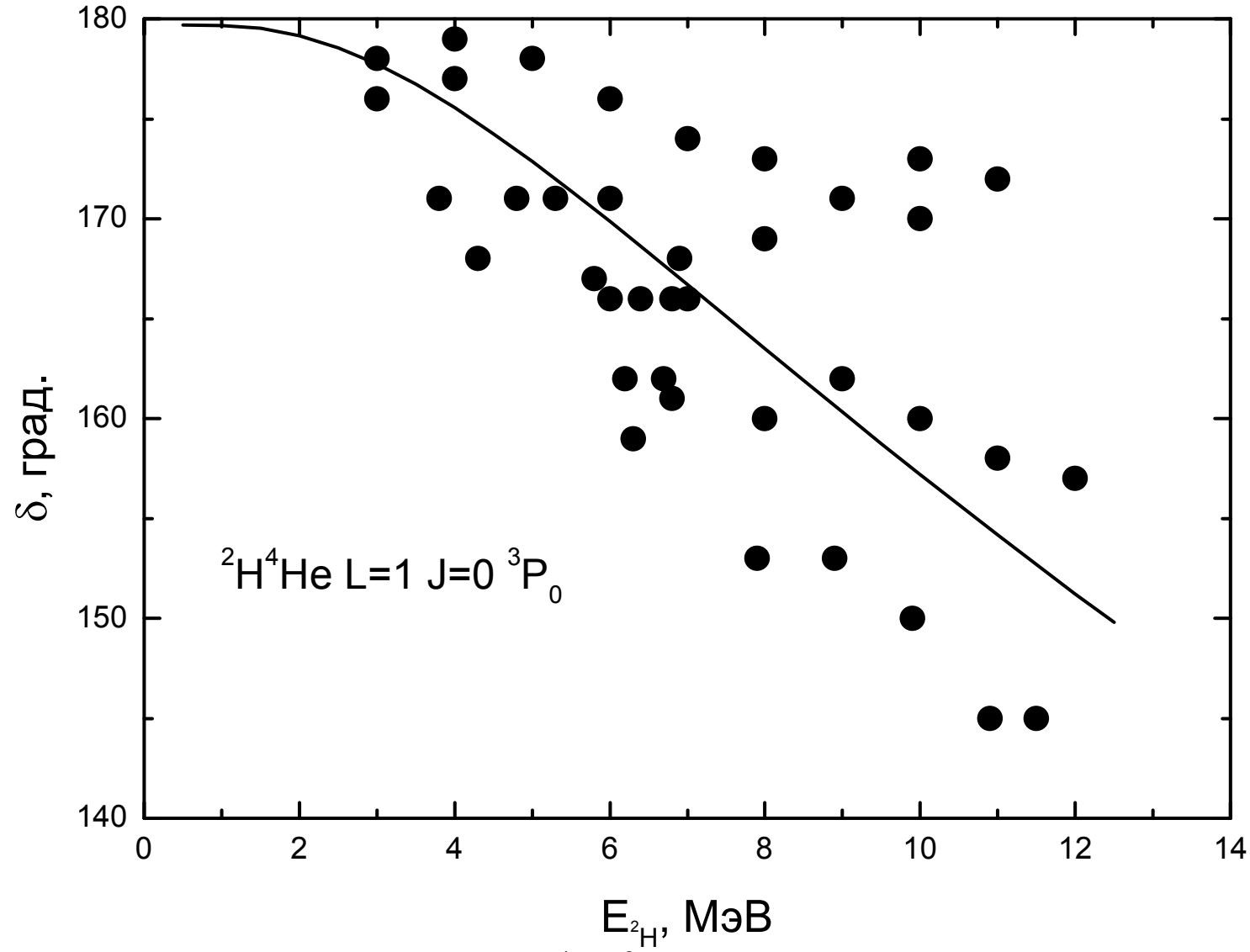

Рис.3.1.3. Фазы упругого $^4$He$^2$H рассеяния для $L$ = 1 и $J$ = 0. Экспериментальные данные из [191-195].

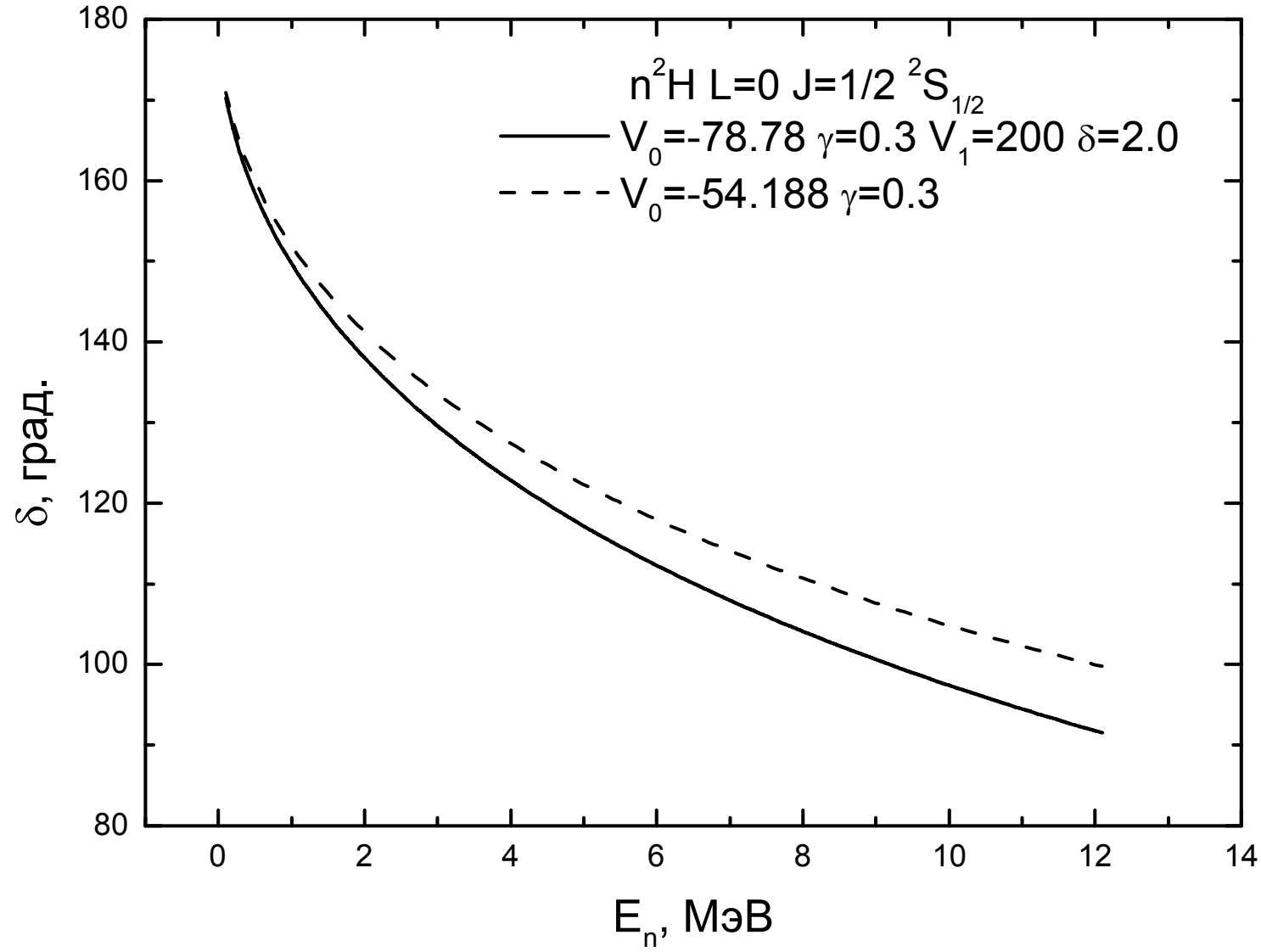

Рис.3.1.4. Фазы упругого n$^2$H рассеяния для $L$ = 0.

Табл.3.1.1 Параметры межкластерных парных потенциалов взаимодействия.

| Пара | Система | ${}^{2S+1}L_J$ | $V_0$, МэВ | $\gamma$ , Фм$^{-2}$ | $V_1$, МэВ | $\delta$ , Фм$^{-2}$ |
|---|---|---|---|---|---|---|
| 12 | $^4$He$^2$H | $^3P_0$ | -10.0 | 0.1 | 72.0 | 0.2 |
| 13 | n$^4$He | $^2S_{1/2}$ | -115.5 | 0.16 | 500 | 1.0 |
| 23 | $^2$Hn | $^2S_{1/2}$ | -78.78 | 0.3 | 200 | 2.0 |

В каждой паре частиц использовано только одно состояние с определенным орбитальным моментом и один потенциал для заданной парциальной волны и спинового состояния. Это представляется вполне оправданным, если состояния и потенциалы для остальных парциальных волн (в каждой паре частиц) вносят меньший вклад и приводят только к небольшим поправкам к расчетным характеристикам рассматриваемого ядра. Такую модель можно назвать одноканальной, поскольку учитывается только одного состояние с определенным орбитальным моментом и его взаимодействием в каждой паре частиц.

Характеристики связанного состояния ядра трития с потенциалом из табл.3.1.1 приведены в табл.3.1.2. Для расчетов характеристик трития в $^2$Hn канале и ядра $^7$Li в трехтельной модели использован радиус протона 0.8768(69) Фм из работы [120], нейтрона – равный нулю, а радиус $^2$H принят равным 1.9660(68) Фм [196]. Для радиуса $^4$He использована величина 1.671(14) Фм, приведенная в [197].

Табл.3.1.2. Характеристики тритона в n$^2$H модели с потенциалом из табл.3.1.1.

| Система | ${}^{2S+1}L_J$ | $E$, МэВ | $R_z$, Фм | $R_m$, МэВ | $C_w$ ($R$ , Фм) |
|---|---|---|---|---|---|
| $^2$Hn | $^2S_{1/2}$ | -6.257 | 2.17 | 2.13 | 2.01(1) (5÷20) |

Величина безразмерной асимптотической константы $C_w$ для $^2$Hn системы, определение которой дано выражением (1.2.3), также показана в табл.3.1.2. Указанная для $C_w$ ошибка определяется усреднением ее величины по приведенному в скобках интервалу расстояний. Экспериментальная величина радиуса $^3H$ по самым современным данным равна 1.7591(363) Фм [198], а его энергия связи в рассматриваемом канале равна -6.257 МэВ [199].

### 3.1.2 Компьютерная программа Computer program

Приведем текст компьютерной программы для вычисления энергии связи и ВФ ядра $^7Li$ в трехтельном $^4He^2Hn$ канале. Программа основана на вариационном методе с разложением ВФ по неортогональному вариационному базису и независимым варьированием всех параметров разложения, подробно описанном в первом разделе и работе [17], где был приведен вариант этой программы на языке Turbo Basic фирмы Borland.

Для написания текста данной программы использовался современный алгоритмический язык Fortran-90, который в своих основных и используемых здесь возможностях мало отличается от описания последней известной нам версии Fortran-2005. Все параметры и переменные пояснены в распечатке программы или аналогичны, описанным ранее параметрам, для других компьютерных программ.

Например, величина Z с номером всегда обозначает заряд кластера, а его масса обозначается, в данном случае, буквой M с номером, RK и RM с цифрой – зарядовый и массовый радиусы кластеров и т.д.

```
PROGRAM THREE_BODY_7LI
! Трехтельная программа для 7Li
IMPLICIT REAL(8) (A-Z)
INTEGER I,J,K,NF, NV, NP,NITER,NP2,IJK,ITER,LA,IA
DIMENSION
FF(0:500000),FU(0:500000),L(0:50,0:50),PH5(0:50)
COMMON /M/
T(0:50,0:50),L1(0:50,0:50),XP(0:50),VN12(0:50,0:50),VN13(0:5
0,0:50),VN121(0:50,0:50),VN131(0:50,0:50),VN122(0:50,0:50),
VN132(0:50,0:50),VN23(0:50,0:50),VN231(0:50,0:50),VN232(0
```

```
:50,0:50),H(0:50,0:50),SV(0:50),VK12(0:50,0:50),VK13(0:50,0:
50),VK23(0:50,0:50),VCB(0:50,0:50)
COMMON /A/
PM0,R122,PM23,A11,V122,M23,R121,V121,R132,M3,M2,M1
COMMON /B/
R131,V131,HC,R232,V232,R231,V231,PI,A23,A13,A12
COMMON /C/ PVC,EPP,ZYS,V132,PNC,NEV
CHARACTER(9) FILI,FILO
BBB(A,AA,K)=K**2/(2.*K+1.)+(2.*K+3.)*A/AA**2-K
Z1=2.0D-000
Z2=1.0D-000
Z3=0.0D-000
M1=4.0D-000
M2=2.0D-000
M3=1.0D-000
Z=Z1+Z2+Z3
M23=M3+M2
PM23=M3*M2/M23
AM0=M1+M2+M3
PM0=M1*M23/AM0
RK1=1.67D-000
RK2=1.96D-000
RK3=0.0D-000
RM1=1.67D-000
RM2=1.96D-000
RM3=0.877D-000
NF=6000
HF=0.005D-000
BB=1.0D+030
LA=1
HC=0.010D-000
PNC=-10.0D-000
PVC=0.0D-000
A11=41.46860D-000
A12=1.4399750D-000*Z1*Z2
A13=1.4399750D-000*Z1*Z3
A23=1.4399750D-000*Z3*Z2
P1=4.0D-000*DATAN(1.0D-000)
```

```
PI=DSQRT(P1)
! 1 - AL; 2 - D; 3 - N;    L  -  N-AL - 0, AL-D - 1, N-D - 0
! N-D
V231=-78.78D-000; ! J= 1/2  ; LAM=0 E=6.257 RZ=2.17
! RM=2.13 Cw=2.01(1) 5-20 Fm
R231=0.3D-000
V232=200.0D-000
R232=2.0D-000
! AL-D
V121=-10.0D-000; ! J=0; L=1
R121=0.1D-000
V122=72.0D-000
R122=0.2D-000
! AL-N
V131=-115.5D-000; ! J=1/2; L=0
R131=0.16D-000
V132=500.0D-000
R132=1.0D-000
! - - - - - - - - - - - - - - - - - - - - - - - - - - - - - - - - - - - -
NP=10
FILI='ALFAA.DAT'
FILO='ALFA2.DAT'
EP=3.0D-015; ! ТОЧНОСТЬ ЭНЕРГИИ
EPP=2.0D-015; ! ТОЧНОСТЬ ДЕТЕРМИНАНТА
NITER=30; ! ЧИСЛО ИТЕРАЦИЙ
NV=1 ; ! ЕСЛИ =0 - НЕТ ВАРЬИРОВАНИЯ, =1 - ВАРЬИРО-
ВАНИЕ ВСЕХ ПАРАМЕТРОВ
PH=0.00001D-000
IA=1
IF(IA==0) THEN
DO I=1,NP
XP(I)=I/10.0D-000
XP(I+NP)=XP(I)*2.0D-000
ENDDO
ELSE
OPEN (1,FILE=FILI)
READ(1,*)
DO I=1,NP
```

```
READ(1,*) J,XP(I),XP(I+NP)
ENDDO
CLOSE(1)
ENDIF
NP2=2*NP
AAA1: DO ITER=1,NITER
PH55=PH/ITER
50 FMIN=BB
DO IJK=1,NP2
60 PH5(IJK)=XP(IJK)*PH55
XP(IJK)=XP(IJK)+PH5(IJK)
IF (XP(IJK)<0.0D-000) THEN
XP(IJK)=XP(IJK)-PH5(IJK)
GOTO 61
ENDIF
CALL MINIM(NP,F,LA,FILI)
CC=BB; BB=F
IF(NV==0) GOTO 7654
IF(F<CC)THEN
PRINT*, ITER,IJK,F,ABS(F-CC)
IF (ABS(F-CC)>=EP) GOTO 60
ELSE
XP(IJK)=XP(IJK)-PH5(IJK)
ENDIF
61 ENDDO
PH55=-PH55/2.0D-000
IF (ABS(CC-F)>=EP/2.0D-000) GOTO 50
ENDDO AAA1
7654 PRINT*,
"****************************************************"
PRINT*, "E = ",F
PRINT*
PRINT*, "          N           ALFA                 BET"
DO I=1,NP
PRINT*,I,XP(I),XP(I+NP)
ENDDO
PRINT*
899  CONTINUE
```

```
ZYS=1.0D-000
ALA=F
CALL MINIM(NP,ALA,LA,FILI)
! - - - - - - - - - - - - - - - - - - НОРМИРОВКА - - -  - - - - - - - - - - -
IF (LA==1) THEN
FA=3.; AF=2.
ENDIF
IF (LA==2) THEN
FA=15.; AF=6.
ENDIF
IF (LA<1 .OR. LA>2) STOP
CALL SVNOR(PI,LA,NP,XP,SV)
!--------------------ПРОВЕРКА НОРМИРОВКИ -------------------
S=0.0D-000
DO I=1,NP
DO J=1,NP
AL=XP(I)+XP(J)
BT=XP(I+NP)+XP(J+NP)
DO K=0,NF
R=HF*K
RR=(R**2)*AL
AB=(R**2)*DEXP(-RR)
FF(K)=AB
ENDDO
CALL SIMPS(NF,HF,FF,S1)
DO K=0,NF
R=HF*K
RR=(R**2)*BT
AC=R**(2*LA+2)*DEXP(-RR)
FU(K)=AC
ENDDO
CALL SIMPS(NF,HF,FU,S2)
S=S+SV(I)*SV(J)*S1*S2
ENDDO
ENDDO
SNOR=S
PRINT*
PRINT*, "   NORM =",SNOR,'        NEV-DET = ',NEV
```

```
PRINT*
PRINT*, '         N              SV'
DO I=1,NP
PRINT*,  I,SV(I)
ENDDO
6622 CONTINUE
! ****************** НОРМИРОВКА ******************
BN=LA+1.5
SS=0.
DO I=1,NP
DO J=1,NP
AL=XP(I)+XP(J)
BT=XP(I+NP)+XP(J+NP)
L(I,J)=1./AL**1.5/BT**BN
SS=SS+SV(I)*SV(J)*L(I,J)
ENDDO
ENDDO
SN=DSQRT(FA*P1/16./2.**LA*SS)
PRINT*
PRINT*, "  N= ", SN
! ******************  RM ****************************
 S=0.
 DO I=1,NP
 DO J=1,NP
 AL=XP(I)+XP(J)
 BT=XP(I+NP)+XP(J+NP)
S=S+SV(I)*SV(J)/BT**BN/AL**1.5*(3.*PM23/AL+(2.*LA+3.)
*PM0/BT)
 ENDDO
 ENDDO
 RMM=P1*FA*S/2.**(LA+1)/16.
RRR=M1/AM0*RM1**2+M2/AM0*RM2**2+M3/AM0*RM3*
*2+RMM/AM0
 RM=DSQRT(RRR)
 PRINT*, " RM = ", RM
! ******************  RZ **************************
CCC=(Z1*M23**2+(Z2+Z3)*M1**2)/AM0**2
DDD=(Z2*M3**2+Z3*M2**2)/M23**2
```

```
EEE=-M1/AM0/M23*(Z2*M3-Z3*M2)
S=0.
SS=0.
SSS=0.
 DO I=1,NP
 DO J=1,NP
 AL=XP(I)+XP(J)
 BT=XP(I+NP)+XP(J+NP)
 S=S+SV(I)*SV(J)*(2.*LA+3.)/BT**(LA+2.5)/AL**1.5
 SS=SS+SV(I)*SV(J)*3./BT**(LA+1.5)/AL**2.5
 SSS=SSS+SV(I)*SV(J)*AF/BT**(LA+2)/AL**2
 ENDDO
 ENDDO
 RMM=FA/2**(LA+1)*P1/16.*(CCC*S+DDD*SS)+EEE*SSS
 RRR=Z1/Z*RK1**2+Z2/Z*RK2**2+Z3/Z*RK3**2+RMM/Z
 RZ=DSQRT(RRR)
 PRINT*, " RZ = ",RZ
! ********************** Q ************************
 QQ=-10.*2./5.*RMM+2.86
 PRINT*, " Q = ",QQ
! ************** CONTROL ENERGY  ****************
SC=0.
SK=0.
SP=0.
SL=0.
S1=0.
S3=0.
S2=0.
SH=0.
DO I=1,NP
DO J=1,NP
S1=S1+SV(I)*SV(J)*VK12(I,J)
S2=S2+SV(I)*SV(J)*VK13(I,J)
S3=S3+SV(I)*SV(J)*VK23(I,J)
SC=SC+SV(I)*SV(J)*VCB(I,J)
SK=SK+SV(I)*SV(J)*T(I,J)
SP=SP+SV(I)*SV(J)*(VN12(I,J)+VN13(I,J)+VN23(I,J))
SL=SL+SV(I)*SV(J)*L1(I,J)
```

```
SH=SH+SV(I)*SV(J)*H(I,J)
ENDDO
ENDDO
SC1=S1*P1/16.0D-000
SC2=S2*P1/16.0D-000
SC3=S3*P1/16.0D-000
SCU=SC1+SC2+SC3
SC=SC*P1/16.0D-000
SK=SK*P1/16.0D-000
SP=SP*P1/16.0D-000
SL=SL*P1/16.0D-000
SH=SH*P1/16.0D-000
ST=SCU+SK+SP+SC
PRINT*
PRINT*, "COUL. ENERGY  VK = ",SCU," 12 = ",SC1," 13 =
",SC2," 23 = ",SC3
PRINT*, "CENTROB. ENERGY =  ",SC
PRINT*, "KINETICH. ENERGY =  ",SK
PRINT*, "M.E. OT L1 = ",SL
PRINT*, "POTENS. ENERGY =  ",SP
PRINT*, "POLNAY ENERGY ST =  ",ST
PRINT*, "POLNAY ENERGY SH =  ",SH
!--------------------------------------------------------------------------
PRINT*, "???"
READ*, AAA
IF(AAA==0) GOTO 2244
OPEN (1,FILE=FILO)
WRITE(1,*) '        N                  ALFA                        BETTA'
DO I=1,NP
WRITE(1,*) I,XP(I),XP(I+NP)
ENDDO
WRITE(1,*)
WRITE(1,*) 'E = ',F
WRITE(1,*)
WRITE(1,*)'SUM(H*SV-E*L*SV) FROM SV=',ALA
WRITE(1,*)
WRITE(1,*) '        N                   SV'
DO I=1,NP
```

```
WRITE(1,*) I,SV(I)
ENDDO
WRITE(1,*)
WRITE(1,*) 'NOR= ',SNOR, SN
WRITE(1,*)
WRITE(1,*) 'NEV-DET=',NEV
WRITE(1,*)
WRITE(1,*) "RM=",RM," RZ=",RZ
WRITE(1,*)
WRITE(1,*) "Q=",QQ
CLOSE(1)
2244 CONTINUE
END
SUBROUTINE SIMPS(N,H,F,S)
IMPLICIT REAL(8) (A-Z)
INTEGER I,N
DIMENSION F(0:500000)
A=0.0D-000;B=0.0D-000
DO I=1,N-1,2
B=B+F(I)
ENDDO
DO I=2,N-2,2
A=A+F(I)
ENDDO
S=H*(F(0)+F(N)+2.0D-000*A+4.0D-000*B)/3.0D-000
END
SUBROUTINE MINIM(NP,ALA,L,FILO)
IMPLICIT REAL(8) (A-Z)
INTEGER   NP,KK,JJ,L,K,I
COMMON /M/
T(0:50,0:50),L1(0:50,0:50),XP(0:50),VN12(0:50,0:50),VN13(0:5
0,0:50),VN121(0:50,0:50),VN131(0:50,0:50),VN122(0:50,0:50),
VN132(0:50,0:50),VN23(0:50,0:50),VN231(0:50,0:50),VN232(0
:50,0:50),H(0:50,0:50),SV(0:50),VK12(0:50,0:50),VK13(0:50,0:
50),VK23(0:50,0:50),VCB(0:50,0:50)
COMMON /A/
PM0,R122,PM23,A11,V122,M23,R121,V121,R132,M3,M2,M1
COMMON /B/
```

```
R131,V131,HC,R232,V232,R231,V231,PI,A23,A13,A12
COMMON /C/ PVC,EPP,ZYS,  V132,PNC,NEV
CHARACTER(9) FILO
BBB(A,AB,K)=(1.*K)**2/(2.*K+1.)+(2.*K+3.)*A/AB**2-
(1.*K)
P1=4.0D-000*DATAN(1.0D-000)
PI=DSQRT(P1)
IF(L==1) THEN
FA=3.; FAA=1.;FFA=1.
ENDIF
IF(L==2) THEN
FA=15.; FAA=3.;FFA=2.
ENDIF
IF(L<1 .OR. L>2) THEN
PRINT*, "STOP"; STOP
ENDIF
A1: DO KK=1,NP
A5: DO JJ=1,NP
AL=XP(KK)+XP(JJ)
AL1=XP(KK)*XP(JJ)
BT=(XP(KK+NP)+XP(JJ+NP))
BT1=XP(KK+NP)*XP(JJ+NP)
H1=FA/2.**L*A11/PM23/BT*BBB(AL1,AL,0)/AL**0.5/BT**(
L+0.5)
H2=FA/2.**L*A11/PM0/AL*BBB(BT1,BT,L)/AL**0.5/BT**(L
+0.5)
T(KK,JJ)=H1+H2
L1(KK,JJ)=FA/2.**L/AL**1.5/BT**(L+1.5)
AA=AL*BT+R121*(AL+BT*(M3/M23)**2)
DD=(M3/M23)**2*R121+AL
VN121(KK,JJ)=FA/2.**L*V121*DD**L/AA**(L+1.5)
AA=AL*BT+R122*(AL+BT*(M3/M23)**2)
DD=(M3/M23)**2*R122+AL
VN122(KK,JJ)=FA/2.**L*V122*DD**L/AA**(L+1.5)
VN12(KK,JJ)=VN121(KK,JJ)+VN122(KK,JJ)
AA=AL*BT+R131*(AL+BT*(M2/M23)**2)
DD=(M2/M23)**2*R131+AL
VN131(KK,JJ)=FA/2.**L*V131*DD**L/AA**(L+1.5)
```

```
AA=AL*BT+R132*(AL+BT*(M2/M23)**2)
DD=(M2/M23)**2*R132+AL
VN132(KK,JJ)=FA/2.**L*V132*DD**L/AA**(L+1.5)
VN13(KK,JJ)=VN131(KK,JJ)+VN132(KK,JJ)
VN231(KK,JJ)=FA/2.**L*V231/BT**(L+1.5)/(AL+R231)**1.5
VN232(KK,JJ)=FA/2.**L*V232/BT**(L+1.5)/(AL+R232)**1.5
VN23(KK,JJ)=VN231(KK,JJ)+VN232(KK,JJ)
VK12(KK,JJ)=2.*FFA*A12/PI/AL**1.5/BT**(L+1)
VK13(KK,JJ)=2.*FFA*A13/PI/AL**1.5/BT**(L+1)
VK23(KK,JJ)=2./2.**L*FA*A23/PI/AL/BT**(L+1.5)
VCB(KK,JJ)=(1.*L)*(1.*L+1.)/2.**L*FAA*A11/AL**1.5/BT**
(L+0.5)/PM0
H(KK,JJ)=T(KK,JJ)+VN23(KK,JJ)+VN12(KK,JJ)+VN13(KK,JJ
)+VCB(KK,JJ)+VK12(KK,JJ)+VK13(KK,JJ)+VK23(KK,JJ)
ENDDO A5
ENDDO A1
CALL MINI(NP,ALA,DETER)
EE=ALA
OPEN (1,FILE=FILO)
WRITE(1,*) '        N              ALFA                    BETTA'
DO I=1,NP
WRITE(1,*) I,XP(I),XP(I+NP)
ENDDO
WRITE(1,*)
WRITE(1,*) 'E = ',EE
CLOSE(1)
IF (ZYS==1.0D-000) THEN
CALL VEC(NP,ALA)
CALL SVNOR(PI,L,NP,XP,SV)
ENDIF
END
SUBROUTINE SVNOR(PI,L,NP,XP,SV)
IMPLICIT REAL(8) (A-Z)
INTEGER NP,L,I,J
DIMENSION SV(0:50),XP(0:50),A(0:50,0:50)
IF (L==1) THEN
FA=3.
ENDIF
```

```
IF (L==2) THEN
FA=15.
ENDIF
SS=0.
DO I=1,NP
DO J=1,NP
AL=XP(I)+XP(J)
BT=XP(I+NP)+XP(J+NP)
A(I,J)=1./AL**1.5/BT**(L+1.5)
SS=SS+SV(I)*SV(J)*A(I,J)
ENDDO
ENDDO
ANOR=DSQRT(16.*2.**L/FA/SS)/PI
DO I=1,NP
SV(I)=ANOR*SV(I)
ENDDO
END
SUBROUTINE MINI(NP,COR,D)
IMPLICIT REAL(8) (A-Z)
INTEGER C,  NP
COMMON /A/
PM0,R122,PM23,A11,V122,M23,R121,V121,R132,M3,M2,M1
COMMON /B/
R131,V131,HC,R232,V232,R231,V231,PI,A23,A13,A12
COMMON /C/ PVC,EPP,ZYS,  V132,PNC,NEV
PN=PNC; PV=PVC; H=HC; E=EPP
IF(PN>PV) THEN
PNN=PV; PV=PN; PN=PNN
ENDIF
A=PN
1 CALL DET(NP,A,D1); B=A+H
2 CALL DET(NP,B,D2)
IF (D1*D2>0.0D-000) THEN
B=B+H; D1=D2
IF (B<=PV .AND. B>=PN) GOTO 2
C=0; RETURN; ELSE
A=B-H; H=H*1.0D-001
IF(ABS(D2)<E .OR. ABS(H)<E) GOTO 3
```

```
B=A+H; GOTO 1
ENDIF
3 C=1; COR=B; D=D2
END
SUBROUTINE DET(NP,LLA,S)
IMPLICIT REAL(8) (A-Z)
INTEGER I,J,K,NP
DIMENSION
LLL(0:50,0:50),B(0:50,0:50),C(0:50,0:50),AAA(0:50,0:50)
COMMON /M/
T(0:50,0:50),L1(0:50,0:50),XP(0:50),VN12(0:50,0:50),VN13(0:5
0,0:50),VN121(0:50,0:50),VN131(0:50,0:50),VN122(0:50,0:50),
VN132(0:50,0:50),VN23(0:50,0:50),VN231(0:50,0:50),VN232(0
:50,0:50),H(0:50,0:50),SV(0:50),VK12(0:50,0:50),VK13(0:50,0:
50),VK23(0:50,0:50),VCB(0:50,0:50)
COMMON /A/
PM0,R122,PM23,A11,V122,M23,R121,V121,R132,M3,M2,M1
COMMON /B/
R131,V131,HC,R232,V232,R231,V231,PI,A23,A13,A12
COMMON /C/ PVC,EPP,ZYS,  V132,PNC,NEV
DO I=1,NP
DO J=1,NP
LLL(I,J)=(H(I,J)-LLA*L1(I,J))
B(I,J)=0.0D-000
C(I,J)=0.0D-000
ENDDO
ENDDO
GOTO 234
PRINT*, "                    МАТРИЦА  LLL=H-E*L1"
PRINT*
DO II=1,NP
DO KK=1,NP
PRINT*, LLL(II,KK)
ENDDO
PRINT*
ENDDO
234 CONTINUE
 ! - - - - LLLL - РАЗЛОЖЕНИЕ НА ТРЕУГОЛЬНЫЕ - - - - - - -
```

```
DO I=1,NP
C(I,I)=1.0D-000
B(I,1)=LLL(I,1)
C(1,I)=LLL(1,I)/B(1,1)
ENDDO
DO I=2,NP
DO J=2,NP
S=0.0D-000
IF (J>I) GOTO 1
DO K=1,I-1
S=S+B(I,K)*C(K,J)
ENDDO
B(I,J)=LLL(I,J)-S
GOTO 2
1 S=0.0D-000
DO K=1,I-1
S=S+B(I,K)*C(K,J)
ENDDO
C(I,J)=(LLL(I,J)-S)/B(I,I)
2 CONTINUE
ENDDO
ENDDO
 ! - - - - - ПРОВЕРКА РАЗЛОЖЕНИЯ МАТРИЦЫ LLL - - - - -
 SS=0.0D-000
 DO I=1,NP
 DO J=1,NP
 S=0.0D-000
 DO K=1,NP
 S=S+B(I,K)*C(K,J)
 ENDDO
 AAA(I,J)=S-LLL(I,J)
 SS=SS+AAA(I,J)
 ENDDO
 ENDDO
 NEV=SS
 GOTO 678
 PRINT*, "              ДЛЯ ДЕТЕРМИНАНТА  N=LLL-B*C =0"
 DO I=1,NP
```

```
 PRINT*
 DO J=1,NP
 PRINT*,AAA(I,J)
 ENDDO
 ENDDO
 PRINT*
678 CONTINUE
S=1.0D-000
DO K=1,NP
S=S*B(K,K)
 ENDDO
GOTO 991
PRINT 22, LLA,S,SS
PRINT*, "       DET=",S
PRINT*, "       NEV=",SS
991 CONTINUE
22 FORMAT(3E15.5)
END
SUBROUTINE VEC(NP,LLA)
IMPLICIT REAL(8) (A-Z)
INTEGER I,J,K,J1,I1,NP
DIMENSION
LLL(0:50,0:50),D(0:50),Y(0:50),B(0:50,0:50),AD(0:50,0:50),X(0
:50),C(0:50,0:50),E2(50)
COMMON /M/
T(0:50,0:50),L1(0:50,0:50),XP(0:50),VN12(0:50,0:50),VN13(0:5
0,0:50),VN121(0:50,0:50),VN131(0:50,0:50),VN122(0:50,0:50),
VN132(0:50,0:50),VN23(0:50,0:50),VN231(0:50,0:50),VN232(0
:50,0:50),H(0:50,0:50),SV(0:50),VK12(0:50,0:50),VK13(0:50,0:
50),VK23(0:50,0:50),VCB(0:50,0:50)
COMMON /A/
PM0,R122,PM23,A11,V122,M23,R121,V121,R132,M3,M2,M1
COMMON /B/
R131,V131,HC,R232,V232,R231,V231,PI,A23,A13,A12
COMMON /C/ PVC,EPP,ZYS,  V132,PNC,NEV
 DO I=1,NP
 DO J=1,NP
 LLL(I,J)=(H(I,J)-LLA*L1(I,J))
```

```
 B(I,J)=0.0D-000
 C(I,J)=0.0D-000
 ENDDO
 ENDDO
 DO I=1,NP-1
 DO J=1,NP-1
 AD(I,J)=LLL(I,J)
 ENDDO
 ENDDO
 I1=1
 I2=NP-1
 J=NP
 DO I=I1,I2
 D(I)=-LLL(I,J)
 ENDDO
 NP=NP-1
CALL TRI(NP,AD,B,C,SOB)
 ! - - - - - - - - - - - - - - - - - - - - - - - - - - - - - - - - - - - - - - - - - - - -
 Y(1)=D(1)/B(1,1)
 DO I=2,NP
 S=0.0D-000
 DO K=1,I-1
 S=S+B(I,K)*Y(K)
 ENDDO
 Y(I)=(D(I)-S)/B(I,I)
 ENDDO
  X(NP)=Y(NP)
 DO I=NP-1,1,-1
 S=0.0D-000
 DO K=I+1,NP
 S=S+C(I,K)*X(K)
 ENDDO
 X(I)=Y(I)-S
 ENDDO
 DO I=1,NP
 SV(I)=X(I)
 ENDDO
 NP=NP+1
```

```
 SV(NP)=1
 S=0.0D-000
 DO I=1,NP
 S=S+SV(I)**2
 ENDDO
 !PRINT*,'S=', S
 SS=0.0D-000
 DO I=1,NP
 SV(I)=SV(I)/DSQRT(ABS(S))
! SS=SS+SV(I)**2
 ENDDO
!AN=1.0D-000/DSQRT(ABS(SS))
!AN=1.0D-000
!PRINT*, "                    H*SV-LA*L*SV=0"
 SSS=0.0D-000
 DO I=1,NP
 S=0.0D-000
 SS=0.0D-000
 DO J=1,NP
 !SV(J)=SV(J)*AN
 S=S+H(I,J)*SV(J)
 SS=SS+LLA*L1(I,J)*SV(J)
 ENDDO
 E2(I)=S-SS
 SSS=SSS+E2(I)
 ENDDO
 LLA=SSS
!DO I=1,NP
!PRINT*,"E2 = ",E2(I)
!ENDDO
PRINT*
PRINT*,'          SUM(H*SV-E*L*SV) FROM SV =',LLA
END
SUBROUTINE TRI(NP,AD,B,C,S)
IMPLICIT REAL(8) (A-Z)
INTEGER I,J,K,NP
DIMENSION
AD(0:50,0:50),B(0:50,0:50),C(0:50,0:50),AAA(0:50,0:50)
```

```
 DO I=1,NP
 C(I,I)=1.0D-000
 B(I,1)=AD(I,1)
 C(1,I)=AD(1,I)/B(1,1)
 ENDDO
 DO I=2,NP
 DO J=2,NP
 S=0.0D-000
 IF (J>I) GOTO 551
 DO K=1,I-1
 S=S+B(I,K)*C(K,J)
 ENDDO
 B(I,J)=AD(I,J)-S
 GOTO 552
551 S=0.0D-000
 DO K=1,I-1
 S=S+B(I,K)*C(K,J)
 ENDDO
 C(I,J)=(AD(I,J)-S)/B(I,I)
552 CONTINUE
ENDDO
ENDDO
 ! - - - - - - - - - - - - - - - - - - - - - - - - - - - - - - - - - - - - - - - - - - -
 SS=0.0D-000
 DO I=1,NP
 DO J=1,NP
 S=0.0D-000
 DO K=1,NP
 S=S+B(I,K)*C(K,J)
 ENDDO
 AAA(I,J)=S-AD(I,J)
 SS=SS+AAA(I,J)
 ENDDO
 ENDDO
 GOTO 578
 PRINT*, "                    NEV = AD - B*C =0"
 DO I=1,NP
 DO J=1,NP
```

```
 PRINT*,AAA(I,J)
 ENDDO
 ENDDO
578 S=1.0D-000
 DO K=1,NP
 S=S*B(K,K)
 ENDDO
! GOTO 9753
! PRINT*, "              DET=",S
 !PRINT*,S
 PRINT*, "              NEV-TRI=",SS
 !PRINT*,SS
9753 PRINT*
 END
 SUBROUTINE WW(SK,L,GK,R,N,H,WH)
 IMPLICIT REAL(8) (A-Z)
 INTEGER I,L,N,NN
 DIMENSION V(50000)
 H=H
 N=N
 SS=DSQRT(ABS(SK))
 AA=GK/SS
 BB=L
 NN=500
 HH=.02D-000
 ZZ=1+AA+BB
 AAA=1.0D-000/ZZ
 NNN=2000
 DO I2=1,NNN
 AAA=AAA*I2/(ZZ+I2)
 ENDDO
 GAM=AAA*NNN**ZZ
 RR=R
 CC=RR*SS*2
 DO I=0,NN
 TT=HH*I
 V(I)=TT**(AA+BB)*(1+TT/CC)**(BB-AA)*DEXP(-TT)
 ENDDO
```

```
CALL SIMPS(NN,HH,V,S)
WH=S*DEXP(-CC/2.0D-000)/(CC**AA*GAM)
END
```

### 3.1.3 Трехтельные результаты
### Three-body results

Приведем теперь результаты счета по этой программе для ядра $^{7}$Li при рассмотренном выше варианте трехтельной конфигурации для девяти членов разложения волновой функции (1.5.4) по гауссойдам

**E = -8.7165 04042795002**

| N | ALFA (α) | BETTA (β) |
|---|---|---|
| 1 | 2.695013648564534E-001 | 5.552519454982658E-002 |
| 2 | 6.073846174799727E-002 | 5.582211053901705E-002 |
| 3 | 1.481076486508074E-001 | 1.500571262902319E-001 |
| 4 | 1.219211094860576E-001 | 2.100593191345530E-001 |
| 5 | 1.583008396850423E-001 | 6.443497149569889E-001 |
| 6 | 1.572092636709496E-001 | 6.485407339616155E-001 |
| 7 | 2.048327219956353E-001 | 5.160938755052593E-001 |
| 8 | 2.920843757329559E-001 | 3.970902201955188E-001 |
| 9 | 1.185843962546213 | 7.762432248877493E-002 |

SUM(H*SV-E*L*SV) FROM SV = -3.048700181196296E-011
NORM = 9.999999999986114E-001
NEV-DET = -9.947598300641403E-014

| N | SV |
|---|---|
| 1 | -2.006282287524374E-002 |
| 2 | -1.041690910372422E-002 |
| 3 | -1.956124177741528E-001 |
| 4 | 1.721055580493311E-001 |
| 5 | -33.347719101844450 |
| 6 | 31.786460544316070 |
| 7 | 2.111064771161292 |
| 8 | -5.187214457242435E-001 |
| 9 | 4.082610488279214E-002 |

RM = 2.771546765454730

Для десяти членов разложения ВФ по гауссойдам получаем

**E = -8.7176 07265169926**

| N | ALFA | BET |
|---|---|---|
| 1 | 2.667953617399743E-001 | 5.601125106169563E-002 |
| 2 | 5.941262333765297E-002 | 5.493848544738812E-002 |
| 3 | 1.393163512886810E-001 | 1.556109489548065E-001 |
| 4 | 1.235101199700397E-001 | 1.918288577203961E-001 |
| 5 | 1.584363057560162E-001 | 6.454611171440240E-001 |
| 6 | 1.578191661203245E-001 | 6.479935485451310E-001 |
| 7 | 2.037134087039333E-001 | 5.101654896420405E-001 |
| 8 | 2.707874198704808E-001 | 4.030486689121375E-001 |
| 9 | 1.211284751587861 | 7.694488373218295E-002 |
| 10 | 4.678503078478220 | 7.811004516393501E-002 |

NEV-TRI = 3.979039320256561E-013
SUM(H*SV-E*L*SV) FROM SV = -3.541195364720196E-009
NORM = 9.999999999949636E-001
NEV-DET = 3.836930773104541E-013

| N | SV |
|---|---|
| 1 | 2.090487646994713E-002 |
| 2 | 9.678149889880678E-003 |
| 3 | 2.816715670580448E-001 |
| 4 | -2.562071190446284E-001 |
| 5 | 65.304218586437270 |
| 6 | -63.510565912605150 |
| 7 | -2.554362633383337 |
| 8 | 7.253463724545455E-001 |
| 9 | -4.040198847629709E-002 |
| 10 | 8.813225618937962E-004 |

NN = 9.999999999997704E-001
RM = 2.792145900219183
RZ = 2.517493303959341
Q = -35.515670530665620

COUL. ENERGY VK = 7.722608537158121E-001
12 = 7.722608537158121E-001
13 = 0.000000000000000E+000
23 = 0.000000000000000E+000
CENTROB. ENERGY = 1.906186930356347
KINETICH. ENERGY = 15.485397205276890
M.E. OT L1 = 9.999999999985132E-001
POTENS. ENERGY = -26.881452254540800
**POLNAY ENERGY ST = -8.717607265191750**
**POLNAY ENERGY SH = -8.717607265145604**

Как видно, результаты для разных *N* практически совпадают – их отличие составляет величину порядка 1 кэВ, что демонстрирует насыщение процесса сходимости расчетной трехтельной энергии. Для сравнения приведем экспериментальное значение трехтельной энергии связи -8.724 МэВ ядра $^7$Li [200]. Она отличается от полученной выше энергии только на 6÷7 кэВ.

Для массового и зарядового радиусов получены величины 2.79 Фм и 2.52 Фм. Последняя из них оказывается заметно больше экспериментальных данных 2.39(3) Фм и 2.35(10) Фм [200]. Однако здесь, как и раньше [13], использовался $^2$Hn потенциал, приводящий к завышенному радиусу трития (см. табл.3.1.2), что могло повлиять и на радиус самого ядра $^7$Li. Заметим, что современный радиус дейтрона, равный 2.1402(28) Фм [120], и ядра $^4$He 1.6753(28) Фм [198], также несколько больше использованных здесь величин 1.97 Фм и 1.67 Фм [121-123,196].

Тем самым дейтронный кластер нужно деформировать, как в ядре трития, так и в $^7$Li, поскольку в свободном состоянии дейтрон очень "рыхлая" система. Для того чтобы получить правильный зарядовый радиус ядра $^7$Li, равный 2.39(3) Фм необходимо уменьшить радиус дейтронного кластера, также как это было сделано раньше для ядра трития [13] и принять его примерно равным 1.4 Фм. Нужно отметить, что наиболее современные значения зарядового радиуса ядра $^7$Li равны 2.4017(281) Фм [198] или 2.4173(280) Фм, как приведено в работе [201]. Однако эти результаты были получены нами еще в конце 90-х годов прошлого века [87], поэтому здесь использовались несколько более старые значения всех радиусов.

Для квадрупольно момента ядра $^7$Li в проведенных выше расчетах получена величина -35.5 мб, которая не намного меньше известных данных -40.7(8) мб [200] и -36.6(3) мб [145]. В распечатке результатов расчета приведены нормировки NORM и NN волновой функции, полученные двумя разными способами и равные в обоих случаях единице с высокой степенью точности. Точность, с которой определяются собственные вектора, не менее $10^{-8}$, а невязки триангуляризации NEV-TRI и вычисления детерминанта NEV-DET находятся на уровне $10^{-12}\div10^{-13}$.

В рассмотренной конфигурации расположения кластеров имеется возможность варьирования параметров $P_0$ потенциала в $^4$He$^2$H канале из-за больших ошибок в фазах упругого рассеяния. Эта возможность позволила нам скомпенсировать неучет других орбитальных конфигураций, т.е. одноканальность модели, и получить правильную энергию связи ядра. В тоже время, именно эта конфигурация явно выделяет $^4$He$^3$H структуру ядра $^7$Li, которая имеет наибольшую вероятность существования [11,13]. Поэтому дополнительное варьирование параметров $P_0$ потенциала позволило уточнить энергию связи $^7$Li, приведя ее в хорошее соответствие с экспериментальной величиной.

В заключение этого параграфа обратим внимание на строки

**E = -8.7176 0726 51 69926**
**POLNAY ENERGY ST = -8.7176 0726 51 91750**
**POLNAY ENERGY SH = -8.7176 0726 51 45604**

которые показывают трехтельную энергию связи ядра $^7$Li в приведенных выше распечатках. Разница в численных значениях этой энергии показывает точность, с которой она определяется разными методами, и которая составляет порядок ε ~ $10^{-10}$ МэВ.

## *3.2 Трехтельная модель ядра $^9$Be*
## *Three-body model of $^9$Be*

Перейдем теперь к рассмотрению ядра $^9$Be в трехтельной кластерной $^4$He$^3$H$^2$H модели. Будем считать, что в основании треугольника из трех частиц находятся $^3$H$^2$H кластеры (частицы 23) с орбитальным моментом относительного движения $\lambda = 0$ и спином 1/2, т.е. рассматривается только дублетное состояние этих кластеров. Ядро $^4$He (частица 1) находится в вершине треугольника и его положение относительно центра масс двухкластерной системы определяется орбитальным моментом $l$.

Полный спин системы трех частиц считается равным 1/2, а орбитальный момент $\boldsymbol{L} = \boldsymbol{l} + \lambda$, равный 1, может быть получен, например, из комбинации $l = 1$ и $\lambda = 0$ ($\boldsymbol{l} = \boldsymbol{l}_{\alpha t} + \boldsymbol{l}_{\alpha d}$). Здесь предполагается, что именно эта орбитальная конфигурация доминирует в рассматриваемой $^4$He$^3$H$^2$H модели, т.е. также рассматривается одноканальная трехтельная модель этого ядра. При такой конфигурации кластеров полный момент системы $\boldsymbol{J} = \boldsymbol{L} + \boldsymbol{S}$ равен $3/2^-$ и $1/2^-$, первый из которых соответствует ОС ядра $^9$Be.

### 3.2.1 Потенциалы и фазы рассеяния
### Potentials and scattering phase shifts

В расчетах использованы бинарные межкластерные потенциалы для $^4$He$^3$H и $^4$He$^2$H систем с отталкивающим кором и запрещенным состоянием в $^3$H$^2$H канале обычного вида (1.2.1). Параметры межкластерных парных потенциалов приведены в 1÷6 столбцах табл.3.2.1. В седьмом столбце табл.3.2.1 приведены канальные энергии связи $^4$He$^3$H системы в ядре $^7$Li и $^4$He$^2$H канала в $^6$Li, в восьмом – среднеквадратичные зарядовые радиусы связанных состояний этих пар частиц и в девятом – безразмерные асимптотические константы связанных состояний в двухчастичных каналах, найденные с функцией Уиттекера (1.2.3).

Табл.3.2.1. Параметры потенциалов в бинарных кластерных системах и основные характеристики связанных состояний.

| Сис-тема | $^{2S+1}L_J$ | $V_0$, МэВ | $\gamma$, Фм$^{-2}$ | $V_1$, МэВ | $\delta$, Фм$^{-2}$ | $E$, МэВ | $R_z$, Фм | $C_w$ |
|---|---|---|---|---|---|---|---|---|
| 1 | 2 | 3 | 4 | 5 | 6 | 7 | 8 | 9 |
| $^4$He$^3$H | $^2P_{1/2}$ | -85.82 | 0.13 | 90.0 | 0.2 | -1.989 | 2.6 | 3.86(1) |
| $^3$H$^2$H | $^2S_{1/2}$ | -44.5887 | 0.15 | 4.5 | 0.015 | – | – | – |
| $^4$He$^2$H | $^3S_1$ | -71.91 | 0.15 | 70.0 | 0.2 | -1.474 | 2.66 | 3.27(1) |

Параметры потенциалов подобраны таким образом, чтобы максимально точно воспроизвести соответствующие экспериментальные фазы упругого рассеяния, которые показаны на рис.3.2.1, 3.2.2 и 3.2.3. В качестве потенциалов кластерной $^3$H$^2$H системы использованы чистые по схемам Юнга взаимодействия [37], а результаты расчета фаз с таким потенциалом показаны на рис.3.2.1 непрерывной линией.

Пунктиром на рис.3.2.1 приведена полоса ошибок определения чистых $^3$H$^2$H фаз [37], которая получается из экспериментальных данных различных исследований, описанных в работе [202]. Точки и квадраты показывают извлеченные из экспериментальных данных фазы рассеяния, также приведенные в работе [202], а кружками и открытыми квадратами приведены МРГ (метод резонирующих групп) вычисления фаз рассеяния [203].

Такой $^4$He$^2$H потенциал с периферическим отталкиванием содержит связанное запрещенное состояние при энергии -11.49 МэВ. Отметим, что нам не удалось найти другие параметры потенциала, т.е. без связанного запрещенного уровня, с которыми можно было бы описать чистую дублетную $^2S_{1/2}$ фазу упругого $^3$H$^2$H рассеяния.

Именно такая форма взаимодействия с очень небольшим уточнением глубины $^3$H$^2$H потенциала на 0.0887 МэВ, предварительно фиксированного по фазам рассеяния [13], позволяет получить правильную величину трехтельной энергии связи ядра $^9$Be. Таким образом, глубина этого потенциала изменялась для наилучшего описания энергии связи $^9$Be в трехтельном канале, и это изменение составило всего 0.0877 МэВ.

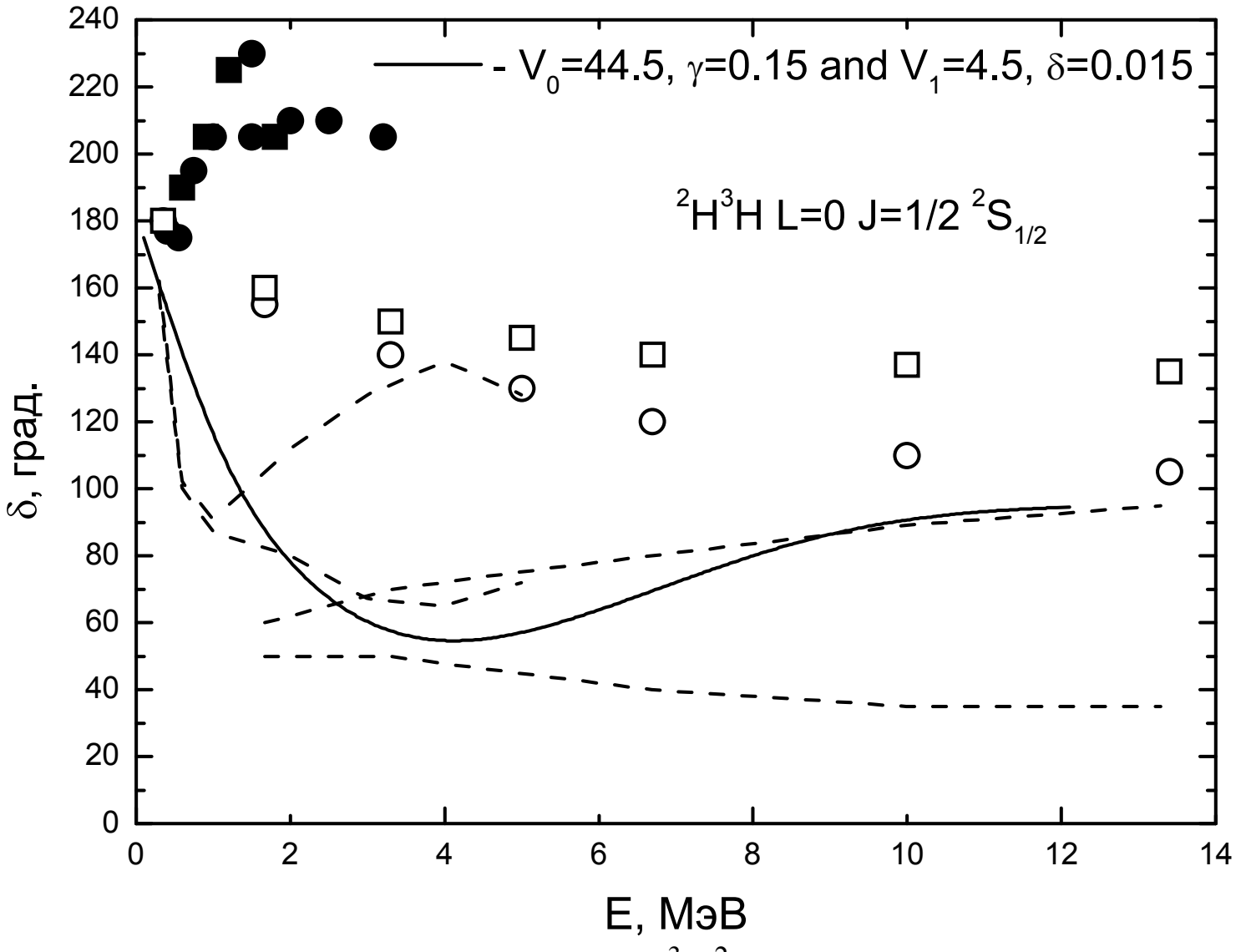

Рис.3.2.1. Чистые фазы упругого $^3$H$^2$H рассеяния для *S* волны. Непрерывная линия – результаты расчета фазы с потенциалом из табл.3.2.1, пунктиром приведена полоса ошибок определения чистых фаз [37], которая получается из данных экспериментальной работы [202]. Точки и квадраты – извлеченные из экспериментальных данных фазы рассеяния [202]. ружки и открытые квадраты – МРГ вычисления фаз рассеяния [203].

Для $^4$He$^3$H системы использован потенциал первого возбужденного $^2P_{1/2}$ состояния без запрещенного связанного уровня, фаза которого показана непрерывной линей на рис.3.2.2. Такой потенциал хорошо описывает фазу рассеяния [204,205] и приводит, по-видимому, к наилучшему описанию характеристик связанного состояния ядра $^9$Be в трехчастичной модели.

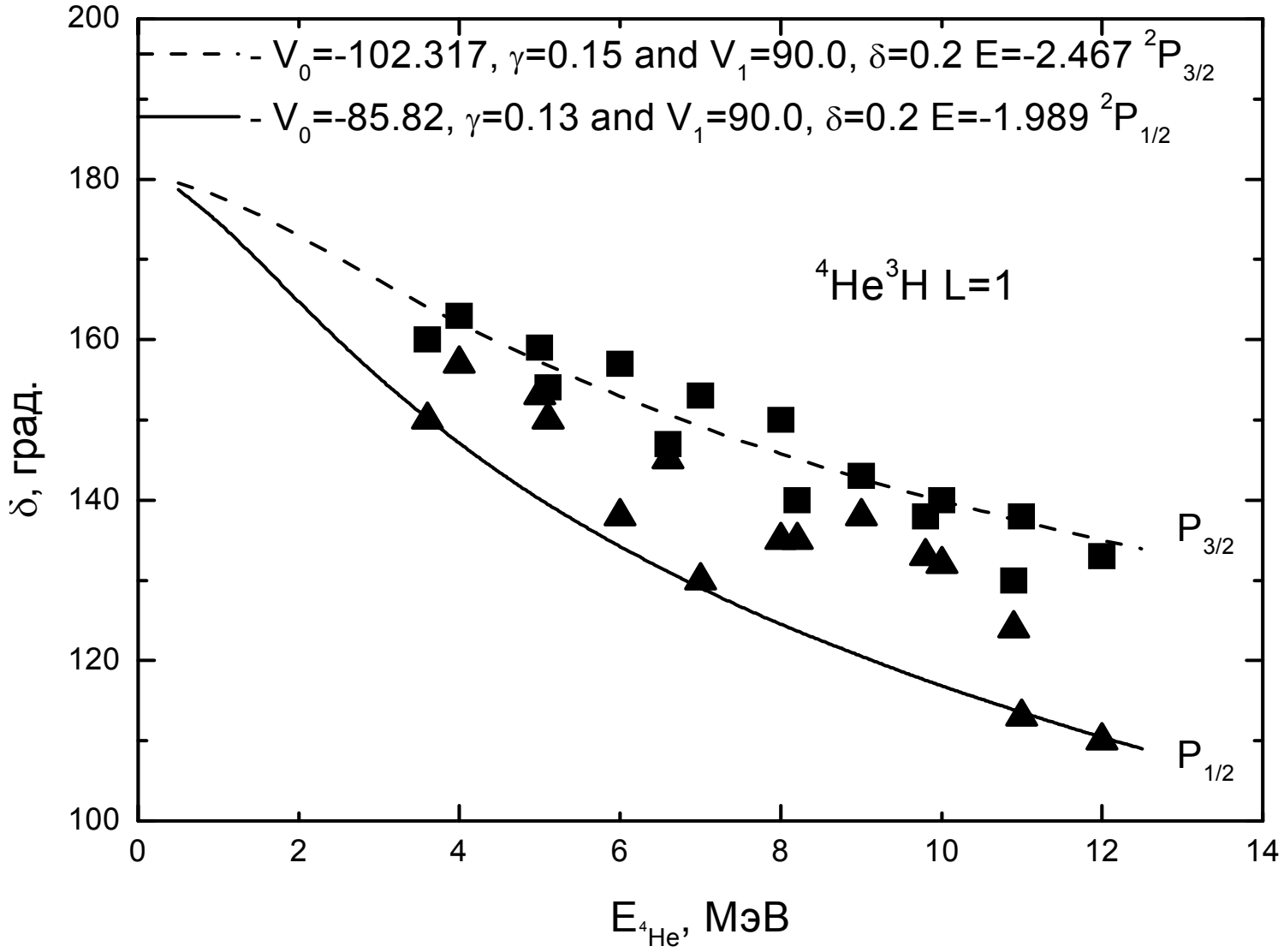

Рис.3.2.2. Фазы упругого $^4$He$^3$H рассеяния для *P* волны. Треугольники и квадраты – извлеченные из экспериментальных данных фазы рассеяния [204,205].

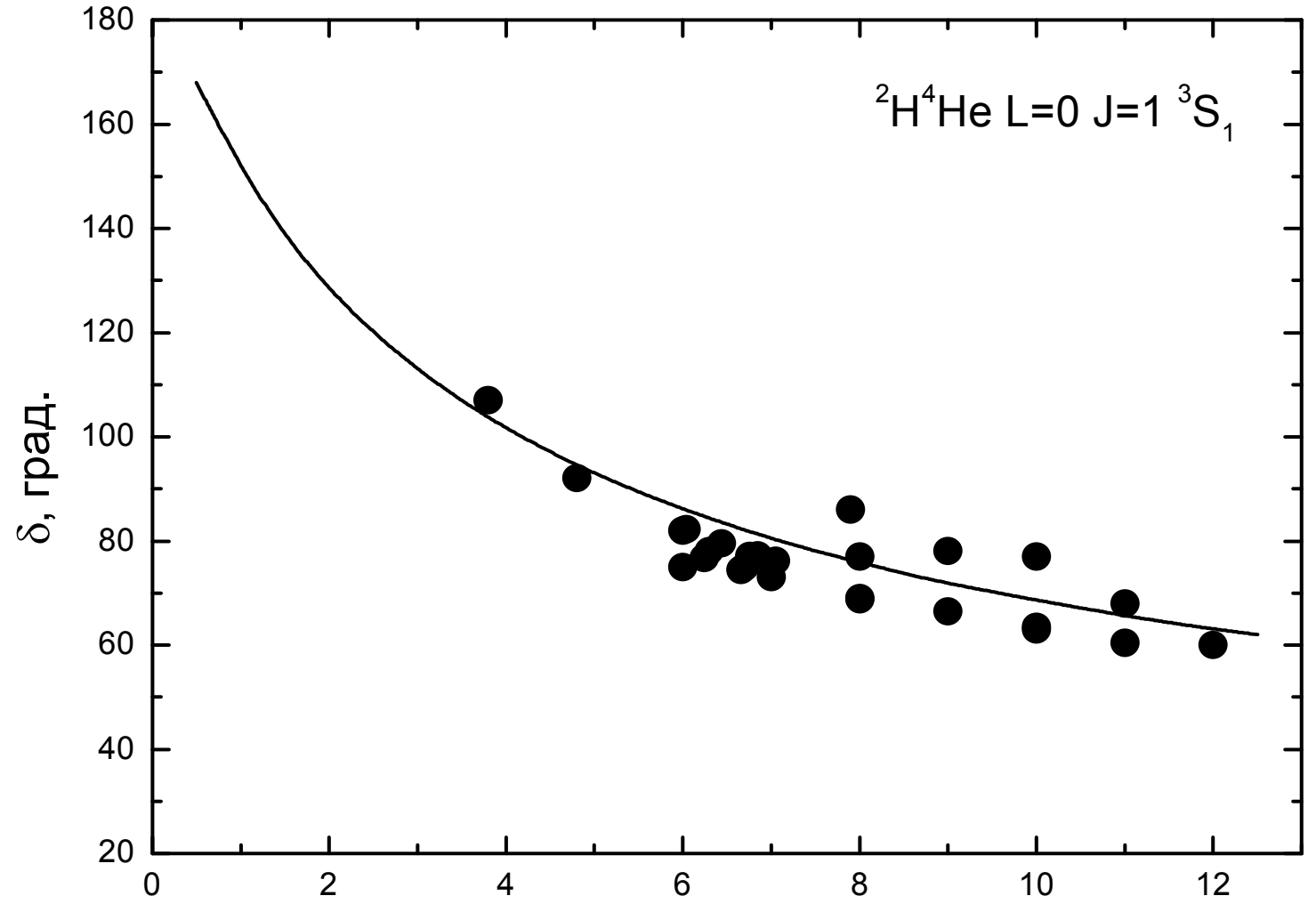

Рис.3.2.3. Фазы упругого $^4$He$^2$H рассеяния при $L$ = 0. Экспериментальные данные из [191-195].

Тем самым, исходя их этих результатов, следует считать, что кластерная $^4$He$^3$H система находится внутри ядра $^9$Be в виртуальном возбужденном $^2P_{1/2}$ состоянии, а не на основном $^2P_{3/2}$ уровне. Возможно, в будущем этот результат можно будет проверить другими, независимыми методами или подходами.

Потенциал в $^4$He$^2$H системе наилучшим образом описывает характеристики связанного состояния ядра $^6$Li, приведенные в табл.3.2.1, и не имеет ЗС, а качество описания $^3S_1$ фазы рассеяния [191-195] показано на Рис.3.2.3 непрерывной линией.

## 3.2.2 Трехтельные результаты и фотосечения Three-body results and photo cross sections

Приведем часть компьютерной программы, которая описана в предыдущем параграфе этого раздела и задает характеристики кластеров и парных потенциалов:

```
! - - - - - - - - - - - - - - - - - - - - - - - - - - - - - - - - - - -
Z1=2.0D-000
Z2=1.0D-000
Z3=1.0D-000
M1=4.0D-000
M2=3.0D-000
M3=2.0D-000
…
RK1=1.670D-000
RK2=1.70D-000
RK3=1.960D-000
RM1=1.670D-000
RM2=1.70D-000
RM3=1.960D-000
…
! 1 - AL; 2 - T; 3 - D;    L  -  AL-T - 1, AL-D -0, D-T - 0
! D-T
 V231=-44.5887D-000      ; ! J= 1  ; L=0
 R231=0.150D-000
```

```
 V232=4.50D-000
 R232=0.0150D-000

! AL-D
 V131=-71.90D-000   ; ! J=1/2 ; L=0
 R131=.150D-000
 V132=70.0D-000
 R132=0.20D-000

! AL-T
 V121=-85.820D-000   ; ! J=1/2 ; L=1
 R121=.130D-000
 V122=90.0D-000
 R122=0.20D-000
 ! - - - - - - - - - - - - - - - - - - - - - - - - - - - - - - - - - - -
```

С описанными выше парными межкластерными потенциалами найдена трехтельная волновая функция, параметры и коэффициенты разложения которой при $N$ = 10 даны в табл.3.2.2 и 3.2.3, энергия связи ядра $^9$Be, нормировка ВФ *Nor* и его зарядовый радиус $R_z$, которые приведены в табл.3.2.4. Как видно их этих таблиц, полученные характеристики ядра $^9$Be хорошо согласуются с имеющимися экспериментальными данными, а нормировка трехтельной ВФ практически равна единице.

Табл.3.2.2. Параметры $\alpha_i$ и $\beta_i$ разложения трехтельной волновой функции (1.5.4) ядра $^9$Be в трехтельной модели.

| *I* | $\alpha_i$ | $\beta_i$ |
|---|---|---|
| 1 | 3.273095667111755E-001 | 8.983828859473847E-002 |
| 2 | 3.172872902170442 | 1.207723068311927E-001 |
| 3 | 1.844963898289113E-001 | 5.507018219497912E-002 |
| 4 | 9.153345511558431E-002 | 1.025123206313353E-001 |
| 5 | 2.213721392830491E-001 | 3.620573882275837E-001 |
| 6 | 2.269904002428514E-001 | 4.017242168731691E-001 |
| 7 | 1.985961231123472E-001 | 1.634388187790707E-001 |
| 8 | 5.073985961415315E-001 | 4.769549654021469E-001 |
| 9 | 5.124708157837092E-001 | 4.828596792375761E-001 |
| 10 | 3.843254062764651E-001 | 2.094550079909185E-001 |

Табл.3.2.3. Коэффициенты $C_i$ разложения трехтельной волновой функции (1.5.4) ядра $^9$Be в трехтельной модели.

| *i* | $C_i$ |
|---|---|
| 1 | 2.564677848067949E-002 |
| 2 | -2.825587985872606E-003 |
| 3 | 3.379570860934880E-003 |
| 4 | 1.204452338550613E-002 |
| 5 | -6.029327933110736E-001 |
| 6 | 4.606318693160417E-001 |
| 7 | 1.677085370227155E-001 |
| 8 | -3.310475484431805 |
| 9 | 3.189208283959322 |
| 10 | 1.415768198190696E-001 |

Табл.3.2.4. Некоторые характеристики ядра $^9$Be в трехтельной кластерной модели.

| Хар-ки | Расчет | Эксперимент |
|---|---|---|
| $E$, МэВ | -19.1632 | -19.1633 [206] |
| $R_z$, Фм | 2.56 | 2.519(12) [206] |
| *Nor* | 9.999999999999867E-001 | |

Как уже говорилось, для достижения лучшего согласия расчетной трехтельной энергии $^9$Be с экспериментом [206], поскольку имеются большие ошибки и неопределенности в $^3$H$^2$H фазовом анализе, приводящие к неоднозначностям по-строения чистых фаз, глубина этого потенциала несколько варьировалась. Первоначальная глубина притягивающей части, с которой рассчитывалась фаза рассеяния, приведенная на рис.3.2.1, равна $V_0$ = -44.5 МэВ, а для поиска трехтельной энергии связи использовалось несколько большее значение $V_0$ = -44.5887 МэВ. На фазы упругого рассеяния такие изменения глубины потенциала влияния практически не оказывают.

На рис.3.2.4а, 3.2.4б показано сравнение радиальных функций $^3$H+$^6$Li относительного движения построенных в $^4$He$^4$Hen и $^4$He$^3$H$^2$H моделях соответственно. Использование потенциалов с отталкиванием в одноканальной $^4$He$^3$H$^2$H модели $^9$Be приводит нас к безузловой 1*P* ВФ относительного движения. В то же время, $^4$He$^4$Hen модель, построенная с парными потенциалами глубокого притяжения, содержащими ЗС, дает узловую 3*P* волновую функцию.

Сплошная кривая на рис.3.2.4а соответствует результирующей волновой функции. Полученные таким образом двухтельных волновые функции $^3$H+$^6$Li канала использованы далее в расчетах сечений фоторазвала ядра $^9$Be(γ,$^3$H)$^6$Li [207], которые представлены на рис.3.2.5а,б вместе с экспериментальными данными работы [208].

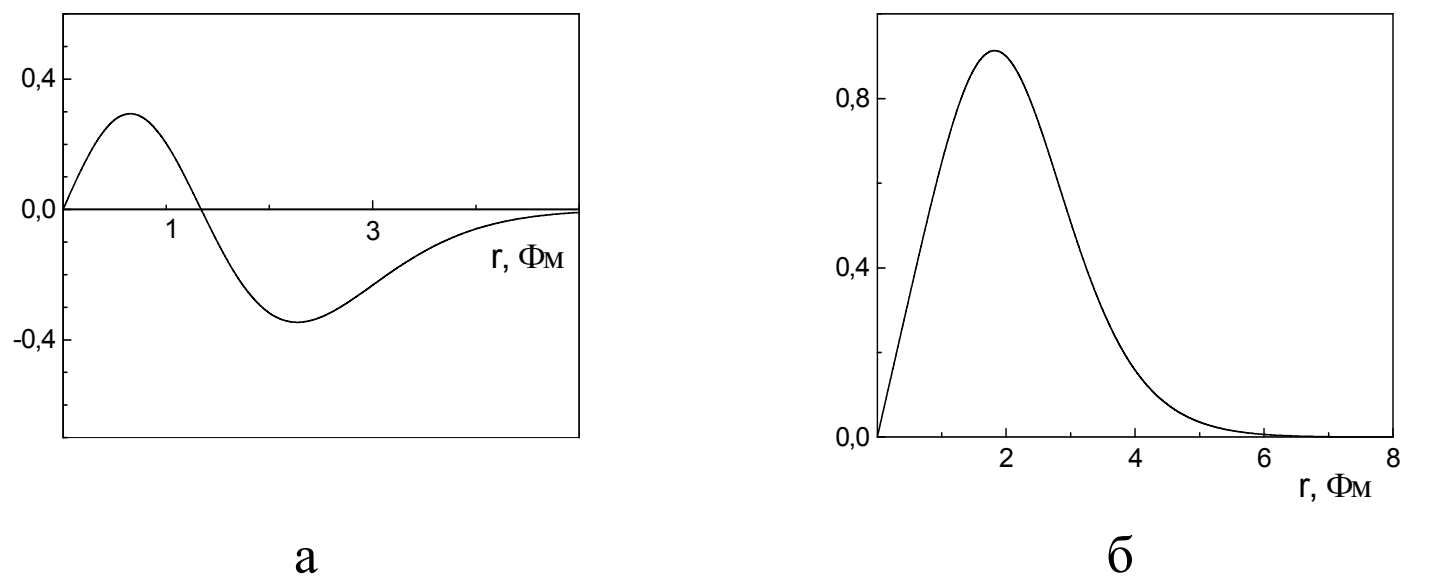

а б

Рис.3.2.4. Радиальные функции относительного движения кластеров в канале $^6$Li+$^3$H ядра $^9$Be.
*а* –$^4$He$^4$Hen модель и б – $^4$He$^3$H$^2$H модель.

Следует отметить, что рассчитанная в этих двух моделях форма сечения фоторазвала практически не отличается, хотя и имеет некоторые количественные расхождения. Это вполне объясняется схожестью «хвоста» двух различных волновых функций на больших расстояниях, что, в свою очередь, приводит к подобности результатов при рассматриваемых, сравнительно малых энергиях процесса фоторазвала $^9$Be в двухчастичный $^3$H$^6$Li канал.

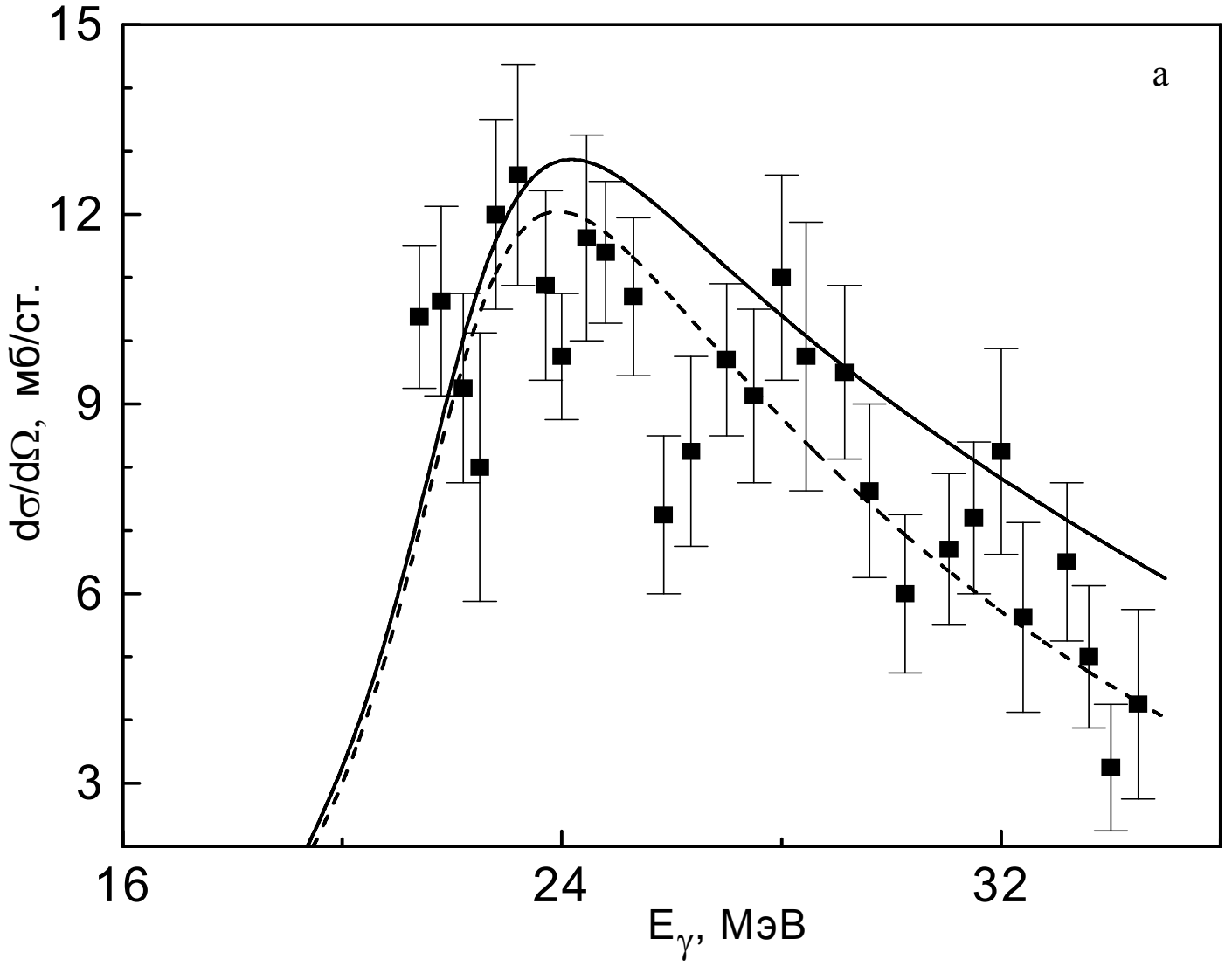

Рис.3.2.5а. Дифференциальные сечения процесса $^9$Be($\gamma$,$^3$H)$^6$Li. Квадраты – эксперимент [208]. Теоретический расчет для n$^4$He$^4$He модели. Пунктир – дипольный $E1$ переход, сплошная кривая – суммарное сечение.

Таким образом, в рамках рассматриваемых вариационных методов, получены новые результаты для трехтельной $^4$He$^3$H$^2$H кластерной модели ядра $^9$Be [207]. В таких расчетах использован неортогональный вариационный базис, независимое варьирование всех параметров разложения ВФ по гауссойдам, межкластерные потенциалы, чистые в некоторых случаях по схемам Юнга и согласованные с фазами упругого рассеяния в двухчастичных системах.

Как видно, именно используемые варианты потенциалов взаимодействия между кластерами приводят к правильной энергии связи ядра $^9$Be, описанию некоторых других его характеристик в трехтельном кластерном канале и разумному объяснению экспериментальных дифференциальных сечений рассмотренной реакции фоторазвала ядра $^9$Be в двухчастичный $^3$H$^6$Li канал.

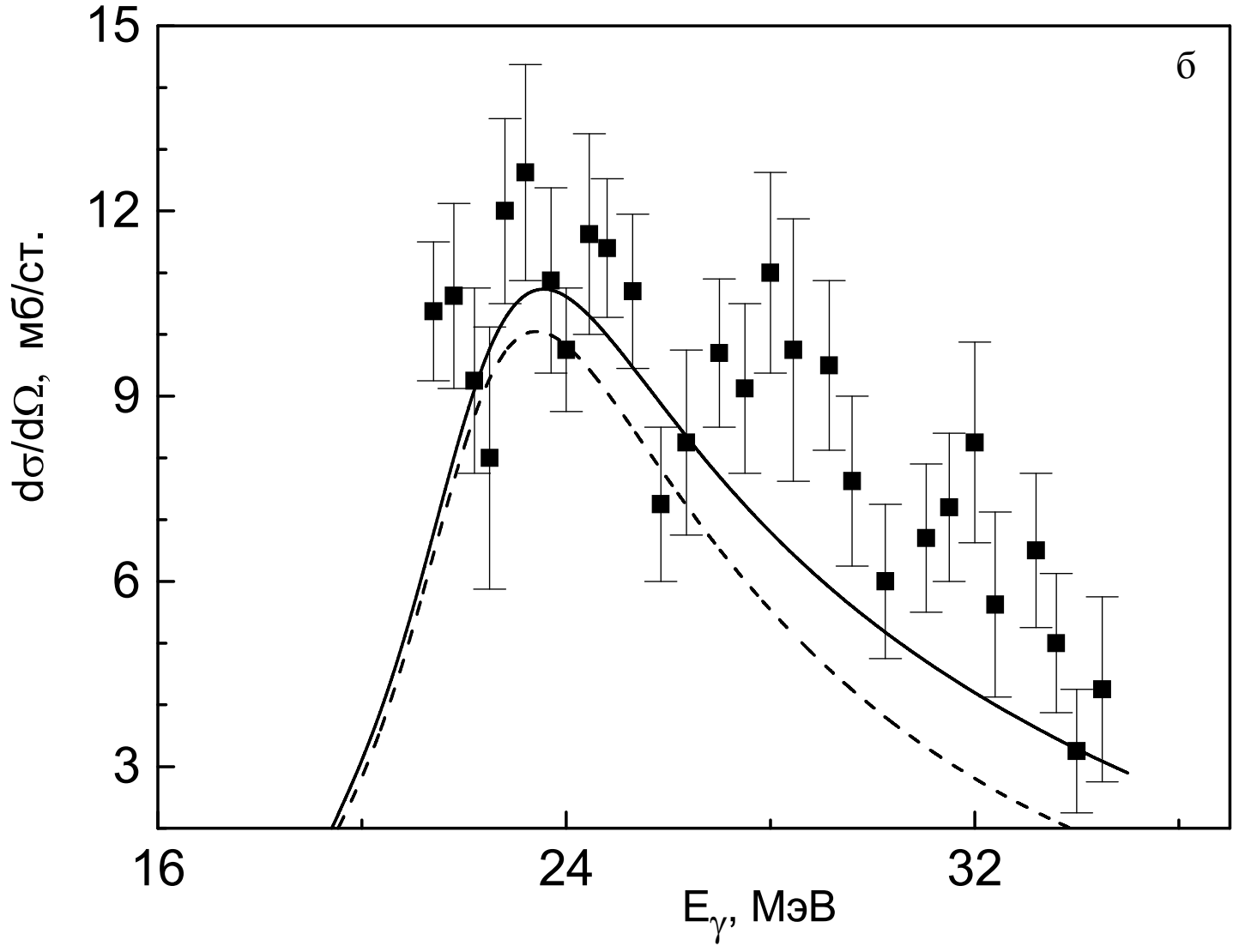

Рис.3.2.5б. Дифференциальные сечения процесса $^9$Be(γ,$^3$H)$^6$Li. Квадраты – эксперимент [208]. Теоретический расчет для $^4$He$^3$H$^2$H модели. Пунктир – дипольный *E*1 переход, сплошная кривая – суммарное сечение.

Дальнейшее развитие теоретических исследований в этом направлении требует решения трехтельной задачи со связью каналов, т.е. учета различных возможных парциальных волн в каждом двухтельном плече трехчастичной системы [207]. Однако, рассмотренный здесь одноканальный подход вполне позволяет правильно получить сечение процесса фоторазвала в области низких энергий. Поэтому вполне естественно в дальнейшем выполнить расчеты и при энергиях γ-кванта стремящихся к нулю. Тем самым рассматривать область астрофизических энергий, которые представляют интерес для задач ядерной астрофизики и термоядерных процессов во Вселенной.

## 3.3 Трехкластерная структура $^{11}$B Three-body structure of $^{11}$B

Рассмотрим теперь возможность использования трехтельной модели для следующего нечетного легкого атомного ядра $^{11}$B, которое можно представить трехтельной $^4$He$^4$He$^3$H структурой. Момент основного состояния $^{11}$B равен $3/2^-$ и может быть образован при $\lambda = 0$ и $l = 1$, поскольку момент $^4$He равен нулю, а $^3$H имеет полуцелый спин 1/2. В используемой модели основание треугольника по-прежнему состоит из частиц 2 и 3, которыми в данном случае являются две $\alpha$-частицы с нулевым относительным моментом $\lambda$. Орбитальный момент $l$, равный 1, может быть получен из комбинации $l_{12} = 1$ и $l_{13} = 1$. Напомним, что здесь $\boldsymbol{l} = \boldsymbol{l}_{12} + \boldsymbol{l}_{13}$, а моменты $l_{12}$ и $l_{13}$ это орбитальные моменты между частицами 12 и 13, причем частицей 1, находящейся в вершине треугольника, является ядро $^3$H.

В качестве межкластерных потенциалов будут использованы $^4$He$^3$H взаимодействия в основном $^2P_{3/2}$ состоянии ядра $^7$Li. Здесь предполагается, что именно эта орбитальная конфигурация доминирует в рассматриваемой одноканальной $^4$He$^4$He$^3$H модели. Конечно, в многоканальном варианте такой трехтельной модели возможен вклад, например, конфигураций $l_{12} = 1$ и $l_{13} = 0$ или $l_{12} = 0$ и $l_{13} = 1$, которые также приводят к $l = 1$.

Напомним (п.п.1.6), что в проведенных расчетах, как и ранее, при каждом значении вариационных параметров $\alpha_i$ и $\beta_i$, которые варьируются независимо друг от друга, находим некоторую энергию системы $E$, которая дает ноль детерминанта, а затем, изменяя эти параметры, проводим поиск минимума трехтельной энергии $E$, которая является собственной энергией вариационной задачи. Затем увеличиваем размерность базиса $N$ и повторяем все вычисления, до тех пор, пока величина собственного значения, т.е. энергия связи $E_N$, при очередной размерности базиса $N$ не станет отличаться от предыдущего значения $E_{N-1}$ на величину $\varepsilon$, которая обычно задается на уровне 1.0÷2.0 кэВ. Эта минимальная энергия и будет реальной энергией связи трехчастичной системы, т.е. энергией связи ядра в такой модели. Причем размерность гауссова базиса обычно не превышает 10÷12 [16].

### 3.3.1 Потенциалы и фазы
### Potentials and phase shifts

В настоящих расчетах для $^4$He$^3$H и $^4$He$^4$He систем использованы бинарные межкластерные потенциалы с отталкивающим кором (1.2.1), параметры которых приведены в табл.3.3.1. Фазы рассеяния, соответствующие таким потенциалам, показаны на рис.3.2.2 и 3.3.1 штриховыми линиями. Экспериментальные данные для упругого $^4$He$^4$He рассеяния взяты из работ [128,130,209,210].

Табл.3.3.1. Параметры парных межкластерных потенциалов.

| Система | $^{2S+1}L_J$ | $V_0$, МэВ | γ, Фм$^{-2}$ | $V_1$, МэВ | δ, Фм$^{-2}$ |
|---|---|---|---|---|---|
| $^4$He$^4$He | $^1S_0$ | -204.0 | 0.2025 | 500.0 | 0.36 |
| $^4$He$^3$H | $^2P_{3/2}$ | -102.317 | 0.15 | 90.0 | 0.2 |

Энергия связи $^7$Li в $^4$He$^3$H канале с потенциалом из табл.3.3.1 и целыми значениями масс частиц -2.467 МэВ точно совпадает с экспериментальной величиной [200]. Зарядовый радиус равен 2.40 Фм, а асимптотическая константа на интервале 7÷16 Фм имеет значение $C_w$ = 3.57(1) [13,89]. В качестве зарядового и массового радиуса тритона принималась величина 1.70 Фм, а для $^4$He 1.67 Фм [121-123]. Напомним, что одно из наиболее современных значений зарядового радиуса ядра $^7$Li равно 2.4017(281) Фм [198].

Для асимптотической двухчастичной константы основного $^2P_{3/2}$ состояния $^4$He$^3$H системы в ядре $^7$Li, например, в работе [211] с уиттекеровской асимптотикой (1.2.3), учитывающей кулоновские эффекты [17], получено, при пересчете к безразмерной величине (1.2.3) с $k_0$ = 0.453 Фм$^{-1}$ [14], следующее значение: 3.87(16). В работе [212] для ОС, также после пересчета к безразмерной величине, приведено 3.73(26), что вполне согласуется с полученными здесь результатами. Такой пересчет значений требуется, поскольку в указанных работах использовалась несколько другая форма определения асимптотических констант, а именно:

$$\chi_L(r) = C_W W_{-\eta L+1/2}(2k_0 r)$$

На рис.3.3.1 непрерывной линией показана $^1S_0$ фаза упругого $^4$He$^4$He рассеяния с потенциалом Али-Бодмера (параметры даны на рис.3.3.1), который приводит к несколько заниженной трехтельной энергии связи $^4$He$^4$He$^3$H системы в ядре $^{11}$B. Требуется немного, примерно на 7%, увеличить глубину его притягивающей части (см. табл.3.3.1) для того чтобы получить практически правильную величину энергии связи, сходимость которой от числа членов разложения волновой функции показана в табл.3.3.2.

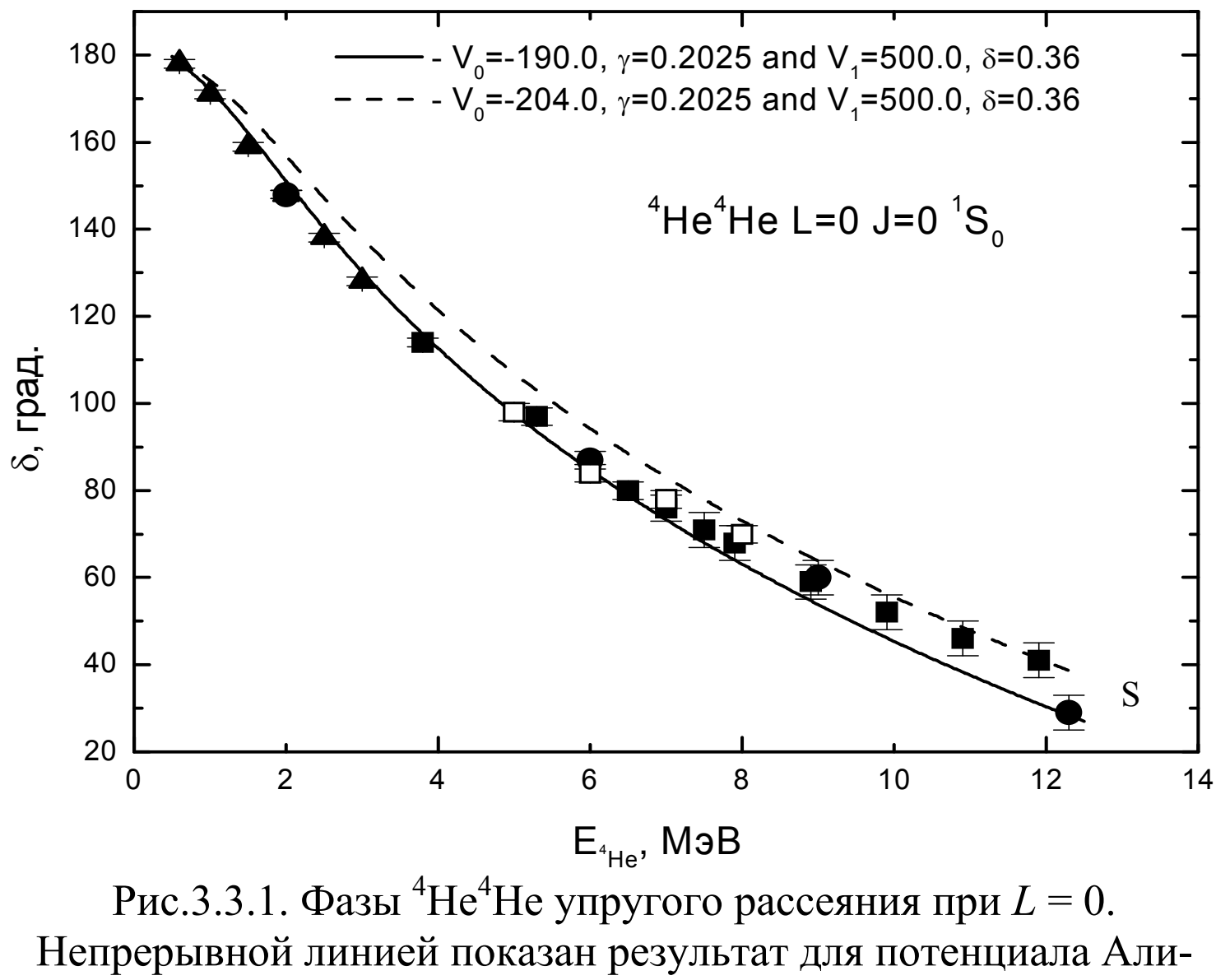

Рис.3.3.1. Фазы $^4$He$^4$He упругого рассеяния при $L = 0$. Непрерывной линией показан результат для потенциала Али-Бодмера [209], штриховой – для его модифицированного варианта из табл.3.3.1. Экспериментальные данные взяты из работ: [209] – •, [128] – ▲, [210] – □, [130] – ■.

Табл.3.3.2. Сходимость трехтельной энергии связи $E$ ядра $^{11}$B в зависимости от числа гауссойд $N$ в разложении ВФ. Экспериментальная величина энергии связи $^{11}$B в этом канале равна -11.131 МэВ [213].

| $N$ | 4 | 6 | 8 | 10 | 12 |
|---|---|---|---|---|---|
| $E$, МэВ | -10.832 | -10.985 | -11.070 | -11.072 | -11.079 |

Фаза такого потенциала показана на рис.3.3.1 штриховой кривой и идет несколько выше извлеченных из эксперимента фаз упругого $^4$He$^4$He рассеяния. Необходимость изменения глубины $^4$He$^4$He потенциала, возможно, связана с одноканальностью используемой здесь модели, в которой учитывается только одна допустимая орбитальная конфигурация с межкластерным $^2P_{3/2}$ потенциалом связанного $^4$He$^3$H состояния при $l$ = 1 для $l_{12}$ = 1 и $l_{13}$ = 1, параметры которого приведены в табл.3.3.1. Учет других $^4$He$^3$H конфигураций также с $l$ = 1, но при $l_{12}$ = 1 и $l_{13}$ = 0 или $l_{12}$ = 0 и $l_{13}$ = 1, по-видимому, мог бы увеличить трехтельную энергию связи, и изменять глубину $^4$He$^4$He потенциала не потребовалось.

## 3.3.2 Трехтельные результаты
## Three-body results

### 3.3.2.1 Первый вариант межкластерных потенциалов
### First variant of intercluster potentials

Приведем часть программы, полностью распечатанной в пп.3.1.2, в которой записаны характеристики кластеров и потенциалы их взаимодействия:

```
!----------------------------------
Z1=1.0D-000
Z2=2.0D-000
Z3=2.0D-000
M1=3.0D-000
M2=4.0D-000
```

```
M3=4.0D-000
…
RK1=1.7D-000
RK2=1.67D-000
RK3=1.670D-000
RM1=1.7D-000
RM2=1.67D-000
RM3=1.670D-000
…
! 1 - T; 2 - AL; 3 - AL;   L  -  AL1-AL2 - 0, AL1-T - 1, AL2-T - 1
! AL-AL
V231=-204.0D-000      ; ! J= 0  ; LAM=0
R231=0.45D-000**2
V232=500.0D-000
R232=0.6D-000**2

! AL-T
V131=-102.317D-000 ;    ! J=3/2; L=1
R131=0.15D-000
V132=90.0D-000
R132=0.2D-000

! AL-T
V121=V131    ;  ! J=3/2; L=1
R121=R131
V122=V132
R122=R132
! - - - - - - - - - - - - - - - - - - - - - - - - - - - - - - - - - - - -
```

Параметры и коэффициенты разложения волновой функции $^4He^4He^3H$ системы для $N$ = 10 в ядре $^{11}B$ приведены ниже в самой распечатке, а результаты для энергии показаны в табл.3.3.2

**E = -11.072136455745790** (N = 10)

| N | ALFA | BET |
|---|---|---|
| 1 | 7.731577265613154E-002 | 1.240840193536184E-001 |
| 2 | 1.641040105068961E-001 | 2.285876535575549E-001 |
| 3 | 4.870741010024022E-001 | 1.536884862684963E-001 |
| 4 | 2.513045621382256E-001 | 3.041700707766410E-001 |
| 5 | 2.033807929862554E-001 | 3.450251748941237E-001 |
| 6 | 6.368494296100554E-001 | 1.898054241645994E-001 |
| 7 | 3.430614492885307E-001 | 1.441809735418989E-001 |
| 8 | 3.520135247366762E-001 | 2.289195177072528E-001 |
| 9 | 2.131971494536152E-001 | 3.314403264542553E-001 |
| 10 | 4.968108631879107E-001 | 2.048954900045679E-001 |

SUM(H*SV-E*L*SV) FROM SV = -1.776356839400251E-013
NORM = 9.999999999998579E-001
NEV-DET = 1.364242052659392E-012

| N | SV |
|---|---|
| 1 | 7.141139314701980E-002 |
| 2 | 6.873184414416353E-001 |
| 3 | 7.416580078033560E-001 |
| 4 | 4.203984406616693 |
| 5 | 3.779698938600395 |
| 6 | -1.470993620114665 |
| 7 | -6.275457436460156E-001 |
| 8 | -3.525509423084515 |
| 9 | -7.047436459053254 |
| 10 | 3.191334856523210 |

N = 9.999999999999705E-001
RM = 2.633871831198857
RZ = 2.630340587104462

COUL. ENERGY VK = 4.106613183539739
12 = 1.219823574100133
13 = 1.219823574100133
23 = 1.666966035339474
CENTROB. ENERGY = 4.025371560588050
KINETICH. ENERGY = 8.030009418604076
POTENS. ENERGY = -27.234130618489780
**POLNAY ENERGY ST = -11.072136455757920**
**POLNAY ENERGY SH = -11.072136455745180**

Подобные результаты для энергии и ВФ при размерности $N = 12$

**E = -11.079033093916390** (N = 12)

| N | ALFA | BET |
|---|---|---|
| 1 | 6.409144140489247E-002 | 1.019587659674352E-001 |
| 2 | 1.258066165364320E-001 | 1.811335435670906E-001 |
| 3 | 5.502163968570325E-001 | 1.563736148998969E-001 |
| 4 | 1.093264054668316E-001 | 2.915855267598159E-001 |
| 5 | 1.333280173011495E-001 | 4.867098271947788E-001 |
| 6 | 8.577841282629973E-001 | 1.559124765668210E-001 |
| 7 | 2.427745646298275E-001 | 9.280370272042079E-002 |
| 8 | 4.374090344872209E-001 | 1.607892257131171E-001 |
| 9 | 2.506356884255520E-001 | 3.365880418027703E-001 |
| 10 | 5.992286208956132E-001 | 1.732142250126302E-001 |
| 11 | 2.513360250891818E-001 | 3.505196617789699E-001 |
| 12 | 9.123756893621239E-001 | 1.549639056362407E-001 |

SUM(H*SV-E*L*SV) FROM SV = 4.089173444299377E-012
NORM = 1.000000000000028
NEV-DET = 1.818989403545857E-012

| N | SV |
|---|---|
| 1 | 2.405128330550160E-002 |
| 2 | 2.838460999953977E-001 |
| 3 | 3.355552135939775 |
| 4 | 6.347242717086900E-002 |
| 5 | -7.943048747668001E-002 |
| 6 | -5.622161796232756 |
| 7 | -1.975249769693670E-002 |
| 8 | -2.913336286105724 |
| 9 | -1.743989588892319 |
| 10 | 9.471131965767997E-001 |
| 11 | 1.677487262418139 |
| 12 | 4.030809374250345 |

N = 1.000000000000014
RM = 2.635529943243490

RZ = 2.631840422205140

COUL. ENERGY VK = 4.103278121515663
12 = 1.218338113251590
13 = 1.218338113251590
23 = 1.666601895012484
CENTROB. ENERGY = 4.011294763849525
KINETICH. ENERGY = 8.028619034600876
POTENS. ENERGY = -27.222225013883560
**POLNAY ENERGY ST = -11.079033093917500**
**POLNAY ENERGY SH = -11.079033093917130**

Из этих распечаток видно, что ошибки поиска детерминанта NEV-DET имеют порядок величины $10^{-12}$, полная ошибка поиска энергии и собственных векторов, определяемая выражением $(H\text{-}EL)C$, оказывается меньше $4.1{\cdot}10^{-12}$, а нормировка полученной волновой функции отличается от единицы только в 12÷14 знаке после запятой.

### 3.3.2.2 Второй вариант межкластерных потенциалов Second variant of intercluster potentials

Однако нужно отметить, что расчетные среднеквадратичные радиусы получаются несколько больше экспериментальной величина для зарядового радиуса $^{11}$B, равного 2.406(29) Фм [198], несмотря на то, что радиус $^{7}$Li в $^{4}$He$^{3}$H канале имеет правильное значение. Обратим теперь внимание на тот факт, что радиусы ядер $^{7}$Li 2.4017(281) Фм [198] и $^{11}$B 2.406(29) Фм [198] практически совпадают и, по-видимому, если ядро $^{7}$Li в $^{4}$He$^{3}$H конфигурации и находится внутри $^{11}$B в трехтельной $^{4}$He$^{4}$He$^{3}$H модели, то оно должно быть несколько сжато.

Поэтому возьмем $^{4}$He$^{3}$H потенциал с параметрами притягивающей части $V_1$ = -121.405 МэВ и $\alpha$ = 0.18 Фм$^{-2}$, без изменения его отталкивания (см. табл.3.3.1 и 3.3.3 ). Он приводит к энергии связи $^{7}$Li с целыми значениями масс частиц -2.467 МэВ, полностью совпадающей с экспериментальной величи-ной [200]. Зарядовый радиус равен 2.24 Фм, а асимптотическая константа на интервале 5÷15 Фм имеет значение $C_w$ = 2.58(1) [13,89]. Как видно, в таком потенциале ядро $^7$Li несколько деформировано, а именно, сжато по сравнению с его свободным состоянием.

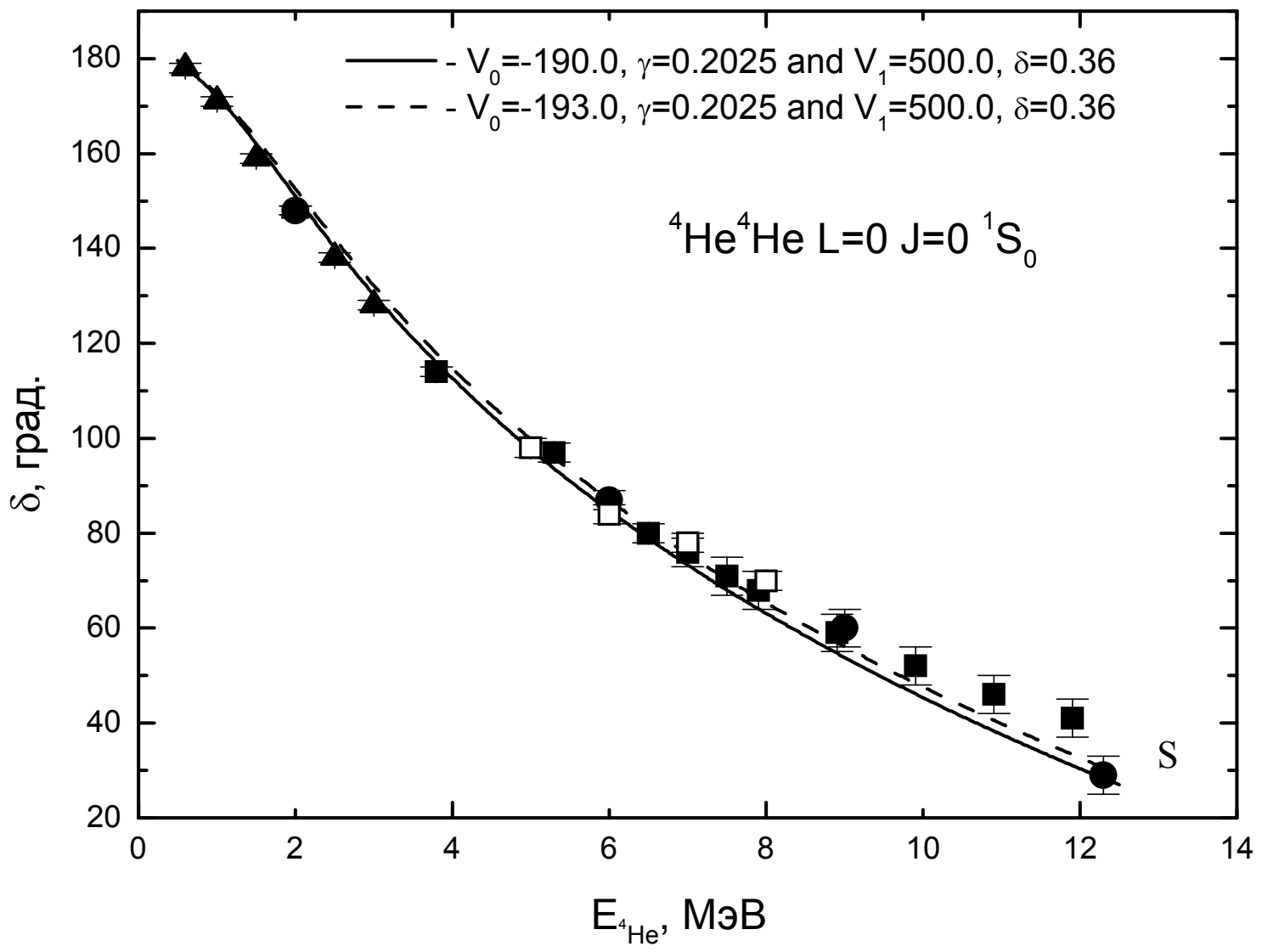

Рис.3.3.2. Фазы $^4$He$^4$He упругого рассеяния при $L$ = 0. Непрерывной линией показан результат для потенциала Али-Бодмера [209], штриховой – для его модифицированного варианта из табл.3.3.3. Экспериментальные данные взяты из работ: [209] – •, [128] – ▲, [210] – □, [130] – ■.

Далее, для получения практически правильной величины трехтельной энергии, как будет показано в распечатке ниже, требуется намного меньше деформировать $^4$He$^4$He потенциал, а именно, принять его глубину -193.0 МэВ, не меняя других его параметров, как показано в табл.3.3.3. Фаза его упругого рассеяния показана на рис.3.3.2 пунктирной кривой и практически не отличается от результатов для стандартного потенциала Али-Бодмера, которые, по-прежнему, показаны непрерывной линией.

Табл.3.3.3. Новый вариант параметров парных межкластерных потенциалов.

| Система | $^{2S+1}L_J$ | $V_0$, МэВ | γ, Фм$^{-2}$ | $V_1$, МэВ | δ, Фм$^{-2}$ |
|---|---|---|---|---|---|
| $^4$He$^4$He | $^1S_0$ | -193.0 | 0.2025 | 500.0 | 0.36 |
| $^4$He$^3$H | $^2P_{3/2}$ | -121.405 | 0.15 | 90.0 | 0.2 |

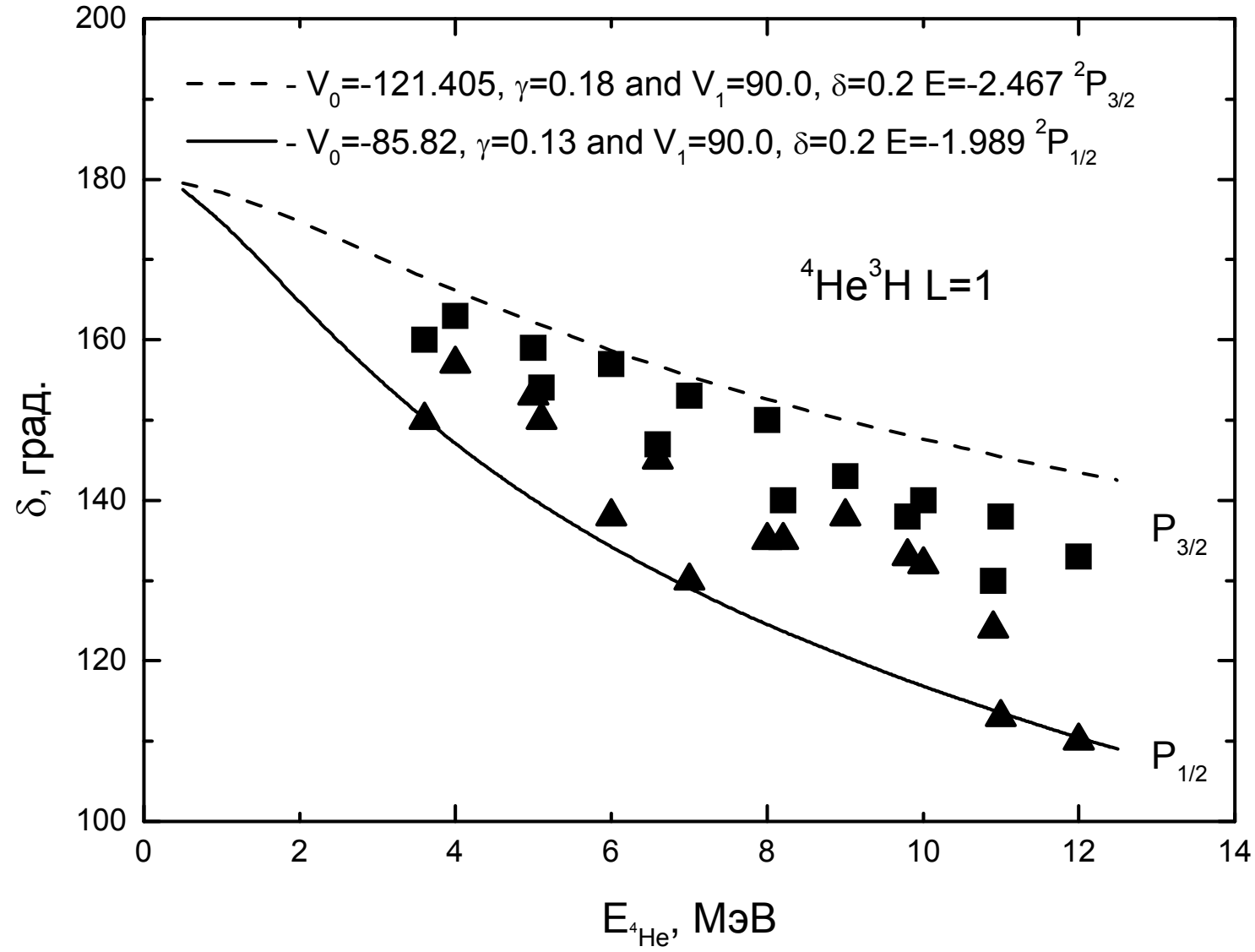

Рис.3.3.3. Фазы упругого $^4$He$^3$H рассеяния для *P* волны. Треугольники и квадраты – извлеченные из экспериментальных данных фазы упругого рассеяния [204,205]. Кривые – расчеты с разными потенциалами, параметры которых показаны на рисунке и в табл.3.3.3.

На рис.3.3.3 приведена фаза используемого далее потенциала $^4$He$^3$H взаимодействия, которая, по сравнению с результатами, представленными на рис.3.2.2, проходит немного выше, а, по сути, по верхней границе фаз упругого $^4$He$^3$H рассеяния, извлеченных из экспериментальных данных [204,205].

Приведем часть программы, в которой записаны эти по-тенциалы взаимодействия кластеров:

```
! - - - - - - - - - - - - - - - - - - - - - - - - - - - - - - - - - -
! 1 - T; 2 - AL; 3 - AL;   L  -  AL1-AL2 - 0, AL1-T - 1, AL2-T - 1
! AL-AL
V231=-193.0D-000      ; ! J= 0  ; LAM=0
R231=0.45D-000**2
V232=500.0D-000
R232=0.6D-000**2

! AL-T
V131=-121.405D-000 ;    ! J=3/2; L=1
R131=0.18D-000
V132=90.0D-000
R132=0.2D-000

! AL-T
V121=V131    ;  ! J=3/2; L=1
R121=R131
V122=V132
R122=R132
! - - - - - - - - - - - - - - - - - - - - - - - - - - - - - - - - - -
```

Теперь приведем результаты расчета трехтельной энергии связи ядра $^{11}B$ в $^4He^4He^3H$ канале с описанными выше потенциалами для 10 членов разложения ВФ по гауссойдам *N*:

**E = -11.031463381093430**

| N | ALFA | BET |
|---|---|---|
| 1 | 6.860823790482709E-002 | 1.372018518013886E-001 |
| 2 | 1.979200374955967E-001 | 1.975207162666516E-001 |
| 3 | 4.887720500875407E-001 | 1.729550202313665E-001 |
| 4 | 1.536806478562412E-001 | 3.328419037585270E-001 |
| 5 | 9.581401187947329E-002 | 4.647401165174852E-001 |
| 6 | 7.235261491934235E-001 | 2.130457228008369E-001 |
| 7 | 2.517628737184591E-001 | 1.646905851358956E-001 |
| 8 | 3.718525859440748E-001 | 2.263181658977888E-001 |

9 2.301200947513070E-001 3.053675821366754E-001
10 5.342732966940103E-001 2.386280212884875E-001

SUM(H*SV-E*L*SV) FROM SV = 1.456612608308205E-012
NORM = 9.999999999999014E-001
NEV-DET = 0.000000000000000E+000

| N | SV |
|---|---|
| 1 | 5.144041160323528E-002 |
| 2 | 6.941910561622631E-001 |
| 3 | 5.424054031163350E-001 |
| 4 | 1.121880192698402 |
| 5 | -1.131115274640460E-001 |
| 6 | -8.746059242237850E-001 |
| 7 | -5.773075026219299E-001 |
| 8 | -1.357323308702923 |
| 9 | -1.431015963101069 |
| 10 | 1.939168998797320 |

N = 9.999999999999950E-001
RM = 2.545788338764759
RZ = 2.554149440825896

COUL. ENERGY VK = 4.549379683245643
12 = 1.434603714056763
13 = 1.434603714056763
23 = 1.680172255132118
CENTROB. ENERGY = 5.629493999635255
KINETICH. ENERGY = 9.387040848724377
POTENS. ENERGY = -30.597377914995690
**POLNAY ENERGY ST = -11.031463383390410**
**POLNAY ENERGY SH = -11.031463383390960**

Далее следуют результаты расчета трехтельной энергии связи для $N = 12$:

**E = -11.032671516388140**

| N | ALFA | BET |
|---|---|---|
| 1 | 6.866999434690753E-002 | 1.308910045752647E-001 |
| 2 | 1.149793056566817E-001 | 1.884387791693870E-001 |
| 3 | 5.539896465720190E-001 | 1.648434683949910E-001 |
| 4 | 1.528193850522205E-001 | 3.151598143556194E-001 |
| 5 | 1.052759959020325E-001 | 3.918444921609474E-001 |
| 6 | 8.827637468448799E-001 | 1.561502493903658E-001 |
| 7 | 2.403581793183441E-001 | 1.111462488829206E-001 |
| 8 | 5.436244156008828E-001 | 1.700073788396006E-001 |
| 9 | 2.569650583366106E-001 | 3.146550134272712E-001 |
| 10 | 5.752686749704641E-001 | 1.756637188474036E-001 |
| 11 | 2.882071283272155E-001 | 3.321865686346414E-001 |
| 12 | 9.046635125615116E-001 | 1.547378005133593E-001 |

SUM(H*SV-E*L*SV) FROM SV = -4.312994406063808E-012
NORM = 1.000000000006423
NEV-DET = 9.094947017729282E-013

| N | SV |
|---|---|
| 1 | 4.219506029644921E-002 |
| 2 | 6.537938394165033E-002 |
| 3 | 10.984160595778020 |
| 4 | 1.332911137207201 |
| 5 | -1.904812220035353E-001 |
| 6 | -16.076855570816770 |
| 7 | -3.480370810845272E-002 |
| 8 | -19.639156270511190 |
| 9 | -3.254715851761231 |
| 10 | 11.029434819668920 |
| 11 | 1.864459157683097 |
| 12 | 13.879575716434890 |

N = 9.999999999999663E-001
RM = 2.545605920214188
RZ = 2.553862814168674
COUL. ENERGY VK = 4.545338040958250
12 = 1.432617674939411
13 = 1.432617674939411
23 = 1.680102691079429

CENTROB. ENERGY = 5.611694081091503
KINETICH. ENERGY = 9.392846392977125
POTENS. ENERGY = -30.582550036090190
**POLNAY ENERGY ST = -11.032671521063310**
**POLNAY ENERGY SH = -11.032671521338680**

Отсюда видно, что даже в этом случае радиусы $^{11}$B, примерно равные 2.55 Фм, оказываются заметно больше их экспериментальной величины 2.406(29) Фм [198], но для получения правильной энергии связи $^{11}$B практически не требуется деформировать $^4$He$^4$He потенциал, а парное $^4$He$^3$H взаимодействие все еще может быть согласовано с фазами рассеяния, как показано на рис.3.3.3.

Конечно, можно еще изменить $^4$He$^3$H потенциал, сделав его более узким, что приведет к уменьшению радиуса $^7$Li и в результате, возможно, удастся получить правильный радиус $^{11}$B. Но, как видно из рис.3.3.3, при таком изменении потенциала уже не удастся правильно описать извлеченную из эксперимента фазу упругого $^4$He$^3$H рассеяния, которая имеет тенденцию к увеличению значений при увеличении глубины потенциала, которые показаны в табл.3.3.1 и табл.3.3.3. Эта тенденция хорошо видна и при сравнении рис.3.2.2 и рис.3.3.3.

В этом смысле рассмотренный здесь второй вариант потенциалов является наиболее оптимальным с точки зрения минимального изменения параметров хорошо определяемого на основе экспериментальных фаз рассеяния $^4$He$^4$He потенциала. Причем, такой вариант потенциалов все еще позволяет сохранить разумную форму расчетных $^4$He$^3$H фаз рассеяния, показанных на рис.3.3.3.

С другой стороны, всегда можно считать, что из-за большой трехтельной энергии связи $^{11}$B, которая больше энергии связи тритонного кластера в 3$N$ канале, вполне возможна его деформация, т.е. небольшое сжатие с уменьшением радиуса. Действительно, в свободном состоянии радиус трития $^3$H оказывается больше радиуса ядра $^4$He и если тритий сильно связать в ядре, его радиус может несколько уменьшиться, уменьшив тем самым, радиус $^{11}$B.

Как видно из приведенных выше результатов трехтельная модель и в этом случае позволяет получить разумное описание некоторых основных характеристик нечетного ядра $^{11}$B. Хотя имеющиеся ошибки фазового анализа приводят к неопределенностям в параметрах межкластерных потенциалов, но практически в их пределах удается построить потенциалы, приводящие к вполне приемлемым результатам при описании основных характеристик ядра $^{11}$B. Кроме того, одноканальность модели также приводит к некоторым неопределенностям в потенциалах взаимодействия, уменьшить которые, как уже говорилось, позволит, по-видимому, только рассмотрение двух- и трехканальных конфигураций (по орбитальному моменту) в некоторых парах частиц.

# Заключение
# Conclusion

Тем самым, использованный альтернативный метод нахождения собственных значений обобщенной матричной задачи, рассматриваемой на основе вариационных методов решения уравнения Шредингера с использованием неортогонального вариационного базиса, избавляет нас от возможных неустойчивостей, возникающих в процессе применения обычных методов решения такой математической модели, т.е. обычного метода ортогонализации по Шмидту [17].

В рамках рассматриваемых вариационных методов, получены новые результаты для трехтельных моделей ядер $^7$Li, $^9$Be и $^{11}$B. Для этого использован неортогональный вариационный базис, независимое варьирование параметров, межкластерные потенциалы, чистые по схемам Юнга и согласованные с фазами упругого рассеяния в двухчастичных каналах. Все эти результаты позволяют правильно воспроизвести некоторые рассмотренные экспериментальные характеристики связанных состояний этих ядер и сечения некоторых фотоядерных процессов.

Еще раз обратим внимание, что рассмотренная трехтельная модель позволяет, в частности, провести определенную проверку построенных по фазам рассеяния парных межкластерных потенциалов. Получаемые на ее основе результаты позволяют убедиться в целесообразности дальнейшего использования таких взаимодействий для расчетов, связанных с рассмотрением некоторых, в том числе, астрофизических характеристик ядерных систем и термоядерных реакций, например, астрофизических $S$-факторов на легких атомных ядрах при низких и сверхнизких энергиях [9,12,13,189].

# ҚОРЫТЫНДЫ

*Көптеген тәжірибелік және теориялық зерттеулер нәтижесінде, ұрылымдары екі немесе үш бөлікті каналдардан тұратын кластерлік модельдеу жеңіл атомдар ядроларының көптеген сипаттамалаларын түсіндіретіндігі айқындалды [12].*

Бірінші тарауда келтірілген үздіксіз және дискретті спектрлердегі ядроның толқындық функцияларының есептеу әдістерін біле отырып, әр түрлі есептеулерді орындауда төмен және өте төмен энергиялардың ядролық физиканың мәселелерінің шешімі және ядролық астрофизика үшін кез келген моделді есептер [17]. Термоядролық процесстердің ядролық мінездемелерінің нақты есептеулерін Шредингер теңдеулерін орындау үшін, жеңіл ядролық бөлшектердің арасындағы өзара байланысатын потенциалдарын білуі керек, олар термоядролық реакцияларда қатысатын кластерлер [14].

Екінші тарауда келтірілген шашырау процесінің кластераралық потенциалдарын құру үшін, фазалық талдаудың нәтижелері қолданылады, фазаларының іздестіруіне іс жүзінде нөлдік энергиядан басталады. Сондықтан потенциалдар, серпімді шашыратудың процесстері үшін, резонанстық күйлердің болуын есепке алатын, төмен және өте төмен энергиялардың ядролық физиканың және астрофизикалық есептерінің кез келген есептеулерінде қолдану мүмкін [14].

Мысалы, электромагнитті өткелдердің үздіксіз спектрді ядроныңбайланыс күйінен күйге өту үшін керек болатын потенциалдары кластераралық каналдарындағы ядролардың негізгі күйлерінің негізгі мінездемелерінің сипаттау негіздерінде беріледі. Олар, тек қана, Юнгтің бір схемасына тәуелді, яғни орбиталық симметрия бойынша мөлдір болу [9,12,13].

Мұндай потенциалдардың кейбір гаустық және Вудс- Саксон байланыс түрлері $p^2H$, ядролық бөлшектерінің жүйелері үшін NN, $p^2H$, $p^3H$, $p^3He$, $p^4He$, $p^6Li$, $p^7Li$, $p^9Be$, $p^{12}C$, $p^{13}C$, $^2H^2H$, $^2H^3He$, $^2H^3H$, $^3H^3He$, $^3H^3H$, $^3He^3He$, $^2H^4He$, $^3H^4He$, $^3He^4He$, $^2H^6Li$, $^4He^4He$, $^4He^{12}C$ және кейбір басқа НИИЯФ МГУ және Алматыдағы авторлардың біртума жұмыстарынан бұрын алынды. Кейіннен, олар жүйеленген, анықталған, қайта тексерілген және кітаптың келесі шолуларында берілген [9-11,12-15,17,29,37]. Тыйым салынған және рұқсат етілген күйлердің, кластерлердің салыстырмалы қозғалысындағы тұжырымдама басқа мағнада бұл кластераралық өзарабайланыстарын құруда есепке алынды [11].

Үшінші тарауда, серпімді шашыратудың фазаларының негізінде жұп кластераралық потенциал-дарын құрудың дұрыстығының қосымша бақылауына бағытталған кейбір жеңіл атом ядроларының бір арналы модельінің үш денелі нәтижелері толық келтірілген [16,87,207]. Өзара байланыстың кейбірі сыннан өткен және тексерілгендері сайып келгенде астрофизикалық *S*-факторларды есептеуге – Күн және жұлдыздардағы термоядролық процесстердің ядролық сиппатамаларын анықтауға қолданылады [14,168].

Қорытындысында, төмен және өте төмен ядролық физиканың нақтылы әдістері көрсетуге талпындық, яғни астрофизикалық энергиялардың термоядролық реакциялардың есептеулері үшін төмен энергияларда қолдануға болатындығын көрсеттік. Көрсетілген, ядролық физиканың негізгі кейбір әдістері біздің Әлемнің ядролық астрофизикасының нақтылы мәселелерін шешуге мүмкіндік береді [168].

# ЗАКЛЮЧЕНИЕ

*В результате многочисленных экспериментальных и теоретических исследований было показано, что именно кластерная модель позволяет успешно объяснять многие характеристики легких атомных ядер, структура которых представляется в двух- или трехчастичных каналах [12].*

Таким образом, зная методы расчета волновых функций ядра в непрерывном и дискретном спектрах, которые были приведены в первом разделе, можно рассматривать любые модельные задачи для выполнения различных вычислений и решения большинства проблем ядерной физики низких и сверхнизких энергий и ядерной астрофизики [17]. Однако для выполнения реальных расчетов ядерных характеристик термоядерных процессов при решении уравнения Шредингера нужно знать потенциалы взаимодействия между легкими ядерными частицами – кластерами, которые участвуют в термоядерных реакциях [14].

Для построения межкластерных потенциалов процессов рассеяния обычно используются результаты фазового анализа, приведенные во втором разделе, в котором поиск фаз рассеяния начинается практически с нулевой энергии. Поэтому потенциалы, получаемые на основе таких фаз для процессов упругого рассеяния, которые учитывают наличие резонансных состояний, можно использовать в любых расчетах связанных с решением ядерно-физических и астрофизических задач низких и сверхнизких энергий [14].

Межкластерные потенциалы связанных в ядре состояний кластеров, которые требуются для расчетов,

например, электромагнитных переходов из связанного состояния ядра в состояния непрерывного спектра, обычно строятся на основе описания ими некоторых характеристик основных состояний ядер в кластерных каналах. Они должны, как правило, зависеть только от одной схемы Юнга, т.е. быть чистыми по орбитальной симметрии [9,12,13].

Некоторые варианты таких потенциалов взаимодействия гауссова и Вудс-Саксоновского типа для систем ядерных частиц NN, $p^2H$, $p^3H$, $p^3He$, $p^4He$, $p^6Li$, $p^7Li$, $p^9Be$, $p^{12}C$, $p^{13}C$, $^2H^2H$, $^2H^3He$, $^2H^3H$, $^3H^3He$, $^3H^3H$, $^3He^3He$, $^2H^4He$, $^3H^4He$, $^3He^4He$, $^2H^6Li$, $^4He^4He$, $^4He^{12}C$ и некоторые другие были получены ранее в оригинальных работах авторов из НИИЯФ МГУ и Алматы. Впоследствии, они были систематизированы, перепроверены, уточнены и приведены в следующих обзорах и книгах [9-11,12-15,17,29,37]. В некоторых случаях при построении таких межкластерных взаимодействий учитывалась концепция запрещенных и разрешенных состояний в относительном движении кластеров, что позволило избавиться от присутствия на малых расстояниях отталкивающего кора [11].

В третьем разделе книги приведено довольно подробное рассмотрение результатов трехтельной одноканальной модели некоторых легких атомных ядер, которое, по сути, направлено на дополнительный контроль правильности построения парных межкластерных потенциалов на основе фаз упругого рассеяния [16,87,207]. В дальнейшем некоторые из полученных и проверенных таким образом взаимодействий используются для расчетов астрофизических *S*-факторов и других ядерных характеристик термоядерных процессов на Солнце и звездах [14,168].

И в заключение еще раз обратим ваше внимание, что в настоящей книге сделана попытка продемонстрировать определенные методы ядерной физики низких и сверхнизких, т.е. астрофизических энергий, которые могут быть использованы для расчетов термоядерных ре-акций на Солнце и звездах. Показано, как на основе некоторых методов ядерной физики можно получить потенциалы, которые позволяют решать в дальнейшем определенные проблемы ядерной астрофизики нашей Вселенной [168].

# CONCLUSION

*It was shown, in the issue of numerous experimental and theoretical studies, that cluster model allows one to successfully describe many characteristics of light atomic nuclei, which structure can be represented in two- or three-body channels [12].*

Thereby, it is possible to consider any model problems for carrying out different calculations and solve the majority problems of low and ultralow energy nuclear physics and nuclear astrophysics [17] if we know the methods of calculation of nuclear wave function in continuous and discrete spectra, which are given in the first section. However, it is necessary to know interaction potentials between light nuclear particles – clusters, taking part in thermonuclear reactions [14], for carrying out real calculations of nuclear characteristics of thermonuclear processes by solving the Schrödinger equation.

The results of the phase shift analysis, given in the second section, where the search of phase shifts of scattering practically begins from zero energy, are used for the construction of intercluster potentials of elastic scattering processes. Therefore, the potentials obtained on the basis of such phase shifts of elastic scattering processes and taking into account the presence of the resonance states, which may be used in any calculations connected with the solving of nuclear-physical and astrophysical problems of low and ultralow energies [14].

The intercluster potentials of the bound states, which are needed for calculations, for example, the electromagnetic transitions from the bound state of nucleus to the states of continuous spectrum, usually are constructed on the basis of description of fundamental characteristics of the ground states of nucleus in cluster channels. They should, as a rule, depend from the only one Young's scheme, i.e. be pure according to orbital symmetry [9,12,13].

Some variants of these potentials of Gaussian and Woods-Saxon types for systems of nuclear particles NN, $p^2H$, $p^3H$, $p^3He$, $p^4He$, $p^6Li$, $p^7Li$, $p^9Be$, $p^{12}C$, $p^{13}C$, $^2H^2H$, $^2H^3He$, $^2H^3H$, $^3H^3He$, $^3H^3H$, $^3He^3He$, $^2H^4He$, $^3H^4He$, $^3He^4He$, $^2H^6Li$, $^4He^4He$, $^4He^{12}C$ and other were obtained earlier in the original works of authors from Institute of Nuclear Physics (MSU) and Almaty. Later, they were systemized, double-checked, specified and given in the next reviews and books [9-11,12-15,17,29,37]. The conception of forbidden and allowed states in relative cluster motion is taken into account in some cases of construction of these intercluster interactions, and this fact allows us to get rid of the existence of repulsive core at low distances [11].

The quite detailed consideration of the results of three-body one-channel model of certain light atomic nuclei is given in the third section of the book; intrinsically, this is the additional control of correctness of the construction of pair intercluster potentials obtained on the basis of elastic scattering [16,87,207]. Further, some of the similarly obtained and checked interactions are used for calculations of the astrophysical *S*-factors and other nuclear characteristics of thermonuclear processes in the Sun and the stars [14,168].

Once more, in conclusion, we are paying your attention that it was done the attempt to demonstrate certain methods of nuclear physics of low and ultralow energy, i.e. astrophysical energies, which can be used for the calculations of thermonuclear reactions in the Sun and the stars. It was shown that results, which allow one to solve some problems of nuclear astrophysics of our Universe [168], can be obtained on the basis of certain methods of nuclear physics.

# АЛҒЫС БІЛДІРУ

# БЛАГОДАРНОСТИ

Автор выражает большую признательность проф. Блохинцеву Л.Д., проф. Ишханову Б.С. и проф. Неудачину В.Г. (НИИЯФ МГУ им. М.В. Ломоносова, Москва, Россия), д.ф.-м.н. Узикову Ю.Н. (ОИЯИ, Дубна, Россия), проф. Дуйсебаеву А.Д. и проф. Буртебаеву Н.Т. (Институт ядерной физики Национального ядерного центра РК, Алматы, Казахстан), проф. Данаеву Н.Т. (Казахский Национальный Университет им. аль-Фараби, Алматы, Казахстан), проф. Чечину Л.М. (Астрофизический институт им. В.Г. Фесенкова, Алматы, Казахстан) за очень ценные обсуждения некоторых, рассмотренных в книге вопросов.

Следует особо отметить неоценимое содействие данной работе со стороны научного консультанта книги Академика НАН РК, д.ф.-м.н., проф. Боос Э.Г. (Физико-технический институт, Алматы, Казахстан).

Хочу отметить огромный вклад научного редактора проф. Бурковой Н.А. (Казахский Национальный Университет им. аль-Фараби, Алматы, Казахстан), которая сделала целый ряд полезных замечаний, правок и дополнений при редактировании книги.

Кроме того, автор выражает большую благодарность Джазаирову-Кахраманову А.В. (Астрофизический институт им. В.Г. Фесенкова, Алматы) за перевод части текста книги на английский язык, а Шамшековой С. (Астрофизический институт им. В.Г. Фесенкова, Алматы) за перевод на казахский язык.

Данная работа частично поддерживалась грантами Программы фундаментальных исследований МОН РК через Астрофизический институт им. В.Г. Фесенкова АО "НЦКИТ" НКА РК.

В связи с этим, выражаю искреннюю благодарность президенту АО "НЦКИТ" НКА РК проф. Жантаеву Ж.Ш. за постоянное содействие и поддержку всей тематики по Ядерной астрофизике, а также директору Астрофизического института Омарову Ч.Т. за большую помощь при издании настоящей книги.

# ACKNOWLEDGMENTS

I would like to express a profound gratitude to Prof. Blokhintsev L.D., Prof. Ishkhanov B.S. and Prof. Neudatchin V.G. (Institute of Nuclear Physics, Moscow State University, Moscow, Russia), Dr. Uzikov Yu.N. (JIRN, Dubna, Russia), Prof. Duisebaev A.D. and Prof. Burtebaev N.T. (Institute of nuclear physics of the National Nuclear Centre of the Republic of Kazakhstan, Almaty), Prof. Danaev N.T. (al-Farabi Kazakh National University, Almaty, Kazakhstan) and Prof. Chechin L.M. (Fessenkov's Astrophysical Institute, Almaty, Kazakhstan), for the very important discussions of some questions which where considered in the book.

It is necessary to note the invaluable assistance for this work from the scientific consultant of this book: Academician of the National Academy of Kazakhstan, Prof. Boos E.G. (Applied-physics Institute, Almaty, Kazakhstan).

I want to mention a special contribution made by science editors Prof. Burkova N.A. (Al-Farabi Kazakh National University, Almaty, Kazakhstan) who wrote a number of useful comments, introduced corrections and amendments while editing the book.

In addition, I want to express great thanks to Dzhazairov-Kakhramanov A.V. for the translation of a part of the book into English and to Shamshekova S. (Fessenkov's Astrophysical Institute, Almaty, Kazakhstan) for the translation into Kazakh.

The work has been partly supported by the MES RK (the Ministry of Education and Science of the Republic of Kazakhstan) Program of Fundamental Research via the Fessenkov V.G. Astrophysical Institute "NCSRT" NSA RK.

Thereby, I would like to express a particular gratitude to the President of "NCSRT" NSA RK Prof. Zhantaev Zh.Sh. for the assistance and the permanent support of the whole nuclear astrophysics themes, as well as to the Director of Astrophysical Institute Omarov Ch.T. for the big assistance in the publication of this book.

# ЛИТЕРАТУРА

*Дубовиченко Сергей Борисович*

# ИЗБРАННЫЕ МЕТОДЫ ЯДЕРНОЙ АСТРОФИЗИКИ

Методы ядерной физики в избранных
задачах ядерной астрофизики

# Избранные методы ядерной астрофизики

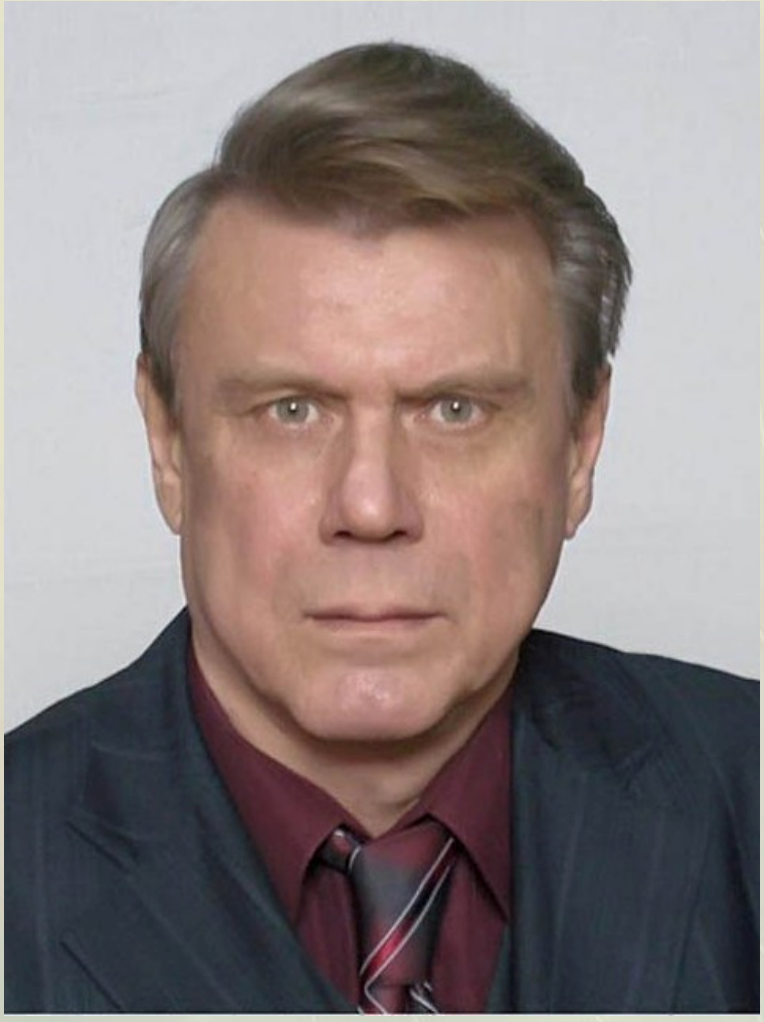

## Дубовиченко Сергей Борисович

Член-корреспондент Российской Академии Естествознания (РАЕ Россия)

Академик Международной Академии Информатизации (МАИН Казахстан)

Член Нью-Йоркской Академии Наук (NYAS USA)

***Доктор физико - математических наук Казахстана и России по специальностям 01.04.16 и 05.13.18***

***Член Европейского Физического Общества***

***Лауреат премии ЛКСМ КазССР***

***Лауреат гранта международного фонда Дж. Сороса***

***Главный научный сотрудник Астрофизического института им. В.Г. Фесенкова НЦКИТ НКА РК***

***Профессор***

**http://www.dubovichenko.ru**
**dubovichenko@gmail.com**
**dubovichenko@mail.ru**